\documentclass[10pt,twocolumn]{article}

\usepackage[T1]{fontenc}
\usepackage[utf8]{inputenc}
\usepackage[margin=2.0cm]{geometry}
\usepackage{microtype}
\usepackage{amsmath,amssymb,amsfonts}
\usepackage{bm}
\usepackage{graphicx}
\usepackage{booktabs}
\usepackage{caption}
\usepackage{subcaption}
\usepackage{float}
\usepackage{placeins}
\usepackage[hidelinks]{hyperref}
\usepackage{doi}    
\usepackage{color}
\usepackage{multirow}
\usepackage{array}
\usepackage{enumitem}
\usepackage{algorithm} 
\usepackage{algpseudocode}

\usepackage{booktabs}
\usepackage{multirow}
\usepackage{graphicx}
\usepackage{caption}

\usepackage{xltabular}  
\usepackage{makecell}
\usepackage{oplotsymbl}
\usepackage{longtable}
\usepackage{changepage}
\usepackage[numbers,square,sort&compress]{natbib}
\usepackage{bibunits}

\defaultbibliographystyle{unsrtnat}
\defaultbibliography{refs}

\definecolor{yellow}{RGB}{252, 211, 3}
\definecolor{green}{RGB}{60,179,113}
\definecolor{blue}{RGB}{3, 165, 252}
\definecolor{gray}{RGB}{70, 80, 75}

\definecolor{lightblue}{rgb}{.90,.95,1}
\definecolor{lightgreen}{rgb}{.90,1,.95}
\definecolor{darkgreen}{rgb}{0,.5,0.5}
\definecolor{darkblue}{rgb}{0.122, 0.467, 0.706}

\newif\ifcomments
\commentsfalse  

\ifcomments
    \usepackage[authormarkup=inline]{changes}
\else
    \usepackage[authormarkup=inline,final]{changes}
\fi
\definechangesauthor[name={Heng}, color={purple}]{hx}
\definechangesauthor[name={Zhuoran}, color={darkblue}]{zr}

\definechangesauthor[name={Reviewer 1}, color = red]{R1}
\definechangesauthor[name={Reviewer 2}, color = brown]{R2}
\definechangesauthor[name={Author}, color = teal]{Author}

\newcolumntype{Y}{>{\raggedright\arraybackslash}X}      
\newcolumntype{C}[1]{>{\centering\arraybackslash}p{#1}} 
\newcolumntype{L}[1]{>{\raggedright\arraybackslash}p{#1}}
\newcolumntype{M}[1]{>{\centering\arraybackslash}m{#1}} 

\newcommand{\cb}{,\allowbreak\ }

\newcommand{\flowcell}[2]{%
  \begin{minipage}[c]{\linewidth}
    \centering
    \vspace{4pt}
    \includegraphics[
      width=0.95\linewidth,
      height=12mm,
      keepaspectratio
    ]{#1}\par
    \footnotesize #2
  \end{minipage}
}

\newcommand{\flowcellTall}[2]{%
  \begin{minipage}[c]{\linewidth}
    \centering
    \vspace{4pt}
    \includegraphics[
      width=0.95\linewidth,
      height=28mm, 
      keepaspectratio
    ]{#1}\par
    \footnotesize #2
  \end{minipage}
}

\newcommand{\flowcellMedium}[2]{%
  \begin{minipage}[c]{\linewidth}
    \centering
    \vspace{4pt}
    \includegraphics[
      width=0.95\linewidth,
      height=20mm, 
      keepaspectratio
    ]{#1}\par
    \footnotesize #2
  \end{minipage}
}

\newcommand{\catcell}[1]{%
  \begin{minipage}[c]{\linewidth}
    \centering
    \itshape #1
  \end{minipage}
}

\title{Towards a unified data-driven turbulence model through multi-objective learning}

\author{Zhuo-Ran Liu\textsuperscript{1,2}, Hao-Chen Wang\textsuperscript{1,2}, Zhuo-Lin Zhao\textsuperscript{1,2}, Heng Xiao\textsuperscript{1,*}}

\date{} 

\begin{document}
\begin{bibunit}
\twocolumn[{
  \begin{center}
    {\LARGE\bfseries Toward a unified data-driven turbulence model through multi-objective learning\par}
    \vspace{1em}

    {\normalsize
      Zhuoran Liu\textsuperscript{1,2},\;
      Haochen Wang\textsuperscript{1,2},\;
      Zhuolin Zhao\textsuperscript{1,2},\;
      Heng Xiao\textsuperscript{1,2,*}\par
    }
    \vspace{0.6em}

    {\small
      \textsuperscript{1}Stuttgart Center for Simulation Science, University of Stuttgart, Stuttgart 70569, Germany\\
      \textsuperscript{2}Institute of Aerospace Thermodynamics, University of Stuttgart, Stuttgart 70569, Germany\\[4pt]
      \textsuperscript{*}Corresponding author: \texttt{heng.xiao@simtech.uni-stuttgart.de}
    }
    \vspace{1em}

    \begin{minipage}{0.9\textwidth}
      \begin{abstract}
Turbulence remains one of the last unresolved problems of classical physics and a major bottleneck to accurate flow prediction in climate, aerospace, and energy systems. Industrial simulations therefore rely on averaged representations of turbulence, which often struggle to predict flows governed by multiple interacting mechanisms.
We present a unified, data-driven turbulence modeling framework designed to learn robustly from sparse, indirect observations across diverse flow regimes. The framework embeds physical consistency into a flexible, frame-invariant closure, automatically selects representative training cases based on similarity of flow-feature distributions, and learns a single, unified model through a multi-objective ensemble strategy that balances competing objectives across flows and quantities of interest.
The resulting \emph{unified foundation model} adapts seamlessly across regimes without manual intervention. It outperforms existing turbulence models across a broad spectrum of canonical flows and maintains improved performance in complex three-dimensional configurations of industrial relevance, including a gas turbine diffuser, a generic car, \added[id=Author]{and a generic aircraft}. When application-specific accuracy is required, the framework further enables \emph{specialist models} through additive fine-tuning on targeted flow datasets.
The results demonstrate the feasibility of a deployable and generalized turbulence modeling approach that unifies multiple flow mechanisms within a single architecture for a broad range of natural and industrial flows.  
    \end{abstract}
    \vspace{0.6cm}
    
    \end{minipage}
    \vspace{0.6cm}
  \end{center}
}]

\section*{INTRODUCTION}
\label{sec:intro}

Turbulence remains a central challenge in classical physics~\cite{kim2024early}. Its presence is ubiquitous in natural and engineered flows from sub-meter to planetary scales. Accurate prediction of turbulent flows has far-reaching impacts on modern society, ranging from reducing uncertainties in climate projections and improving fidelity in extreme weather events~\cite{trok2024machine} to designing safer and more efficient mission-critical systems such as aircraft engines and power plants~\cite{spalart2016role}. The chaotic and multi-scale nature of turbulence~\cite{pope2000} is responsible for critical physics in these systems, such as momentum and heat transport in planetary boundaries and flow through engine turbines and pollutant transport in rivers and oceans. Despite the importance of turbulent transport in such processes, our current modeling of such effects in state-of-the-art simulations is still crude. For example, in weather forecast models, the vertical turbulent mixing is still represented as one-dimensional \emph{parameterization} schemes known as the planetary boundary layer scheme~\cite{zhu2025toward}. Similarly, in industrial computational fluid dynamics, Reynolds-averaged Navier--Stokes (RANS) solvers are still the workhorse tool for routine simulations~\cite{slotnick2014cfd}, where \added[id=R2]{the RANS equations govern the mean flow field and} the ``averaged effects'' of the turbulent fluctuations, \added[id=R2]{introduced through averaging as the Reynolds stress,} are represented with \emph{turbulence models}~\cite{spalart2015philosophies,xiao2019quantification}. This is because such applications require either long-time integration or many simulations for design optimization or uncertainty quantification. It becomes prohibitively expensive to fully resolve all the turbulence scales (as in Direct Numerical Simulations) or even to partially resolve them (as in Large Eddy Simulations). Despite the growth of computational resources, flow solvers based on pure modeling or parameterization of the turbulence (referred to  as \textit{turbulence modeling} collectively hereafter) will remain as the backbone in many fields in the years to come. Therefore, the theoretical foundation and practical development of turbulence will continue to have societal and scientific importance.

\subsection*{Pursuit of a universal constitutive relation}
More than half a century ago, Lumley~\cite{lumley1970toward} postulated the existence of a turbulent constitutive relation that maps the mean velocities to the Reynolds stresses. From a phenomenological perspective, such a relation shall be inferred from data and constrained by mathematical invariance and symmetry, which in principle enables transferability across similar flows. Mathematically, it defines the Reynolds stress at a given location as a functional of the mean velocity field evaluated over a surrounding neighborhood, which may in the limiting case reduce to a purely local dependence. Using experimental data, Lumley established that such a local constitutive relation exists for homogeneous shear flows and homogeneous strain, both of which behave analogously to viscoelastic fluids with an increasing relaxation timescale~\cite{lumley1970toward}. Despite its conceptual depth, this hypothesis has had limited impact on the subsequent development of turbulence modeling. Although numerous turbulence models have since been proposed, none has achieved the universality across a broad range of flows envisioned by Lumley. \deleted[id=R1]{suggesting that such universality may not be a complex objective under practical modeling constraints.}
\added[id=R1]{Recent analyses have further emphasized fundamental limitations of single-point, local closures in representing non-local transport and non-equilibrium effects in turbulence~\cite{girimaji2024turbulence}, as purely local constitutive relations cannot capture the history dependence and spatial interactions inherent in turbulent flows. Consequently, the existence of a universal, local constitutive relation remains uncertain under practical modeling constraints.}

\subsection*{From universality to unification}

Rather than pursuing a universal constitutive relation of uncertain existence, this work focuses on developing a unified turbulence model that can represent multiple flow mechanisms within a single framework. Here, a \textit{unified model} denotes a closure capable of handling multiple flow mechanisms, such as attached boundary layers, separated flows, and secondary flows, without manual zoning, switching, or blending~\cite{rumsey2022nasa, spalart2023old}. Such unification is particularly attractive for industrial applications such as aircraft aerodynamics and turbine flows, where multiple flow mechanisms coexist, interact, and cannot be clearly separated spatially. Developing a \emph{unified foundation model} of this type is the central objective of this work. Nevertheless, even a more limited degree of unification, referred to here as a \emph{specialist model}, would already offer substantial practical value~\cite{rumsey2022search, wu2023enhancing}.

\subsection*{Efforts toward a unified turbulence model} 
Achieving a unified turbulence model relies on training with diverse flows that capture multiple flow mechanisms.
A number of strategies have been pursued with varying degrees of success.
For example, multi-case training using symbolic model representations has produced models that perform well across several wall-bounded flows, representing some of the earliest steps toward unified turbulence modeling~\cite{fang2023toward,waschkowski2022multi}.
\added[id=R2]{Nevertheless, the restrictive functional forms and reliance on derivative-free optimization can limit scalability, particularly when many competing objectives arise.}
An alternative strategy, which we refer to as ``unification through aggregation", relies on expert models tailored to individual flow mechanisms, with a classifier assigning one or more models to each spatial location during prediction~\cite{cherroud2025space,buchanan2025data}.
\added[id=R2]{Such approaches, however, require multiple model evaluations and can struggle when different flow mechanisms strongly interact within the same region.}
Yet another strategy toward a unified model is to provide users with a generalized model that can be tuned for individual flows, exemplified by the generalized \(k\)--\(\omega\) model (GEKO)~\cite{menter2025generalized}.
\added[id=R2]{This model relies on manual tuning for different flows, which limits its ability to achieve true unification.}
Similar efforts towards a generalizable closure have also emerged in wall-modeled large eddy simulations, such as the knowledge-integrated additive learning~\cite{zhang2025knowledge} and building block approaches~\cite{lozano2023machine}.
However, to this day, a unified RANS turbulence model that can seamlessly handle coexisting, vastly different flow mechanisms in a single flow (e.g., flows with massive separation and swirling simultaneously) does not yet exist. 
The difficulty lies in the fact that different flow regimes impose conflicting requirements on the model: adjustments that improve one regime can degrade another. 
\added[id=R2]{This conflict fundamentally limits the transferability of existing models across diverse flow regimes.}
As noted in the NASA Langley Turbulence Modeling Workshop summary, after a decade of intense research, the community has yet to produce a data-driven turbulence model that exceeds current models in terms of predictability, generality, and robustness~\cite{rumsey2022nasa}.

\subsection*{Unification via multi-objective learning}
\begin{figure*}[!htb]
  \makebox[0pt][l]{
    \begin{minipage}{17cm}
      \includegraphics[width=\textwidth]{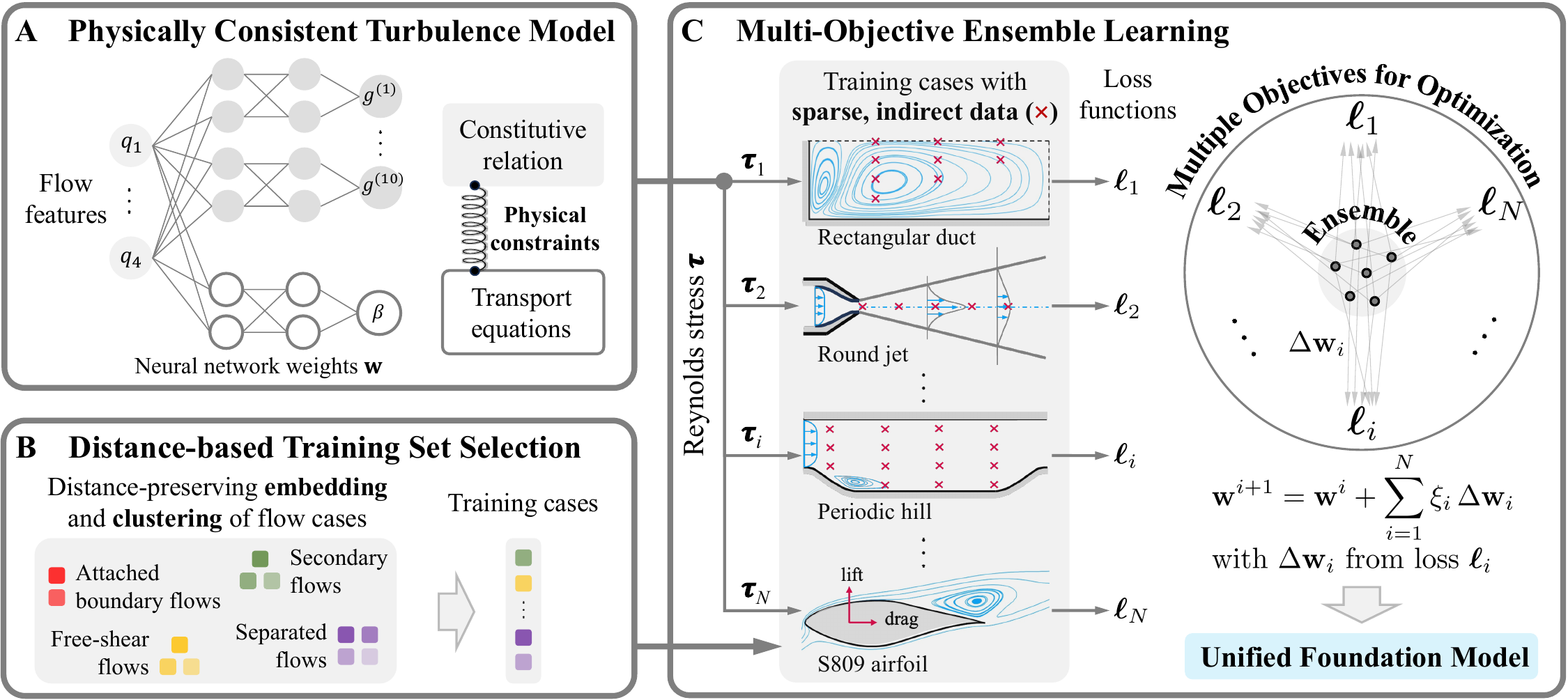}
      \caption{The proposed framework proceeds through three steps to construct a unified turbulence model. The learning is formulated as a multi-objective optimization problem, yielding a single neural-network-based model that reconciles competing objectives across flows and quantities of interest. (A) First, a physically consistent and frame-invariant model representation is employed, in which the turbulent constitutive relation and transport equations are learned in a coupled and internally consistent manner under physical constraints, enabling a single model to adapt to different flow mechanisms without manual switching. (B) Second, a comprehensive dataset of flows is compiled, and a distance-based training-set selection strategy is used to automatically identify a compact and representative set of training cases by comparing probability distributions of local, frame-invariant flow features, thereby spanning relevant flow physics while avoiding redundancy. (C) Finally, an ensemble-based, multi-objective learning framework is applied to learn a unified model from diverse flows and sparse, indirect observations, balancing competing objectives across flows and quantities of interest. Taken together, these three components yield a unified foundation turbulence model that captures multiple flow regimes within a single set of learned network weights. For application-specific accuracy, the unified foundation model can be adapted into a specialist model through additive fine-tuning (see Fig.~S1 in Supplementary Material).}
      \label{fig:method-overview}
    \end{minipage}
  }
\end{figure*}

\added[id=R2]{In this work, we develop a unified data-driven turbulence modeling framework that learns a single closure capable of representing multiple flow mechanisms. The unification is achieved through a strategy that balances conflicting objectives and branches across non-conflicting ones.}
\deleted[id=R2]{In this work, we pursue a strategy of unification through balancing and branching to construct a unified data-driven turbulence model that performs robustly across diverse flow regimes.} As illustrated in Fig.~\ref{fig:method-overview}, the proposed machine-learning framework proceeds through three steps that together yield a unified turbulence model. First, we employ a physically consistent, frame-invariant model representation that enables a single turbulence model to adapt to different flow mechanisms without manual switching (Fig.~\ref{fig:method-overview}A)~\cite{ling2016reynolds,taghizadeh2020turbulence}. Second, we use a distance-based selection strategy to automatically identify a compact and representative set of training flows that spans the relevant flow physics while avoiding redundancy (Fig.~\ref{fig:method-overview}B). Finally, we devise an ensemble-based, multi-objective learning strategy to balance competing objectives across flow regimes and, through this process, learn a unified turbulence model from heterogeneous flows and sparse, indirect observations~\cite{zhang2022ensemble} (Fig.~\ref{fig:method-overview}C). The resulting foundation model captures multiple flow mechanisms within a single set of network weights and is seamlessly integrated into a Reynolds-averaged Navier–Stokes solver in OpenFOAM.

Building on prior work that enables learning a turbulence model from indirect data~\cite{zhang2022ensemble}, this study introduces several advances that together enable robust unification across flow regimes. First, we develop a flexible and physically consistent turbulence model representation in which the turbulent constitutive relation and transport equations are learned in a coupled manner, ensuring coherent closure behavior and turbulence scales~\cite{taghizadeh2020turbulence}. Second, we introduce an automatic, distribution-based strategy for selecting representative training flows, coupled with the compilation of a comprehensive library of 36 canonical-to-complex flows—the most extensive dataset used to train a single turbulence model to date. Finally, and most importantly, we formulate turbulence model learning as a multi-objective optimization problem, allowing competing objectives arising from different flows and quantities of interest to be reconciled within a single unified model. 
\added[id=R2]{
In particular, this formulation naturally accommodates multiple quantities of interest from the same flow (e.g., drag and lift for an airfoil), which can introduce competing demands on the turbulence model (see Supplementary Material, \S~3.2.2).
}

\paragraph{Flexible and physically consistent model representation}
We construct a turbulence model that combines high expressive power with physical consistency and numerical robustness by embedding invariance, structural hierarchy, and physics-based constraints directly into the model representation (see Fig.~\ref{fig:method-overview}A). The tensor basis neural network (TBNN) framework~\cite{ling2016reynolds} provides a flexible and frame-invariant realization of the general eddy-viscosity model, making it well suited for unified turbulence modeling across diverse flow regimes. 
In its standard form, the TBNN predicts coefficients for tensor bases of all polynomial orders through a single, shared network, implicitly weighting linear and higher-order nonlinear terms equally. This often introduces unnecessary nonlinearity that degrades generalization and numerical stability. To impose a clear physical structure, we introduce a parallel TBNN architecture (see Fig.~S2b in Supplementary Material),  in which shared invariant features are mapped to separate, order-specific network branches that predict coefficients for low-order and higher-order tensor bases independently, enabling physics-informed regularization of nonlinear contributions. Beyond the constitutive relation alone, turbulence modeling requires internal consistency with the turbulence transport equations that determine turbulent time and length scales. Rather than modifying either component in isolation~\cite{ling2016reynolds}, we treat coefficients in both the constitutive relation and the transport equations as learnable fields within a unified network architecture and optimize them in a coupled manner~\cite{taghizadeh2020turbulence}. Finally, we impose physics-based constraints to ensure correct behavior in canonical limits, including decaying homogeneous isotropic turbulence, equilibrium homogeneous shear, compatibility with the logarithmic law of the wall, and edge of turbulence region~\cite{wilcox1998turbulence,spalart2023old}. 
Together, these design choices establish a turbulence model representation that is flexible yet structured, enforces internal consistency between constitutive and transport closures, and embeds physical constraints to ensure correct behavior in canonical flows.
\added[id=R1]{Despite these improvements, the model remains a single-point, local closure and thus inherits fundamental limitations. Such closures approximate inherently multi-point and non-local turbulence dynamics using local flow quantities and cannot distinguish between different flow structures that produce similar Reynolds stresses~\cite{reynolds1995one}. The present work aims to improve predictive accuracy within this framework rather than to overcome these intrinsic limitations.}

\paragraph{Distribution-based training set selection}
Developing a unified turbulence model requires training data that span diverse flow mechanisms while keeping the training cost manageable. To this end, we compiled a comprehensive library of benchmark flows from the literature, including both classical datasets~\cite{kim1987turbulence} and recent datasets designed for data-driven turbulence modeling~\cite{xiao2020flows}. The resulting library comprises~36 flow cases spanning canonical to complex configurations; nine representative cases are selected for training using a hierarchical clustering strategy based on probability-distribution distances (described below), while the remaining cases are reserved for testing (see Fig.~\ref{fig:unified-foundation-model}a). 
\added[id=R1]{As extrapolation beyond the training feature space remains challenging for data-driven turbulence models~\cite{taghizadeh2021turbulence, srivastava2021generalizable}, the evaluation focuses on generalization within the feature space spanned by the training flows.}
For each flow, appropriate quantities of interest (e.g., velocity profiles, drag, and lift; see Table~\ref{tab:benchmark_flows}) are identified based on the underlying physics and modeling challenges, providing a rigorous basis for model evaluation. Unlike data-rich fields such as weather forecasting or genomics, turbulence modeling encompasses many distinct flow mechanisms but offers limited accessible data, making principled case selection essential and requiring an objective notion of similarity between flows. To define such distances systematically, we refrain from heuristic judgment, which lacks objectivity and reproducibility, and from pointwise flow comparisons, which are sensitive to mesh resolution.  Instead, we assess flow similarity by comparing probability distributions of local, frame-invariant flow features (see Fig.~\ref{fig:method-overview}B). Together, this dataset compilation and distribution-based selection strategy yields a training set that spans dominant flow mechanisms while preserving physical representativeness.

\paragraph{Multi-objective ensemble learning from sparse, indirect observations} 
To learn a unified turbulence model from sparse, indirect observations across multiple flows, we develop a learning framework that combines ensemble-based inference with multi-objective optimization (see Fig.~\ref{fig:method-overview}C). In practice, most experimental and engineering datasets provide only indirect measurements, such as velocity profiles or integral forces, rather than full-field turbulence quantities~\cite{zhang2022ensemble,zhang2023combining}. Learning from such data using gradient-based methods typically requires adjoint-enabled or fully differentiable solvers~\cite{holland2019field,duraisamy2021perspectives,strofer2021end}, which are often unavailable in complex engineering applications. We therefore employ a regularized ensemble Kalman learning framework~\cite{zhang2020regularized}, which enables non-intrusive parameter updates through ensemble-based covariances between model predictions and observations while constraining deviations from a physically meaningful baseline model~\cite{parish2016paradigm}. Beyond learning from indirect data, unified turbulence modeling requires balancing performance across multiple flows and quantities of interest. Rather than optimizing a single aggregated loss, we formulate training as a multi-objective optimization problem and seek Pareto-optimal solutions that reconcile competing objectives across flow regimes~\cite{sener2018multi,desideri2012multiple}. Although originally developed for gradient-based learning, this framework naturally extends to ensemble-based updates by interpreting objective-specific ensemble corrections as generalized descent directions. Together, this ensemble-based, multi-objective formulation enables robust learning of a single turbulence model across diverse flows and observational constraints.
\added[id=R1]{Sensitivity analyses of the training set selection and model size further confirm this robustness (Supplementary Material,~\S3).}

When application-specific accuracy is prioritized, the unified foundation model can be further adapted into a specialist model through additive fine-tuning, in which a compact correction module is trained while all foundation parameters remain fixed. This minimal-modification strategy reallocates model capacity toward targeted flow mechanisms while preserving the generalization and robustness inherited from the unified foundation model.

In summary, the proposed framework integrates a flexible and physically consistent model representation with representative training-case selection and multi-objective ensemble learning. 
\added[id=R1]{The framework targets generalization within the feature space spanned by the training flows, while regularization toward the baseline model ensures stable behavior outside this regime.}
As demonstrated in the results section, this integrated design enables robust and generalizable predictions across diverse flow phenomena without the need for manual zoning or handcrafted blending. Together, these elements advance turbulence modeling toward solutions that are both predictive and practical for large-scale engineering applications. More importantly, the methodology itself is scalable to a large number of flows and target quantities of interest---up to 40 objectives~\cite{sener2018multi}---offering a viable path toward unifying diverse benchmark flow mechanisms and enabling application to truly complex industrial configurations.

\section*{RESULTS}
\label{sec:results}

\subsection*{Unified foundation model}
We developed a unified turbulence modeling framework that is trained once and generalizes well across a broad range of flows, including attached boundary layers, separated, secondary, and free-shear flows. Based on this framework, we construct a unified foundation model that integrates multiple diverse flow mechanisms into a single formulation. Unlike most existing data-driven models, which are trained for narrow flow categories and tend to degrade outside their training regimes~\cite{rumsey2022search}, our model performs well on the training flows and improves predictions for unseen flows exhibiting similar flow mechanisms. This directly addresses long-standing challenges emphasized in recent reviews and community discussions~\cite{spalart2023old, rumsey2022nasa, rumsey2022search}, particularly the need for general turbulence models that deliver consistent, broadly applicable improvements while remaining robust and free from manual zoning. Such requirements were central to the NASA 2022 ``Collaborative Testing Challenge"~\cite{rumsey2022nasa}. Our unified foundation model makes substantial progress toward this goal, delivering robust and accurate predictions across multiple flow regimes. The NASA challenge cases, included in our evaluation, also demonstrate clear performance improvements (Supplementary Material,~\S5).

\begin{figure*}[!htb]
    \centering
    \begin{minipage}{\textwidth}
        \begin{minipage}{17cm}
            \centering
            \begin{subfigure}[b]{8.3cm}
                \centering
                \includegraphics[width=8.3cm]{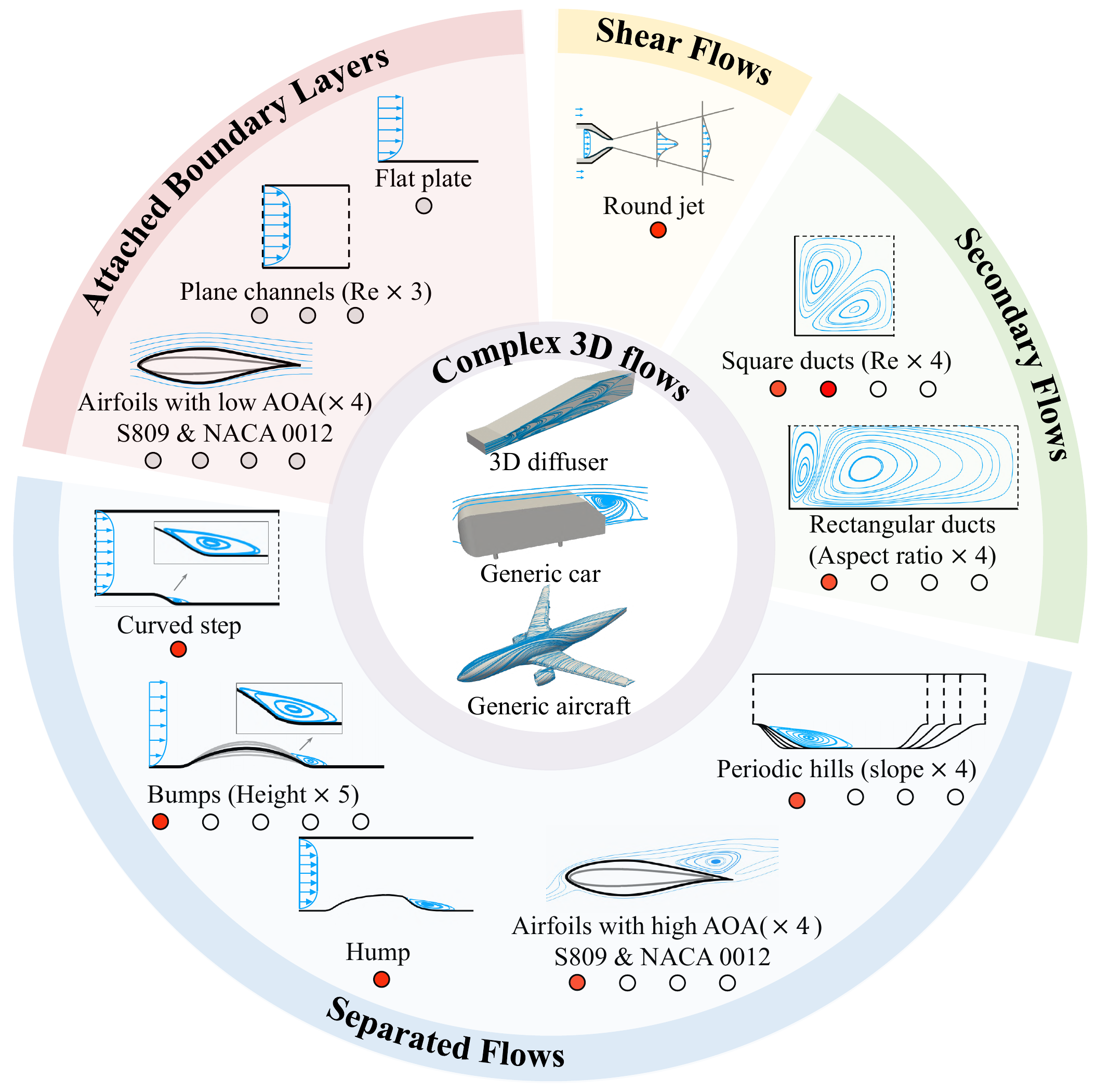}
                \caption{\added[id=Author]{Case library for model training and evaluation.}}
                \label{fig:case-setup-foundation}
            \end{subfigure}
            \hfill
            \begin{subfigure}[b]{8.3cm}
                \centering        
                \includegraphics[width=8.3cm]{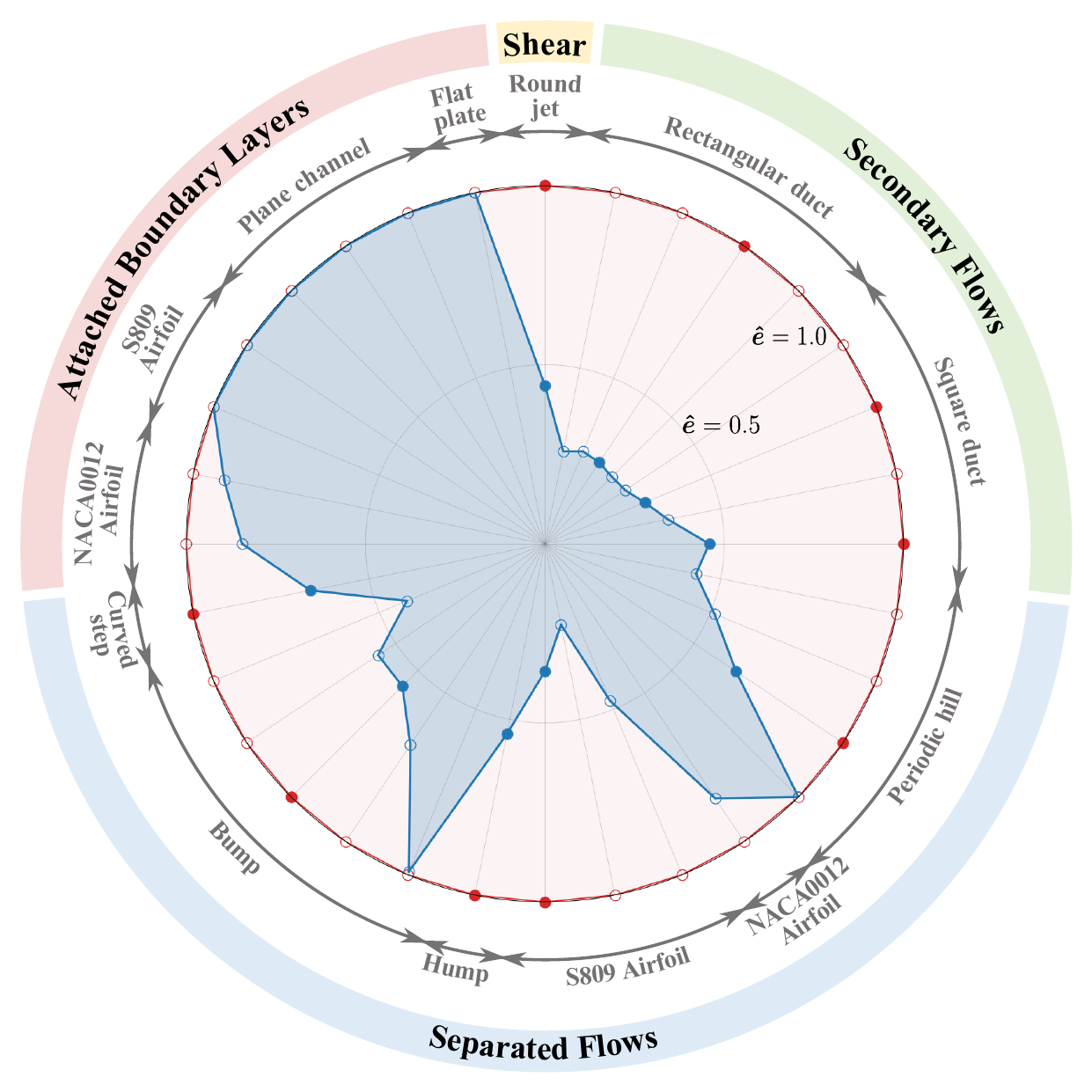}
                \caption{Performance of unified foundation model.}
                \label{fig:radar-foundation}
            \end{subfigure}

            \caption{
            Overview of the training and evaluation cases and performance of the unified foundation turbulence model.
            (a) Library of canonical and complex flows used for training and evaluation, comprising 36 cases in total. Canonical flows are grouped into four categories—attached boundary layers, free-shear flows, secondary flows, and separated flows—while two complex three-dimensional flows involving multiple mechanisms are shown in the inner circle. Nine representative cases (highlighted with red circles) form the training set, and the remaining 27 cases are used for testing.
            (b) Radar chart comparing normalized misfits of the baseline and unified foundation models across flow categories. Smaller radial distances indicate improved performance. The unified foundation model demonstrates robust performance across all categories while maintaining accuracy comparable to the baseline for bump flows, for which only one representative case is shown.
            }
            \label{fig:unified-foundation-model}
        \end{minipage}
    \end{minipage}
\end{figure*}

\paragraph{Model training and evaluation setup}
The unified foundation turbulence model is trained on nine flows representing distinct flow mechanisms, identified using the distribution-based training set selection method. These include a curved step and a periodic hill for internal separated flows; a bump, a hump, and an S809 airfoil at high angle of attack for external separated flows; two square ducts at different Reynolds numbers and a rectangular duct for secondary flows; and a round jet for free-shear flows. Each case provides sparse, indirect, and heterogeneous observations tailored to its nature: sparse velocity measurements for the curved step, periodic hill, bump, hump, square and rectangular ducts, and round jet; and aerodynamic forces for the S809 airfoil. In total, ten training objectives are defined, as the S809 airfoil case involves two competing objectives: lift and drag. Once trained, the model’s generalization is tested on 25 unseen cases without manual zoning or parameter tuning, as shown in Fig.~\ref{fig:case-setup-foundation} and Table~\ref{tab:benchmark_flows} (Supplementary Material,~\S4).

\begin{table*}[t]
  \begingroup
  \small
  \centering
  \setlength{\tabcolsep}{6pt} 
  \renewcommand{\arraystretch}{1.3}
  \begin{tabular}{
    p{0.17\textwidth}   
    p{0.20\textwidth}   
    p{0.2\textwidth}    
    p{0.17\textwidth}   
    >{\centering\arraybackslash}p{0.06\textwidth}
    >{\centering\arraybackslash}p{0.06\textwidth}
  }

    \textbf{Categories} & \textbf{Flows} & \textbf{\makecell{Control \\ Parameter}} & \textbf{Observations} & \textbf{\makecell{No. \\ Train}} & \textbf{\makecell{No. \\ Test}} \\
    \specialrule{1pt}{0pt}{0pt}

    \multirow{3}{*}[-1ex]{\makecell{\textit{Attached} \\ \textit{Boundary} \\ \textit{Layers}}}
      & Flat plate 
      &  & Velocities, friction
      &  & 1 \\
    &
      Plane channel 
      & Reynolds number & Velocities
      &  & 3 \\
    & Airfoil (attached)
      & Shape, angle of attack & Lift, drag
      &  & 4 \\ \hline

    \multirow{4}{*}{\textit{Separated Flows}}
    & Curved step
      &  & Velocities
      & 1 &  \\
    & Hump
      &  & Velocities
      & 1 &  \\
    & Bump 
      & Height & Friction
      & 1 & 4 \\
    & Periodic hill
      & Slope steepness & Velocities
      & 1 & 3 \\
    & Airfoil (separated)
      & Shape, angle of attack & Lift, drag
      & 1 & 3 \\ \hline

    \multirow{2}{*}{\textit{Secondary Flows}}
    & Square duct
      & Reynolds number & Velocities
      & 2 & 2 \\
    & Rectangular duct
      & Aspect ratio & Velocities
      & 1 & 3 \\ \hline

    \multirow{1}{*}{\textit{Free-shear Flows}}
      & Round jet
      &  & Velocities
      & 1  \\ \hline

    \multirow{4}{*}{\textit{Complex 3D Flows}}
    & Generic car
      &  & Drag
      &  & 1 \\
    & 3D diffuser
      &  & Velocities, friction
      &  & 1 \\ 
    & \multirow{2}{*}{\added[id=Author]{Generic aircraft}}
      & \multirow{2}{*}{Angle of attack}  & \multirow{2}{*}{\shortstack[l]{Lift, drag,\\ friction, pressure}}
      &  & \multirow{2}{*}{2} \\ 
    &  &  &  &  &  \\
      \midrule
      
    \multicolumn{4}{r!}{\textbf{Totals}} & \textbf{9} & \textbf{27} \\
    \specialrule{1pt}{0pt}{0pt}

  \end{tabular}
  \vspace{6pt}
    \caption{Categorized benchmark turbulent flows for evaluating the unified foundation model. The benchmark covers five categories: attached boundary layers, separated flows, secondary flows, free-shear flows, and complex three-dimensional flows. Most cases vary by control parameters such as Reynolds number, angle of attack, or geometric features (e.g., bump height, hill slope steepness), while cases without such parameters correspond to a single configuration. Available observations include velocity, wall friction, lift, and drag. ``No. Train" and ``No. Test" denote the numbers of training and test cases, respectively. The datasets include flat plate~\cite{jespersen2016overflow}, plane channel~\cite{kim1987turbulence}, S809/NACA 0012 airfoils~\cite{somers1997design,ladson1988effects}, curved step~\cite{bentaleb2012large}, bump~\cite{matai2019large}, hump~\cite{greenblatt2006experimental}, periodic hill~\cite{xiao2020flows}, square duct~\cite{pinelli2010reynolds}, rectangular duct~\cite{vinuesa2018secondary}, round jet~\cite{bridges2010establishing}, generic car~\cite{ashton2024ahmedml}, three-dimensional diffuser~\cite{cherry2008geometric} and generic aircraft~\cite{evans2020test}.}
  \label{tab:benchmark_flows}
  \endgroup
\end{table*}

\paragraph{Evaluation metric}
The unified foundation model performs well across all training cases and generalizes effectively to unseen cases, including both those within the training flow categories and complex three-dimensional configurations. Its performance is evaluated on diverse test cases spanning attached boundary layers, separated flows, secondary flows, free-shear flows, and complex three-dimensional flows. 
The model's performance across training flow categories is shown in Fig.~\ref{fig:radar-foundation}. It is evaluated using the normalized misfit~$\hat{e} = \frac{e_{\text{unified}} - e_{\text{single}}}{e_{\text{base}} - e_{\text{single}}}$, where $e_{\text{unified}}$ is the misfit of the unified foundation model relative to the ground truth, $e_{\text{base}}$ is the misfit of the baseline model, \added[id=R2]{i.e., Wilcox (1988) \(k\)--\(\omega\) model~\cite{wilcox1988reassessment}}, and $e_{\text{single}}$ is the misfit of a single-case trained model, which defines the upper performance limit.
A value of~$\hat{e} < 1$ indicates improved performance over baseline, while~$\hat{e} > 1$ indicates degradation. 

\begin{figure*}[!htb]
    \centering
    \begin{minipage}{\textwidth}
        \begin{minipage}{17cm}
            \centering
            \includegraphics[width=17cm]{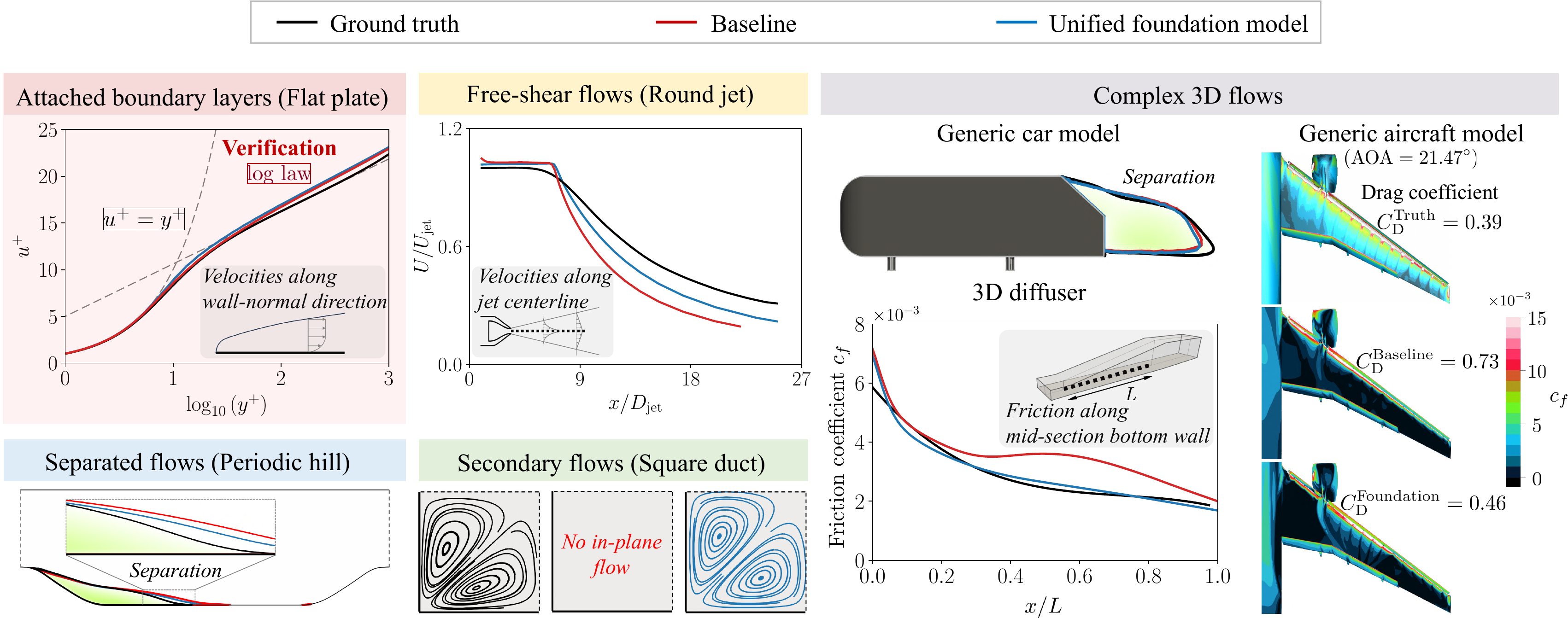}
            \caption{Performance evaluation of the unified foundation model across representative cases from training flow categories and two complex three-dimensional flows. The model is compared with the ground truth and the baseline model. For attached boundary layers, the flat plate recovers the law of the wall. For free-shear flows, the round jet shows improved streamwise velocity decay along the jet centerline. For separated flows, the periodic hill with a lower slope steepness demonstrates better prediction of separation size and reattachment location. For secondary flows, the square duct at a higher Reynolds number shows accurate recovery of corner vortices. For complex three-dimensional flows, the generic car shows separation predictions comparable to the baseline, \deleted[id=Author]{while} the three-dimensional diffuser matches the ground-truth skin-friction coefficients along the bottom-wall midsection, \added[id=Author]{and the generic aircraft shows improved skin-friction and drag predictions at large angles of attack}.}
            \label{fig:detail-foundation}
        \end{minipage}
    \end{minipage}
\end{figure*}

\paragraph{Performance evaluation of the unified foundation model} Overall, the unified foundation model generalizes well, achieving lower misfits than the baseline in most cases and comparable performance otherwise.
For \textbf{attached boundary layers}, such as the zero pressure gradient flat plate, the model matches the baseline model in regimes where linear eddy-viscosity models are already reliable, recovering the law of the wall and maintaining accuracy across geometries and Reynolds numbers.
In \textbf{free-shear flows}, represented here by a round jet, the model provides more accurate predictions of centerline velocity decay, \added[id=R2]{a key metric that reflects the jet spreading rate. Since this behavior is strongly influenced by the turbulent transport equations, the improved prediction highlights the importance of maintaining physical consistency between the constitutive relation and the transport equations}.
For \textbf{secondary flows} in ducts with different Reynolds numbers and aspect ratios, the model better captures in-plane flows, \added[id=R2]{which require a nonlinear eddy-viscosity model to represent the anisotropic stress driving the secondary motion, and maintains} accuracy under both geometric and Reynolds number extrapolation. 
For \textbf{separated flows}, the model substantially improves predictions of massive separation, such as periodic hills with varying slope steepness, bumps with different heights, the hump, the curved step, and S809 airfoil at high angles of attack. 
\added[id=R2]{These improvements mainly arise from the spatial variation of the leading tensor-basis coefficient $g^{(1)}$ shown in Fig.~\ref{fig:method-overview-SI}A, which represents the eddy-viscosity contribution and directly governs separation and reattachment.}
For \textbf{complex three-dimensional flows}, \added[id=R2]{including the generic car, three-dimensional diffuser, and generic aircraft, the model generally improves predictions across different flow mechanisms, while remaining comparable to the baseline otherwise.} \deleted[id=Author]{The generic car shows improved separation prediction, consistent with its dominant separated-flow characteristics} \added[id=Author]{The generic car shows separation predictions comparable to the baseline}; the three-dimensional diffuser recovers the ground-truth skin-friction coefficients along the bottom-wall midsection, \added[id=R2]{where secondary flow interacts with separation}; \added[id=Author]{the generic aircraft exhibits improved skin-friction and drag predictions under large angles of attack, where multiple interacting flow mechanisms are present with separation being the dominating effect.}
The overall improvements for complex three-dimensional flows remain limited, as the separation location in the diffuser is still not predicted accurately. Detailed performance of the unified foundation model on seven representative test cases is shown in Fig.~\ref{fig:detail-foundation}.
The unified foundation model demonstrates improved generalization across diverse flow regimes, achieving better, or at least comparable, accuracy to the baseline model in all test cases. 

\subsection*{Specialist model}
\label{sec:specialist}
The specialist model further improves prediction accuracy for targeted flows through additive fine-tuning \added[id=R2]{of the unified foundation model} on three representative cases with related flow mechanisms. Building on the unified foundation model, it preserves much of the foundation performance while reallocating model capacity toward the target mechanisms. Although some generalization capability may be reduced, this trade-off yields substantially improved accuracy for the targeted flows. Such specialization reflects common industrial practice, where high fidelity for specific applications is prioritized over uniform performance.

\paragraph{Model training and evaluation setup}
The specialist model targets flows dominated by separation and secondary flows. Fine-tuning is performed using three canonical cases: the curved step representing separated flows, and rectangular ducts with aspect ratios of 3 and 10 representing secondary flows. These cases are selected as the three closest matches to the target flows using the distribution-based training case selection method. This combination enables the model to improve its performance for both secondary and separated flow mechanisms.
The specialist model is evaluated on all benchmark canonical flows and further tested on an asymmetric three-dimensional diffuser flow, which involves interactions between the fine-tuned flow mechanisms.
The three-dimensional diffuser is a challenging validation case featuring incompressible, asymmetric internal flow with strong adverse pressure gradients and significant Reynolds-stress anisotropy. The resulting three-dimensional separation closely mirrors practical diffuser behavior, making it an ideal benchmark for evaluating separation and secondary flow prediction capabilities. 

\begin{figure*}[t!]
    \centering
    \begin{minipage}{\textwidth}
        \begin{minipage}{17cm}
            \centering
            \includegraphics[width=17cm]{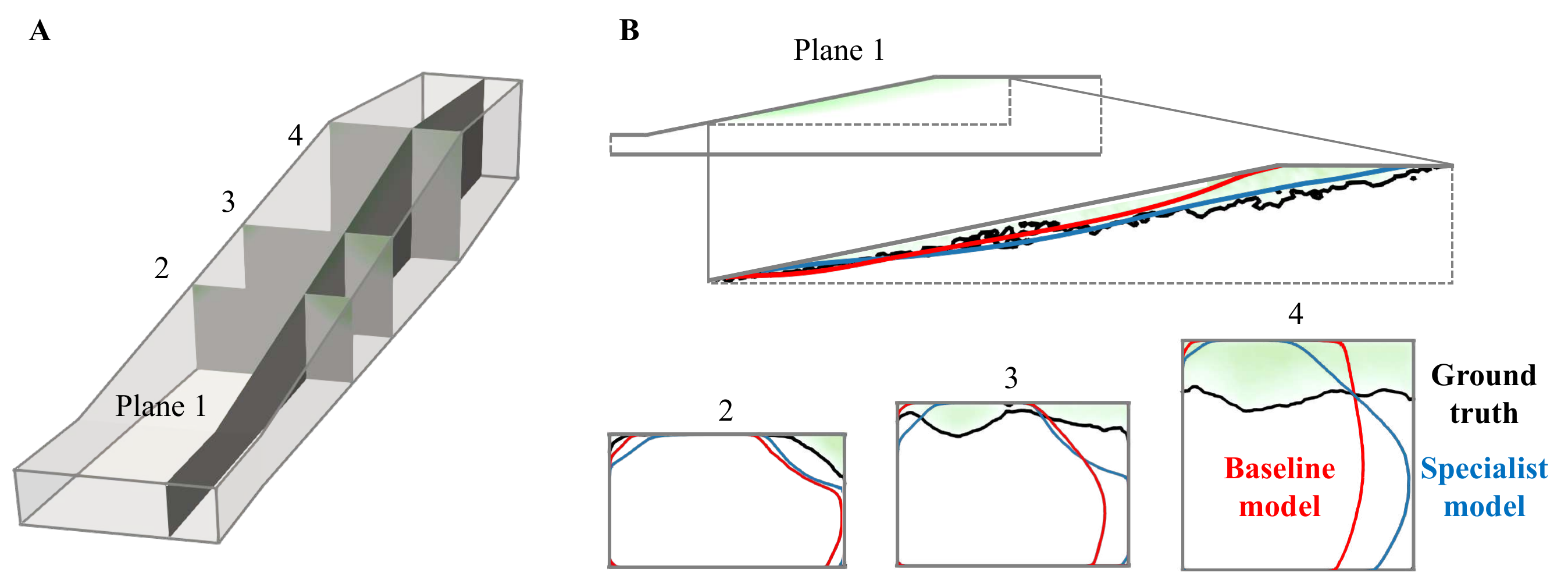}
            \caption{Significant performance improvement of the specialist turbulence model for a complex three-dimensional diffuser flow. Panel~A shows the diffuser geometry and the sampling planes used for evaluation, including a vertical spanwise plane near the inclined sidewall (Plane~1) and three streamwise cross-sectional planes within the expansion region (Planes~2--4).
            \added[id=R2]{The ground truth on Plane~1 come from experiments~\cite{cherry2008geometric} and contain noticeable irregularities, whereas the data on Planes~2--4 are obtained from the DNS mean flow field~\cite{ohlsson2010dns}.}
            Panel~B compares flow separation predicted by the baseline model, the specialist model, and the ground truth.
            The baseline model fails to capture the correct separation behavior, producing spurious sidewall separation and missing the dominant separation pattern.
            In contrast, the specialist model accurately reproduces both the location and extent of separation observed in the ground truth, correctly capturing the upper-wall separation across all planes. These results demonstrate a large and systematic improvement in predictive accuracy achieved by the specialist model for complex three-dimensional flows.}
            \label{fig:specialist-diffuser}
        \end{minipage}
    \end{minipage}
\end{figure*}

\paragraph{Performance evaluation of the specialist model}
The specialist model maintains good generalization across all benchmark canonical flows, while delivering clear performance enhancements for the fine-tuning flow categories and, in particular, for the complex three-dimensional diffuser case.
With additive fine-tuning, the model reduces the over-prediction of separation in the curved step and accurately recovers in-plane secondary flows in rectangular ducts. While the unified foundation model improves predictions for both flows, fine-tuning delivers substantially higher accuracy.
Moreover, the specialist model captures diffuser separation more accurately, correcting the baseline~\(k\)--\(\omega\) model’s sidewall separation and aligning it with the ground truth, where separation occurs along the upper wall in the expansion region. The evaluation results are summarized as follows.
\begin{enumerate}
    \renewcommand{\labelenumi}{(\arabic{enumi})}
    \item \textit{Generalization in benchmark flows}.
    The specialist model remains robust across all benchmark canonical flow cases. 
    Within the fine-tuning flow categories, including secondary and separated flows, most cases show clear performance improvements relative to the unified foundation model. 
    Outside these categories, predictive accuracy is mildly degraded and remains comparable to, or slightly below, the baseline model. 
    Nevertheless, the specialist model continues to satisfy essential physical constraints, including consistency with the logarithmic law of the wall (Supplementary Material,~\S5).

    \item \textit{Generalization to complex 3D flow}.
    In the three-dimensional diffuser, where secondary flows interact with separation, the specialist model suppresses spurious sidewall separation and correctly predicts the primary upper-wall separation observed in the ground truth, as shown in Fig.~\ref{fig:specialist-diffuser}. Cross-sectional analyses show that the specialist model accurately captures the growth and reattachment of the separated region, while the baseline model misplaces both the location and extent of separation (Supplementary Material,~\S5). The specialist model's success on this complex configuration demonstrates strong generalization beyond the training manifold and highlights its ability to capture interacting flow mechanisms in industrial applications.
\end{enumerate}

\subsection*{Interpretation for model unification}
\label{sec:interpretation}
The unified turbulence model combines two complementary strategies for flow mechanism unification: automatic balancing of overlapping input features and internal branching for case-specific features across multiple flows. In overlapping regions, where identical features occur in different flows, the model resolves conflicts through multi-objective learning. In case-specific regions, it learns distinct mappings for each flow, acting as a piecewise function with internal branching. Together, these mechanisms enable the model to generalize across flows governed by distinct yet interacting flow mechanisms. We next use the unified foundation model to interpret how these strategies enable generalization.

\begin{figure*}[t!]
    \centering
    \begin{minipage}{\textwidth}
        \begin{minipage}{17cm}
            \centering
            \includegraphics[width=17cm]{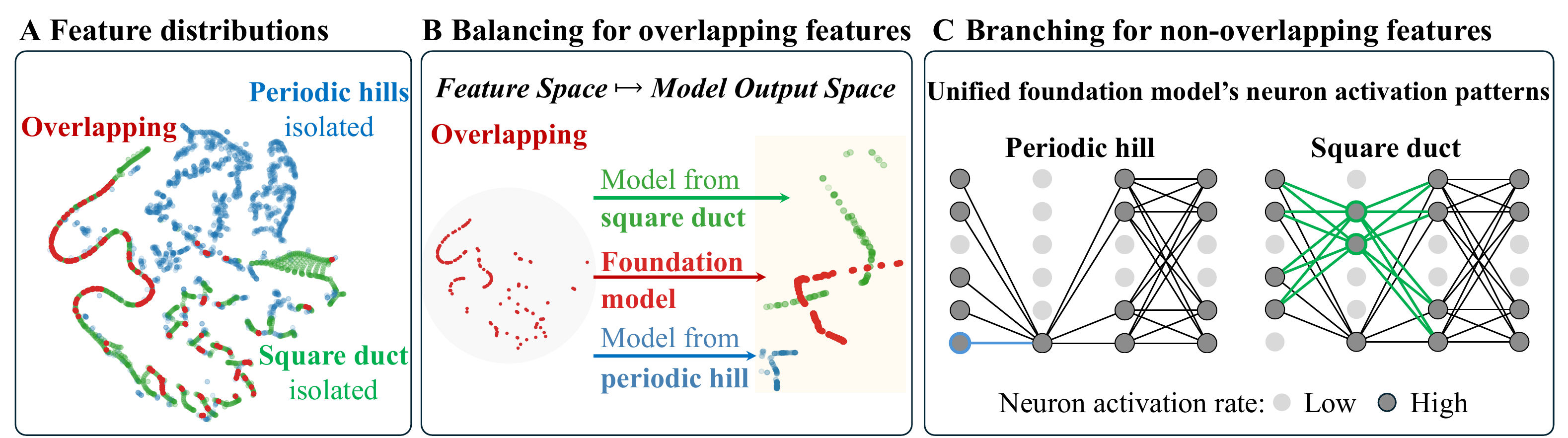}
            \caption{
            Feature-space overlap and conflict resolution enabled by the unified foundation model.
            Panel~A shows a t-SNE projection of the four-dimensional feature space, highlighting overlapping and case-specific regions for the periodic hill (blue) and the square duct (green), with red indicating overlap.
            Panel~B compares model responses in the overlapping feature regions, demonstrating that single-case trained models map similar features to different outputs, which limits generalization, whereas the unified foundation model resolves these conflicts and produces balanced predictions.
            Panel~C shows neuron activation patterns in the sub-network associated with the quadratic coefficient~$g^{(2)}$, where periodic hill features activate fewer neurons, while square duct features induce stronger activation and rely more heavily on $g^{(2)}$, reflecting differences in the underlying turbulence mechanisms.}
            \label{fig:interpretation}
        \end{minipage}
    \end{minipage}
\end{figure*}

The unified foundation model's input feature analysis reveals that different flows exhibit overlapping features while also maintaining distinct case-specific structures. We project the four frame-invariant input features into two dimensions using t-SNE~\cite{maaten2008visualizing} for visualization. Fig.~\ref{fig:interpretation}A presents the results for two representative flows: periodic hill and square duct. The plot shows overlapping regions, as well as distinct regions specific to each case. This demonstrates that the unified foundation model captures common structures shared across flows while preserving case-specific characteristics.

\paragraph{Balancing in overlapping regions}
Overlapping regions in the feature space serve as critical zones where the model must resolve conflicts among cases to enable joint learning of interacting mechanisms. Because identical features in these regions can correspond to different outputs across cases, deterministic models alone cannot eliminate the resulting discrepancies. Multi-objective learning addresses this by adjusting the contribution of each case during training, promoting balanced optimization and compelling the model to reconcile conflicting objectives. As shown in Fig.~\ref{fig:interpretation}B, the single-case trained models often map overlapping features to significantly different outputs, creating distinct regions in the output space and limiting generalization. The unified foundation model resolves these conflicts by maintaining a balanced representation, achieving an intermediate and more generalizable state.
This balancing not only enhances generalization but also preserves the ability to capture genuine interactions where multiple mechanisms coexist.

\paragraph{\added[id=R2]{Origins of feature space overlapping}}
\added[id=R2]{Conflicts can originate not only from different flow mechanisms across different flows, but also from conflicting quantities of interest within the same flow.
Quantities of interest from the same flow are not necessarily aligned: while some probe the same dominant mechanism and respond consistently to model corrections, others reflect different physical sensitivities. When training relies on only a subset of these quantities, the resulting model can neglect unresolved mechanisms, leading to degraded predictions for the others and the appearance of conflicting objectives.
For example, in the curved-step flow, the velocity field and skin-friction coefficient are aligned, both governed by separation and reattachment dynamics. In contrast, for the S809 airfoil, lift and drag impose conflicting corrections on the dominant closure coefficient. This conflict is reflected not only at the level of integrated quantities but also in the spatial structure of the learning signals: the gradients with respect to the turbulent eddy viscosity induced by lift and drag exhibit opposing patterns across the flow domain (see Fig.~S10 in Supplementary Material), indicating that these two objectives drive the model toward fundamentally different flow adjustments. These competing directions correspond to distinct physical pathways, with lift favoring enhanced separation and drag favoring increased wake mixing (Supplementary Material, \S~3.2). Together, these observations provide a concrete physical basis for the conflicts observed in overlapping feature regions and motivate the need for multi-objective learning to reconcile them within a unified framework.}

\paragraph{Branching in case-specific regions}
In non-overlapping regions of the feature space, the network learns case-specific mappings, and effectively behaves like a piecewise function. To study this behavior, we extract case-specific features and pass them through the unified foundation model. For each neuron, we compute its activation rate, defined as the fraction of inputs producing a positive output. For rectified linear units (ReLU), this is equivalent to the proportion of non-zero outputs. The analysis is performed separately for the periodic hill and square duct.
The activation patterns of the sub-network for the quadratic coefficient $g^{(2)}$ are shown in Fig.~\ref{fig:interpretation}C. Dark neurons and connections represent pathways activated in both the periodic hill and square duct cases, highlighting their similarities. In the periodic hill, most first-layer neurons are active, with one additional neuron uniquely activated (blue), but the second layer channels information almost entirely through a single pathway. In contrast, in the square duct, second-layer activations are more widely distributed, with several unique pathways remaining active (green). These distinct connection patterns demonstrate how the model reconfigures its internal branching to adapt to the statistical and structural characteristics of each flow.

\section*{DISCUSSION AND CONCLUSIONS}
\label{sec:Discussion-Conclusions}

An accurate, predictive turbulence model that performs reliably across a wide range of flow mechanisms, often simultaneously present in industrial flows, has long been a central goal in industrial computational fluid dynamics. 
In this work, we demonstrate that multi-objective learning combined with a physically consistent model representation and distribution-based training set selection enables the training of a unified foundation model on a heterogeneous dataset consisting of nine canonical flows (e.g., periodic hill for separated flows, rectangular duct for secondary flows, round jet for free-shear flows) and various target quantities (e.g., velocities, drag, lift). 
With the strategy of ``unification through balancing and branching", this unified foundation model achieves Pareto-optimal performance across a broad range of test cases, including previously unseen and complex scenarios involving multiple interacting flow mechanisms. Learning from canonical flows provides interpretability, ease of verification, and a scalable route to generalization. In contrast to training directly on application-specific flows, which are often inaccessible or lack clear mechanism labeling, our approach provides a structured path to unification by expanding coverage with canonical flows through multi-objective learning, balancing trade-offs in parallel while avoiding catastrophic forgetting~\cite{zhang2025knowledge}.

While the unified foundation model yields consistent improvements over the baseline model, not all test cases show dramatic gains. This is expected, as many industrial flows involve additional complexities beyond the training flows and limited objective set. Nevertheless, the framework proves highly effective in application-specific scenarios where a small number of flow mechanisms dominate, as additive fine-tuning reallocates model capacity toward these mechanisms and delivers substantially improved accuracy.
We demonstrated this through two specialist models: one for three-dimensional diffuser flow, where adverse pressure-gradient-induced separation interacts with secondary motions, and the other for external vehicle aerodynamics, trained on flows with separations of varying severity (Supplementary Material,~\S5). 
These examples show that selecting representative canonical flows with related mechanisms and applying fine-tuning enables efficient training of specialist models that achieve high accuracy for the target mechanisms while maintaining strong generalization within their respective domains.

The long-term goal is to develop a unified turbulence model that performs robustly across an entire device, rather than in isolated regions. For instance, while a three-dimensional diffuser represents a single component in a gas turbine, a full engine comprises many more interacting components with distinct flow characteristics. Switching among specialist models within a single simulation would be cumbersome and potentially unstable. Ideally, a single model should adapt seamlessly across the system. Achieving this will require scaling the current framework to accommodate a significantly larger set of objectives. Recent advances in multi-objective learning suggest that neural networks can balance up to 40 objectives~\cite{sener2018multi}, making full-device unification technically feasible for configurations such as entire gas turbines or aircraft frames.

To achieve more consistent improvements across diverse flows and quantities, several technical enhancements are necessary. First, gradient accuracy for each objective must be improved. While our ensemble-based method is robust and broadly applicable, adjoint-based or fully differentiable solvers~\cite{bezgin2025jax} could provide more precise gradients, enabling more sophisticated architectures. Additionally, feature augmentation could better utilize the network’s representational capacity by increasing the input space dimensionality, helping reduce objective conflicts through internal branching. However, the averaging process that maps turbulent fluctuations to mean fields inherently causes information loss, meaning some mechanisms may remain indistinguishable in the input space. In such cases, effective balancing, supported by accurate gradients, remains essential. 
\added[id=R2]{Furthermore, this framework can be extended to unsteady flows by incorporating time-resolved observations over a finite time window~\cite{luo2021novel}, enabling more accurate modeling of inherently unsteady phenomena such as airfoil dynamic stall~\cite{corke2015dynamic}.}
The presented framework, with its physics-informed architecture and gradient-aware training, provides a promising foundation for further advances in unified turbulence modeling.

\section*{MATERIALS AND METHODS}

\subsection*{Physically consistent model representation}
For constant-density, incompressible turbulent flows, the mean flow satisfies the Reynolds-averaged Navier--Stokes equations,
\begin{subequations}
\begin{align*}
\nabla \cdot \bm{U} &= 0, \\
\frac{\mathrm{D}\bm{U}}{\mathrm{D}t} - \nu \nabla^2 \bm{U} + \frac{1}{\rho}\nabla P &= \nabla \cdot \bm{\tau},
\end{align*}
\end{subequations}
with the Reynolds stress $\bm{\tau} = -2k\!\left(\mathbf{b} + \tfrac{1}{3}\mathbf{I}\right)$.
The transport of turbulence scales (the turbulent kinetic energy $k$ and the specific dissipation rate $\omega$) is described by partial differential equations (PDE)~$\bm{\mathcal{R}}=0$,
\[
\bm{\mathcal{R}}\left[k,\omega; \, \boldsymbol{\beta}(\bm{q})\right] =  \mathbf{0} ,
\]
where $\boldsymbol{\beta}$ consists of the constant coefficients in the transport PDE, while the Reynolds stress anisotropy $\mathbf{b}$ is modeled using the general eddy-viscosity constitutive relation~\cite{pope1975more, ling2016reynolds}:
\[
\mathbf{b} = \sum_{i=1}^{10} g^{(i)}(\bm{q})\,\mathbf{T}^{(i)} .
\] 
The transport equations and constitutive relation are modified simultaneously by formulating the coefficients $\boldsymbol{\beta}$ and $g^{(i)}$ in both relations as function of local features $\bm{q}$   using parallel tensor basis neural networks with essential physical constraints~\cite{taghizadeh2020turbulence, wilcox1998turbulence}. More details can be found in the Supplementary Material,~\S2.

\subsection*{Distribution-based training set selection}
Each flow case is represented by the empirical distribution of its local, frame-invariant flow features.
At each grid point, an invariant feature vector $\bm{q}(x)\in\mathbb{R}^4$ is extracted, and the collection of these samples induces an empirical probability measure in feature space.
The training case selection procedure proceeds as follows:
\begin{enumerate}
    \renewcommand{\labelenumi}{(\arabic{enumi})}
    
    \item \textit{Computing pairwise flow similarity.}
    Similarity between two flow cases with feature distributions $\pi_1$ and $\pi_2$ is quantified using the Wasserstein-2 distance.
    The squared Wasserstein-2 distance is defined as
    \[
    W_2^2(\pi_1,\pi_2)
    =
    \min_{\gamma \in\Gamma(\pi_1,\pi_2)}
    \int \|\bm{q}-\bm{q}^{\star}\|^2 \,\mathrm{d}\gamma(\bm{q},\bm{q}^{\star}),
    \]
    where $\bm{q}$ and $\bm{q}^{\star}$ denote features sampled from these two flow cases, and $\Gamma(\pi_1,\pi_2)$ denotes the set of admissible transport plans.

    \item \textit{Clustering and embedding.}
    Pairwise Wasserstein distances between all candidate flow cases are assembled into a symmetric distance matrix.
    Hierarchical clustering is applied to identify groups of statistically similar flows without prescribing the number of clusters \emph{a priori}.
    To aid interpretation of the clustering structure, multi-dimensional scaling is used to embed the flow cases into a 2D plane that approximately preserves the pairwise distribution distances.

    \item \textit{Selecting representative training cases.}
    Representative training cases are selected as cluster medoids, defined as the cases that minimize the sum of distances to all other members within the same cluster.
\end{enumerate}
This procedure ensures coverage of dominant flow mechanisms while avoiding redundant cases and limiting the overall training cost.

\subsection*{Multi-objective ensemble learning}
We couple regularized ensemble learning from indirect data with a multi-objective strategy to balance competing training cases and quantities of interest.
Given $N$ objectives, each defined by a training case and a quantity of interest, the update at each iteration proceeds as follows:
\begin{enumerate}
    \renewcommand{\labelenumi}{(\arabic{enumi})}
    
    \item \textit{Computing regularized EnKF updates for each objective.}
    For objective $j$, sparse observations $\mathbf{y}_j$ are related to the model parameters $\mathbf{w}$ through an observation operator $\mathcal{H}_j$.
    An objective-specific ensemble update $\Delta\mathbf{w}_j$ is computed using the regularized ensemble Kalman filter method, which leverages ensemble covariances and avoids explicit adjoint gradients.
    The update is obtained by minimizing the cost function with regularization:
    \begin{align*}
        \ell_j =\;& \|\mathbf{w}^{i+1}_j - \mathbf{w}^i_j\|_{\mathsf{P}^{-1}}^2 
        + \|\mathbf{y}_j - \mathcal{H}_j[\mathbf{w}^{i+1}_j]\|_{\mathsf{R}^{-1}}^2  \\
        & + \|\mathcal{G}[\mathbf{w}^{i+1}_j]\|_{\mathsf{Q}^{-1}}^2,
    \end{align*}
    where $\mathcal{G}$ penalizes deviations from a physically meaningful baseline model and stabilizes learning from limited, indirect data.

    \item \textit{Reconciling conflicting objectives by adaptive weighting.}
    The objective-specific updates $\{\Delta\mathbf{w}_j\}_{j=1}^N$ can point in competing directions.
    To balance these conflicts, we seek nonnegative weights $\bm{\xi} = (\xi_1,\dots,\xi_N)$, with $\sum_{j=1}^N \xi_j = 1$, such that the combined update direction $\sum_{j=1}^N \xi_j \,\Delta\mathbf{w}_j$ minimizes interference among objectives.
    Treating each $\Delta\mathbf{w}_j$ as a generalized gradient direction, we normalize the updates to reduce sensitivity to heterogeneous scales and determine $\bm{\xi}$ by minimizing
    \[
        \min_{\bm{\xi}} \left\| \sum_{j=1}^N \xi_j \,\Delta\mathbf{w}_j \right\|^2,
        \quad \text{s.t. } \xi_j \ge 0,\;\sum_{j=1}^N \xi_j = 1.
    \]
    This auxiliary problem is solved using the Frank--Wolfe algorithm, which iteratively identifies the most conflicting objective and adjusts the weights via an analytic line search, yielding a compromise direction that balances progress across all objectives.

    \item \textit{Forming the unified update.}
    The final model update blends the objective-specific EnKF updates using the learned weights,
    \[
        \mathbf{w} \leftarrow \mathbf{w} + \eta \sum_{j=1}^{N} \xi_j \,\Delta\mathbf{w}_j,
    \]
    where $\eta$ is a step size.
\end{enumerate}

\section*{DATA AVAILABILITY}
The source code for our model and training datasets are available at GitHub: \url{https://github.com/ITLR-DDSim/unified-turbulence-modeling}.

\section*{SUPPLEMENTARY DATA}
Supplementary data are available at NSR online.

\section*{ACKNOWLEDGMENTS}
We thank Dr. Xin-Lei Zhang and Dr. Xu-Hui Zhou for their insightful discussions, and Dr. Zhou for providing the Ahmed body evaluation case. We also acknowledge Mr. Jiaqi Li and Dr. Xiang Yang for their assistance in setting up the CRM-HL case.

\section*{FUNDING}
Z.L., H.W., and H.X. acknowledge support from the Deutsche Forschungsgemeinschaft (DFG, German Research Foundation) under Germany's Excellence Strategy - EXC 2075 - 390740016. Z.Z. acknowledges funding from the DFG (Project Number 551388164) and the Carl Zeiss Foundation (CZS Project Number P2021-04012). The authors also acknowledge support from the Stuttgart Center for Simulation Science (SimTech).

\section*{AUTHOR CONTRIBUTIONS}
Z.-R.L. and H.X. designed research;
Z.-R.L., H.-C.W., and Z.-L.Z. performed research;
Z.-R.L. and H.X. analyzed data;
and all authors contributed to writing the paper.

\renewcommand{\refname}{References}
\putbib

\end{bibunit}

\clearpage
\onecolumn
\begin{bibunit}
\section*{\LARGE\bfseries Supplementary Material}
\setcounter{figure}{0}
\setcounter{table}{0}
\setcounter{equation}{0}
\renewcommand{\thefigure}{S\arabic{figure}}
\renewcommand{\thetable}{S\arabic{table}}
\renewcommand{\theequation}{S\arabic{equation}}

\tableofcontents
\section{Introduction}
Turbulence is ubiquitous in natural and industrial flows and remains a central challenge in classical physics due to its chaotic and multiscale nature. These characteristics govern essential transport processes, such as momentum, heat, and pollutant transport, yet make it prohibitively expensive to resolve all relevant scales in practical simulations. As a result, most predictions rely on turbulence modeling within Reynolds-averaged Navier--Stokes solvers~\cite{slotnick2014cfd}. In such approaches, the averaged effects of unresolved turbulent fluctuations on the mean flow must be represented through closure models that use simplified, typically local, flow information. Despite decades of development and their central role in simulations, existing turbulence models remain crude approximations of complex, nonlinear, and multiscale turbulent dynamics. Consequently, improving the theoretical foundations and practical performance of turbulence modeling remains both scientifically fundamental and societally important.

\subsection{Turbulence modeling: from universality to unification}
Early efforts in turbulence modeling were shaped by Lumley’s proposal that turbulent momentum transport might admit a constitutive relation, analogous to those used in continuum mechanics, linking the mean velocity field to the Reynolds stress~\cite{lumley1970toward}. In that work, Lumley argued that such a relation should respect fundamental symmetry and invariance properties and remain independent of boundary conditions imposed on the Reynolds stresses, thereby providing a structured framework for their representation. Using experimental data, he further demonstrated that a local constitutive relation can be identified in highly idealized settings, such as homogeneous shear and homogeneous strain flows, where turbulence exhibits viscoelastic-like behavior.
However, these successes remained confined to canonical configurations and had little influence on practical turbulence modeling. In complex industrial flows, multiple mechanisms such as attached boundary layers, flow separation, curvature effects, and secondary motions often coexist and interact within the same configuration, making strict universality unlikely under practical modeling constraints. Specifically, industrial applications strongly favor local closures, in which the Reynolds stress depends only on velocity gradients at the same spatial location. In contrast, a truly universal constitutive relation---if it exists at all---would likely be nonlocal, reflecting the inherently nonlocal nature of turbulent physics.
\added[id=R1]{This discrepancy has also been emphasized in recent studies of machine-learning-based closures, which highlight that single-point constitutive relations cannot, in general, capture non-local transport and non-equilibrium effects through purely local mappings from mean-flow quantities to Reynolds stresses~\cite{girimaji2024turbulence}.} 
This realization motivated a gradual shift away from the pursuit of universal closures toward the more pragmatic objective of unification, in which a single model is expected to represent multiple flow mechanisms without manual zoning, switching, or retuning. As a result, robustness, simplicity, and computational efficiency became dominant considerations, leading to the widespread adoption of empirical eddy-viscosity models such as the Spalart--Allmaras and \(k\)--\(\omega\) SST models~\cite{xiao2019quantification}. Decades of experience with such models suggest that a simple, local universal constitutive relation is unlikely to exist~\cite{duraisamy2019turbulence}. Nevertheless, a \emph{unified turbulence model} that can handle multiple coexisting mechanisms remains highly desirable for industrial applications~\cite{rumsey2022nasa}.

Eddy-viscosity models provide a natural foundation for a unified turbulence model.
This choice is practical because eddy-viscosity models are local, robust, and computationally efficient, making them reliable for industrial solvers and a strong basis for data-driven extensions~\cite{xiao2019quantification}. Conventional eddy-viscosity models combine turbulence transport equations for turbulent scales with a linear constitutive relation to estimate the Reynolds stress. This framework can be generalized to nonlinear forms, offering greater flexibility than the standard Boussinesq assumption~\cite{speziale1987nonlinear,gatski1993explicit}. The general eddy-viscosity model~\cite{pope1975more} has been realized through tensor basis neural networks~\cite{ling2016reynolds} and symbolic formulations~\cite{weatheritt2016novel,schmelzer2018data}. Alternatively, Reynolds stress modeled from the eddy viscosity models can also be modified directly through eigen-perturbations~\cite{wang2017physics,wu2018physics}. Other studies have instead focused on improving the turbulence transport equations, for example by introducing data-driven multiplicative correction fields~\cite{parish2016paradigm}.  
Beyond eddy-viscosity models, nonlocal approaches offer an alternative. These include Reynolds stress transport models~\cite{launder1975progress} and data-driven nonlocal constitutive relations~\cite{han2023equivariant,zhou2022frame,luther2025non}, which could capture additional physical effects such as non-equilibrium effects in the turbulence anisotropy. However, nonlocal models face major challenges: they are computationally expensive, less robust, and difficult to treat solid boundaries (wall models). Machine learning has also been used to refine specific terms in Reynolds stress transport models, such as the pressure--strain correlation~\cite{alaya2022evolutionary}, but these models themselves remain fragile. Considering these trade-offs, the tensor basis neural network formulation of the general eddy-viscosity model provides a well-balanced compromise among accuracy, flexibility, and computational efficiency.

\subsection{Toward a unified turbulence model: existing approaches and key challenges}
Despite the merits of eddy-viscosity models as a foundation for unified data-driven turbulence modeling, a predictive and generalizable model cannot be achieved by modifying individual closure components in isolation.
Instead, success critically depends on preserving internal consistency between the turbulence transport equations and the constitutive relation for the Reynolds stress, as well as satisfying additional physical constraints. Nevertheless, most existing data-driven eddy-viscosity models modify only one of these tightly coupled components, which fundamentally limits their ability to achieve unification.

Eddy-viscosity models consist of two tightly coupled components: turbulence transport equations define the turbulent length and time scales, determining Reynolds stress amplitude, while the constitutive relation characterizes its anisotropy. Many studies focus exclusively on refining the constitutive relation or the turbulence anisotropy using tensor basis neural networks, symbolic regression, or random forests~\cite{ling2016reynolds,schmelzer2018data,wang2017physics,wu2018physics}, while others introduce corrections to the transport equations~\cite{parish2016paradigm}. Although both approaches have achieved improved performance over baseline models, treating these components independently can break the physical coupling between turbulent scale evolution and the Reynolds stress anisotropy. Recent analyses have shown that such decoupled corrections can lead to severe inconsistencies, even in simple canonical flows~\cite{taghizadeh2020turbulence,taghizadeh2021turbulence}. These observations highlight the importance of preserving the internal consistency in the efforts to achieve a unified turbulence model.

Beyond internal consistency, a unified eddy-viscosity model must also respect physical constraints arising from behaviors observed in canonical flows (e.g., homogeneous isotropic decaying turbulence and homogeneous shear turbulence), near-wall dynamics (law of the wall), and free-stream limits, among other~\cite{spalart2023old}. These flows have traditionally been used to calibrate turbulence models, imposing additional physical constraints among model coefficients in the transport equations. These coefficients, in turn, rely on implicit assumptions embedded in the constitutive relation.
Existing data-driven studies rarely address this full set of coupled constraints in a systematic and coherent manner; instead, different works tend to modify different subsets of model components, leading to patchwise improvements that do not generalize across flow regimes.
Ignoring the dependence between transport equations and constitutive relation results in inconsistent Reynolds-stress definitions and incorrect equilibrium behavior, undermining robustness and generalizability. Resolving this physical consistency issue systematically is therefore a prerequisite for unified turbulence modeling.

Within the eddy-viscosity framework, several strategies have been explored to advance turbulence unification, with varying degrees of success. 
Multi-case training using symbolic model representations has demonstrated promising generalization across several wall-bounded flows, representing some of the earliest steps toward unified turbulence modeling~\cite{fang2023toward}. However, the restrictive functional forms of symbolic models and their reliance on derivative-free optimization methods, such as non-dominated sorting genetic algorithm II, limit the scalability of training, particularly when balancing many competing objectives required for unification.
Another line of research pursues unification through aggregation, in which expert models tailored to individual flow mechanisms are combined into a composite framework using mixture-of-experts or classifier-driven model selection~\cite{cherroud2025space,buchanan2025data}. While this strategy enables mechanism-specific modeling, it typically requires multiple model evaluations and can struggle when strong interactions between flow mechanisms occur within the same region.
A different approach employs adjoint-based field inversion and machine learning to infer multiplicative correction fields to existing turbulence models~\cite{rumsey2022search,wu2025development}. These methods have proven effective for flows dominated by similar mechanisms~\cite{wu2025development}, but their extension to flows with fundamentally different coexisting mechanisms (e.g., separated and secondary flows, as will be shown in the three-dimensional diffuser or the wing-body junction~\cite{gand2010flow}) remains challenging, often leading to non-convergent training and limited generalizability~\cite{rumsey2022search}. Finally, the generalized \(k\)--\(\omega\) model provides a pragmatic alternative by introducing tunable closure terms that allow systematic adjustments for different flow conditions~\cite{menter2025generalized}. Although attractive for practical use in commercial solvers, its performance still relies heavily on manual tuning and therefore falls short of true unification.
Despite these advances, a unified turbulence model capable of seamlessly handling multiple, strongly interacting flow mechanisms within a single configuration does not yet exist~\cite{rumsey2022nasa}. Achieving such a model remains a central open challenge, as improvements targeting one flow regime often degrade performance in others. Taken together, these limitations indicate that unification cannot be achieved through isolated improvements or post hoc aggregation alone, but instead requires a learning framework that jointly enforces physical consistency, balances competing objectives, and operates directly on sparse, indirect observations, since full-field Reynolds stress data are rarely available in realistic applications.

\subsection{Scope and contributions of this work}
In this work, we propose a unified, data-driven turbulence modeling framework that learns a single, neural-network-based eddy-viscosity model that is robust and accurate across diverse flow regimes. The approach combines a physically consistent model representation, a distribution-based selection of representative training flows, and a multi-objective ensemble learning strategy to balance competing objectives arising from sparse, indirect observations in different flow mechanisms. The resulting unified turbulence model delivers consistent, across-the-board improvements over a wide range of canonical and complex flows, while remaining compatible with existing Reynolds-averaged Navier--Stokes solvers, thereby offering a viable pathway toward industrial deployment.

\FloatBarrier
\section{Methodology}
\label{sec:method}

\begin{figure}[!htb]
    \centering
    \includegraphics[width=0.99\textwidth]{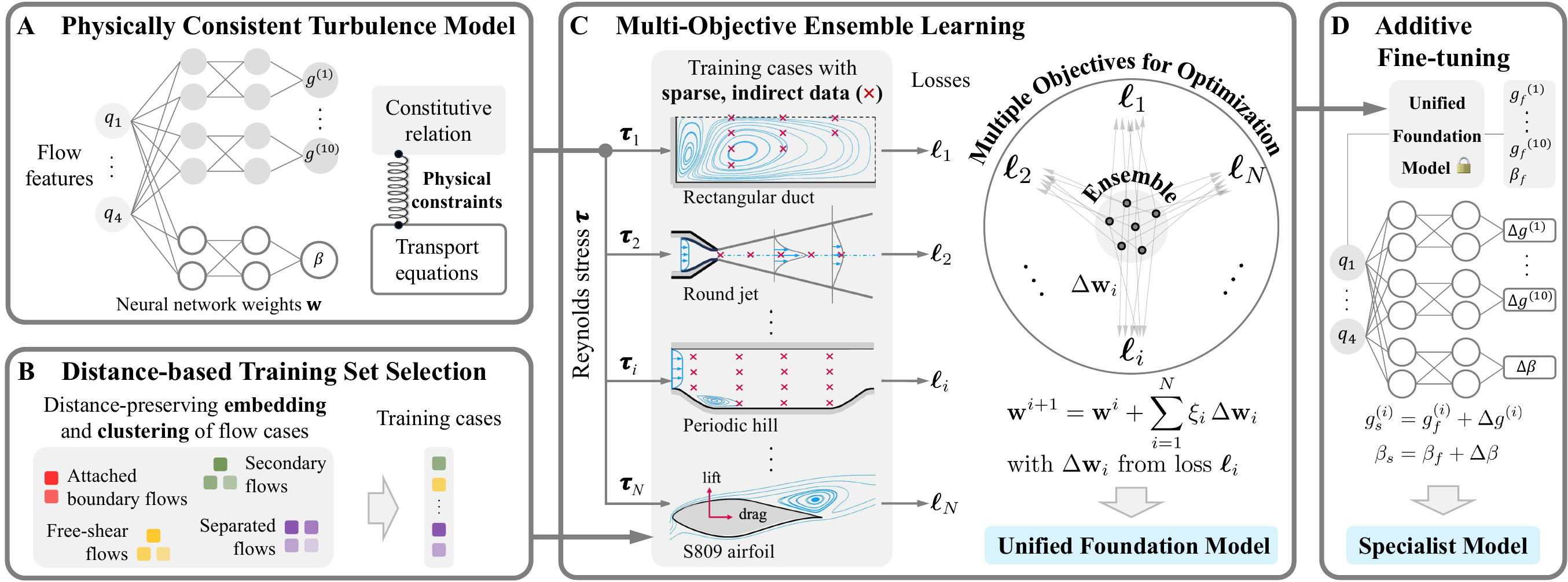}
    \caption{
    The proposed methodology establishes a unified foundation turbulence model that captures diverse flow regimes and supports application-specific adaptation through fine-tuning.
    (A) A physically consistent and frame-invariant model representation is adopted, in which a flexible turbulent constitutive relation is learned in a coupled manner with turbulence transport equations under established physical constraints, enabling consistent prediction of Reynolds stresses across different flow regimes.
    (B) A representative and compact training dataset is automatically identified by comparing probability distributions of local, frame-invariant flow features; distance-preserving embedding and clustering are used to select flows that span distinct flow mechanisms while avoiding redundancy.
    (C) The unified model is learned using an ensemble-based, multi-objective optimization framework, in which Reynolds stresses predicted by the model are propagated through the Reynolds-averaged Navier--Stokes equations and compared against sparse, indirect observations from the selected training cases, with adaptive weighting used to balance heterogeneous objectives across flows.
    Together, these steps yield a single set of network weights that captures multiple flow regimes within a unified foundation model.
    (D) For application-specific accuracy, the foundation model can be adapted into a specialist model through additive fine-tuning, in which compact correction modules target specific flow mechanisms while preserving the learned foundation behavior.}
    \label{fig:method-overview-SI}
\end{figure}

The proposed framework integrates several complementary components into a coherent learning pipeline that advances data-driven turbulence modeling from formulation to deployment (Fig.~\ref{fig:method-overview-SI}). 
\textbf{First}, a flexible and physically consistent model structure is established, in which the constitutive relation and turbulence transport equations are treated in a coupled, frame-invariant manner and constrained by established physical requirements, ensuring both expressiveness and physical admissibility. 
\textbf{Second}, representative training cases are selected using a distribution-based strategy that compares flows through distances between their flow-feature probability distributions. This selection provides broad coverage of distinct flow mechanisms while avoiding redundant information.
\textbf{Third}, building on the resulting model structure and training set, learning is formulated as a multi-objective problem and carried out using a regularized ensemble Kalman framework that infers turbulence closures from sparse and indirect observations. The ensemble-based formulation enables gradient-free inference and jointly learns multiple flow cases and quantities of interest, allowing competing objectives to be reconciled and yielding a unified foundation model with balanced performance across flow regimes.
When application-specific accuracy is required, the unified foundation model can be further adapted through an \textbf{optional} additive fine-tuning procedure. In this step, compact correction modules focus model capacity on target flow mechanisms while preserving the physical structure and generalization properties learned in the unified foundation model.

\subsection{Flexible and physically consistent model representation}
The turbulence model is designed to be frame-invariant, flexible, and physically consistent through its model representation. Frame invariance is enforced by expressing the Reynolds stress closure in a tensor-basis form, which provides a robust foundation for data-driven modeling. To improve numerical stability and explicitly regularize model nonlinearity, linear and nonlinear tensor-basis contributions are treated separately, ensuring that dominant low-order effects are captured first while higher-order corrections are introduced in a controlled manner. Physical consistency between the constitutive relation and the turbulence transport equations is enforced as an essential component of the model. Their coefficients are therefore modeled in a coupled framework as learnable fields subject to established physical constraints, including consistency with near-wall behavior such as the logarithmic law of the wall. This coupled and constrained formulation prevents unphysical Reynolds stress predictions and further enhances numerical stability.

\subsubsection{General eddy-viscosity formulation}
For constant-density, incompressible turbulent flows, the mean flow satisfies the Reynolds-averaged Navier--Stokes (RANS) equations:
\begin{align*}
\nabla \cdot \bm{U} &= 0, \\
\frac{\mathrm{D} \bm{U}}{\mathrm{D} t} - \nu \nabla^2 \bm{U} + \frac{1}{\rho} \nabla P &= \nabla \cdot \bm{\tau},
\end{align*}
where $\bm{U}$ denotes the mean velocity, $P$ is the mean pressure normalized by density $\rho$, and $\bm{\tau}$ is the Reynolds stress tensor that accounts for the effect of unresolved velocity fluctuations. Since these fluctuations are not resolved in RANS equations, the Reynolds stress requires a closure model.

The Reynolds stress tensor is decomposed into deviatoric and isotropic components as
\begin{equation*}
\boldsymbol{\tau} = -\left(\mathbf{a} + \frac{2}{3}k\mathbf{I}\right),
\end{equation*}
where $k$ is the turbulence kinetic energy and $\mathbf{I}$ is the second-order identity tensor. The normalized anisotropy tensor is defined as $\mathbf{b} = \mathbf{a}/(2k)$.
According to the general eddy-viscosity formulation~\cite{pope1975more}, the normalized anisotropy tensor~$\mathbf{b}$ can be expressed as a linear combination of ten tensor bases:
\begin{equation}
\mathbf{b} = \sum_{i=1}^{10} g^{(i)}(\theta_1, \dots, \theta_5)\,\mathbf{T}^{(i)},
\label{eq:b}
\end{equation}
where $g^{(i)}$ are scalar coefficient functions of five invariants $\theta_i$, and $\mathbf{T}^{(i)}$ are tensor bases. Both $\mathbf{T}^{(i)}$ and $\theta_i$ are constructed from the normalized mean strain-rate and rotation-rate tensors.
The normalized strain-rate and rotation-rate tensors are commonly defined as
\begin{equation}
\hat{\mathbf{S}} = \tau_s \mathbf{S}, \quad \hat{\mathbf{W}} = \tau_s \mathbf{W},
\label{eq:timescale-norm}
\end{equation}
where $\mathbf{S} = \frac{1}{2}\left[\nabla\boldsymbol{U} + (\nabla\boldsymbol{U})^\top\right]$ and $\mathbf{W} = \frac{1}{2}\left[\nabla\boldsymbol{U} - (\nabla\boldsymbol{U})^\top\right]$ are the mean strain-rate and rotation-rate tensors, respectively. The turbulence time scale $\tau_s$ is estimated as $\tau_s = k/\varepsilon = 1/(C_{\mu}\omega)$, where $C_{\mu}$ is the turbulent-viscosity constant.
The first two scalar invariants and corresponding tensor bases in Eq.~\eqref{eq:b} are defined as:
\[
\theta_1 = \{\hat{\mathbf{S}}^2\}, \quad \theta_2 = \{\hat{\mathbf{W}}^2\}, \quad \mathbf{T}^{(1)} = \hat{\mathbf{S}}, \quad \mathbf{T}^{(2)} = \hat{\mathbf{S}}\hat{\mathbf{W}} - \hat{\mathbf{W}}\hat{\mathbf{S}},
\]
where $\{\cdot\}$ denotes the tensor trace. Definitions of the remaining tensor bases and invariants are available in~\cite{pope1975more, ling2016reynolds}. For two-dimensional turbulent flows, only the first two tensor bases and invariants are required to form a complete representation.

The tensor basis neural network (TBNN)~\cite{ling2016reynolds} provides a data-driven realization of the general eddy-viscosity model by learning the coefficient functions $g^{(i)}$ as functions of the frame-invariant scalars $\theta_1$, ..., $\theta_5$. These learned coefficients are then combined with the corresponding tensor basis $\mathbf{T}^{(i)}$ according to~Eq.~\eqref{eq:b} (see Fig.~\ref{fig:parTBNN}). This formulation preserves frame invariance by construction while providing a flexible neural-network representation of the constitutive relation.
In classical formulations, the time-scale normalization in Eq.~\eqref{eq:timescale-norm} is applied to the strain-rate and rotation-rate tensors to render them dimensionless; however, it does not bound their magnitude and may yield excessively large values in regions of strong gradients, which is undesirable for machine learning. To ensure stable and efficient learning, the flow features derived from these tensors are required to lie within a bounded range.
We therefore avoid conventional global min--max normalization, which depends on dataset extrema and may cause feature clustering near singular points.
Instead, we adopt a local time-scale normalization~\cite{wu2018physics,zhang2023physical} applied directly to the strain-rate and rotation-rate tensors, which combines non-dimensionalization with bounded scaling and preserves local flow physics.
Specifically,
\[
\hat{\mathbf{S}} = \frac{\mathbf{S}}{|\mathbf{S}| + 1/\tau_s},
\quad
\hat{\mathbf{W}} = \frac{\mathbf{W}}{|\mathbf{W}| + 1/\tau_s},
\]
from which the scalar invariants and other flow features are subsequently derived.

\subsubsection{Parallel tensor basis neural network representation}
\begin{figure}[!htb]
  \centering
  \begin{subfigure}[t]{0.45\textwidth}
    \centering
    \includegraphics[width=0.95\textwidth]{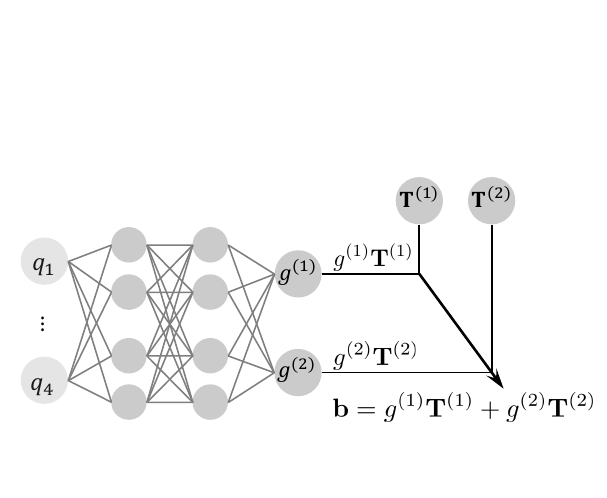}
    \caption{Original TBNN}
    \label{fig:tbnn-orig}
  \end{subfigure}
  \hspace{-0.01\textwidth}
  \begin{subfigure}[t]{0.26\textwidth}
    \centering
    \includegraphics[width=0.95\textwidth]{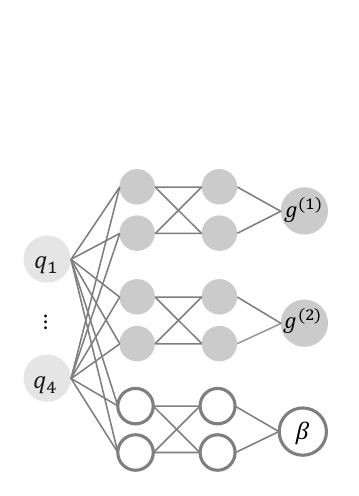}
    \caption{Consistent parallel TBNN}
    \label{fig:tbnn-parallel}
  \end{subfigure}
  \hspace{-0.01\textwidth}
  \begin{subfigure}[t]{0.26\textwidth}
    \centering
    \includegraphics[width=0.95\textwidth]{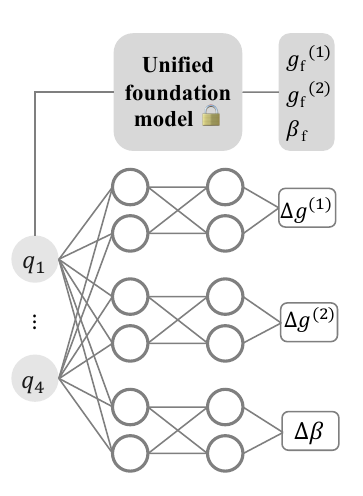}
    \caption{Additive fine-tuning}
    \label{fig:tbnn-finetune}
  \end{subfigure}

  \caption{
    Architectural extensions of the tensor basis neural network for unified and specialist turbulence modeling.
    (a) Original tensor basis neural network (TBNN). A single neural network predicts the coefficient functions $g^{(i)}$ from the flow features $\bm{q}$, which are combined with the predefined tensor basis $\mathbf{T}^{(i)}$ through the constitutive relation $\sum_i g^{(i)} \mathbf{T}^{(i)}$ as in Eq.~\eqref{eq:b}.
    (b) Proposed consistent parallel TBNN used for the unified foundation model. Separate sub-networks predict the linear and quadratic coefficients $g^{(1)}$ and $g^{(2)}$, while an additional sub-network jointly predicts the turbulence transport-equation coefficient $\beta$, enabling enforcement of physical consistency between the constitutive relation and the transport equations. For clarity, the tensor-basis combination shown in (a) is not repeated.
    (c) Proposed additive fine-tuning strategy for specialist models. A compact set of additional trainable network weights is introduced to capture flow-specific corrections, while all network weights of the unified foundation model remain fixed. The underlying tensor-basis formulation is unchanged.}
    
  \label{fig:parTBNN}
\end{figure}

We extend the original TBNN (Fig.~\ref{fig:parTBNN} a) to the parallel TBNN (Fig.~\ref{fig:parTBNN} b) to enhance flexibility and control over model nonlinearity. Unlike the original TBNN, which predicts all coefficients using a single network, the parallel TBNN employs multiple independent sub-networks to predict individual coefficient functions. In this work, we construct a quadratic eddy-viscosity model based on the first two tensor bases, $\mathbf{T}^{(1)}$ and $\mathbf{T}^{(2)}$. The model input consists of four flow features $\mathbf{q}=\{q_i\}_{i=1}^4$. The first two are scalar invariants derived from the velocity gradient, denoted by $\theta_1$ and $\theta_2$. The third feature, $q_3 = \nu_t/(100\nu + \nu_t)$, represents the viscosity ratio, and the fourth feature, $q_4 = \min\!\left(\sqrt{k}d/(50\nu),\, 2\right)$ corresponds to the turbulent Reynolds number~\cite{ling2015evaluation}. These features are summarized in Table~\ref{tab:feature-definitions}.

\begin{table}[h!]
    \centering
    \renewcommand{\arraystretch}{1.2}
    \begin{tabular}{m{0.1\linewidth} >{\raggedright\arraybackslash}m{0.2\linewidth} >{\raggedright\arraybackslash}m{0.25\linewidth}}
    \specialrule{1pt}{0pt}{0pt}
    \textbf{Feature} & \textbf{Description} & \textbf{Formula} \\
    \specialrule{1pt}{0pt}{0pt}
    
    $q_1$ & First scalar invariant & $q_1 = \theta_1 = \{\hat{\mathbf{S}}^2\}$ \\
    
    $q_2$ & Second scalar invariant & $q_2 = \theta_2 = \{\hat{\mathbf{W}}^2\}$ \\
    
    $q_3$ & Viscosity ratio & $q_3 = \tfrac{\nu_t}{100\nu + \nu_t}$ \\
    
    $q_4$ & Turbulent Reynolds number & $q_4 = \min\!\left(\tfrac{\sqrt{k}\,d}{50\nu},\, 2\right)$ \\
    \specialrule{1pt}{0pt}{0pt}
    \end{tabular}
    \caption{Summary of model input features, where curly brackets $\{\cdot\}$ indicate trace of a tensor; $\nu$ and $\nu_t$ are the laminar and eddy viscosities, $\hat{\mathbf{S}}$ and $\hat{\mathbf{W}}$ the normalized strain- and rotation-rate tensors, $k$ the turbulent kinetic energy, and $d$ the wall distance.}
    \label{tab:feature-definitions}
\end{table}

Separate sub-networks are used to predict the linear coefficient $g^{(1)}$ and the quadratic coefficient $g^{(2)}$, while sharing the same invariant input features. This parallel architecture imposes a clear structural hierarchy by treating linear and higher-order nonlinear contributions independently, rather than implicitly weighting them equally within a single network. By emphasizing linear contributions and introducing nonlinear corrections in a controlled manner, the design enables physics-informed regularization, improving interpretability, numerical robustness, and generalization.

\subsubsection{Physical consistency requirement within the turbulence model}
To ensure physical consistency, the constitutive relation and turbulence transport equations are treated in a coupled manner and learned simultaneously under essential physical constraints.
We adopt the $k$--$\omega$ (1988) model as the baseline,
\begin{align*}
\frac{\mathrm{D}k}{\mathrm{D}t} &= P_k - \beta^* k \omega 
+ \nabla \cdot \!\left[(\nu + \sigma^* \nu_t)\nabla k \right], \\[0.6em]
\frac{\mathrm{D}\omega}{\mathrm{D}t} &= \alpha\, \frac{\omega}{k} P_k - \beta\, \omega^2 
+ \nabla \cdot \!\left[(\nu + \sigma \nu_t)\nabla \omega \right],
\end{align*}
where the transport coefficients $(\alpha, \beta, \beta^*, \sigma, \sigma^*)$ are traditionally calibrated using canonical flows under the assumption of a linear eddy-viscosity model.
Such fixed calibration limits generalization to complex industrial flows and becomes internally inconsistent when the constitutive relation is extended to nonlinear forms.

To address these limitations, only the dissipation coefficient $\beta$ in the turbulent specific dissipation rate $\omega$ equation is treated as a spatially varying quantity and is learned jointly with the constitutive relation. The parallel TBNN directly predicts the dissipation coefficient $\beta$ together with the tensor-basis coefficients $g^{(i)}$ (Fig.~\ref{fig:parTBNN} b). 
Each sub-network is implemented as a fully connected neural network: the sub-network for $g^{(1)}$ uses four hidden layers with four neurons per layer, the sub-network for $g^{(2)}$ uses four hidden layers with six neurons per layer, and the sub-network for $\beta$ uses four hidden layers with four neurons per layer; all sub-networks employ the ReLU activation function, resulting in a total of 333 trainable network weights.
The remaining transport coefficients are then derived sequentially using established physical constraints~\cite{wilcox1998turbulence, pope2000, kok2000resolving, cazalbou1994behavior}. Specifically, the following constraints are enforced to ensure equilibrium behavior, near-wall asymptotics, and freestream robustness:
\begin{enumerate}[leftmargin=3em,label=({\arabic*})]
\item $\beta^* / \beta = 1.25$, based on homogeneous isotropic decaying turbulence;
\item $(Sk/\varepsilon)^2 = -\beta / (g^{(1)} \alpha \beta^*) = 17.06$, enforcing equilibrium homogeneous shear turbulence;
\item $\sigma = \sqrt{\beta^*}\, (\beta/\beta^* - \alpha) / \kappa^2$, ensuring compatibility with the logarithmic law of the wall;
\item $\sigma > \sigma^* > 0.5$ and $\sigma - \sigma^* \le \sigma \sigma^*$, mitigating dependence on the prescribed freestream turbulence levels.
\end{enumerate}
Here, $\varepsilon$ denotes the turbulence dissipation rate, $\kappa$ is the von K\'arm\'an constant, and $S=\sqrt{2 S_{ij} S_{ij}}$ is the magnitude of the strain-rate tensor $S_{ij}$.
Together, these conditions define the physical limits that the coupled model must satisfy in the learned model. This coupled formulation allows turbulence scales and Reynolds stress anisotropy to adapt coherently to the local flow conditions while remaining anchored to well-established turbulence physics. As a result, the proposed representation provides an expressive, frame-invariant, and physically consistent foundation for learning nonlinear eddy-viscosity models.
\added[id=R1]{Within this formulation, the model remains a single-point, local closure. While it does not explicitly represent non-local transport or history effects, the framework can be extended to incorporate such physics. In particular, additional transported quantities can be introduced to encode Reynolds-stress anisotropy, providing a limited form of non-locality. This extension enables improved representation of non-equilibrium behavior, while retaining the robustness and efficiency of eddy-viscosity formulations~\cite{wu2025data}.}

\subsection{Distribution-based training set selection}

Unified data-driven turbulence modeling requires training cases that span diverse flow mechanisms while avoiding redundant information and keeping training cost manageable. 
We address this requirement by formulating training-set selection as a distribution-based problem, in which candidate flows are compared using distances between probability distributions of physically meaningful flow features. This strategy promotes diversity in the training data and ensures that distinct flow regimes contribute complementary information to the learning of the unified model.

\begin{figure}[!htb]
  \centering
  \includegraphics[width=0.75\textwidth]{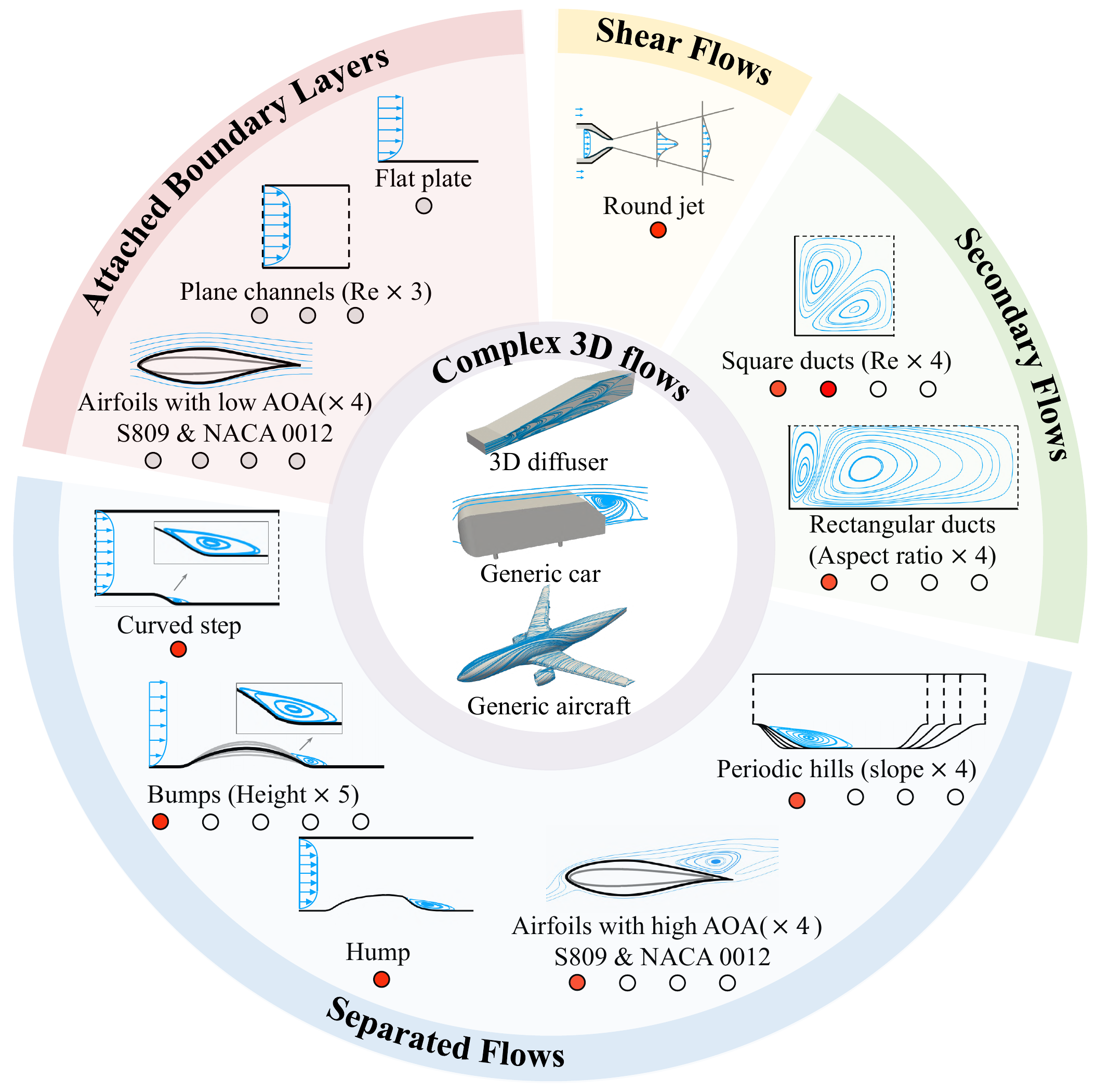}
  \caption{Canonical and complex flows used to train and evaluate the unified foundation model, comprising 36 cases in total. The canonical flows are grouped into four categories: attached boundary layers, free-shear flows, secondary flows, and separated flows, each shown as a sector of the ring, while the three complex 3D flows are placed in the inner circle. Nine representative training flows are highlighted with red circles, identified through the distribution-based training set selection, while the remaining flows are reserved for testing. The testing set includes 23 canonical flows within the training categories and three complex 3D flows. Note that the plane channels and flat plate case are reserved for validation and are thus intentionally excluded from the training set. Detailed specifications are provided in Table~\ref{tbl:foundation-test-cases}. }
  \label{fig:case-setup-foundation-paper}
\end{figure}

As a starting point, we compiled a library of 36 benchmark flow cases from the literature and used a distance-based similarity measure to quantify their distances. The compiled dataset includes both classical datasets~\cite{kim1987turbulence} and recent datasets developed for data-driven turbulence modeling~\cite{xiao2020flows} (Fig.~\ref{fig:case-setup-foundation-paper}). Because turbulence modeling involves many distinct flow regimes but offers limited accessible data compared with data-rich fields such as weather forecasting or genomics~\cite{bi2023accurate,deepmind2025alphagenome}, principled comparison and selection of flow cases are essential. Direct pointwise comparison of flow fields is ill-suited for this purpose, as it depends strongly on mesh resolution, geometry, and grid alignment and therefore does not provide an objective basis for comparing different flows. Instead, each flow is represented statistically through the distribution of its local, frame-invariant flow features, allowing similarity to be assessed at the distribution level. Specifically, local flow features are treated as realizations of a multivariate random variable, so that each flow case induces an empirical probability distribution in feature space. At each spatial location \(x\), the local feature vector is denoted by $\bm{q}(x) \in \mathbb{R}^4$ (see Table~\ref{tab:feature-definitions}), with each grid point providing one sample of $\bm{q}$. The resulting empirical measure $\mu$ summarizes the statistical structure of the flow and enables direct comparison between different flows independent of their geometries and mesh resolutions.

\begin{figure}[!htb]
    \centering
    \includegraphics[width=0.625\textwidth]{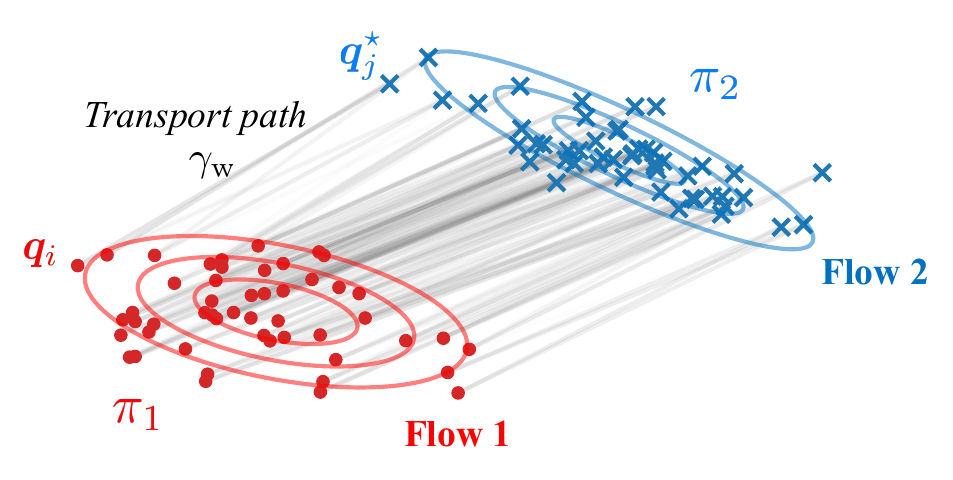}
    \caption{Illustration of similarity between two flow cases based on their feature distributions. Feature samples extracted from the two flows (represented by \textcolor{red}{\textbullet} and \textbf{\textcolor{blue}{\texttimes}}) are compared, and gray lines indicate the optimal matching between samples under the optimal transport formulation. The overall dissimilarity between the flows is quantified by aggregating the transport distances between matched feature samples, yielding the distance between their feature distributions.}
    \label{fig:OT}
\end{figure}

We employ a distance defined directly on probability distributions to quantify the similarity between two flow cases with feature distributions $\pi_1$ and $\pi_2$ (see Fig.~\ref{fig:OT}). 
Specifically, we use the Wasserstein-2 distance from optimal transport theory, which measures how much ``effort'' is required to transform one feature distribution into another. 
This transformation is interpreted as transporting probability mass from one distribution to the other, with the cost proportional to the squared distance traveled in feature space.
Mathematically, the squared Wasserstein-2 distance is defined as
\begin{equation}
\label{eq:wasserstein2}
W_2^2(\pi_1,\pi_2)
=
\min_{\gamma_{\text{w}} \in \Gamma(\pi_1,\pi_2)}
\int\|\bm{q} - \bm{q}^{\star}\|^2 \, \mathrm{d}\gamma_{\text{w}}(\bm{q}, \bm{q}^{\star}),
\end{equation}
where $\Gamma(\pi_1,\pi_2)$ denotes the set of admissible transport plans whose marginals are $\pi_1$ and $\pi_2$. 
Each transport plan~$\gamma_{\text{w}}(\bm{q}, \bm{q}^{\star})$ specifies how much probability mass is moved from feature state $\bm{q}$ in the first flow to feature state $\bm{q}^{\star}$ in the second flow. 
The Wasserstein distance is obtained by selecting the transport plan that minimizes the total transport cost. 
Intuitively, two flows are considered similar if their feature distributions can be matched with only small displacements in feature space.
In practice, the distributions $\pi_1$ and $\pi_2$ are approximated by empirical measures constructed from finite samples,
\[
\pi_1 \approx \frac{1}{N_{\pi_1}} \sum_{i=1}^{N_{\pi_1}} \delta_{\bm{q}_i},
\qquad
\pi_2 \approx \frac{1}{N_{\pi_2}} \sum_{j=1}^{N_{\pi_2}} \delta_{\bm{q}^{\star}_j},
\]
where $\{\bm{q}_i\}$ and $\{\bm{q}^{\star}_j\}$ are flow features extracted from the corresponding flow fields, and $\delta_{\bm{q}_i}$ denotes the Dirac delta (unit point-mass) measure concentrated at $\bm{q}_i$. The resulting empirical measures assign equal probability to each sampled feature vector. The Wasserstein distance between empirical distributions is computed numerically using standard optimal transport solvers and converges to the population distance as the number of samples increases under mild regularity assumptions.

Building on the distribution-based distance measure defined by the Wasserstein-2 distance, we quantify and interpret flow similarity using hierarchical clustering and distance-preserving low-dimensional embedding. An illustrative example of this procedure is shown in Fig.~\ref{fig:foundation-case-selection}.
We first apply hierarchical clustering to the pairwise Wasserstein distances between flow-feature distributions. Starting from individual flow cases, the algorithm iteratively merges the closest clusters, producing the dendrogram shown in Fig.~\ref{fig:foundation-case-selection}a. The resulting hierarchy reveals clear groupings corresponding to secondary flows, separated flows, and free-shear flows, without requiring the number of clusters to be specified \emph{a priori}.
To further aid interpretation, we embed the same distance information into a two-dimensional space using multi-dimensional scaling. The distance-preserving embedding retains the relative similarities defined by the Wasserstein-2 distance and provides an intuitive visualization of how flow cases relate to one another in feature-distribution space, as shown in Fig.~\ref{fig:foundation-case-selection}b. This embedding is invariant under rigid transformations and complements the hierarchical structure by revealing intra- and inter-cluster proximity.
Finally, we select representative training cases from each cluster using the cluster medoid, defined as the flow case that minimizes the sum of Wasserstein distances to all other cases within the same cluster. This strategy ensures coverage of dominant flow mechanisms while avoiding redundant cases and limiting training cost. This distribution-based selection provides a principled foundation for joint training and multi-objective learning in the unified turbulence modeling framework.

\subsection{Multi-objective ensemble learning from sparse, indirect data}

A central contribution of this work is the formulation of turbulence model learning as a multi-objective optimization problem, in which competing objectives arising from different flows and quantities of interest are reconciled within a single unified model. To enable this formulation under the practical constraint of sparse and indirect observations, we combine multi-objective optimization with a regularized ensemble Kalman learning framework, applied to a neural-network–based turbulence model. Within this framework, ensemble-based updates provide robust, gradient-free parameter estimation in the absence of full-field Reynolds-stress data, while the multi-objective formulation balances performance across heterogeneous flow regimes. Together, these elements yield a scalable and physically grounded strategy for constructing unified turbulence models from diverse data sources.

\subsubsection{Multi-objective learning for unified turbulence modeling}
A unified turbulence model should deliver robust and accurate predictions across diverse flows without manual intervention. To this end, we train a single model using heterogeneous data, such as sparse velocity measurements and integral forces, drawn from flows with distinct physical mechanisms, including attached, separated, secondary, and free-shear flows. This formulation naturally gives rise to multiple objectives that must be optimized simultaneously.
Accordingly, the learning problem is formulated as a multi-objective optimization problem, where each flow case and quantity of interest defines an individual objective with loss function $\ell_j$:
\[
\min_{\mathbf{w}}
\left[
\ell_1(\mathbf{w}),\,
\ell_2(\mathbf{w}),\,
\dots,\,
\ell_N(\mathbf{w})
\right]^{\top},
\]
where $\mathbf{w}$ denotes the weights of the neural-network-based turbulence model (Fig.~\ref{fig:parTBNN}).
This formulation explicitly recognizes that improving performance for one objective may degrade performance for others, leading to conflicting objectives during optimization.

\added[id=R2]{Such conflicts arise when different quantities of interest (QoIs) impose different constraints on the turbulence closure. Importantly, QoIs from the same flow are not necessarily aligned and can also be conflicting, as will be demonstrated in \S\ref{subsec:conflicting QoIs}. The nature and extent of this conflict are also model-dependent, since different baseline models can exhibit different physical biases and thus different QoI sensitivities. This motivates a multi-objective learning framework for handling competing QoI constraints within a unified turbulence model.}

In multi-objective learning, Pareto optimality refers to solutions for which no objective can be improved without worsening at least one other objective.
For unified turbulence modeling, a Pareto-optimal solution corresponds to a parameter vector $\mathbf{w}$ that achieves a balanced compromise across multiple flows, rather than optimal performance on a single case.
A common approach to multi-objective learning is to combine objectives into a single weighted loss; however, fixed-weight formulations suffer from two major limitations.
First, suitable weights are not known \emph{a priori} and often require costly grid searches or heuristic tuning~\cite{kendall2018multi,chen2018gradnorm}.
Second, fixed weighting provides no principled notion of optimality in the presence of conflicting objectives, frequently resulting in solutions that favor some flows at the expense of others.

\begin{figure}[!htb]
  \centering

  \begin{subfigure}[t]{0.26\textwidth}
    \centering
    \includegraphics[width=0.8\textwidth]{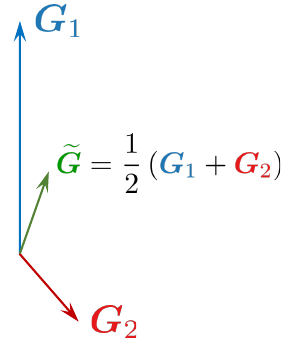}
    \caption{}
    \label{fig:fw-mean}
  \end{subfigure}%
  \hspace{0.004\textwidth}%
  \begin{subfigure}[t]{0.26\textwidth}
    \centering
    \includegraphics[width=0.8\textwidth]{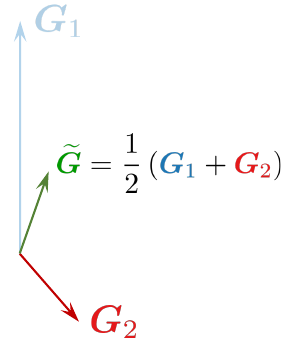}
    \caption{}
    \label{fig:fw-conflict}
  \end{subfigure}%
  \hspace{0.004\textwidth}%
  \begin{subfigure}[t]{0.33\textwidth}
    \centering
    \includegraphics[width=0.8\textwidth]{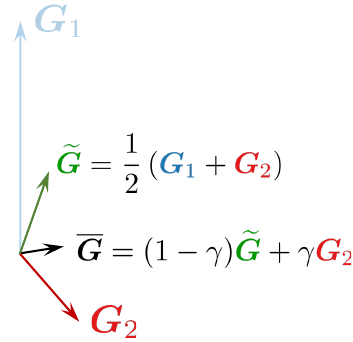}
    \caption{}
    \label{fig:fw-relax}
  \end{subfigure}

  \caption{Illustration of the Frank--Wolfe algorithm for two conflicting objectives $\bm{G}_1$ and $\bm{G}_2$: 
    (a) averaging objective gradients to obtain the mean gradient $\widetilde{\bm{G}}$, 
    (b) identifying the most conflicting gradient $\bm{G}_2$, and 
    (c) under-relaxing the mean gradient toward the conflicting gradient using a step size $\gamma$. 
    The procedure iterates until conflicts are resolved or a stopping criterion is met.}
    
  \label{fig:fw-algorithm}
\end{figure}

To overcome these limitations, we adopt a multi-gradient descent framework with dynamically adjusted weights~\cite{desideri2012multiple}.
Network weights of the model are updated according to
\[
\mathbf{w} \leftarrow \mathbf{w} - \eta \sum_{j=1}^{N} \xi_j \nabla_{\mathbf{w}} \ell_j(\mathbf{w}),
\]
where $\eta$ is the learning rate and $\xi_j$ denotes the weight associated with objective $j$.
Pareto optimality can be characterized through Pareto stationarity using the Karush--Kuhn--Tucker (KKT) conditions~\cite{KuhnTucker1951}.
A network weight vector $\mathbf{w}$ is Pareto stationary if there exists a nonnegative weight vector
$\bm{\xi} = [\xi_1, \xi_2, \dots, \xi_N]^{\top}$,
with $\sum_{j=1}^N \xi_j = 1$, such that
\[
\sum_{j=1}^{N} \xi_j \nabla_{\mathbf{w}} \ell_j(\mathbf{w}) = \mathbf{0}.
\]
This condition implies that the objective gradients are in conflict unless all gradients vanish simultaneously.
The central task of multi-objective optimization is therefore to identify the weight vector $\bm{\xi}$ that reconciles these conflicts.
In practice, vector $\bm{\xi}$ can be determined by solving an auxiliary optimization problem that minimizes the norm of the weighted sum of objective gradients.
A zero norm indicates Pareto stationarity, while a nonzero norm indicates that objectives are locally aligned and can be improved simultaneously.
Each gradient is first normalized by the product of its loss value and its $L^2$ norm to ensure meaningful weighting across heterogeneous objectives.
After normalization, objectives with smaller losses produce larger normalized gradients and therefore receive smaller weights.
This mechanism automatically emphasizes underperforming objectives while preventing well-optimized objectives from dominating the update.
The resolution of conflicts in multi-objective optimization is illustrated in Fig.~\ref{fig:fw-algorithm}.
We use $\bm{G}_j = \nabla_{\mathbf{w}} \ell_j(\mathbf{w})$ to denote the gradient of objective $j$, and $\overline{\bm{G}} = \sum_{j=1}^{N} \xi_j \bm{G}_j$ is their weighted sum.
The Frank--Wolfe algorithm~\cite{jaggi2013revisiting} is used to optimize $\bm{\xi}$ by identifying the objective gradient most misaligned with $\overline{\bm{G}}$ and increasing its weight, while maintaining the normalization constraints.
Compared with derivative-free methods such as non-dominated sorting genetic algorithms, this gradient-based approach scales efficiently to problems with many objectives~\cite{sener2018multi}, making it well suited for unified turbulence model training.

\subsubsection{Regularized ensemble learning from sparse, indirect observations}

In practical turbulence modeling, most experimental and industrial datasets provide only sparse and indirect observations, such as velocities or integral forces, rather than full-field Reynolds stress data. Learning from such indirect data with gradient-based methods typically requires adjoint-enabled or fully differentiable solvers, which are often unavailable in complex industrial applications.
To enable stable and efficient learning under these conditions, we adopt a regularized ensemble Kalman filter (REnKF) method~\cite{zhang2020regularized}.

Turbulence model learning is formulated as a Bayesian inverse problem: given observations $\mathbf{y}$ and a baseline model as the prior, the goal is to infer the posterior distribution of the turbulence model, parameterized by neural network weights $\mathbf{w}$,
\[
\mathbb{P}(\mathbf{w} \mid \mathbf{y}) \propto \mathbb{P}(\mathbf{w}) \, \mathbb{P}(\mathbf{y} \mid \mathbf{w}).
\]
Under Gaussian and linear assumptions, the maximum a posteriori estimate reduces to the optimization problem
\begin{equation}
\ell_{\text{EnKF}}(\mathbf{w}) =
\|\mathbf{w}^{i+1} - \mathbf{w}^i\|_{\mathsf{P}^{-1}}^2
+ \|\mathbf{y} - \mathcal{H}[\mathbf{w}^{i+1}]\|_{\mathsf{R}^{-1}}^2,
\label{eq:loss-enkf}
\end{equation}
where $\mathcal{H}$ is the observation operator, and $\mathsf{P}$ and $\mathsf{R}$ denote the covariance matrices of the prior model and observation errors, respectively.
Iterative EnKF tends to push the solution away from the baseline, effectively reducing the update to maximum likelihood estimation and often leading to overfitting and poor generalization.

To mitigate this issue, we introduce a regularization term into Eq.~\eqref{eq:loss-enkf} that penalizes deviations from physically meaningful baseline model coefficients.
The resulting regularized loss function is
\[
\ell_{\text{REnKF}}(\mathbf{w}) =
\ell_{\text{EnKF}}(\mathbf{w})
+ \|\mathcal{G}[\mathbf{w}^{i+1}]\|_{\mathsf{Q}^{-1}}^2,
\]
where $\mathsf{Q}$ denotes the covariance matrix of the regularization constraints.
The regularization operator $\mathcal{G}$ penalizes deviations of selected model coefficients from their baseline values.
For the constitutive relation, the baseline linear eddy-viscosity coefficients are $g^{(1)}=-0.09$ and $g^{(2)}=0$, while for the turbulence transport equations the dissipation coefficient $\beta$ has a baseline value of $0.072$.
Accordingly, the regularization operator is defined as
\[
\bm{\mathcal{G}} =
\left[
(g^{(1)} + 0.09)^2,\;
(g^{(2)} - 0)^2,\;
(\beta - 0.072)^2
\right].
\]

Minimizing the regularized cost function $\ell_{\text{REnKF}}$ yields a two-stage update scheme~\cite{zhang2020regularized,zhang2023combining}.
First, a pre-correction step accounts for the regularization term,
\[
\widetilde{\mathbf{w}}_j^i = \mathbf{w}_j^i - \mathsf{P}\, \mathcal{G}'[\mathbf{w}_j^i]^\top
\mathsf{Q}^{-1} \mathcal{G}[\mathbf{w}_j^i],
\]
followed by the standard ensemble Kalman update,
\[
\mathbf{w}_j^{i+1} = \widetilde{\mathbf{w}}_j^i + \mathsf{K} \big( \mathbf{y}_j - \mathcal{H}[\widetilde{\mathbf{w}}_j^i] \big), \quad \mathsf{K} = \mathsf{P}\mathsf{H}^\top \left( \mathsf{H}\mathsf{P}\mathsf{H}^\top + \mathsf{R} \right)^{-1}.
\]
Overall, this method stabilizes learning of nonlinear eddy-viscosity models, preserves physical consistency between the constitutive relation and transport equations, and yields turbulence models that are both robust and generalizable.

Within this regularized ensemble Kalman framework, the remaining question is how multiple, potentially competing objectives are incorporated consistently into the ensemble updates.
This is achieved naturally, as each objective $j$ yields an ensemble-based update $\bm{G}_j$ which can be interpreted as a generalized gradient direction.
These objective-specific updates are combined using adaptive weights $\bm{\xi}$ obtained from the multi-objective optimization,
\[
\mathbf{w} \leftarrow \mathbf{w} + \sum_{j=1}^{N} \xi_j \bm{G}_j,
\]
thereby coupling multi-objective learning with the regularized ensemble Kalman method. The complete multi-objective ensemble learning procedure is summarized in Algorithm~\ref{alg:mtl-renkf}.

\begin{algorithm}[!htb]
\caption{Multi-objective ensemble learning ($T=100$)}
\label{alg:mtl-renkf}
\begin{algorithmic}[1]
    \For{$t = 1, 2, \dots, T$}
        \State Compute individual objective $j$'s update $\bm{G}_j$ by regularized ensemble Kalman method
        \State Normalize each update:
        \begin{align*}
            \mathsf{G}_j = \frac{\bm{G}_j}{\ell_j \cdot \|\bm{G}_j\|}
        \end{align*}
        \State Initialize task weights: $\bm{\xi} \gets [1/N, \dots, 1/N]$
        \Repeat
            \State Compute weighted mean gradient: 
            \begin{align*}
                \overline{\mathsf{G}} = \sum_{j=1}^{N} \xi_j \mathsf{G}_j
            \end{align*}
            \State Identify the most conflicting task:
            \begin{align*}
                J = \arg\min_{j} \mathsf{G}_j^{\top} \overline{\mathsf{G}}
            \end{align*}
            \State Compute step size $\gamma$ via line search:
            \begin{align*}
                \gamma = \arg\min_{\gamma \in [0, 1]} \left\| (1 - \gamma) \overline{\mathsf{G}} + \gamma \mathsf{G}_J \right\|^2
            \end{align*}
            \State Update task weights (where $\mathbf{e}_J \in \mathbb{R}^N$ denotes the $J^{\text{th}}$ unit vector):
            \begin{align*}
                \bm{\xi} \gets (1 - \gamma) \bm{\xi} + \gamma \mathbf{e}_J
            \end{align*}
        \Until{Convergence of $\bm{\xi}$ or maximum steps reached}
        \State Update network weights $\mathbf{w}$ with learning rate $\eta$:
        \begin{align*}
            \mathbf{w} \gets \mathbf{w} + \eta \left( \sum_{j=1}^{N} \xi_j \bm{G}_j \right)
        \end{align*}
    \EndFor
    \State \Return Final network weights $\mathbf{w}$
\end{algorithmic}
\end{algorithm}

\subsection{Additive fine-tuning for specialist model}

When application-specific accuracy is required, the unified foundation model can be adapted into a specialist model via additive fine-tuning.
A unified foundation model is designed to perform robustly across diverse flow mechanisms and typically improves upon baseline closures over a wide range of conditions. 
However, this broad objective inevitably trades off accuracy in flow regimes that dominate a specific device or operating condition. 
Enforcing uniform performance across many regimes can dilute model capacity for the mechanisms of primary interest, such as large-scale separation in Ahmed body flows or strong secondary motions in three-dimensional diffusers. 
In these settings, higher predictive accuracy is often achieved by relaxing performance requirements on less relevant regimes and concentrating model capacity on the targeted mechanisms. 
This motivates the introduction of a \emph{specialist model}, which improves accuracy within a specific mechanism class while retaining limited generalizability. 
Crucially, such specialization must preserve essential physical properties inherited from the unified foundation model, such as the logarithmic law of the wall.

The unified foundation model is treated as a strong prior that already encodes broad extrapolation capability across canonical flows through multi-objective training. 
Specialization is therefore formulated as a controlled adaptation that enhances performance on a targeted application domain while introducing the smallest possible deviation from the unified foundation model. 
To this end, an additive parameterization is adopted, in which the foundation network is frozen and only a compact correction module is trained (Fig.~\ref{fig:parTBNN} c).
Let the unified foundation model predict the closure quantities $\{g^{(1)}_{\text{f}}, \cdots, g^{(10)}_{\text{f}},\, \beta_{\text{f}}\}$ using the physically consistent parallel TBNN architecture. The specialist model augments these predictions by learning small additive corrections
$\{\Delta g^{(1)}, \cdots, \Delta g^{(10)},\, \Delta \beta\}$,
such that
\begin{equation}
g^{(i)}_{\text{s}} = g^{(i)}_{\text{f}} + \Delta g^{(i)},
\quad
\beta_{\text{s}} = \beta_{\text{f}} + \Delta\beta.
\label{eq:additive-ft}
\end{equation}
Only the correction networks are trainable and the unified foundation model remains fixed. 
This design ensures that the specialist model (i) starts from a well-validated foundation and (ii) modifies the closure only where required by the new data. 
Additive learning thus provides a natural mechanism for capturing flow-specific effects while retaining the robustness of the unified foundation model.
Each correction term is parameterized by a fully connected neural network. The sub-networks for $\Delta g^{(1)}$, $\Delta g^{(2)}$, and $\Delta \beta$ each use four hidden layers with four neurons per layer and employ the ReLU activation function. In total, the additive fine-tuning architecture introduces 255 trainable network weights.

In the fine-tuning process, the physical consistency is still preserved.
The specialist model continues to predict $\beta_{\text{s}}$ together with the constitutive coefficients $g^{(i)}_{\text{s}}$, while the remaining transport coefficients $(\beta^*, \alpha, \sigma, \sigma^*)$ are computed through the same constraint chain used for the foundation model, including equilibrium homogeneous shear behavior and logarithmic-layer compatibility. 
As a result, key physical properties, such as the logarithmic velocity profile, are preserved by construction. 
This guarantees that fine-tuning improves performance on targeted flows without introducing unphysical inconsistencies.

The specialist model is fine-tuned on a set of targeted flows, supplemented by closely related canonical cases, using the same multi-objective learning framework adopted for foundation training. 
Each flow and quantity of interest defines an individual objective, and conflicts among objectives are resolved through adaptive weighting. 
Denoting the trainable correction network weights by $\mathbf{w}_{\text{s}}$ and the frozen unified foundation model network weights by $\mathbf{w}_{\text{f}}$, the fine-tuning problem is formulated as
\[
\min_{\mathbf{w}_{\text{s}}} \;\sum_{j=1}^{N_s} \xi_j\, \ell_j(\mathbf{w}_{\text{f}}, \mathbf{w}_{\text{s}}),
\]
with the objective-specific loss
\[
\ell_j(\mathbf{w}_\text{f},\Delta\mathbf{w}) =
\| \mathbf{w}_\text{s}^{i+1} - \mathbf{w}_\text{s}^i \|_{\mathsf{P}^{-1}}^2
+ \| \mathbf{y} - \mathcal{H}[\mathbf{w}_\text{f}, \mathbf{w}_\text{s}^{i+1}] \|_{\mathsf{R}^{-1}}^2
+ \| \mathcal{G}[\mathbf{w}_\text{s}^{i+1}] \|_{\mathsf{Q}^{-1}}^2,
\]
where $\ell_j$ follows the same regularized ensemble learning formulation introduced earlier. 
The regularization operator~$\bm{\mathcal{G}}$ penalizes deviations of the additive corrections from zero and is defined as
\[
\bm{\mathcal{G}} =
\left[
\left(\Delta g^{(1)}\right)^2,\cdots,
\left(\Delta g^{(10)}\right)^2,\;
\left(\Delta \beta\right)^2
\right].
\]
This formulation enforces a \emph{minimal-modification} principle: additive corrections are activated only when they yield significant improvements in the objectives.

When multiple candidate fine-tuning cases are available, priority is given to those that are statistically closest to the target application in the model input feature space. 
Similarity between flows is quantified using probability distribution distances between their feature distributions, and only cases with the smallest distances to the target flows are selected. 
This strategy promotes mechanism-consistent specialization—for example, selecting curved-step, periodic-hill, or bump flows to support separation-dominated targets—and empirically improves generalization within closely related configurations.

\FloatBarrier
\section{Verification and validation}
\label{sec:validation}
Physical consistency and multi-objective learning are essential for constructing unified turbulence models that are both reliable and broadly applicable.
Physical consistency is verified by demonstrating that the learned model preserves the logarithmic law of the wall and is validated by improved predictions in a round jet flow, where turbulence transport coefficients strongly govern flow development. \added[id=R2]{We also demonstrate that quantities of interest for model training can be conflicting even within the same flow, highlighting the necessity of multi-objective learning.} The effectiveness of multi-objective learning is demonstrated by its ability to reconcile conflicting objectives in periodic hill flows, maintain performance in non-conflicting square duct and airfoil cases, and scale robustly to an increasing number of training objectives while achieving Pareto-improved solutions.

\subsection{Effectiveness of physical consistency}
We examine both the enforcement and practical impact of physical consistency in the turbulence model. Correct implementation of the imposed constraints is first verified by assessing whether the learned model preserves fundamental theoretical behavior, namely the logarithmic law of the wall. The influence of physical consistency on predictive accuracy is then evaluated by comparing physically consistent and inconsistent models on a round jet flow, a configuration in which turbulence transport coefficients strongly govern flow development.

\subsubsection{Verification of physical consistency within turbulence model}
\begin{figure}
    \centering
    \includegraphics[width=0.75\linewidth]{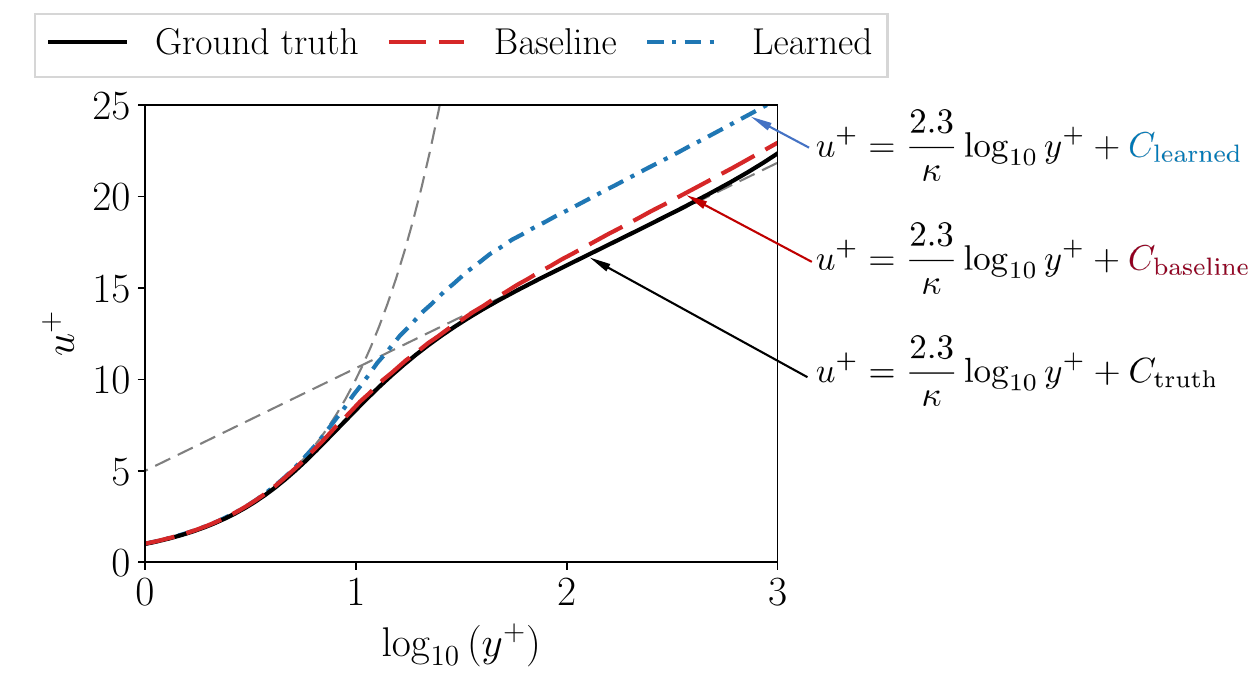}
    \caption{Verification of physical consistency of the learned model using a turbulent flat plate. Mean velocity profiles \(u^+\) are shown versus the wall-normal coordinate \(y^+\) on a logarithmic scale for the learned model (the specialist model for secondary flows with separation), baseline \(k\)--\(\omega\) model, and ground truth. The learned model exhibits the same logarithmic slope in the log layer as the baseline model and the ground truth, consistent with the imposed von K\'arm\'an constant~$\kappa=0.41$. Differences arise only in the additive constant (log-law intercept), resulting in a vertical shift of the logarithmic velocity profile.}
    \label{fig:verification-physical-consistency}
\end{figure}

\added[id=R2]{Physical consistency is assessed in both the inner and outer regions of the boundary layer. In the inner region, we verify whether the learned model preserves the logarithmic law of the wall.}
\deleted[id=R2]{
Physical consistency is verified by examining whether the learned model preserves the logarithmic law of the wall in equilibrium boundary layers.} 
As derived from the physical constraint based on equilibrium log-layer analysis, the logarithmic relationship
\[
u^+ = \frac{1}{\kappa} \ln y^+ + C \approx \frac{2.3}{\kappa} \log_{10} y^+ + C
\]
is enforced through the fixed slope ($1/\kappa$ for $\ln y^+$, equivalently $2.3/\kappa$ for $\log_{10} y^+$), where $\kappa$ is the von K\'arm\'an constant, while the additive constant $C$ is not explicitly constrained. 
To verify correct enforcement, the learned model (the specialist model for secondary flows with separation) is evaluated on a turbulent flat plate. 
Within the logarithmic region, the predicted relationship between $u^+$ and $y^+$ remains logarithmic and preserves the correct slope, confirming that the imposed constraint is respected (see Fig.~\ref{fig:verification-physical-consistency}). 
A vertical shift of the logarithmic profile is observed, which is as expected because the intercept is not specified by the constraint. 
This behavior is consistent with conventional turbulence modeling practice~\cite{wilcox1998turbulence,bin2024constrained}, where both the slope and intercept are calibrated empirically rather than derived from first principles, and therefore confirms correct and reasonable enforcement of the law of the wall. 
If enforcing the intercept is desired, the turbulent flat plate can be included explicitly as a training case to constrain the additive constant.

\begin{figure}
    \centering
    \includegraphics[width=0.55\linewidth]{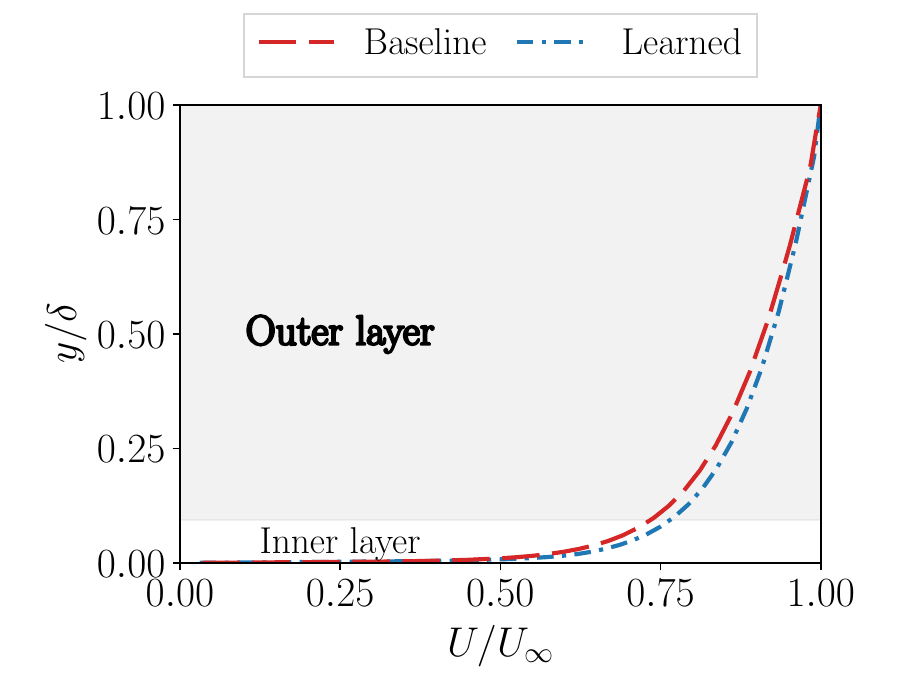}
    \caption{\added[id=R2]{Outer-scaled streamwise velocity profiles for the turbulent flat plate, shown as $y/\delta$ versus $U/U_{\infty}$, where~$U$ is the streamwise velocity, $U_{\infty}$ is the freestream velocity, $y$ is the wall distance, and $\delta$ is the $99\%$ boundary-layer thickness.}}
    \label{fig:verification-physical-consistency-outer-layer}
\end{figure}

\added[id=R2]{To further assess the outer region, we examine the mean streamwise velocity profiles using outer scaling, $y/\delta$ versus $U/U_{\infty}$, where $U_{\infty}$ is the freestream velocity and $\delta$ is the $99\%$ boundary-layer thickness. A small deviation is observed between the learned and baseline profiles, arising from differences in the logarithmic-law intercept $C$ in the inner layer. Despite this offset, the profiles exhibit consistent trends, indicating that the learned model preserves the global boundary-layer structure (see Fig.~\ref{fig:verification-physical-consistency-outer-layer}). Together, the inner- and outer-layer analyses demonstrate that the model maintains physical consistency across the entire boundary layer.}

\subsubsection{Validation of physically consistent model performance}
\begin{figure}
    \centering
    \includegraphics[width=0.6\linewidth]{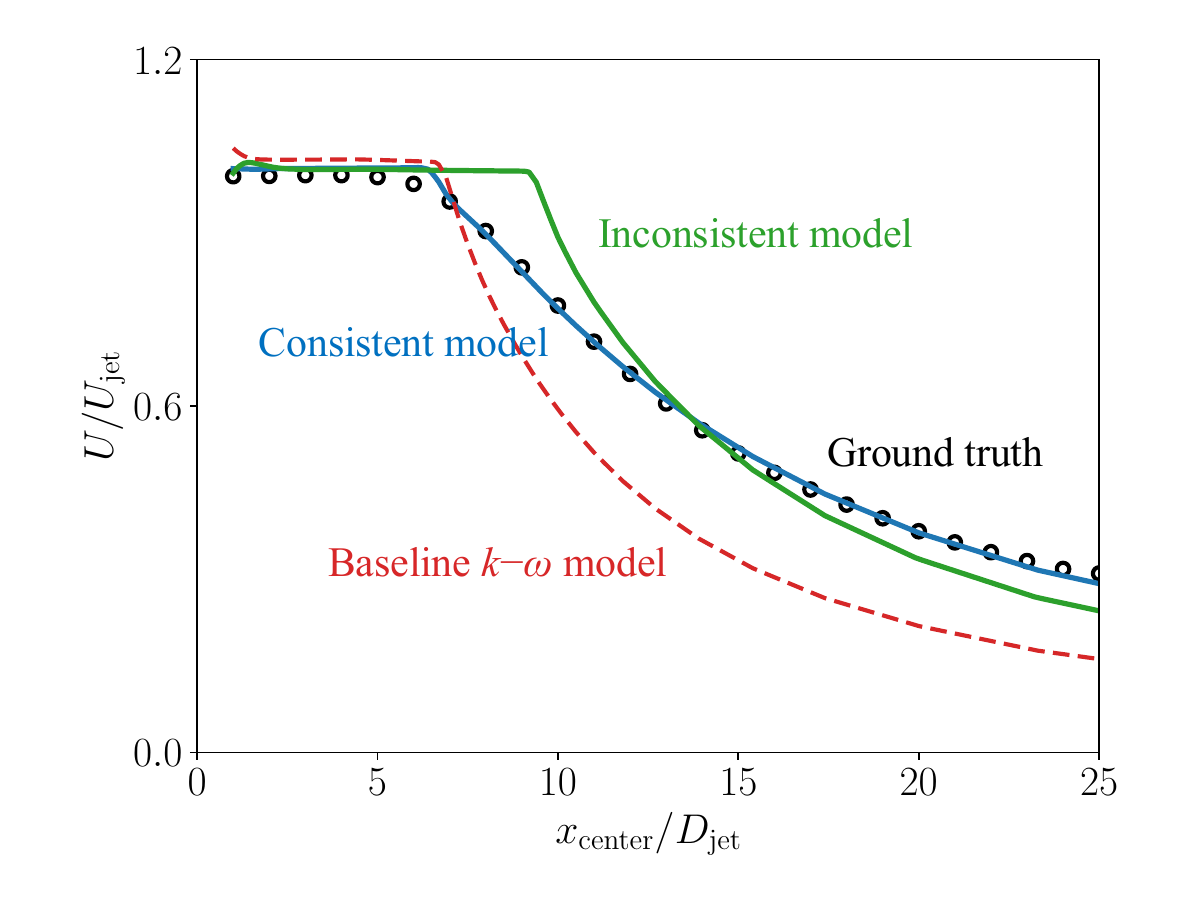}
    \caption{Verification of physical consistency using a round jet flow. The centerline velocity \( U \), normalized by the jet exit velocity \( U_{\mathrm{jet}} \), is plotted against the normalized downstream distance \( x_{\mathrm{center}} / D_{\mathrm{jet}} \), where \( D_{\mathrm{jet}} \) is the jet diameter, for the physically consistent model, physically inconsistent model, baseline \( k\)--\(\omega\) model, and ground truth. The physically consistent model closely reproduces the decay trend of the ground truth along the jet centerline, whereas the physically inconsistent and baseline models deviate downstream.}
    \label{fig:validation-jet-consistency}
\end{figure}

The impact of physical consistency on predictive performance is validated using a round jet flow, a canonical configuration known to be sensitive to turbulence transport coefficients. 
For the baseline $k$--$\omega$ model, the dissipation coefficient $\beta$ in the $\omega$ equation strongly influences the jet spreading rate and velocity decay. 
To isolate the effect of physical consistency, we mainly compare predictions from two models: one that enforces coupled learning of the constitutive relation and turbulence transport equations under physical constraints, and one that does not. 
The physically consistent model closely matches the ground truth, yielding an improved prediction of centerline velocity decay in the jet flow (see Fig.~\ref{fig:validation-jet-consistency}). In contrast, the physically inconsistent model shows pronounced discrepancies downstream of the jet exit, where the velocity decay is noticeably delayed over an extended region. Its decay trend closely follows that of the baseline model, as the two velocity profiles remain nearly parallel. The superior agreement achieved by the physically consistent model underscores the importance of enforcing consistent coupling between the constitutive relation and the turbulence transport equations in jet flows.

\subsection{\added[id=R2]{Characteristics of aligned and conflicting quantities of interest}}
\added[id=R2]{Quantities of interest (QoIs) from the same flow are not necessarily aligned and can be conflicting. Such conflicts arise because different QoIs probe different physical sensitivities and therefore favor different model corrections. The S809 airfoil provides a representative example, where lift and drag induce opposing corrections to the same closure term. By contrast, the curved-step flow shows that QoIs can also remain aligned, with the velocity field and skin-friction coefficient improving consistently under the learned model. Together, these cases show that QoI compatibility depends on both the flow physics and the deficiencies of the baseline model.}

\subsubsection{\added[id=R2]{Aligned quantities of interest: velocity field and friction coefficient of curved step}}
\added[id=R2]{Aligned quantities of interest are quantities that are governed by the same dominant flow mechanism and therefore respond to model changes in a consistent manner. When two quantities depend on the same physical process, improvements in the prediction of that process are expected to affect both quantities in a similar way. In this sense, aligned QoIs do not compete but instead provide complementary views of the same underlying flow physics.}
\begin{figure}
      \centering
      \includegraphics[width=0.72\textwidth]{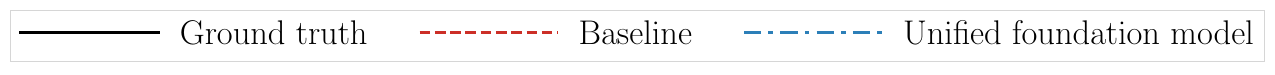}
      
  \begin{subfigure}[t]{0.43\textwidth}
    \centering
    \includegraphics[width=1\textwidth]{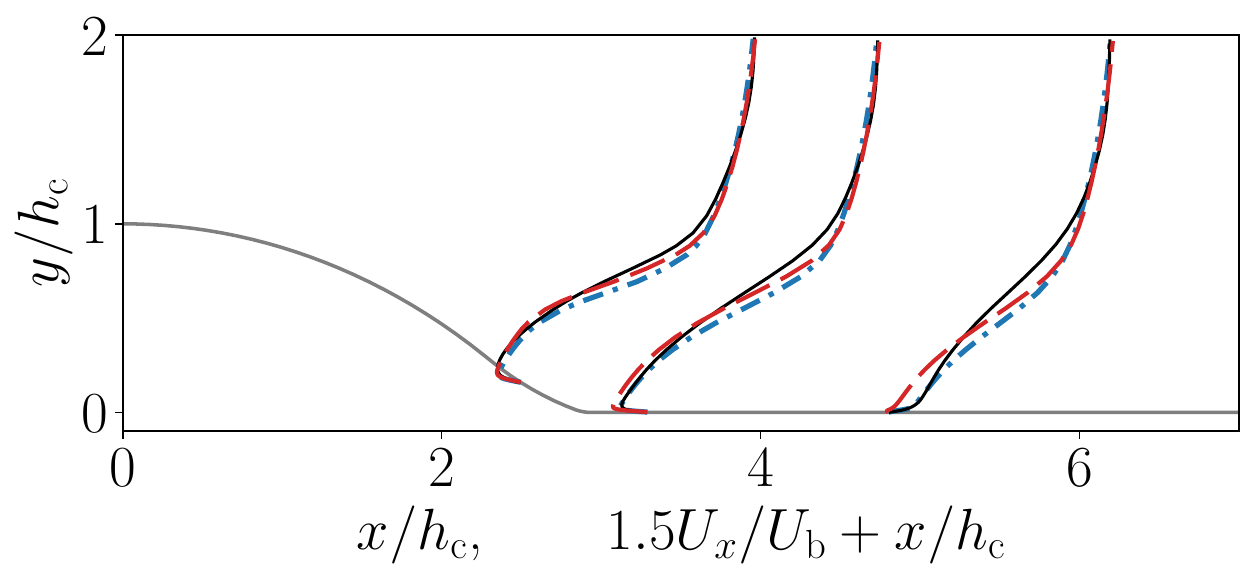}
    \caption{Streamwise velocity profiles}
    \label{fig:drag}
  \end{subfigure}
  \begin{subfigure}[t]{0.54\textwidth}
    \centering
    \includegraphics[width=1\textwidth]{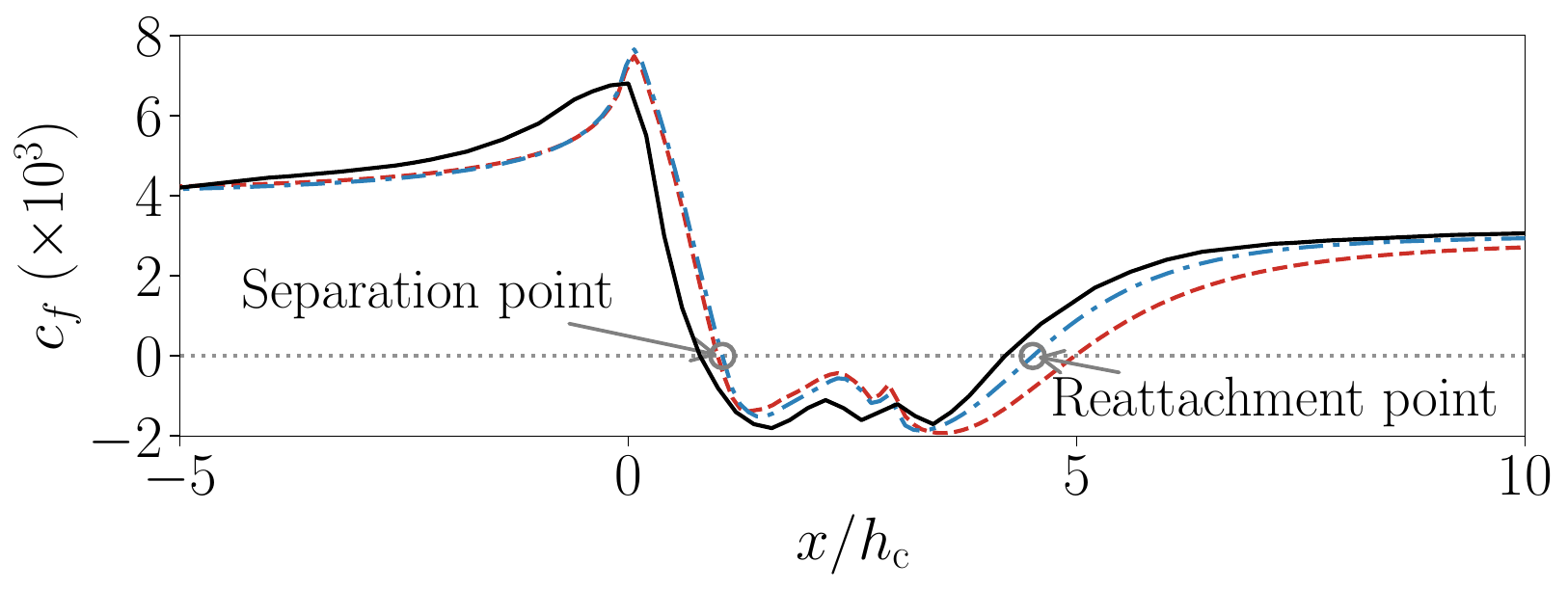}
    \caption{Friction coefficient}
  \end{subfigure}
  \caption{\added[id=R2]{Performance evaluation of the \textbf{unified foundation model} in a \textbf{curved step flow} against the ground truth and the baseline $k$--$\omega$ model. (a) Streamwise velocity profiles \(U_x\), normalized by the bulk velocity \(U_{\mathrm{b}}\), are shown at multiple streamwise locations downstream of the curved step. The profiles $U_x$ versus the wall-normal coordinate \(y\) normalized by the curved-step height \(h_\mathrm{c}\). (b) Skin-friction coefficient distributions $c_f$, are shown along the streamwise direction $x/h_{\mathrm{c}}$. Separation and reattachment points are marked using circles.}}
  \label{fig:curved-step-1}
\end{figure}

\added[id=R2]{The curved step flow provides a representative example. In this case, the dominant mechanism is separation and reattachment induced by the curved geometry and the associated adverse pressure gradient. The key modeling challenge is therefore to predict the development of the separated shear layer, the extent of the recirculation region, and the downstream recovery of the flow. To constrain these features, we use streamwise velocities sampled at multiple downstream locations as observational data. As shown in Fig.~\ref{fig:curved-step-1}a, the unified foundation model yields velocity profiles that are generally closer to the ground truth than the baseline, particularly in the separated and recovery regions, where the shape of the shear layer and the progression toward reattachment are better represented.}

\added[id=R2]{The alignment between the velocity field and the skin-friction coefficient $c_f$ is reflected in the corresponding wall-friction distribution shown in Fig.~\ref{fig:curved-step-1}b. The unified model produces a $c_f$ distribution that more closely follows the ground-truth trend compared to the baseline, particularly in the separated and recovery regions. In particular, it shows a more consistent representation of the negative region associated with reversed flow and a smoother recovery downstream of reattachment, although discrepancies remain near the peak and separation onset. Overall, the changes observed in both the velocity field and $c_f$ follow similar trends. This behavior is consistent with the fact that both quantities are controlled by the same separation and reattachment physics, suggesting that improvements in the learned model primarily act through correcting this shared mechanism.}

\subsubsection{\added[id=R2]{Conflicting quantities of interest: lift and drag of S809 airfoil}}
\label{subsec:conflicting QoIs}

\added[id=R2]{Conflicting quantities of interest (QoIs) arise when different target quantities favor different corrections to the turbulence closure, even within the same flow configuration. In such situations, improving one QoI requires adjusting the flow in a direction that does not improve, and can even deteriorate, another QoI. The source of the conflict therefore does not necessarily lie in using different flows; rather, it originates from the different physical sensitivities encoded by distinct QoIs within the same flow. Importantly, the manifestation of such conflicts is also model-dependent: different baseline turbulence models may produce different deficiencies in the predicted flow field, and therefore induce different QoI sensitivities and preferred correction directions. As a result, whether two QoIs appear aligned or conflicting, and how strongly that conflict is expressed, depends not only on the flow physics but also on the baseline model from which the correction is learned.}

\begin{figure}
      \centering
      \includegraphics[width=0.97\textwidth]{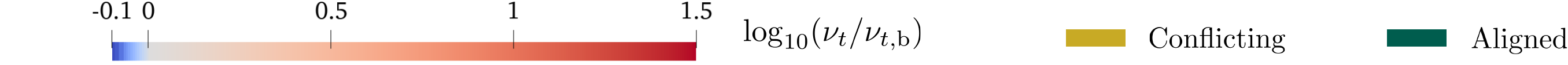}
      
  \begin{subfigure}[t]{0.32\textwidth}
    \centering
    \includegraphics[width=1\textwidth]{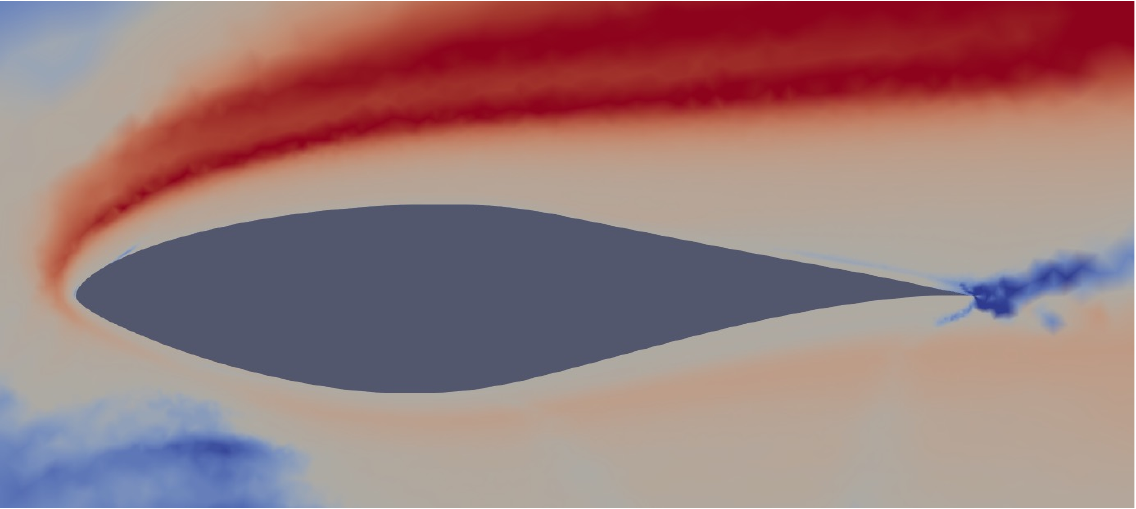}
    \caption{drag-only trained model}
    \label{fig:drag}
  \end{subfigure}
  \begin{subfigure}[t]{0.32\textwidth}
    \centering
    \includegraphics[width=1\textwidth]{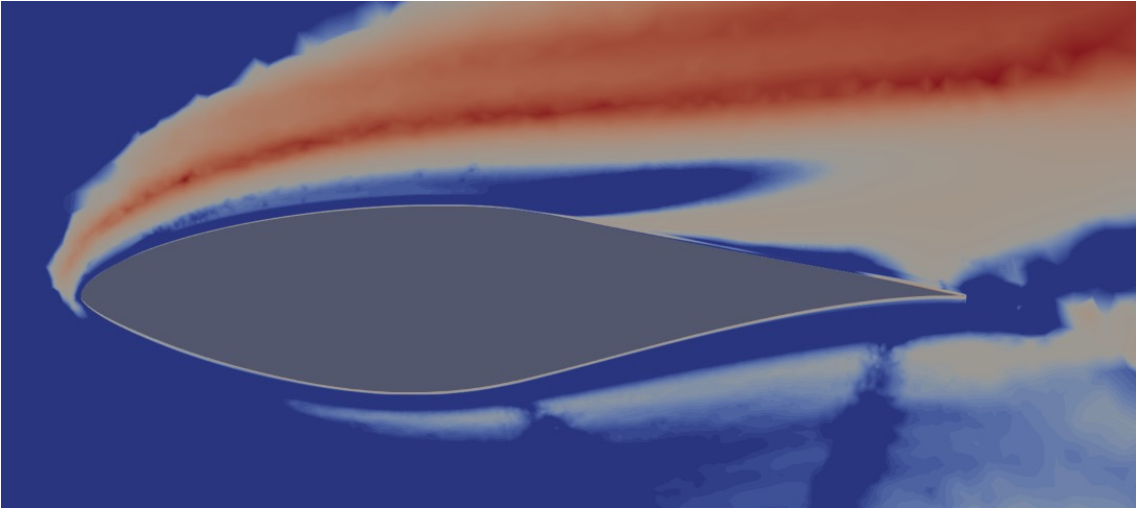}
    \caption{lift-only trained model}
  \end{subfigure}
    \begin{subfigure}[t]{0.32\textwidth}
    \centering
    \includegraphics[width=1\textwidth]{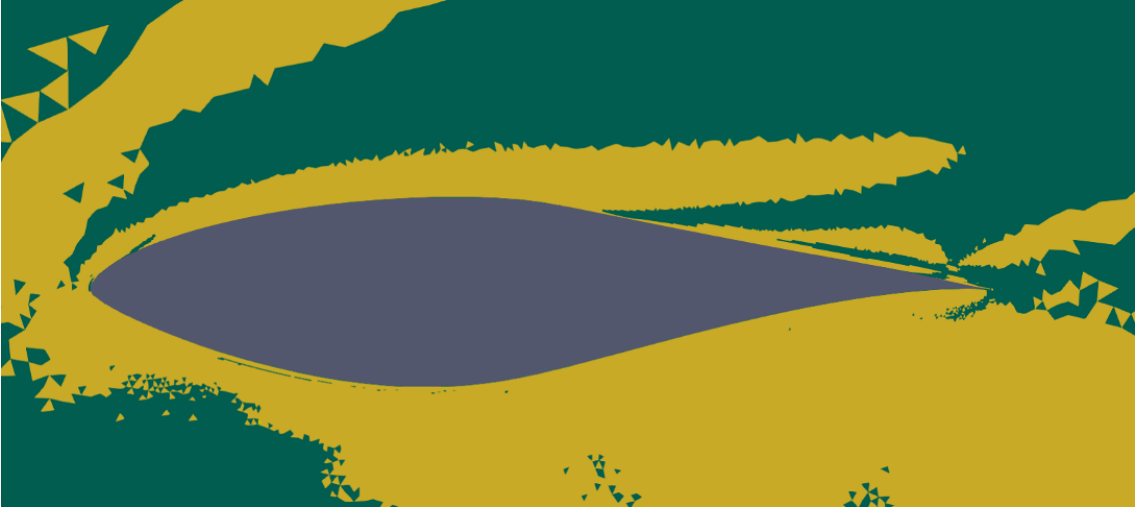}
    \caption{alignment of corrections}
  \end{subfigure}
  \caption{\added[id=R2]{Comparison of eddy-viscosity corrections obtained from single-objective training and their mutual consistency. Panels (a) and (b) show the predicted eddy viscosity, $\log_{10}(\nu_t/\nu_{t,\mathrm{b}})$, from drag-only and lift-only training, respectively, where $\nu_{t,\mathrm{b}}$ denotes the baseline eddy viscosity. Panel (c) highlights regions of alignment and conflict of these corrections.}}
  \label{fig:conflicting-nut}
\end{figure}

\begin{figure}
      \centering
      \includegraphics[width=0.55\textwidth]{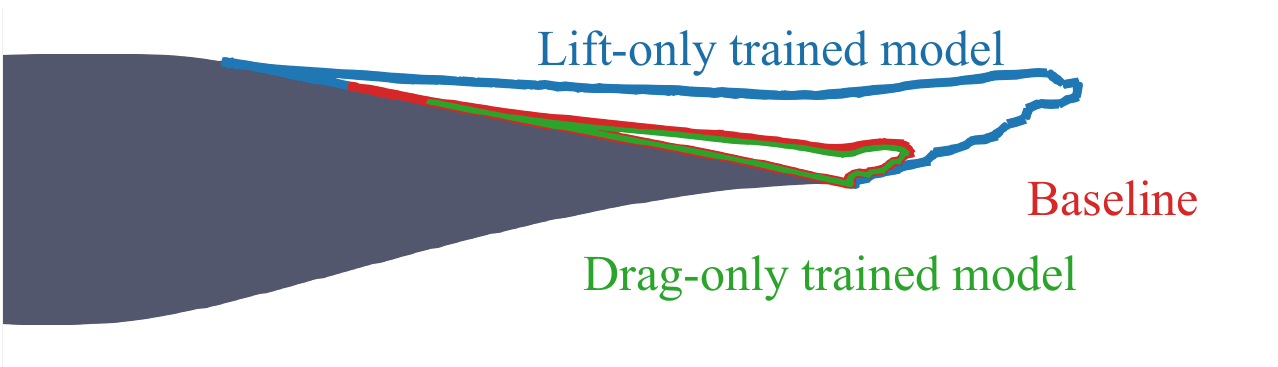}
  \caption{\added[id=R2]{Comparison of the predicted separation bubble for the S809 airfoil obtained from three models: drag-only trained model, baseline model and lift-only trained model.}}
  \label{fig:separation-airfoil}
\end{figure}

\added[id=R2]{In the present S809 airfoil case, the baseline \(k\)--\(\omega\) model overpredicts the lift coefficient and underpredicts the drag coefficient. The baseline solution remains overly attached on the suction side, maintaining excessive suction and thus yielding too much lift. This result is consistent with the interpretation presented in~\cite{zhang2023physical}. The baseline model also underpredicts the wake deficit, which leads to an underprediction of drag, reflecting a known limitation of \(k\)--\(\omega\) type models in capturing mixing and shear-layer development in separated and free-shear flows.}

\added[id=R2]{The results of the single-objective learning cases show that lift and drag drive the model toward opposite eddy-viscosity corrections, as shown in Fig.~\ref{fig:conflicting-nut}c. In particular, the lift-only training case yields a much smaller eddy viscosity especially near the airfoil compared to the baseline (Fig.~\ref{fig:conflicting-nut}b), whereas the drag-only training case yields a much larger eddy viscosity $\nu_t$ (Fig.~\ref{fig:conflicting-nut}a). This indicates that the two objectives favor different turbulence levels, and therefore different flow-adjustment mechanisms.}

\added[id=R2]{The learned behavior is dominated by the linear coefficient $g^{(1)}$, since the quadratic coefficient $g^{(2)}$ remains negligible in both cases and contributes little to the closure. As a result, the difference between the learned models can be interpreted primarily as different corrections to the turbulent eddy viscosity. In the present formulation, this relationship is given by
\[
\nu_t = -\frac{g^{(1)} k}{C_\mu \omega}.
\]
Changes in coefficient $g^{(1)}$ therefore translate directly into changes in turbulent eddy viscosity. The opposite eddy viscosity $\nu_t$ corrections obtained from lift-only and drag-only training (Fig.~\ref{fig:conflicting-nut}c) thus imply opposing adjustments to the dominant closure coefficient $g^{(1)}$, providing strong evidence that the optimization directions associated with these two QoIs are conflicting. These opposite corrections correspond to two distinct physical pathways.}

\added[id=R2]{In the lift-only case, the learned model predicts earlier separation and a substantially larger separation bubble than the baseline solution (Fig.~\ref{fig:separation-airfoil}), which reduces the suction-side pressure loading and brings the lift coefficient closer to the ground truth. This behavior arises from a redistribution of eddy viscosity. Specifically, the learned model decreases eddy viscosity $\nu_t$ in the near-wall region on the suction side, while increasing it in the detached shear layer above the separated region. The reduced near-wall eddy viscosity $\nu_t$ weakens turbulent momentum transfer into the boundary layer under the adverse pressure gradient, making the flow less resistant to separation. At the same time, the elevated eddy viscosity $\nu_t$ in the detached shear layer enhances mixing after separation and helps regulate the downstream development of the enlarged separated flow. The correction therefore acts through a spatial redistribution of turbulent mixing: weaker near-wall mixing to permit stronger separation, and stronger shear-layer mixing once the flow has lifted off. This mechanism remains qualitatively consistent with the separation-based interpretation discussed in~\cite{zhang2023physical}, while indicating that, in the present case, the learned correction is better understood as a spatial reorganization of $\nu_t$ relative to the baseline.}

\begin{figure}
      \centering
      \includegraphics[width=0.65\textwidth]{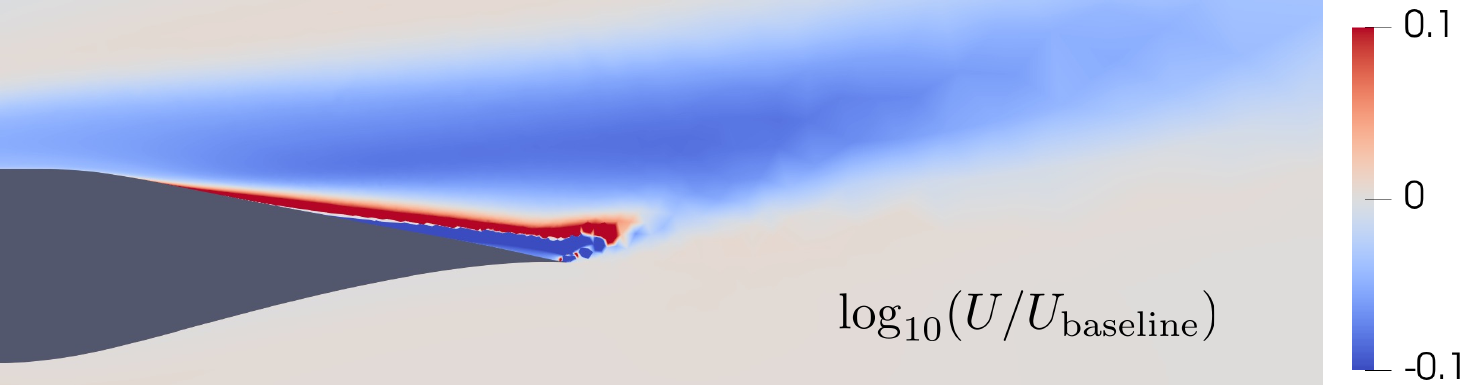}
  \caption{\added[id=R2]{Comparison of the wake deficit between the model trained with only drag and the baseline model for the S809 airfoil. The  velocity ratio $\log_{10}(U/U_{\mathrm{baseline}})$ highlights the regions of velocity deficit and acceleration compared to the baseline model.}}
  \label{fig:different-wakes}
\end{figure}

\added[id=R2]{The drag-only training delays separation and produces a slightly smaller separation bubble than the baseline (Fig.~\ref{fig:separation-airfoil}). At the same time, it generates a stronger and more extended wake deficit, as indicated by the broad blue region of negative $\log_{10}(U/U_{\mathrm{baseline}})$ that persists farther downstream in Fig.~\ref{fig:different-wakes}. This pattern indicates that the drag-only model predicts lower wake velocity than the baseline over much of the wake, with both greater streamwise extent and wider spreading. A localized red region is also observed near the suction-side shear layer and close to the airfoil, indicating limited regions where the drag-only model predicts slightly higher velocity than the baseline. Overall, this behavior is associated with a more global increase in $\nu_t$, including the upper shear layer, the wake, and much of the near-airfoil region, which enhances turbulent mixing and downstream dissipation. Force decomposition further shows that the drag increase arises entirely from pressure drag rather than viscous drag. The model therefore increases drag not by enhancing wall shear, but by modifying the pressure field through stronger wake dissipation. In this sense, separation and drag are not directly equivalent: the drag increase is achieved primarily through wake modification rather than stronger separation.}

\added[id=R2]{Taken together, these results demonstrate that lift and drag are intrinsically conflicting quantities of interest in the present airfoil problem. The lift-only objective drives the dominant closure coefficient $g^{(1)}$, and thus eddy viscosity $\nu_t$, in a direction that promotes earlier separation and enlarges the recirculation zone. In contrast, the drag-only objective drives $g^{(1)}$ in the opposite direction, increasing eddy viscosity $\nu_t$ in the shear layer and wake and thereby increasing pressure drag without requiring a larger separation bubble. These two QoIs therefore favor opposing corrections to the same dominant closure mechanism, leading to distinct physical behaviors of the flow.}

\added[id=R2]{More generally, each quantity of interest encodes specific aspects of the underlying flow physics. In canonical flows, a QoI often reflects a single dominant mechanism, whereas in complex flows, it can combine multiple interacting mechanisms. When model calibration relies on a single QoI, the resulting model is exposed to only a subset of the underlying physics. As a result, it may capture only partial mechanisms and exhibit restricted generalizability, reminiscent of the parable of the six blind men describing an elephant from incomplete perspectives.}

\subsection{Effectiveness of multi-objective learning}
We validate the proposed multi-objective learning framework under two distinct scenarios: a conflict scenario and a non-conflict scenario. In the conflict scenario, the framework automatically resolves competing objectives arising from two periodic hill cases with different separation characteristics, as shown in Fig.~\ref{fig:pehill-conflicting-results}. In the non-conflict scenario, which includes an S809 airfoil and a square duct flow, the framework preserves optimization performance for both cases without introducing trade-offs, as shown in Fig.~\ref{fig:non-conflict-results}.

\subsubsection{Conflict flows: periodic hills with different separation characteristics}
\begin{figure}[!htb]
  \centering

  \begin{subfigure}[t]{0.47\textwidth}
    \centering
    \includegraphics[width=0.96\textwidth]{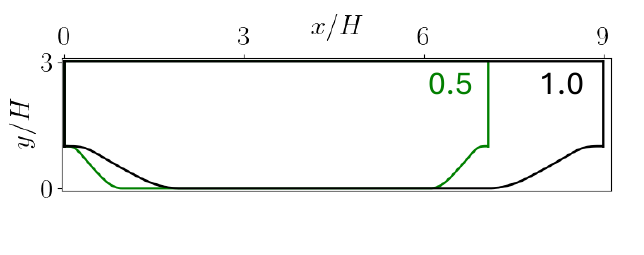}
    \caption{Geometries}
    \label{fig:geom}
  \end{subfigure}
  \begin{subfigure}[t]{0.49\textwidth}
    \centering
    \includegraphics[width=0.96\textwidth]{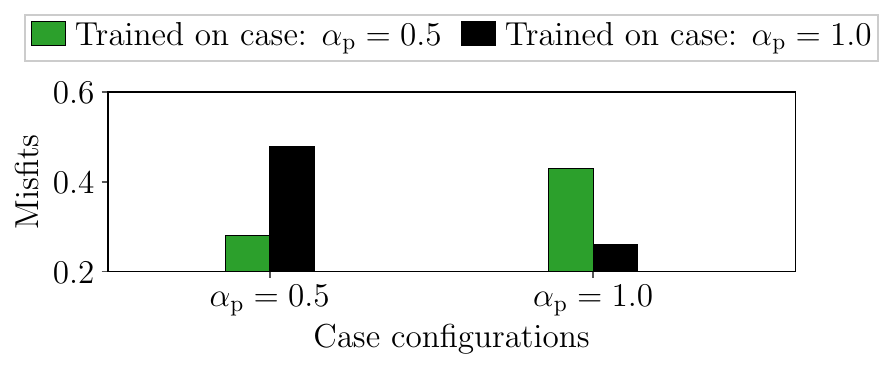}
    \caption{Misfit results}
    \label{fig:misfits}
  \end{subfigure}

  \caption{Comparison of periodic hill geometries and single-case training misfit results.
    (a) Geometries of the two periodic hill configurations with slope parameters $\alpha_{\text{p}} = 0.5$ and $\alpha_{\text{p}} = 1.0$.
    (b) Misfit values obtained from single-case training, where models are trained on one configuration ($\alpha_{\text{p}} = 0.5$ or $\alpha_{\text{p}} = 1.0$) and evaluated on the other configuration.}
  
  \label{fig:geometry_and_misfits}
\end{figure}

The multi-objective framework automatically resolves competing objectives arising from two periodic hill cases with different separation characteristics, arising from differences in hill slope steepness controlled by the slope scaling factor $\alpha_{\text{p}}$. The \(\alpha_{\text{p}} = 0.5\) case is characterized by massive separation, with the recirculation region extending upstream of the hill crest, whereas the \(\alpha_{\text{p}} = 1.0\) case exhibits only mild separation confined to the leeward foot (see Fig.~\ref{fig:geometry_and_misfits}a).

Models trained on a single case fail to generalize across regimes. In particular, a model trained only on the mildly separated case (\(\alpha_{\text{p}} = 1.0\)) incurs large errors when applied to the strongly separated case (\(\alpha_{\text{p}} = 0.5\)), as quantified by the normalized misfits in Fig.~\ref{fig:geometry_and_misfits}b. This behavior is consistent with earlier findings that models trained on mildly separated periodic hill flows struggle to extrapolate to strongly separated configurations~\cite{zhou2022frame,zhang2022ensemble}. These results illustrate the conflicting nature of the two objectives and motivate the use of a principled multi-task optimization strategy.

Fixed-weight loss aggregation cannot reliably reconcile the competing objectives because the optimal balance between cases is unknown a priori and requires manual tuning. To overcome this limitation, the Frank--Wolfe algorithm is employed to adaptively determine the weights by minimizing the norm of the weighted sum of the objective gradients. Two training strategies are compared: grid search over fixed-weight combinations and adaptive weighting using the Frank--Wolfe algorithm.

\begin{figure}[!htb]
  \centering

  \begin{subfigure}[t]{0.48\textwidth}
    \centering
    \includegraphics[width=0.95\textwidth]{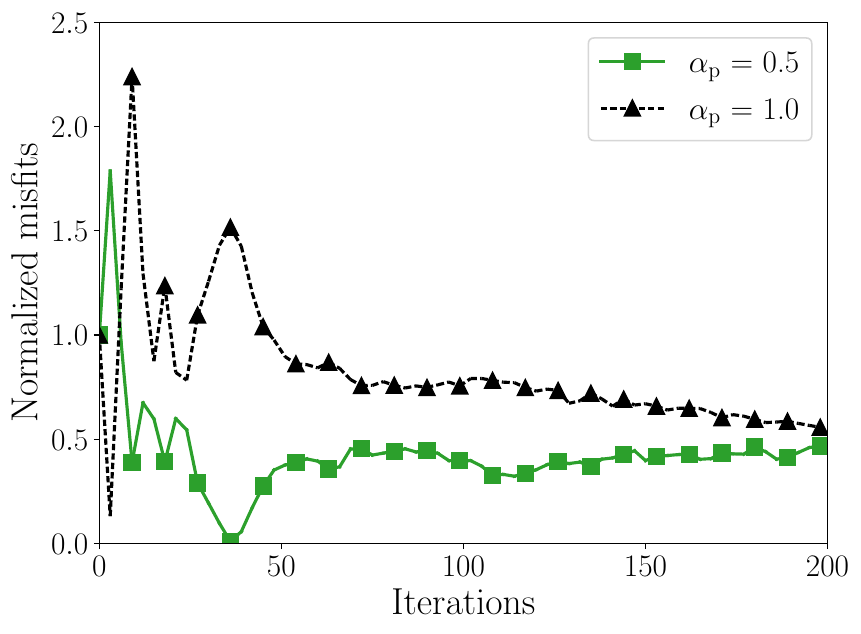}
    \caption{Training evolution of normalized misfits}
    \label{fig:pehill-evolution}
  \end{subfigure}
  \hspace{-0.03\textwidth}
  \begin{subfigure}[t]{0.48\textwidth}
    \centering
    \includegraphics[width=0.95\textwidth]{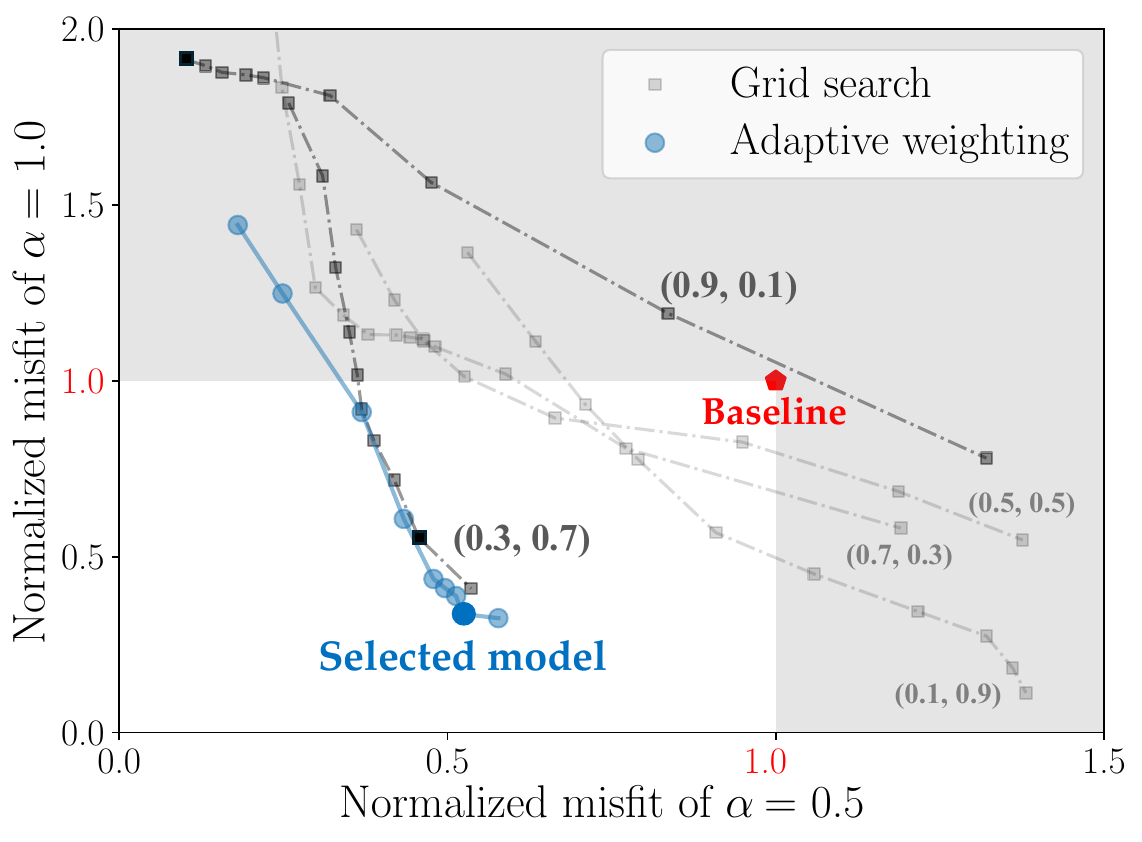}
    \caption{Trade-off between normalized misfits}
    \label{fig:pehill-tradeoff}
  \end{subfigure}

  \caption{Training behavior and trade-off analysis for two periodic hill cases with different slope parameters.
    (a) Evolution of normalized misfits during training for the two cases with $\alpha_{\text{p}} = 0.5$ and $\alpha_{\text{p}} = 1.0$.
    (b) Tradeoff between the two normalized misfits for models obtained using grid search and adaptive weighting. Five symmetric weight pairs spanning 0.1–0.9 are selected and only iterations 150--200 are shown for clarity. The baseline $k$--$\omega$ model (\pentagofill[red]) is located at $(1.0,\,1.0)$.}
  
\label{fig:pehill-conflicting-results}
\end{figure}

Adaptive weighting yields a characteristic training evolution in which the normalized misfits evolve jointly, as shown in Fig.~\ref{fig:pehill-conflicting-results}a. At early stages, improvements in one case degrade the performance of the other, indicating strong competition between objectives. As training progresses, the adaptive weighting reconciles this conflict and drives the optimization toward a balanced solution, in which both cases reach normalized misfits of approximately \(0.5\). Fixed-weight optimization yields an uneven tradeoff that often improves one periodic hill case at the expense of the other in Fig.~\ref{fig:pehill-conflicting-results}b. Grid search over fixed weights explores only a limited set of combinations and frequently produces models that favor one case while degrading the other, with several outcomes performing worse than the baseline. In contrast, the Frank--Wolfe–based adaptive strategy automatically traces a smooth tradeoff curve and converges toward a region in which both normalized misfits are simultaneously reduced, as shown in Fig.~\ref{fig:pehill-conflicting-results}b.

\subsubsection{Non-conflict flows: square duct and airfoil flows}
The proposed framework preserves optimization performance in the non-conflict scenario because the training objectives are compatible. The training cases include a square duct flow and a high--angle-of-attack S809 airfoil flow, which impose distinct constraints on the turbulence model. The square duct case primarily requires accurate modeling of turbulence-induced secondary flows and mainly activates the quadratic term $g^{(2)}$, which governs anisotropy-driven secondary motions. In contrast, the S809 airfoil case focuses on capturing separation under strong adverse pressure gradients and predominantly constrains the linear term $g^{(1)}$, which controls separation behavior. Because each case emphasizes a different model component, their optimization directions remain aligned rather than conflicting.

\begin{figure}[!htb]
  \centering

  \begin{subfigure}[t]{0.5\textwidth}
    \centering
    \includegraphics[width=0.95\textwidth]{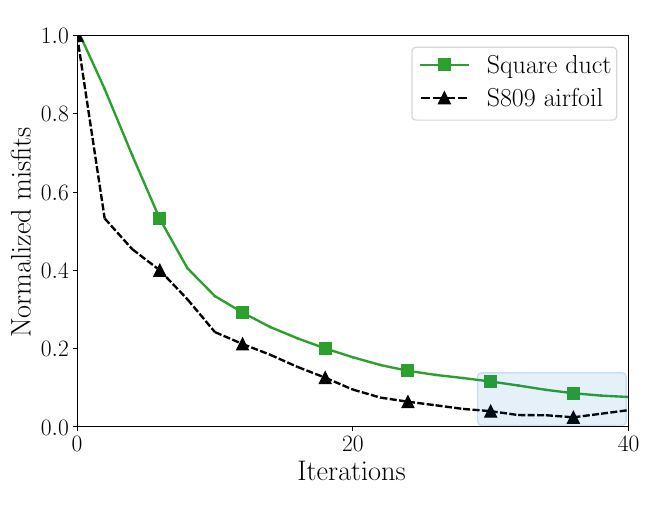}
    \caption{Training evolution of normalized misfit}
    \label{fig:nonconf-evolution}
  \end{subfigure}
  \begin{subfigure}[t]{0.44\textwidth}
    \centering
    \includegraphics[width=0.95\textwidth]{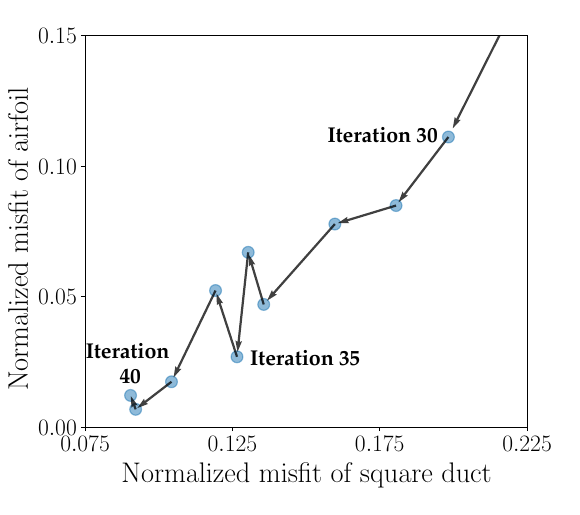}
    \caption{Trade-off between normalized misfit}
    \label{fig:nonconf-tradeoff}
  \end{subfigure}
  \caption{Training behavior and misfit evolution for the square duct and S809 airfoil cases.
    (a) Evolution of normalized misfits during training for the square duct case and the S809 airfoil case using the Frank--Wolfe algorithm.
    (b) Tradeoff of the normalized misfit values between two non-conflict cases obtained through Frank--Wolfe--based training, with selected iterations highlighted.}
  
  \label{fig:non-conflict-results}
\end{figure}

The optimization results confirm the absence of conflict between the objectives. As shown in Fig.~\ref{fig:non-conflict-results}a, the algorithm simultaneously reduces the misfits for both cases during training, indicating stable and consistent convergence. No pronounced Pareto front is observed in Fig.~\ref{fig:non-conflict-results}b, indicating that no meaningful trade-off exists between the objectives. This behavior confirms that, in the non-conflict scenario, reweighting the objectives has negligible influence on the optimal solution.

\subsubsection{Scalability evaluation for multi-objective learning}
\begin{figure}[!htb]
  \centering

  \begin{subfigure}[t]{0.45\textwidth}
    \centering
    \includegraphics[width=0.8\textwidth]{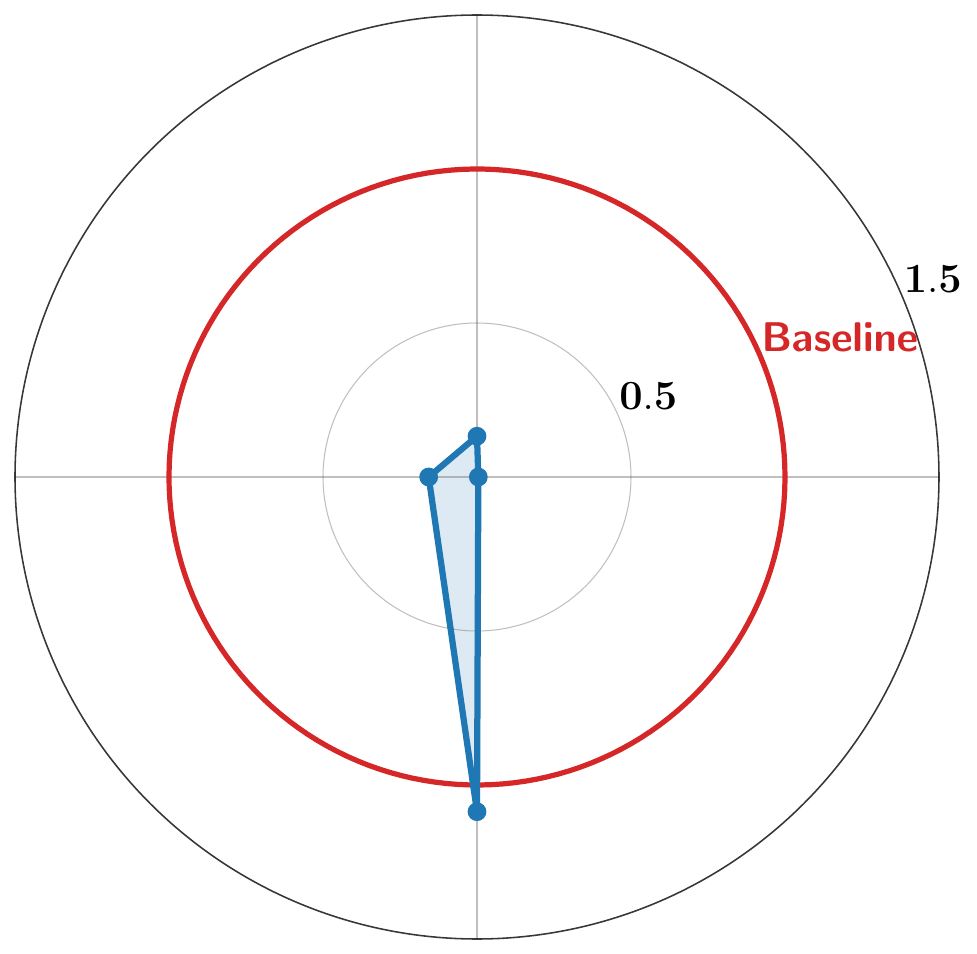}
    \caption{4 objectives}
    \label{fig:scale-a}
  \end{subfigure}
  \hspace{0.02\textwidth}
  \begin{subfigure}[t]{0.45\textwidth}
    \centering
    \includegraphics[width=0.8\textwidth]{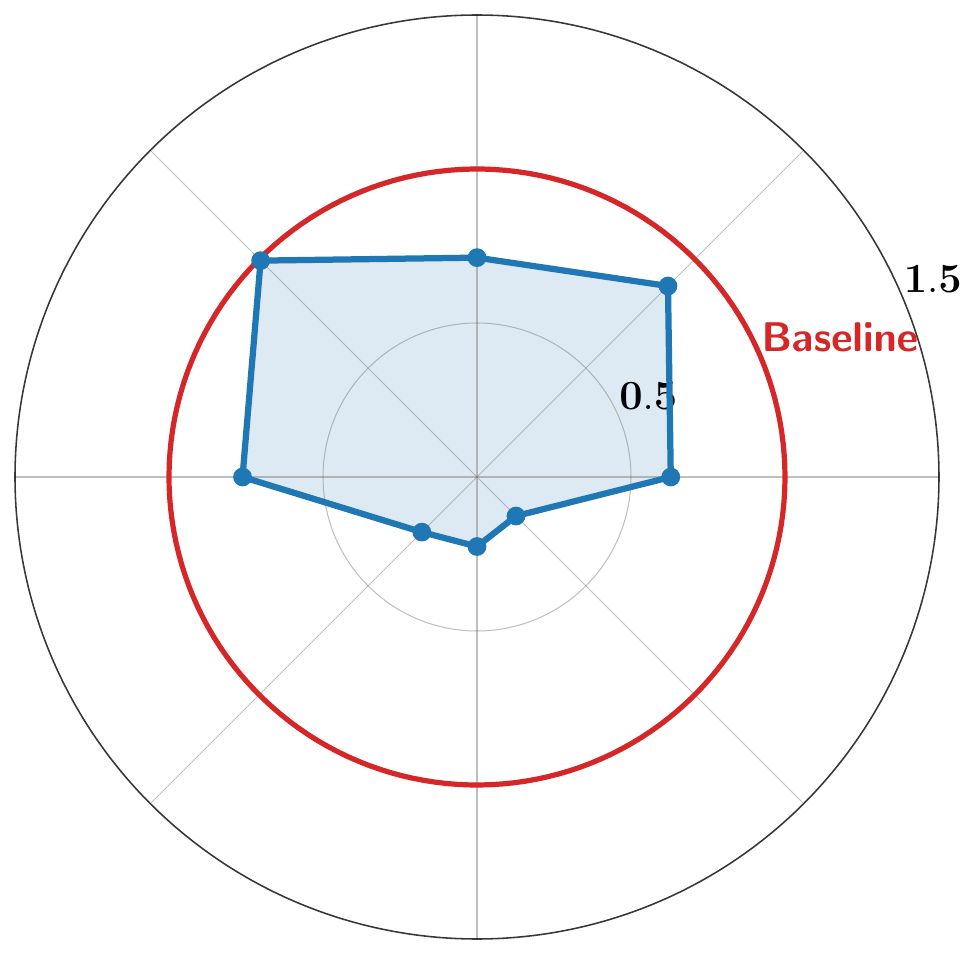}
    \caption{8 objectives}
    \label{fig:scale-b}
  \end{subfigure}

  \begin{subfigure}[t]{0.45\textwidth}
    \centering
    \includegraphics[width=0.8\textwidth]{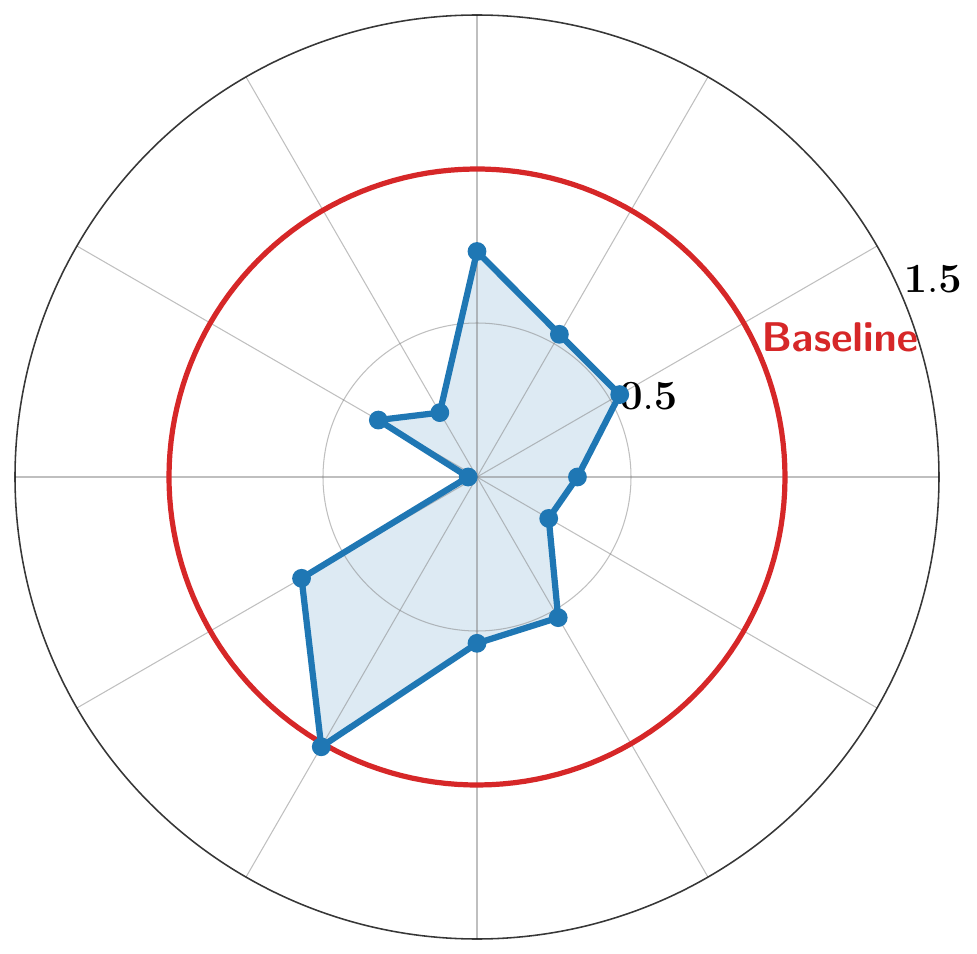}
    \caption{12 objectives}
    \label{fig:scale-c}
  \end{subfigure}
  \hspace{0.02\textwidth}
  \begin{subfigure}[t]{0.485\textwidth}
    \centering
    \includegraphics[width=0.8\textwidth]{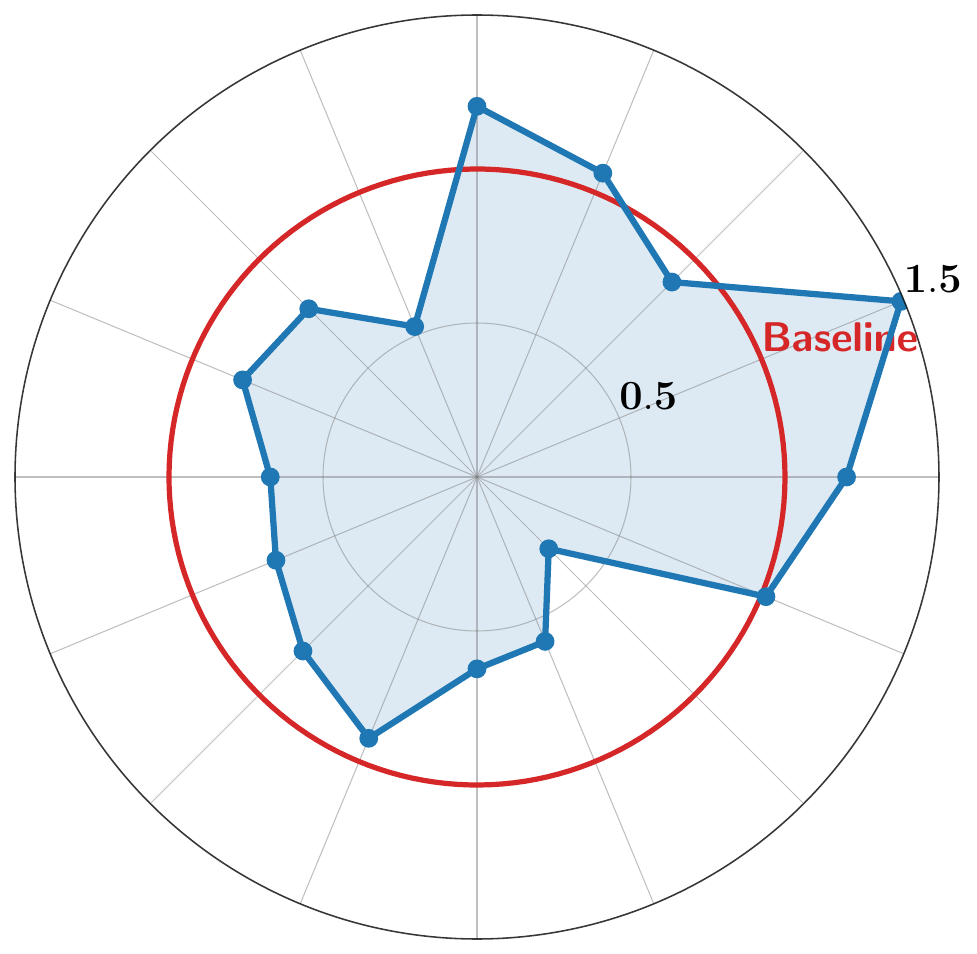}
    \caption{16 objectives}
    \label{fig:scale-d}
  \end{subfigure}

  \caption{Scalability of the multi-objective ensemble learning framework. The analysis compares models trained with 4, 8, 12, and 16 objectives selected through the distribution-based case selection strategy. Each objective~(\textcolor{blue}{\textbullet}) is evaluated by its normalized misfit, with the filled blue polygon representing the collective optimized solution. The baseline \(k\)--\(\omega\) model performance is denoted by the red circle. The results show clear Pareto improvement for 4, 8, and 12 objectives, where all optimized outcomes lie well within the baseline boundary. When scaling to 16 objectives, improvements become less uniform, with some objectives approaching or slightly exceeding the baseline.}
  
  \label{fig:validation-scale}
\end{figure}

We assess the scalability of the proposed multi-objective ensemble learning framework by increasing the number of training objectives selected through the distribution-based case selection strategy. 
Models are trained with 4, 8, 12, and 16 objectives, with the corresponding flows summarized in Table~\ref{tab:dataset_cases} and the resulting normalized misfits shown in Fig.~\ref{fig:validation-scale}.
For 4, 8, and 12 objectives, the optimized solutions consistently outperform the baseline \(k\)--\(\omega\) model, indicating clear Pareto improvement across all objectives. Notably, some objectives achieve performance levels close to those obtained from singly trained models. When the number of objectives increases to 16, optimization becomes more challenging. Several objectives approach or slightly underperform the baseline, and most objectives exhibit more limited improvement, with only one objective reaching approximately half of its singly trained performance. These results indicate stronger trade-offs among competing objectives as the optimization problem becomes more complex, suggesting that additional training iterations or a more expressive model representation may be required to further enhance performance. Nevertheless, the framework maintains stable behavior overall, highlighting its robustness and scalability under highly multi-objective settings.

\begin{table}
  \centering
  \small
  \renewcommand{\arraystretch}{1.2}
  \setlength{\tabcolsep}{8pt}
  \begin{tabular}{lcccc}
    \specialrule{1pt}{0pt}{0pt}
    Flows & 4 objectives & 8 objectives & 12 objectives & 16 objectives \\
    \specialrule{1pt}{0pt}{0pt}

    NACA 0012 airfoil (AOA = 18$^\circ$, drag \& lift) 
    &  &  &  & \checkmark \\

    S809 airfoil (AOA = 14$^\circ$,  drag \& lift) 
    &  & \checkmark & \checkmark & \checkmark \\

    Square duct (Re = 1100) 
    &  & \checkmark & \checkmark & \checkmark \\
    
    Square duct (Re = 3500) 
    & \checkmark & \checkmark & \checkmark & \checkmark \\

    Rectangular duct (AR = 5) 
    &  & \checkmark & \checkmark & \checkmark \\

    Rectangular duct (AR = 10) 
    &  &  &  & \checkmark \\

    Channel (Re = 3300) 
    &  &  &  & \checkmark \\

    Round jet 
    & \checkmark & \checkmark & \checkmark & \checkmark \\

    Hump 
    & \checkmark & \checkmark & \checkmark & \checkmark \\

    Curved step 
    &  &  & \checkmark & \checkmark \\

    Bump (height = 31) 
    &  &  & \checkmark & \checkmark \\

    Periodic hills ($\alpha = 0.8$) 
    &  &  & \checkmark & \checkmark \\

    Periodic hills ($\alpha = 1$) 
    & \checkmark & \checkmark & \checkmark & \checkmark \\

    Periodic hills ($\alpha = 1.5$) 
    &  &  & \checkmark & \checkmark \\

    \specialrule{1pt}{0pt}{0pt}
  \end{tabular}
  \caption{\added[id=R1]{Canonical flows used as training objectives for different objective set sizes (4, 8, 12, and 16). A checkmark indicates the inclusion of each flow. Detailed flow definitions are provided in Table~\ref{tab:cases-setup}.}}
  \label{tab:dataset_cases}
\end{table}

\subsubsection{\added[id=R1]{Sensitivity analysis on model size}}

\begin{figure}[!htb]
  \centering

  \begin{subfigure}[t]{0.45\textwidth}
    \centering
    \includegraphics[width=0.8\textwidth]{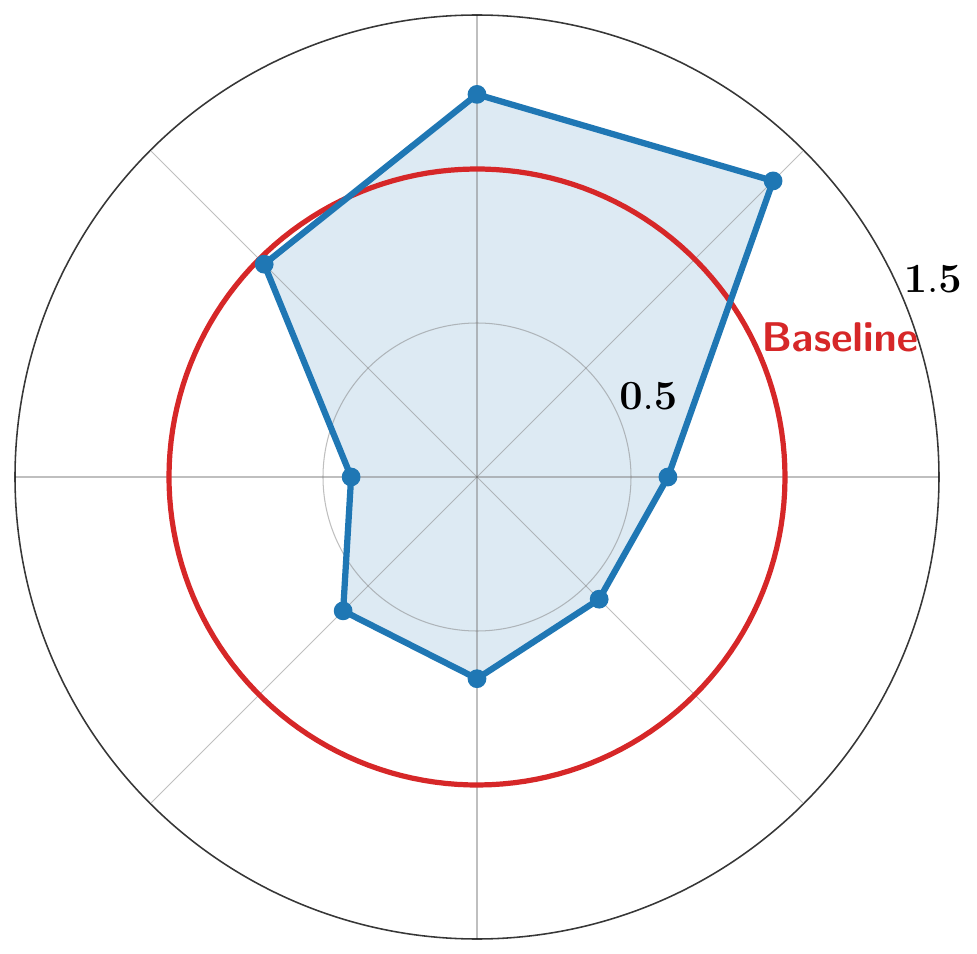}
    \caption{Compact model}
    \label{fig:NNsize-a}
  \end{subfigure}
  \hspace{0.02\textwidth}
  \begin{subfigure}[t]{0.45\textwidth}
    \centering
    \includegraphics[width=0.8\textwidth]{figs/radar_8_objs.pdf}
    \caption{Reference model}
    \label{fig:NNsize-b}
  \end{subfigure}

  \begin{subfigure}[t]{0.45\textwidth}
    \centering
    \includegraphics[width=0.8\textwidth]{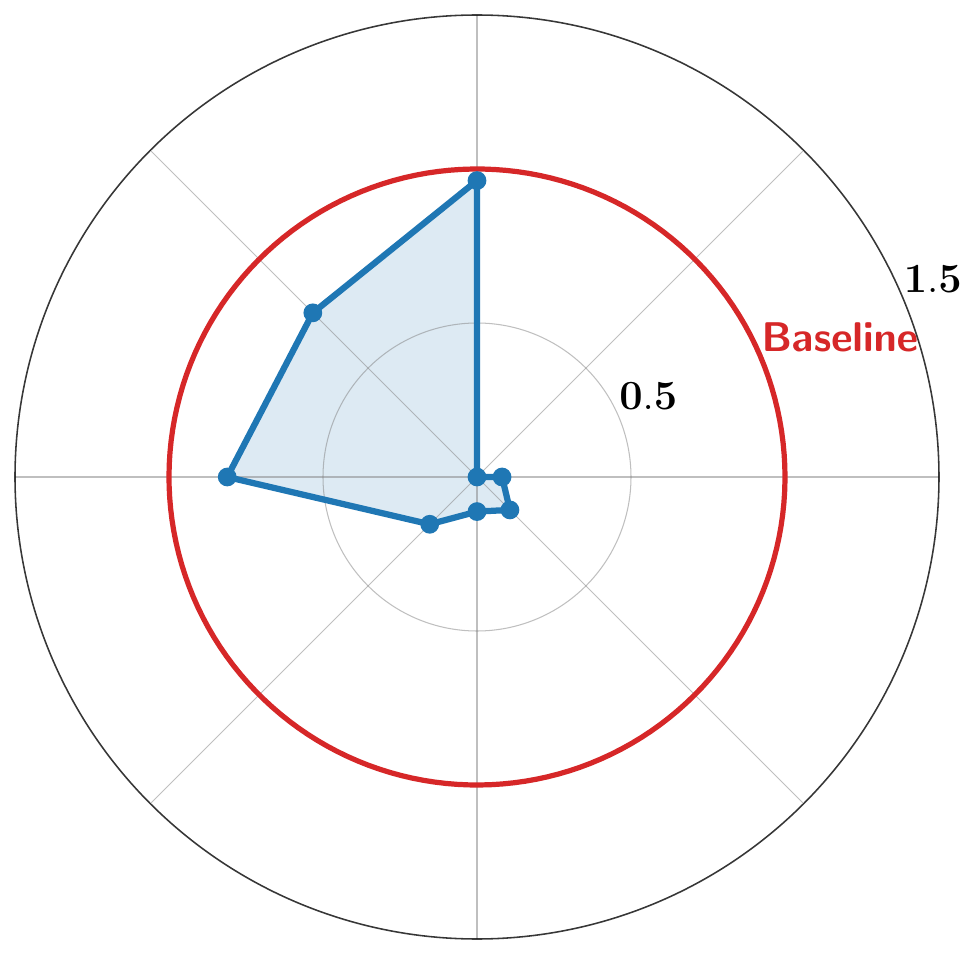}
    \caption{Expanded model}
    \label{fig:NNsize-c}
  \end{subfigure}
  \hspace{0.02\textwidth}
  \begin{subfigure}[t]{0.45\textwidth}
    \centering
    \includegraphics[width=0.8\textwidth]{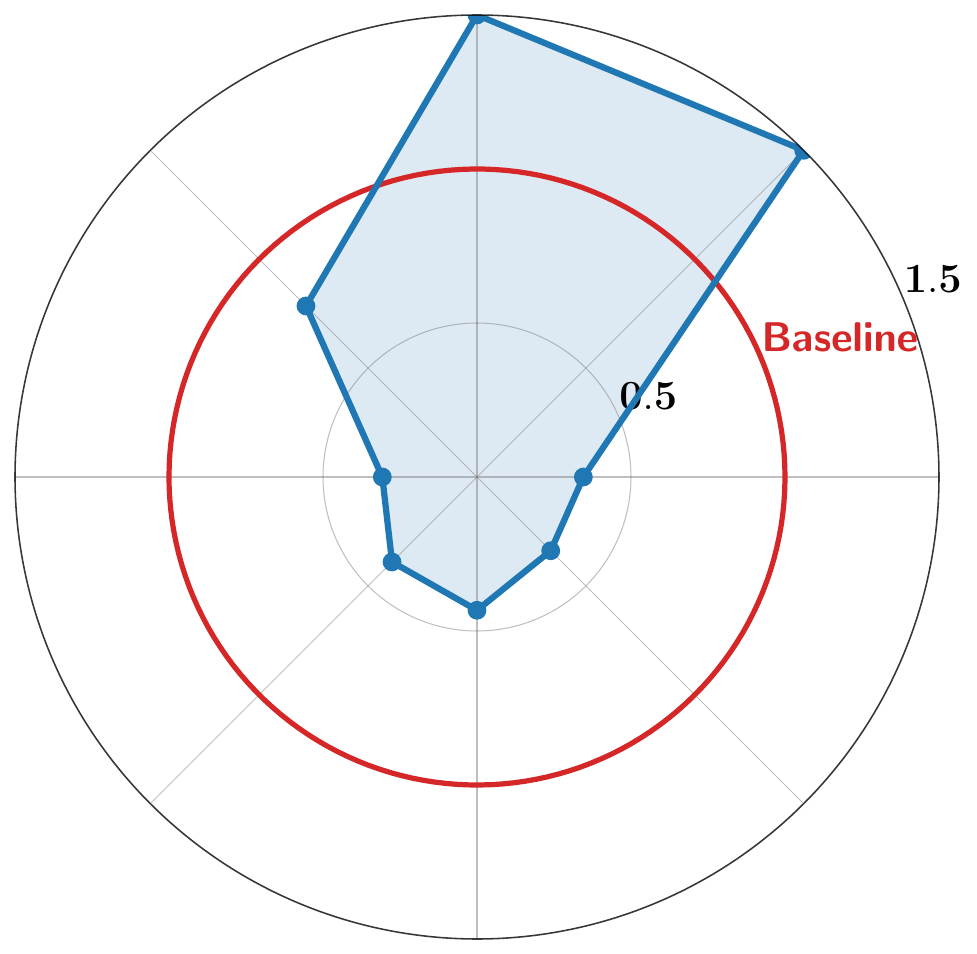}
    \caption{Over-parameterized model}
    \label{fig:NNsize-d}
  \end{subfigure}

  \caption{\added[id=R1]{Sensitivity of the turbulence model to neural network size, shown as radar plots for configurations with different numbers of trainable parameters relative to the reference model. The reference model (b) contains 333 parameters. The other configurations are a compact model (a) with 195 parameters, a moderately expanded model (c) with 651 parameters, and an over-parameterized model (d) with 1275 parameters. Model size is varied by adjusting the number of neurons per layer (width) in each sub-network while keeping the overall architecture unchanged. The radial coordinate represents the normalized misfit $\hat{e}$ for each objective. The red circle denotes the baseline model, and the blue polygon represents the learned model.}}
  \label{fig:Parametric-studies}
\end{figure}

\added[id=R1]{
We assess the sensitivity of the turbulence model to neural network size by varying the number of neurons per layer (width) in each sub-network, while keeping the overall architecture and depth fixed. This analysis focuses on the scenario with 8 objectives (see Table~\ref{tab:dataset_cases}).
}
\added[id=R1]{The study shows that the reference configuration provides the best balance between accuracy and robustness within a many-objective, ensemble-based training setting. Eight training objectives define a many-objective learning scenario, with an ensemble size of 20 samples chosen for computational efficiency. This setup establishes a practical regime for ensemble-based training, where the number of trainable parameters must remain compatible with the ensemble size. Within this regime, four model sizes are examined, ranging from a compact configuration to a heavily over-parameterized one (see Table~\ref{tab:NN-size}). The comparison (Fig.~\ref{fig:Parametric-studies}) shows that the reference configuration achieves a good balance between accuracy and robustness, the compact model exhibits slightly degraded performance due to limited representational capacity, and larger models do not yield consistent improvements, with performance degrading as model size increases.}

\added[id=R1]{The performance degradation observed for larger models arises from a mismatch between model size and ensemble size. In ensemble-based learning, parameter updates are confined to a low-rank subspace determined by the ensemble size~\cite{evensen2022data}. Although modern localization techniques (e.g., noise-informed covariance estimation~\cite{vishny2024high}) are employed, effective learning in high-dimensional parameter spaces remains limited by the ensemble size. As the number of parameters grows relative to the ensemble size, training becomes less efficient and less stable.}

\added[id=R1]{These results indicate that relatively small neural networks are sufficient to represent the underlying turbulence physics, consistent with the low-complexity functional forms of traditional turbulence models. For larger networks to be effective, the ensemble size must be increased accordingly.}

\begin{table}[htbp]
\centering
\begin{tabular}{lcccc}
\specialrule{1pt}{0pt}{0pt}
Configuration & $g^{(1)}$ width & $g^{(2)}$ width & $\beta$ width & Total parameters \\
\specialrule{1pt}{0pt}{0pt}
Compact model & 3 & 4 & 3 & 195 \\
Reference model & 4 & 6 & 4 & 333 \\
Expanded model & 6 & 9 & 6 & 651 \\
Over-parameterized model & 9 & 13 & 9 & 1275 \\
\specialrule{1pt}{0pt}{0pt}
\end{tabular}
\caption{\added[id=R1]{Neural network configurations used in the sensitivity analysis on model size. All models use 4 hidden layers. The model size is varied by changing the number of neurons per layer in each sub-network.}}
\label{tab:NN-size}
\end{table}

\FloatBarrier
\section{Setup of training and test cases}
\label{sec:case}

We present the training and test cases used to evaluate three turbulence models: the unified foundation model, the specialist model for separated flows, and the specialist model for secondary flows with separation. The setup of all cases together with the corresponding observation data are summarized in Table~\ref{tab:cases-setup}.

\small
\setlength{\tabcolsep}{3pt}
\renewcommand{\arraystretch}{1.2}

\begin{xltabular}{\textwidth}{
  M{20mm}     
  L{42mm}     
  Y           
  C{10mm}     
  Y           
  Y           
  C{18mm}     
}

\specialrule{1pt}{0pt}{0pt}
\textbf{Categories} &
\textbf{Flows} &
\textbf{Observations} &
\textbf{No. obs} &
\textbf{Control parameter} &
\textbf{Value} &
\textbf{Label} \\
\specialrule{1pt}{0pt}{0pt}
\endfirsthead

\multicolumn{7}{c}{\small\textit{Table \thetable\ continued from previous page}}\\
\specialrule{1pt}{0pt}{0pt}
\textbf{Categories} &
\textbf{Flows} &
\textbf{Observations} &
\textbf{No. obs} &
\textbf{Control parameter} &
\textbf{Value} &
\textbf{Label} \\
\specialrule{1pt}{0pt}{0pt}
\endhead

\midrule
\multicolumn{7}{r}{\small\textit{Continued on next page}}\\
\endfoot

\endlastfoot

\catcell{Attached\\Boundary\\Layers} &
\flowcell{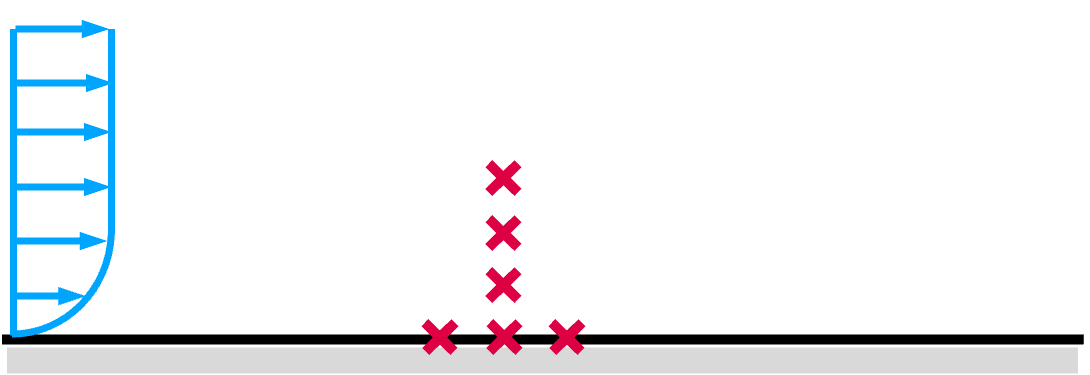}{Zero pressure gradient flat plate} &
Velocities, friction &
34 &
Reynolds number &
$5\times 10^{6}$ &
BL1 \\

& \flowcell{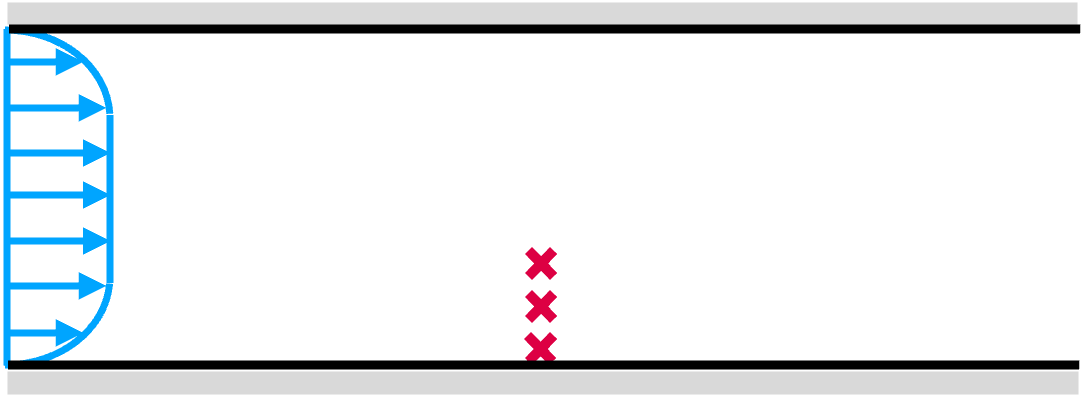}{Plane channel} &
Velocities &
4 &
Reynolds number &
\makecell[l]{3300\\7890\\$8\times 10^{7}$} &
\makecell[c]{BL2.1\\BL2.2\\BL2.3} \\

& \flowcell{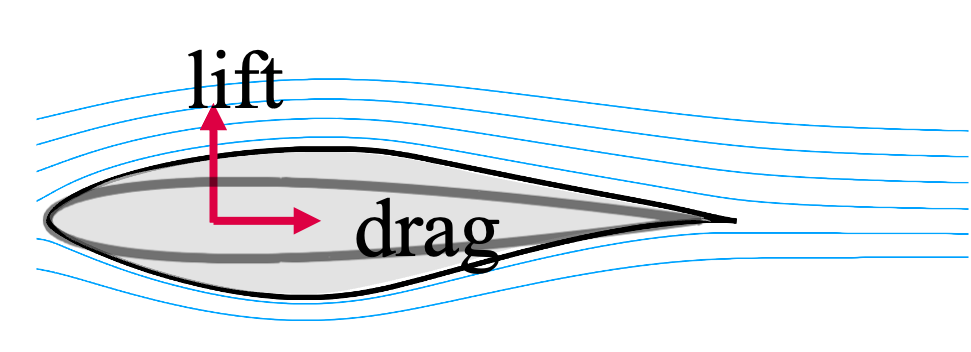}{Airfoils (low AOA)} &
Lift, drag &
2 &
Angle of attack &
\makecell[l]{\scriptsize
S809: $1^\circ$\cb $5^\circ$\\ \scriptsize NACA 0012: \scriptsize $10^\circ$\cb $15^\circ$} &
\makecell[c]{BL3.1\cb 3.2\\ BL3.3\cb 3.4 \\ } \\

\midrule

\catcell{Separated\\Flows} &
\flowcell{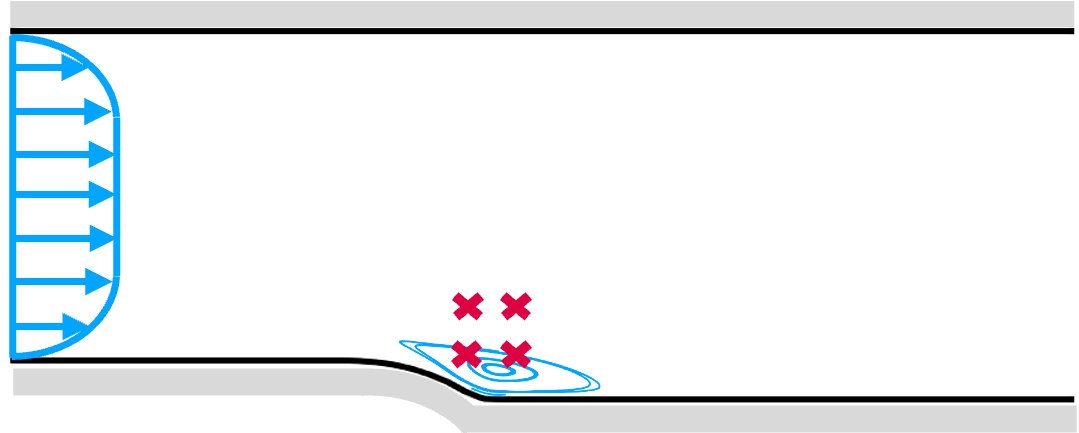}{Curved step} &
Velocities &
27 &
Reynolds number &
13700 &
SP1 \\

& \flowcell{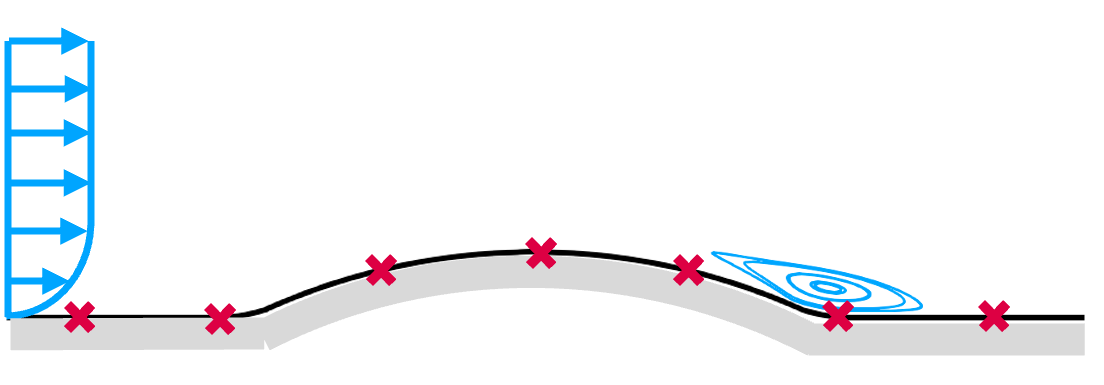}{Bump} &
Friction &
42 &
Height &
\makecell[l]{20 mm\\26 mm\\31 mm\\38 mm\\42 mm} &
\makecell[c]{SP2.1\\SP2.2\\SP2.3\\SP2.4\\SP2.5} \\

& \flowcell{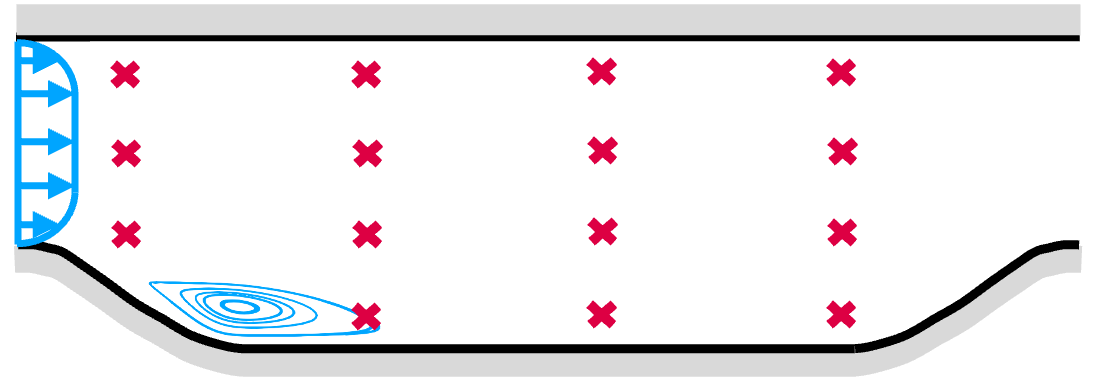}{Periodic hill} &
Velocities &
66 &
Slope steepness &
\makecell[l]{0.8\\1.0\\1.2\\1.5} &
\makecell[c]{SP3.1\\SP3.2\\SP3.3\\SP3.4} \\

& \flowcell{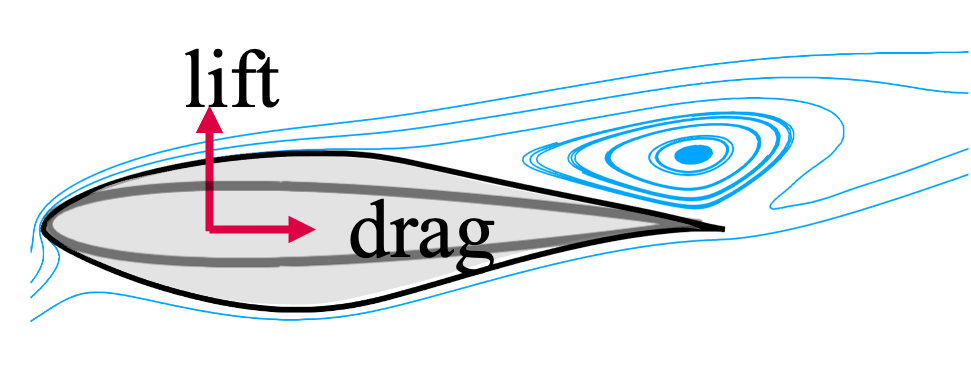}{Airfoils (high AOA)} &
Lift, drag &
2 &
Angle of attack &
\makecell[l]{\footnotesize S809: $14^\circ$\cb $16^\circ$\cb $18^\circ$\\ \footnotesize NACA 0012: $18^\circ$} &
\makecell[c]{SP4.1--4.3\\ SP4.4} \\

& \flowcell{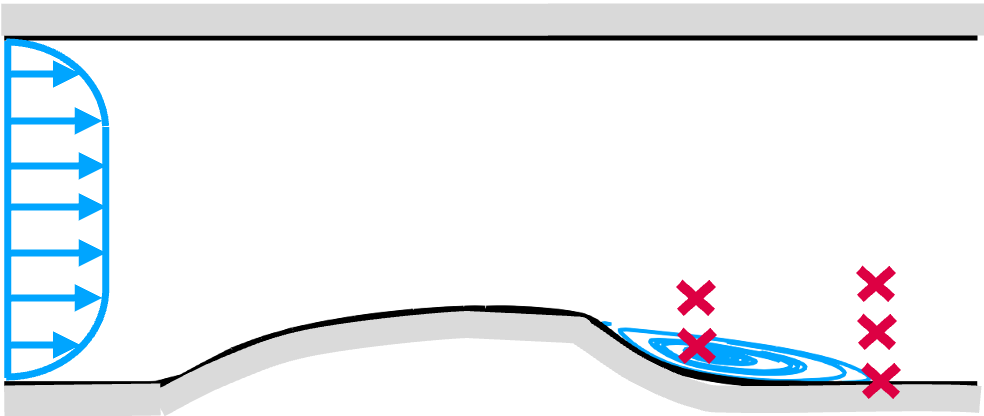}{Hump} &
Velocities &
200 &
Reynolds number &
936000 &
SP5 \\

\midrule

\catcell{Secondary\\Flows\\(Ducts)} &
\flowcellTall{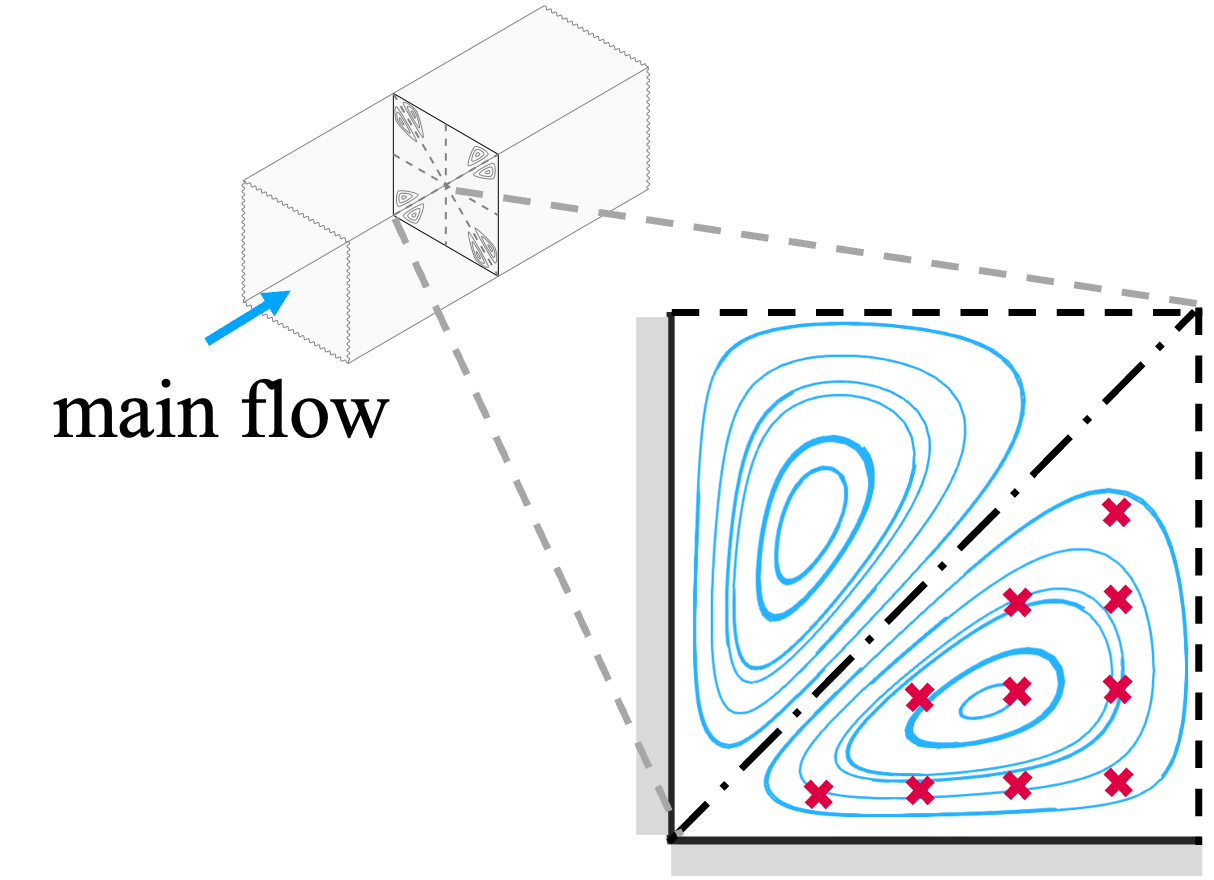}{Square duct} &
Velocities &
193 &
Reynolds number &
\makecell[l]{1100\\1800\\2600\\3500} &
\makecell[c]{DU1.1\\DU1.2\\DU1.3\\DU1.4} \\

& \flowcell{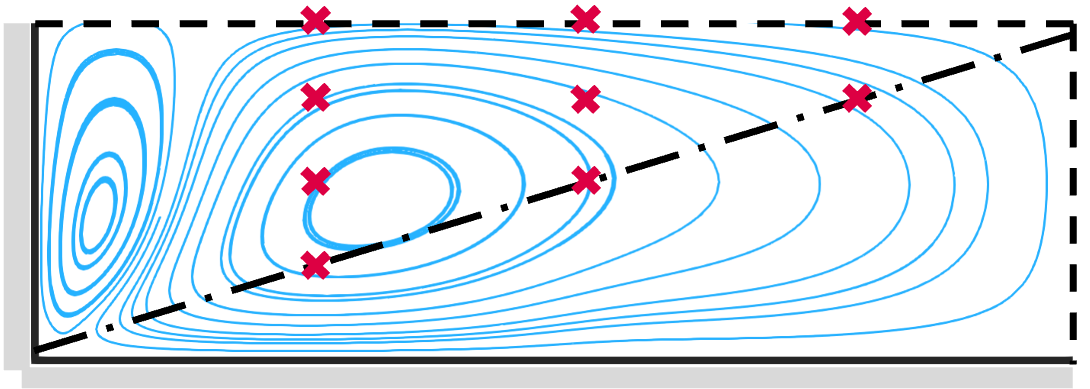}{Rectangular duct} &
Velocities &
35 &
Aspect ratio &
\makecell[l]{3\\5\\7\\10} &
\makecell[c]{DU2.1\\DU2.2\\DU2.3\\DU2.4} \\

\midrule

\catcell{Free-shear\\Flow} &
\flowcell{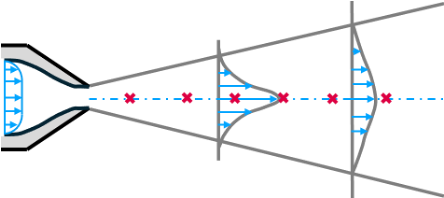}{Round jet} &
Velocities &
25 &
Reynolds number &
5601 &
FS1 \\

\midrule

\catcell{Complex\\3D Flows} &
\flowcell{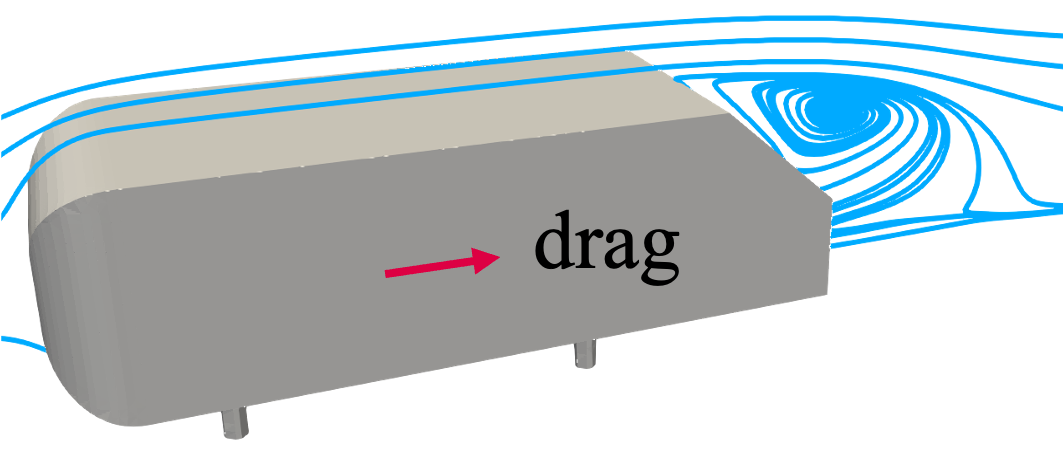}{Generic car (Ahmed body)} &
Drag &
- &
Slant angle &
$40^\circ$ &
CF1 \\

& \flowcell{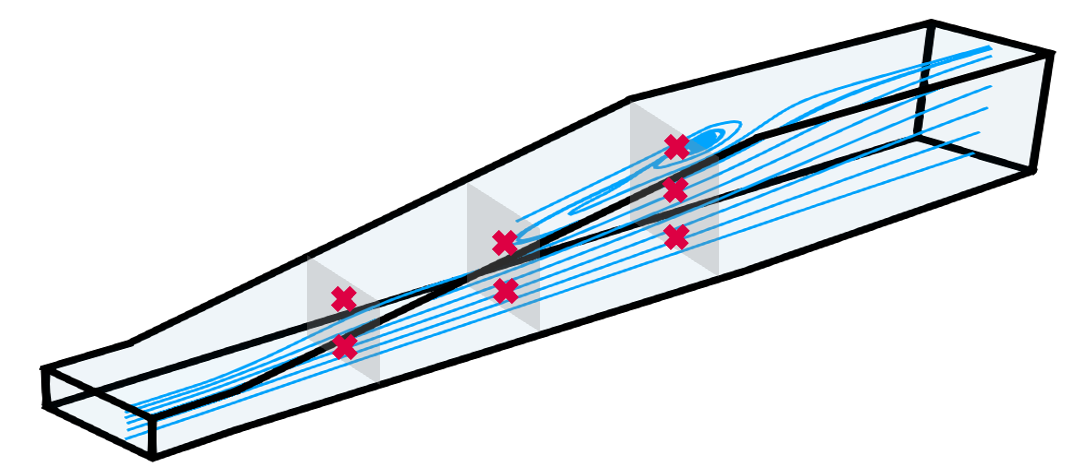}{3D diffuser} &
Velocities, friction &
- &
Reynolds number &
10000 &
CF2 \\

& \flowcellMedium{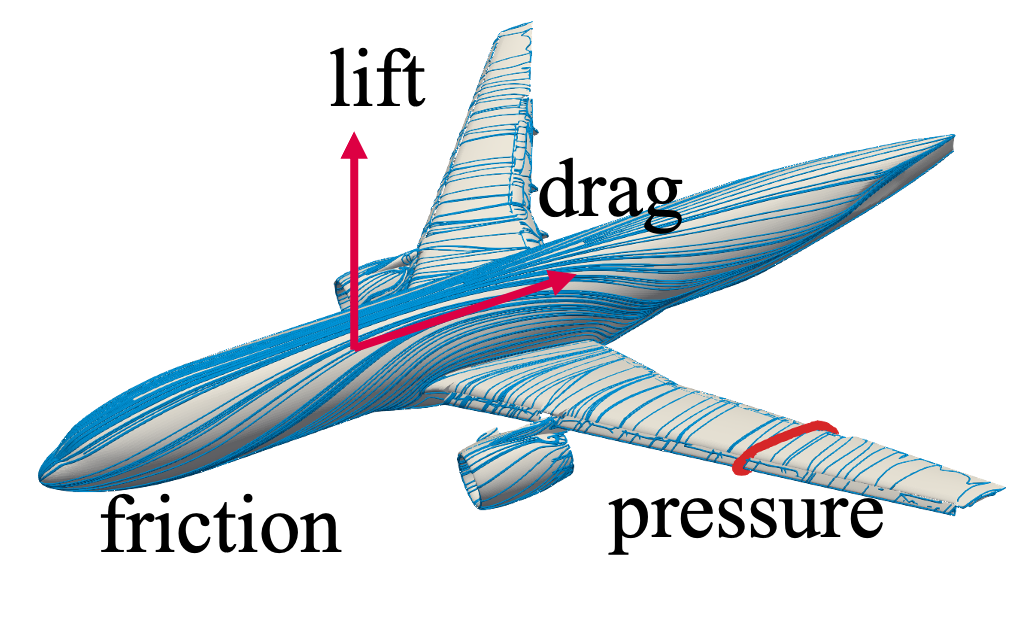}{Generic aircraft (CRM-HL)} &
Lift, drag, friction, pressure &
- &
Angle of attack &
\makecell[l]{$7.05^\circ$\\$21.47^\circ$} &
\makecell[c]{CF3\\CF4} \\

\midrule

\multicolumn{6}{r}{\textbf{Totals}} &
\textbf{36} \\
\bottomrule

\caption{Overview of flow cases and associated observation data used in the present study. The table summarizes the flow categories, configurations, observed quantities (\textbf{\textcolor{red}{\texttimes}} denotes sparse observations), number of observations, control parameters, parameter values, and case labels. Attached boundary-layer flows include a flat plate~\cite{jespersen2016overflow}, plane channels~\cite{kim1987turbulence}, and S809~\cite{somers1997design} and NACA 0012~\cite{ladson1988effects} airfoils at low angles of attack. Separated flows include a curved step~\cite{bentaleb2012large}, bumps~\cite{matai2019large}, a hump~\cite{seifert2002active}, S809 and NACA 0012 airfoils at high angles of attack, and periodic hills~\cite{xiao2020flows}. Secondary flows include square ducts~\cite{pinelli2010reynolds} and rectangular ducts~\cite{vinuesa2018secondary}. Free-shear flow includes a round jet~\cite{bridges2010establishing}, and complex 3D flows include a generic car (Ahmed body)~\cite{lienhart2003flow}, a 3D diffuser~\cite{cherry2008geometric} and \added[id=Author]{a generic aircraft (NASA High-Lift Common Research Model, referred to as CRM-HL)~\cite{evans2020test}.}}
\label{tab:cases-setup}
\end{xltabular}

\subsection{Training cases for unified foundation model}
\label{subsec:foundation}
\begin{figure}[!htb]
  \centering

  \begin{subfigure}[t]{1\textwidth}
    \centering
    \includegraphics[width=0.85\textwidth]{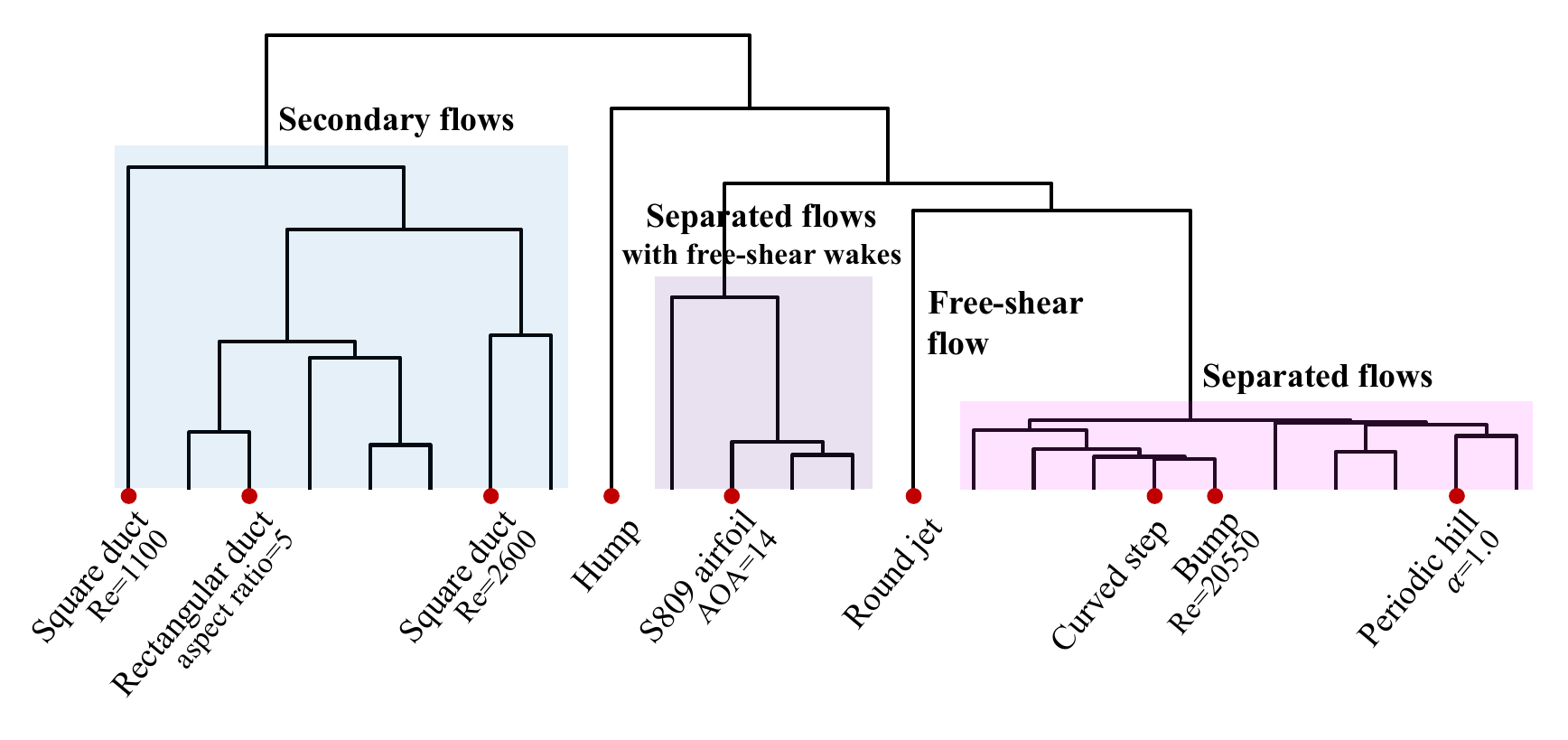}
    \caption{Hierarchical clustering}
    \label{fig:case-selection-dendrogram}
  \end{subfigure}

  \vspace{0.1em}

  \begin{subfigure}[t]{1\textwidth}
    \centering
    \includegraphics[width=0.85\textwidth]{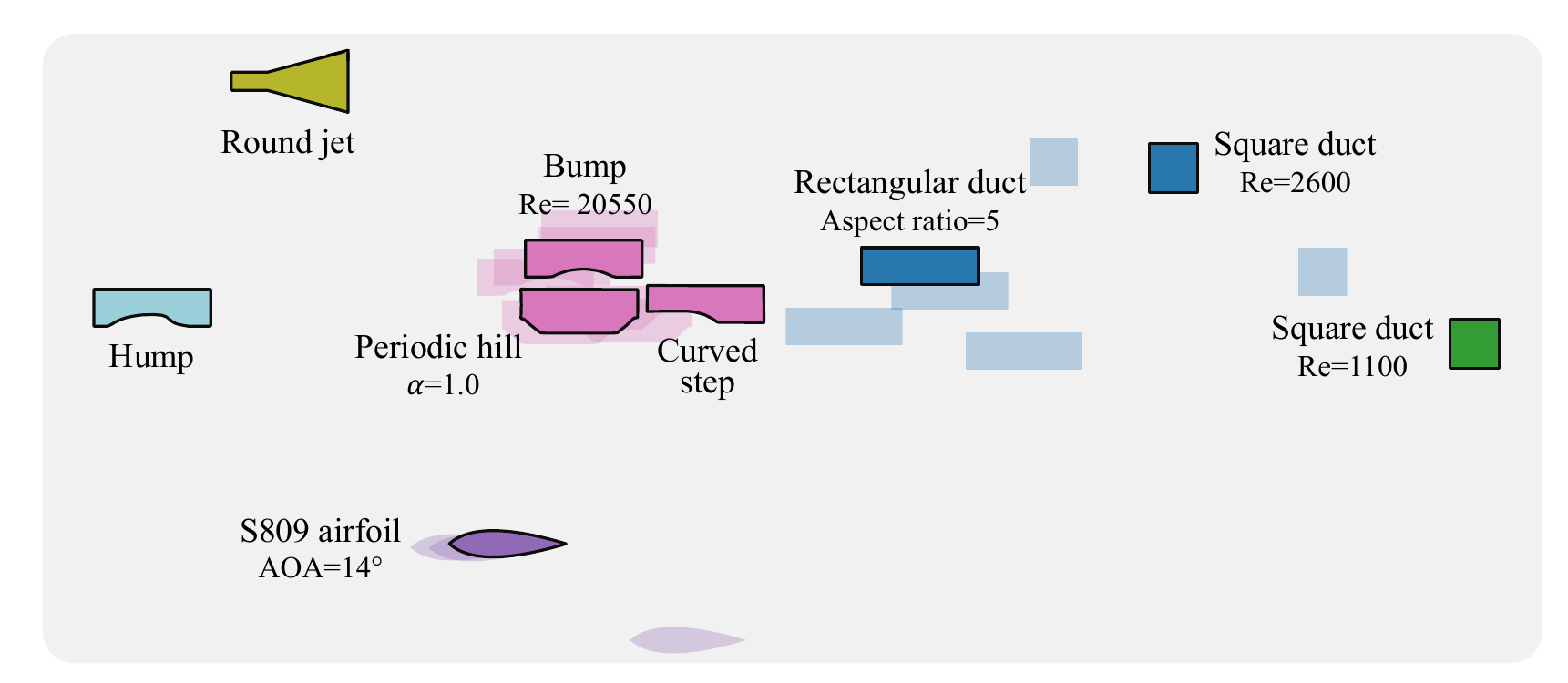}
    \caption{2D Distance-Preserving Embedding Space}
    \label{fig:case-selection-mds}
  \end{subfigure}

  \caption{Distribution-based selection of training cases for the \textbf{unified foundation model}. Flow cases (excluding attached boundary layers reserved for verification) are compared using flow-feature distribution distances to identify representative training cases that span distinct flow mechanisms. (a) Hierarchical clustering dendrogram constructed from pairwise Wasserstein distances between flow-feature distributions, revealing clear groupings corresponding to secondary flows, separated flows, and free-shear flows. (b) Two-dimensional distance-preserving embedding obtained via multi-dimensional scaling, which visualizes the clustering structure and highlights relative similarities among flow cases. Representative cases are selected to cover key mechanisms and are highlighted with solid lines and opaque colors, including rectangular and square ducts for secondary flows, the S809 airfoil, periodic hill, bump, hump, and curved step for separated flows, and the round jet for free-shear flow. The bump case is labeled by its Reynolds number (Re$=20550$), corresponding to a bump height of 31~mm. In total, nine flow cases are selected; for the S809 airfoil, lift and drag are treated as separate objectives, resulting in ten training objectives.}
  \label{fig:foundation-case-selection}

\end{figure}

\begin{figure}[!htb]
  \centering

  \begin{subfigure}[t]{0.2\textwidth}
    \centering
    \includegraphics[width=0.95\textwidth]{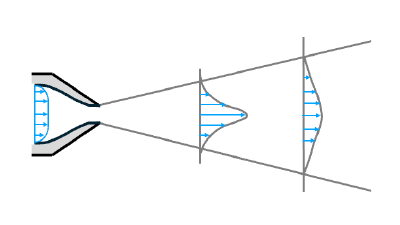}
    \caption{Round jet}
    \label{fig:foundation-setup-jet}
  \end{subfigure}%
  \hfill
  \begin{subfigure}[t]{0.25\textwidth}
    \centering
    \includegraphics[width=0.95\textwidth]{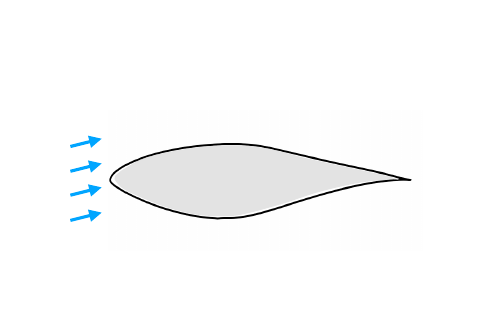}
    \caption{S809 airfoil (AOA $14^\circ$)}
    \label{fig:foundation-setup-airfoil}
  \end{subfigure}%
  \hfill
  \begin{subfigure}[t]{0.26\textwidth}
    \centering
    \includegraphics[width=0.95\textwidth]{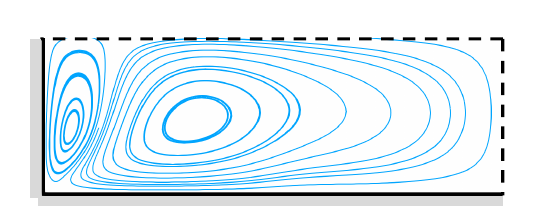}
    \caption{Rectangular duct \\ (Aspect ratio 3)}
    \label{fig:foundation-setup-rectduct}
  \end{subfigure}%
  \hfill
  \begin{subfigure}[t]{0.24\textwidth}
    \centering
    \includegraphics[width=0.95\textwidth]{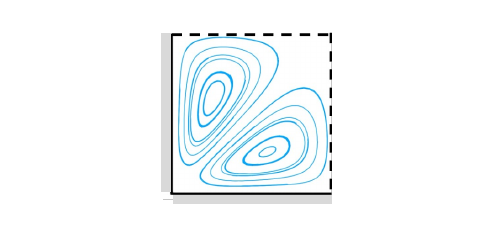}
    \caption{Square duct \\ (Re 1100 and 2600)}
    \label{fig:foundation-setup-squareduct}
  \end{subfigure}

  \vspace{0.8em}

  \begin{subfigure}[t]{0.2\textwidth}
    \centering
    \includegraphics[width=0.95\textwidth]{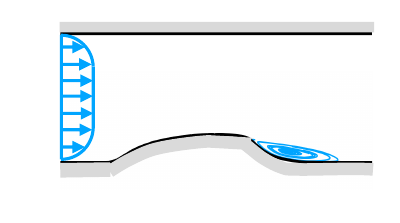}
    \caption{Hump}
    \label{fig:foundation-setup-hump}
  \end{subfigure}%
  \hfill
  \begin{subfigure}[t]{0.25\textwidth}
    \centering
    \includegraphics[width=0.95\textwidth]{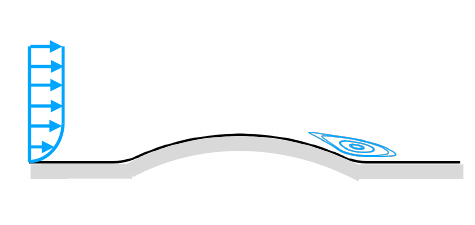}
    \caption{Bump (Height 31 mm)}
    \label{fig:foundation-setup-bump}
  \end{subfigure}%
  \hfill
  \begin{subfigure}[t]{0.26\textwidth}
    \centering
    \includegraphics[width=0.95\textwidth]{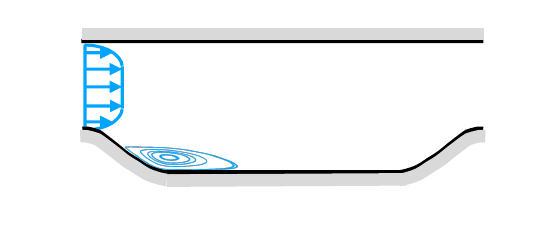}
    \caption{Periodic hill \\ (Slope scaling factor 1.0)}
    \label{fig:foundation-setup-hill}
  \end{subfigure}%
  \hfill
  \begin{subfigure}[t]{0.24\textwidth}
    \centering
    \includegraphics[width=0.95\textwidth]{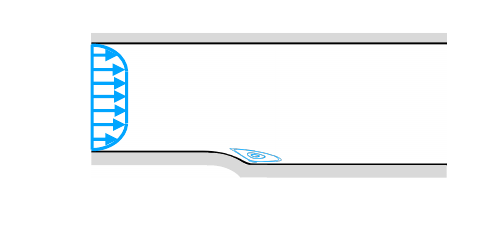}
    \caption{Curved step}
    \label{fig:foundation-setup-step}
  \end{subfigure}

  \caption{Training configurations used for the \textbf{unified foundation model}, selected to span a diverse set of flow mechanisms. The training set comprises nine representative cases, including free-shear flow (round jet), external separated flows (S809 airfoil at $\mathrm{AOA}=14^\circ$, periodic hill with slope scaling factor $\alpha_{\text{p}}=1.0$ and bump with height 31~mm), secondary flows in ducts (rectangular duct with $\mathrm{AR}=3$ and square ducts at $\mathrm{Re}=1100$ and 2600), and internal separated flows (hump and curved step).}
  \label{fig:foundation-setup}
\end{figure}

Training cases for the unified foundation model are selected using a distribution-based strategy that balances flow-mechanism diversity, representativeness, and training cost. To this end, flow cases are embedded into a two-dimensional distance-preserving space based on flow-feature distribution distances and grouped into six clusters, while attached boundary layers are excluded from training and reserved for verification. As shown in Fig.~\ref{fig:foundation-case-selection}b, cases belonging to the same cluster share similar flow characteristics, and the highlighted markers indicate the selected training cases.
Representative cases are then selected to cover distinct physical behaviors (see Fig.~\ref{fig:foundation-setup}). For secondary flows, three cases are selected from two clusters to represent both rectangular and square duct geometries at different Reynolds numbers. 
For separated flows, three cases are manually chosen from one cluster to capture distinct pressure-gradient effects, including internal and external separation under adverse pressure gradients (periodic hill and curved step) and separation with relaminarization under a favorable pressure gradient (bump). Additional separated flow cases, including the hump and a high-angle-of-attack S809 airfoil, are selected from two other clusters. A free-shear flow is represented by a round jet from the remaining cluster. In total, nine representative flow cases are selected for training. For the S809 airfoil, lift and drag are treated as separate, conflicting objectives, yielding ten training objectives in the multi-objective learning framework. 
The unified foundation model trained on these cases is evaluated across flows within the training categories, complex three-dimensional flows, and attached boundary layers as an additional fundamental flow category (Fig.~\ref{fig:case-setup-foundation-paper}).

\paragraph{External separated flows} Flow over the \textbf{S809 airfoil} at high angles of attack ($\mathrm{AOA} > 12^\circ$) is dominated by massive flow separation, making it challenging for traditional turbulence models, which often overestimate the lift force beyond stall~\cite{singh2016using}. At $\mathrm{Re} = 2\times10^{6}$ based on chord length and inflow velocity, the suction side experiences a strong adverse pressure gradient, causing early separation. The case at $\mathrm{AOA} = 14^\circ$ is selected for training, using experimentally measured lift and drag coefficients as observations. Flow over a \textbf{bump} is a widely used two-dimensional test case for turbulence models because the curved surface first accelerates the flow (favourable pressure gradient) and then decelerates it (adverse pressure gradient), strongly disturbing the boundary layer and, for sufficiently large bumps, bringing it close to separation. The configuration consists of a wall-mounted circular arc smoothly blended into a flat plate by convex fillets at both ends. At the selected operating condition, the inflow Reynolds number based on the momentum thickness and freestream velocity is $\mathrm{Re}_{\theta_{\infty}}=2500$. The imposed curvature generates a strong adverse pressure gradient downstream of the crest, which causes primary separation near the trailing fillet and forms a short recirculation bubble. The inlet velocity and turbulence profiles are obtained from a precursor large-eddy simulation of a zero-pressure-gradient turbulent boundary layer at the same $\mathrm{Re}_{\theta_{\infty}}$. Zero-gradient conditions are applied at the top and outlet boundaries, and the lower boundary is treated as a no-slip wall. The training observations consist of the wall skin-friction coefficient $c_f$ sampled at 42 streamwise locations from the large-eddy simulation results.

\paragraph{Internal separated flows} Flow over \textbf{periodic hill} is a benchmark for internal separated flows, characterized by turbulence in the presence of periodic smooth hills that induce flow separation and reattachment. 
The setup operates at a Reynolds number $\mathrm{Re}_{h_{\text{p}}} = 10595$, based on the bulk velocity and hill height $h_{\text{p}}$, with a slope scaling factor $\alpha_{\text{p}} = 1.0$. Smaller values of $\alpha_{\text{p}}$ correspond to steeper hills, while larger values indicate gentler slopes.
The domain applies streamwise-periodic boundary conditions, no-slip walls, and a specified bulk velocity. Velocities from DNS data are extracted at four cross-sections ($x/h_{\text{p}} = 1.0, 3.0, 5.0, 7.0$), yielding 66 velocity components as observations. The \textbf{curved step} case evaluates turbulence model performance for separation caused by abrupt streamline curvature acting on a fully developed turbulent boundary layer. The flow consists of an upstream convex ramp that terminates in a sudden downward step, which forces detachment without prior boundary-layer acceleration. Unlike periodic-hill and bump flows, where separation follows an accelerating boundary layer, this configuration isolates curvature-induced separation physics. The setup resolves both the attached upstream boundary layer and the separated region downstream of the step. The flow operates at $\mathrm{Re}=13700$, and the training data comprise 27 velocity measurements sampled along several wall-normal lines. The \textbf{hump} case serves as a validation benchmark to assess a turbulence model’s ability to reproduce adverse--pressure-gradient--induced separation, reattachment, and boundary-layer recovery in a nominally two-dimensional flow. The flow is fully turbulent upstream, with a freestream velocity of approximately $34.6~\mathrm{m/s}$ and a Reynolds number of $\mathrm{Re}=9.36\times10^{5}$ based on the hump chord $c=420~\mathrm{mm}$. The observation data consist of velocity measurements at three distinct streamwise locations along the hump surface. This configuration has proven challenging for RANS models, which typically underpredict turbulent shear stress in the separated shear layer and consequently overpredict the separation length and delay reattachment.

\paragraph{Secondary flows} Two \textbf{square duct} flows and one \textbf{rectangular duct} flow are included as training cases to expose the model to turbulence-induced anisotropy across different Reynolds numbers and geometric variations. Square duct flow produces secondary motions driven by anisotropy in a non-circular cross-section, and the simulations assume fully developed conditions with periodic streamwise boundaries and no-slip walls. The computations reduce cost by modeling one-quarter of the cross-section using symmetry. Two square-duct cases at different Reynolds numbers ($\mathrm{Re}=1100, \, 2600$) isolate Reynolds-number effects on secondary-flow intensity and anisotropy. The rectangular duct case, with an aspect ratio of 5, introduces a strong geometric asymmetry that alters the structure, relative strength, and spatial distribution of the secondary vortices while preserving the same underlying mechanism. The training data include mean velocities in all three directions at four streamwise locations, totaling around 200 velocity components for each case.

\paragraph{Free-shear flows} The \textbf{round jet} evolves from a potential core into a turbulent shear layer before mixing with the ambient fluid. The total pressure is imposed at the jet inlet, pressure-outlet conditions are applied at the outlet, velocity is specified for the freestream inflow, and no-slip conditions are applied on the walls. Model validation requires reproducing the centerline velocity decay. Accordingly, we used 25 streamwise velocities along the centerline as training observations.

\subsection{Test cases for unified foundation model}
\label{subsubsec:foundation-test}
We evaluate the unified foundation model on 27 turbulent flows, divided into three groups: flows within the training categories, complex three-dimensional flows, and one additional fundamental flow category: the attached boundary layer (Table~\ref{tbl:foundation-test-cases}).
The additional flow category constitutes an asymptotic consistency check for compliance with the law of the wall.

\begin{table}[h!]
\centering

\renewcommand{\arraystretch}{1.2}
\begin{tabular}{m{0.27\linewidth} >{\centering}m{0.18\linewidth} >{\centering\arraybackslash}m{0.16\linewidth} >{\centering\arraybackslash}m{0.12\linewidth}}
\specialrule{1pt}{0pt}{0pt}
\textbf{Case} & \textbf{Control parameter} & \textbf{Value} & \textbf{No. of cases} \\
\specialrule{1pt}{0pt}{0pt}

Flat plate & $\text{Reynolds number}$ & $5\times10^{6}$ & 1 \\
\midrule

\multirow{3}{*}{Plane channel} & \multirow{3}{*}{$\text{Reynolds number}$}   & $3300$ & \multirow{3}{*}{3} \\
                               &                                & $7890$ & \\
                               &                                & $8\times10^{7}$ & \\
\midrule

\multirow{4}{*}{S809 airfoil} & \multirow{4}{*}{$\text{Angle of attack}$}  & $1^\circ$  & \multirow{4}{*}{4} \\
                              &                                & $5^\circ$  & \\
                              &                                & $16^\circ$ & \\
                              &                                & $18^\circ$ & \\
\midrule

\multirow{3}{*}{NACA 0012 airfoil} & \multirow{3}{*}{$\text{Angle of attack}$}  & $10^\circ$  & \multirow{3}{*}{3} \\
                              &                                & $15^\circ$  & \\
                              &                                & $18^\circ$  & \\
\midrule
\multirow{3}{*}{Periodic hill} & \multirow{3}{*}{\shortstack{Slope scaling \\ factor}} & $0.8$ & \multirow{3}{*}{3} \\
 &  & $1.2$ & \\
 &  & $1.5$ & \\
\midrule

\multirow{4}{*}{Bump} & \multirow{4}{*}{\shortstack{Height}} & $20\,\mathrm{mm}$ & \multirow{4}{*}{4} \\
 &  & $26\,\mathrm{mm}$ & \\
 &  & $38\,\mathrm{mm}$ & \\
 &  & $42\,\mathrm{mm}$ & \\
\midrule

\multirow{2}{*}{Square duct} & \multirow{2}{*}{$\text{Reynolds number}$} & $1800$ & \multirow{2}{*}{2} \\
 &  & $3500$ & \\
\midrule

\multirow{3}{*}{Rectangular duct} & \multirow{3}{*}{\shortstack{Aspect ratio}} & $3$ & \multirow{3}{*}{3} \\
 &  & $7$ & \\
 &  & $10$ & \\
\midrule

\multirow{1}{*}{Generic car (Ahmed body)} & Slant angle & \multirow{1}{*}{$40^{\circ}$} & \multirow{1}{*}{1} \\
\midrule

3D diffuser & $\text{Reynolds number}$ & $10^4$ & 1 \\
\midrule

\multirow{2}{*}{\added[id=Author]{Generic aircraft (CRM-HL)}} & \multirow{2}{*}{Angle of attack}& $7.05^{\circ}$ & \multirow{2}{*}{2} \\
 &  & $21.47^{\circ}$ & \\
\midrule

\multicolumn{2}{c}{} & \textbf{Totals} & \textbf{27} \\
\specialrule{1pt}{0pt}{0pt}
\end{tabular}
\caption{Summary of \textbf{test cases} used to evaluate the \textbf{unified foundation model}, including flow configuration, control parameter, parameter values, and the number of cases for each configuration, for a total of 27 test cases.}
\label{tbl:foundation-test-cases}
\end{table}

\paragraph{Attached boundary layers}  
The attached boundary layer category contains eight test cases: a flat plate, channel flows at three Reynolds numbers, S809 airfoils at two low angles of attack and NACA 0012 airfoils also at two low angles of attack. The flat plate case employs uniform inflow and no-slip boundary conditions at the wall, with periodicity applied in the spanwise direction. Two velocity values are sampled along the wall-normal direction to capture near-wall behavior, and the case represents the canonical development of a turbulent boundary layer, where viscous effects dominate close to the wall and gradually transition to turbulent momentum transport farther away. 
The channel flows at three different Reynolds numbers ($\mathrm{Re} = 3300, \, 7890, \, 8 \times 10^7$) use periodic streamwise boundaries and no-slip walls, exhibit the classical viscous, buffer, and logarithmic layers, and use four streamwise velocity samples along the wall-normal direction to probe Reynolds-number effects under otherwise identical conditions.  
The attached S809 airfoil test cases at $\mathrm{AOA} = 1^\circ$ and $5^\circ$ follow the same setup as the training cases, differing only in angle of attack. Lift and drag coefficients are calculated over the entire airfoil surface. These low angles maintain attached flow without massive separation, ensuring they remain representative of attached boundary layers.  
NACA 0012 airfoil at $\mathrm{AOA}=10^\circ$ and $15^\circ$ are also included here. Despite the higher angles of attack, these configurations remain predominantly attached at the selected Reynolds number and therefore also represent attached boundary layers. Their inclusion extends the evaluation to a different airfoil geometry and a higher loading regime while preserving the same underlying flow category.

\paragraph{Separated flows}  
The separated flow category contains four types of test cases: bump with different heights, S809 airfoil at two high angles of attack, periodic hills at three slope scaling factors and NACA 0012 airfoil also at one high angle of attack. 
The bump cases with different heights are simulated with the similar setup as the training case, with the wall friction coefficient used to assess the predictions of separation.
The S809 airfoil test cases at $\mathrm{AOA} = 16^\circ$ and $18^\circ$ share the same setup and sampling strategy as the training case, with lift and drag coefficients extracted; at these higher angles, separation and stall dominate the flow behavior. 
The NACA 0012 airfoil case at $\mathrm{AOA}=18^\circ$ is likewise included as a separated-flow test case. At this angle, the flow exhibits strong separation over a large portion of the suction side, placing it firmly in the stalled regime and making it representative of highly non-equilibrium boundary-layer behavior.
The periodic hill cases at three slope scaling factors (\(\alpha_{\text{p}}=0.8\), \(1.2\), \(1.5\)) impose periodic streamwise and spanwise boundaries with no-slip walls at top and bottom, and streamwise velocity is sampled along wall-normal lines downstream of the hill crest, with points taken every five computational cells, to determine separation and reattachment lengths. 
Together, these cases form a diverse testbed of separated turbulent flows, ranging from localized recirculation to fully stalled aerodynamic configurations.  

\paragraph{Secondary flows}  
The secondary flow category contains two types of test cases: square ducts at two Reynolds numbers (\(\mathrm{Re} = 1800, \, 3500\)) and rectangular ducts at three aspect ratios (AR = $3, 7, 10$). In both geometries, a fully developed turbulent profile is prescribed at the inlet, with no-slip walls along all duct boundaries and a constant mass flux condition to sustain the flow. The domain is periodic in the streamwise direction, ensuring statistical stationarity. Velocity observations are taken in cross-sectional planes, identical to the sampling strategy used in the training case, to capture mean velocity distortion and secondary vortex development. The square duct flows emphasize corner-induced vortices, while the rectangular duct flows show how aspect ratio affects secondary circulations. 
Together, these cases test the ability of turbulence models to capture inherently secondary motions in duct flows.

\paragraph{Complex 3D flows}
The generic car challenges turbulence models with bluff-body wake dynamics, including shear-layer vortices, corner vortices, and a large base recirculation zone. Observation data consist of the surface-integrated drag coefficient, which reflects the accuracy of predicted separation and wake structures.
The asymmetric 3D diffuser tests turbulence models under adverse pressure gradients and Reynolds stress anisotropy that drive asymmetric separation. Observation data include velocity profiles across the separated shear layer, and wall friction trends that capture onset and reattachment.
\added[id=Author]{
The generic aircraft configuration, based on the high-lift Common Research Model, represents a realistic transport aircraft with deployed slat, wing, and flap elements and is studied under high-Reynolds-number conditions ($M=0.2$, $\mathrm{Re}=5.49\times10^{6}$). It generates complex three-dimensional flow characterized by interactions among boundary layers, wakes, and coherent vortical structures, leading to strong Reynolds stress anisotropy and vortex--boundary-layer--driven separation. Two representative conditions are considered: a pre-stall case at $\mathrm{AoA}=7.05^\circ$, corresponding to mildly separated flow before stall and a post-stall case at $\mathrm{AoA}=21.47^\circ$, where stall occurs and large-scale flow separation dominates. Experimental and large-eddy simulation datasets provide integrated forces (lift and drag) and surface measurements (pressure and skin friction).}

\subsection{Specialist turbulence model for separated flows}
Conventional turbulence models often exhibit poor performance in flows with massive separation. To address this limitation, we fine-tune a specialist model tailored for separated flows. This section describes the training configurations and the test cases selected to evaluate its predictive performance.

\begin{figure}[!htb]
  \centering

  \begin{subfigure}[t]{0.45\textwidth}
    \centering
    \includegraphics[width=0.8\textwidth]{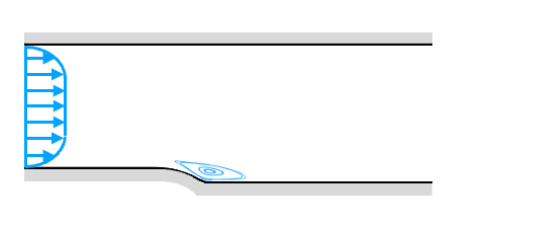}
    \caption{Curved step}
    \label{fig:expert-setup-curved}
  \end{subfigure}
  \hspace{-0.08\textwidth}
  \begin{subfigure}[t]{0.45\textwidth}
    \centering
    \includegraphics[width=0.8\textwidth]{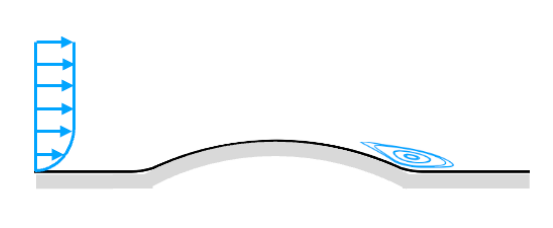}
    \caption{Bump (height = 42 mm)}
    \label{fig:expert-setup-bump}
  \end{subfigure}

  \caption{\textbf{Training configurations} used for the \textbf{specialist model for separated flows}. The training set consists of two benchmark configurations, including a curved-step flow and a bump flow with height 42~mm.}
  
  \label{fig:expert-1-setup}
\end{figure}

\paragraph{Training cases} 
To obtain a specialist model for flows with complex separation, we fine-tune the model using two benchmark separated flows: flow over a bump with $42$ mm height and flow over a curved step as shown in Fig.~\ref{fig:expert-1-setup}. The bump’s convex curvature induces an adverse pressure gradient that produces an incipient separation bubble with reattachment several bump heights downstream, while the curved step’s sudden streamline curvature generates a large separation region followed by gradual reattachment along the lower wall.

\paragraph{Test cases}
Two groups of cases for evaluating the trained specialist model for separated flows are chosen and summarized in Table~\ref{tab:test-cases-separation}, namely test cases within the training categories: bumps with different heights ($20, 26, 31, 38$\,mm) and complex three-dimensional case: the generic car shown in Table~\ref{tab:cases-setup}.

\begin{table}[h!]
\centering

\renewcommand{\arraystretch}{1.2}
\begin{tabular}{m{0.16\linewidth} >{\centering}m{0.2\linewidth} >{\centering\arraybackslash}m{0.08\linewidth} >{\centering\arraybackslash}m{0.13\linewidth}}
\specialrule{1pt}{0pt}{0pt}
\textbf{Case} & \textbf{ Control parameter} & \textbf{Value} & \textbf{No. of cases} \\
\specialrule{1pt}{0pt}{0pt}

\multirow{4}{*}{Bump} & \multirow{4}{*}{$\text{Height}$} & $20\,\mathrm{mm}$ & \multirow{4}{*}{4} \\
                      &                                  & $26\,\mathrm{mm}$ & \\
                      &                                  & $31\,\mathrm{mm}$ & \\
                      &                                  & $38\,\mathrm{mm}$ & \\
\midrule
\multirow{1}{*}{Generic car} & Slant angle & \multirow{1}{*}{$40^{\circ}$} & \multirow{1}{*}{1} \\
\midrule

\multicolumn{2}{c}{} & \textbf{Totals} & \textbf{5} \\
\specialrule{1pt}{0pt}{0pt}
\end{tabular}
\caption{\textbf{Test cases} used to evaluate the \textbf{specialist model for separated flows}, including flow configuration, control parameter, parameter values, and the number of cases for each configuration, for a total of 5 test cases.}
\label{tab:test-cases-separation}
\end{table}

The bumps retain the same configuration as in training while varying the bump height, which systematically alters the adverse pressure gradient and separation strength. These cases extend the training set into controlled parametric variations, enabling assessment of model robustness.

The generic car is a standard benchmark in automotive aerodynamics, used to evaluate turbulence models for bluff-body flows. The model is placed in a wind tunnel domain and a uniform inlet yields a Reynolds number $\mathrm{Re} = 2.8\times10^{6}$. A zero-pressure outlet boundary condition is applied at the downstream end, no-slip conditions are imposed on all solid surfaces, and symmetry conditions are used on the top and lateral boundaries to exploit geometric symmetry and reduce computational cost. The drag coefficient is calculated over the entire generic car surface as the observation data.
The $40^{\circ}$ slanted rear surface produces a complex three-dimensional wake, including top-shear-layer vortices, corner vortices, and a large base recirculation bubble. These flow structures make the generic car a stringent test for assessing turbulence model capability in predicting complex three-dimensional separation and wake dynamics. 

\subsection{Specialist turbulence model for secondary flows with separation}

We fine tune a specialist turbulence model aimed at accurately capturing secondary flows with separation (multiple mechanisms). Recognizing that practical turbulent flows rarely involve a single dominant mechanism, but instead arise from complex interactions among multiple phenomena, the model is trained on configurations where both effects are prominent. This section outlines the training setups and test cases used to evaluate the model’s predictive performance.

\begin{figure}[!htb]
  \centering

  \begin{subfigure}[t]{0.45\textwidth}
    \centering
    \includegraphics[width=0.8\textwidth]{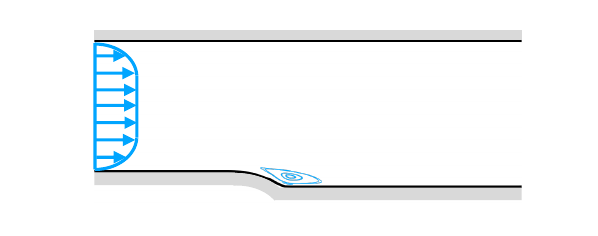}
    \caption{Curved step}
    \label{fig:expert2-curved}
  \end{subfigure}
  \hspace{-0.08\textwidth}
  \begin{subfigure}[t]{0.45\textwidth}
    \centering
    \includegraphics[width=0.8\textwidth]{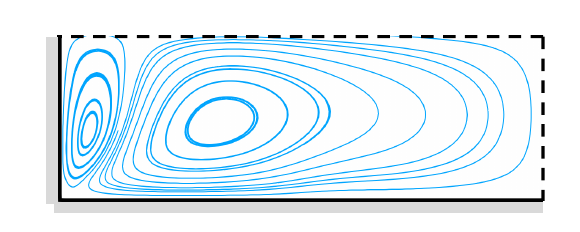}
    \caption{Rectangular ducts (Aspect ratio 3 and 10)}
    \label{fig:expert2-ducts}
  \end{subfigure}

  \caption{\textbf{Training configurations} used for the \textbf{specialist model for secondary flows with separation}. The training set consists of three benchmark configurations, including a curved-step flow and rectangular-duct flows with aspect ratios $\mathrm{AR}=3$ and 10.}
  
  \label{fig:expert-2-setup}
\end{figure}

\paragraph{Training cases} 
We select three benchmark cases to fine tune the specialist model for secondary flows with separation shown in Fig.~\ref{fig:expert-2-setup}, targeting large flow separation and secondary flow mechanisms: flow over a curved step and flows through two rectangular ducts with aspect ratio $\text{AR}=3, \, 10$. These duct cases are detailed in \S\ref{subsec:foundation}. 
The rectangular duct flow exhibits asymmetric corner vortices, with the vortex on the horizontal wall expanding spanwise and dominating while the vertical wall vortex is compressed, producing a skewed concave flow pattern unlike the symmetric structure in square ducts. Training on these cases enhances the model's ability to capture secondary flows, large separations, and their interactions.

\paragraph{Test cases}
We select three groups to evaluate the trained specialist model for secondary flows with separation shown in Table~\ref{tab:test-cases-secondary}: (i) secondary flow cases: square ducts at $\mathrm{Re}=1100, 1800, 2600, 3500$; rectangular ducts with aspect ratio 5 and 7; (ii) separated cases: bumps with heights  $20, 26, 31, 38, 42$ mm; hump and periodic hills with slope scaling factor $\alpha_{\text{p}}=0.8, 1.0, 1.2, 1.5$; (iii) a complex three-dimensional diffuser case. All the test cases within the training categories are detailed in Section.~\ref{subsubsec:foundation-test} and we now focus on the 3D diffuser.

\begin{table}[h!]
\centering

\renewcommand{\arraystretch}{1.2}
\begin{tabular}{m{0.16\linewidth} >{\centering}m{0.2\linewidth} >{\centering\arraybackslash}m{0.16\linewidth} >{\centering\arraybackslash}m{0.16\linewidth}}
\specialrule{1pt}{0pt}{0pt}
\textbf{Flow} & \textbf{Control parameter} & \textbf{Value} & \textbf{No. of cases} \\
\specialrule{1pt}{0pt}{0pt}

\multirow{4}{*}{Square duct} & \multirow{4}{*}{$\text{Reynolds number}$} & $1100$ & \multirow{4}{*}{4} \\
                             &                               & $1800$ & \\
                             &                               & $2600$ & \\
                             &                               & $3500$ & \\   
\midrule

\multirow{2}{*}{Rectangular duct} & \multirow{2}{*}{$\text{Aspect ratio}$} & $5$  & \multirow{2}{*}{2} \\
                                  &                               & $7$  & \\
\midrule

\multirow{5}{*}{Bump} & \multirow{5}{*}{\shortstack{Height}} & $20\,\mathrm{mm}$ & \multirow{5}{*}{5} \\
 &  & $26\,\mathrm{mm}$ & \\
 &  & $31\,\mathrm{mm}$ & \\
 &  & $38\,\mathrm{mm}$ & \\
 &  & $42\,\mathrm{mm}$ & \\
\midrule

Hump & $\text{Reynolds number}$ & $9.36\times 10^5$ & 1 \\
\midrule

\multirow{4}{*}{Periodic hill} & \multirow{4}{*}{\shortstack{Slope scaling \\factor}} & $0.8$ & \multirow{4}{*}{4} \\
                               &                           & $1.0$ & \\
                               &                           & $1.2$ & \\
                               &                           & $1.5$ & \\
\midrule

3D diffuser & $\text{Reynolds number}$ & $10^4$ & 1 \\
\midrule

\multicolumn{2}{c}{} & \textbf{Totals} & \textbf{17} \\
\specialrule{1pt}{0pt}{0pt}
\end{tabular}
\caption{\textbf{Test cases} used to evaluate the \textbf{specialist model of secondary flows with separation}, including flow configuration, control parameter, parameter values, and the number of cases for each configuration, for a total of 17 test cases.}
\label{tab:test-cases-secondary}
\end{table}

The 3D diffuser~\cite{cherry2008geometric} is a challenging validation case for turbulence models. It features incompressible, asymmetric internal flow with strong adverse pressure gradients and significant Reynolds stress anisotropy. The resulting  complex three-dimensional separation closely mirrors the flow behavior in practical diffusers, such as those found between compressors and combustors in jet engines. This case is therefore a valuable benchmark for evaluating a model’s ability to predict corner separation and secondary flow dynamics.
The diffuser consists of an inlet channel, a diffuser section, and a straight outlet section. Both the upper wall and the sidewall expand. Boundary conditions prescribe a fully developed turbulent inflow with a bulk velocity $U_{\mathrm{b}} = 1$~m/s, leading to a Reynolds number $\mathrm{Re} = 10^4$ based on inlet channel height. The diffuser test case is designed to capture three-dimensional separation and reattachment driven by wall expansion. Velocities are sampled at multiple wall-normal locations downstream of the expansion, as indicated by the red markers in Table~\ref{tab:cases-setup}, to resolve the separated shear layer and recirculation bubble. In addition to velocity profiles, the wall friction coefficient is monitored along the bottom wall to quantify separation onset and reattachment. In total, 64 observations are used, providing detailed information on both the mean flow field and wall-bounded response. 

\FloatBarrier
\section{Results}
\label{sec:results}
\subsection{Unified foundation model}

\begin{figure}
  \centering
  \includegraphics[width=0.9\textwidth]{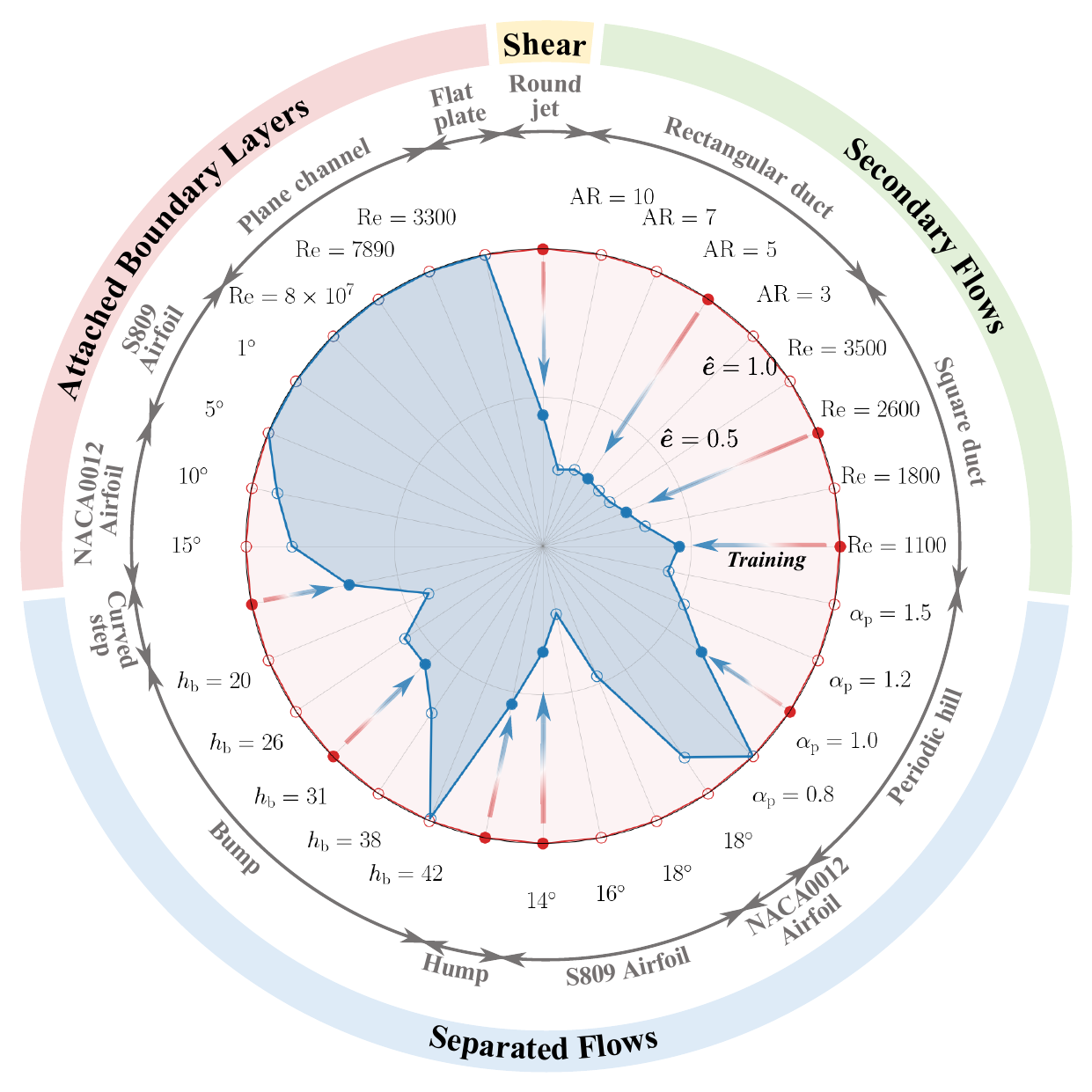}
  \caption{Performance evaluation of the \textbf{unified foundation model} across \textbf{four benchmark flow categories}, shown as a radar chart with comparison to the baseline $k$--$\omega$ model. The plot includes attached boundary layers, secondary flows, separated flows, and free-shear flows and spans multiple configurations. The normalized misfit \(\hat{e}\) is shown as the radial coordinate. The baseline model corresponds to \(\hat{e}=1\) on all axes and is shown as a red curve with round markers. The unified foundation model is shown as a blue filled polygon with markers. Individual test cases are shown along the radial axes and are annotated by parameters such as Reynolds number, aspect ratio, and angle of attack. Across most test cases, the blue filled polygon lies inside the red ring.}
\label{fig:radar-unified-foundation-model}
\end{figure}

We train a unified foundation model on nine canonical benchmark cases and evaluate it on 25 diverse test cases. The unified foundation model demonstrates strong generalization across all test categories and consistently reduces the misfit relative to the baseline \(k\)--\(\omega\) model in Fig.~\ref{fig:radar-unified-foundation-model}.

\begin{figure}
  \centering
  \includegraphics[width=0.75\textwidth]{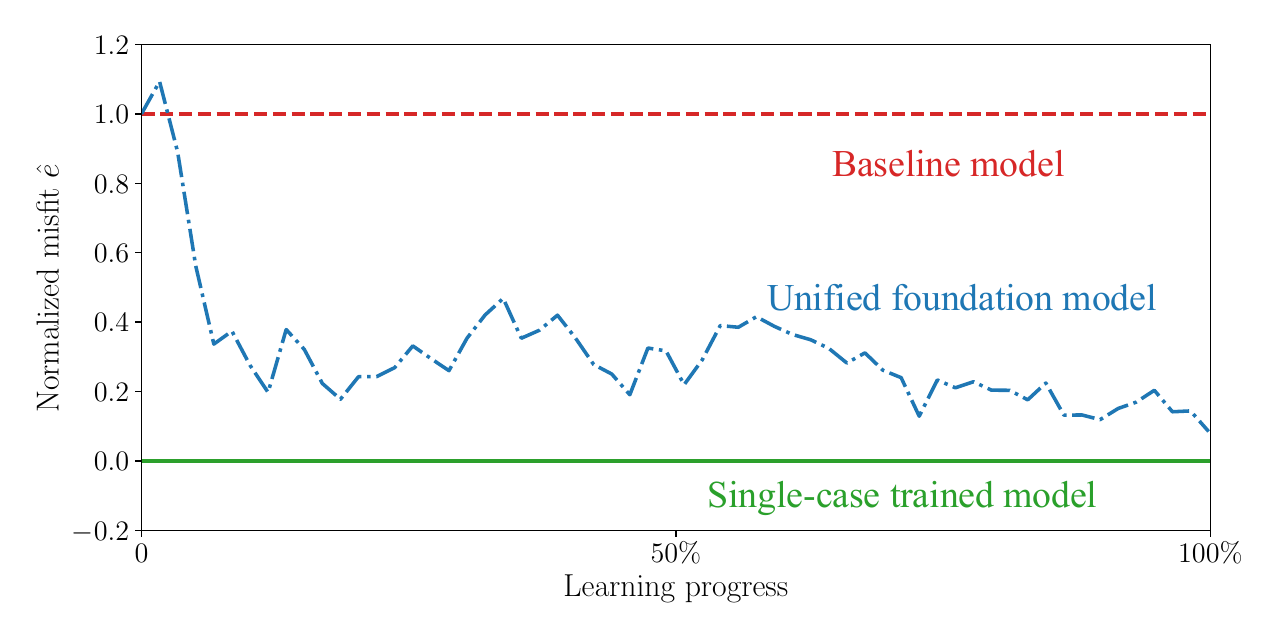}
  \caption{Illustration of the normalized misfit metric $\hat{e}$ used to compare turbulence model performance. The single-case trained model, the unified foundation model, and the baseline $k$--$\omega$ model are shown for reference. The vertical axis shows the normalized misfit $\hat{e}$, defined such that the single-case trained model is fixed at $\hat{e}=0$ and the baseline model is fixed at $\hat{e}=1$, and the horizontal axis indicates learning progress.}
\label{fig:normalized-misfit}
\end{figure}

We introduce the normalized misfit, denoted $\hat{e}$, to quantify the improvement of the unified foundation model relative to the single-case trained model and to the baseline \(k\)--\(\omega\) model. Let $e_{\mathrm{unified}}$, $e_{\mathrm{single}}$, and $e_{\mathrm{base}}$ denote the misfits of the unified foundation model, the single-case trained model, and the baseline model, respectively, and define
\begin{equation}
\hat{e}=\frac{e_{\mathrm{unified}}-e_{\mathrm{single}}}{e_{\mathrm{base}}-e_{\mathrm{single}}}.
\end{equation}
By construction, $\hat{e}=0$ corresponds to performance equal to the single-case trained model and $\hat{e}=1$ corresponds to the baseline; values $\hat{e}<1$ indicate improvement relative to the baseline while values $\hat{e}>1$ indicate degraded performance.
For the rectangular duct case shown in Fig.~\ref{fig:normalized-misfit}, the unified foundation model attains $\hat{e}$ well below unity, indicating better performance relative to the baseline model; the baseline model has $\hat{e}=1$ and the single-case trained model has $\hat{e}=0$ by definition.

The normalized misfit $\hat{e}$ across all nine training cases and 25 test cases summarizes the good generalization behavior of the unified foundation model across four flow categories, as shown in Fig.~\ref{fig:radar-unified-foundation-model}. For attached boundary layers, the unified foundation model attains performance comparable to the baseline on non-trained cases, reflecting the effect of the imposed physical consistency constraints. For free-shear and secondary flows, the unified foundation model yields consistently lower misfits than the baseline $k$--$\omega$ model, indicated by the blue polygon lying inside the baseline ring. For separated flows, the unified foundation model substantially reduces misfit relative to the baseline model in most cases, while achieving only comparable performance for the most challenging configurations, including the bump with the greatest height ($42\,\mathrm{mm}$) and the periodic hill with the steepest slope (\(\alpha_{\text{p}} = 0.8\)). Overall, the normalized misfit distribution demonstrates broad generalization across all four flow categories and simultaneously highlights specific cases of separated flows where targeted refinement is most needed.

\paragraph{Attached boundary layers}

\begin{figure}[!htb]
  \centering

  \includegraphics[width=0.8\textwidth]{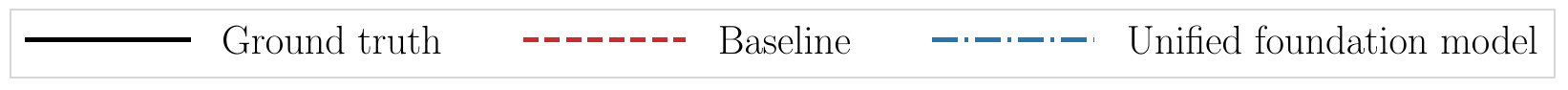}

  \begin{subfigure}[t]{0.48\textwidth}
    \centering
    \includegraphics[width=\textwidth]{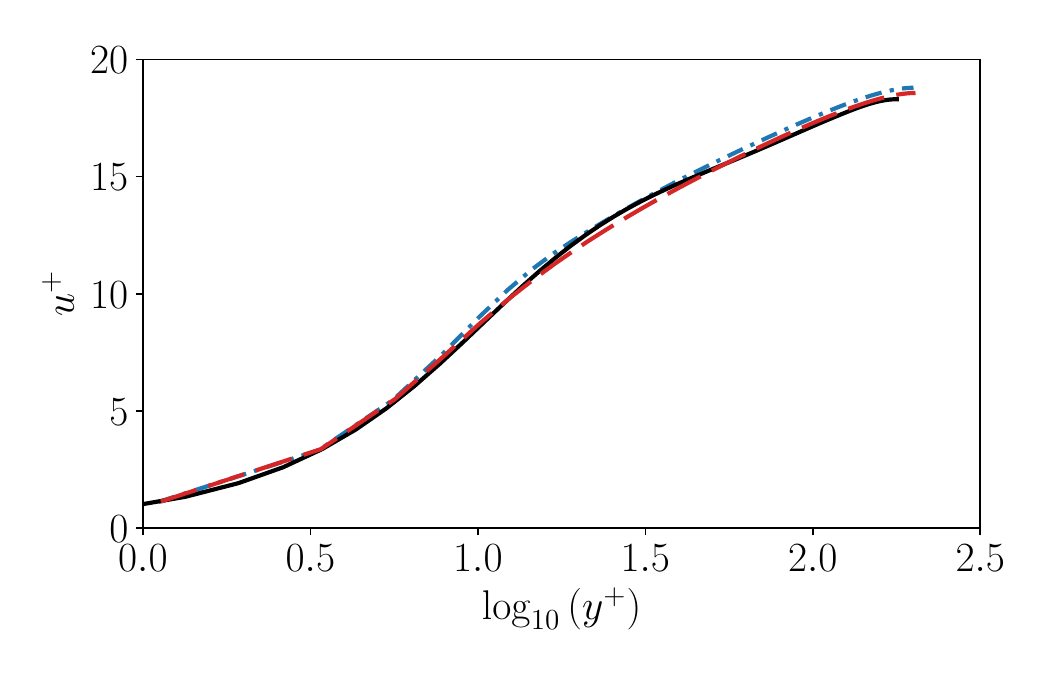}
    \caption{Plane channel: $\mathrm{Re}=3300$}
    \label{fig:canonical-a}
  \end{subfigure}
  \hspace{0.01\textwidth}
  \begin{subfigure}[t]{0.48\textwidth}
    \centering
    \includegraphics[width=\textwidth]{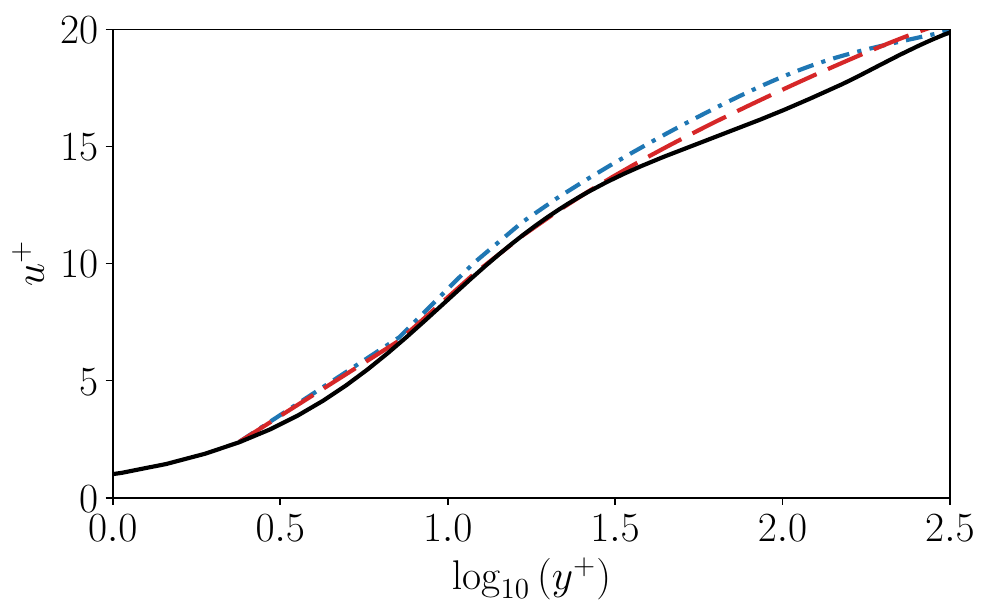}
    \caption{Plane channel: $\mathrm{Re}=7890$}
    \label{fig:canonical-b}
  \end{subfigure}

  \begin{subfigure}[t]{0.48\textwidth}
    \centering
    \includegraphics[width=\textwidth]{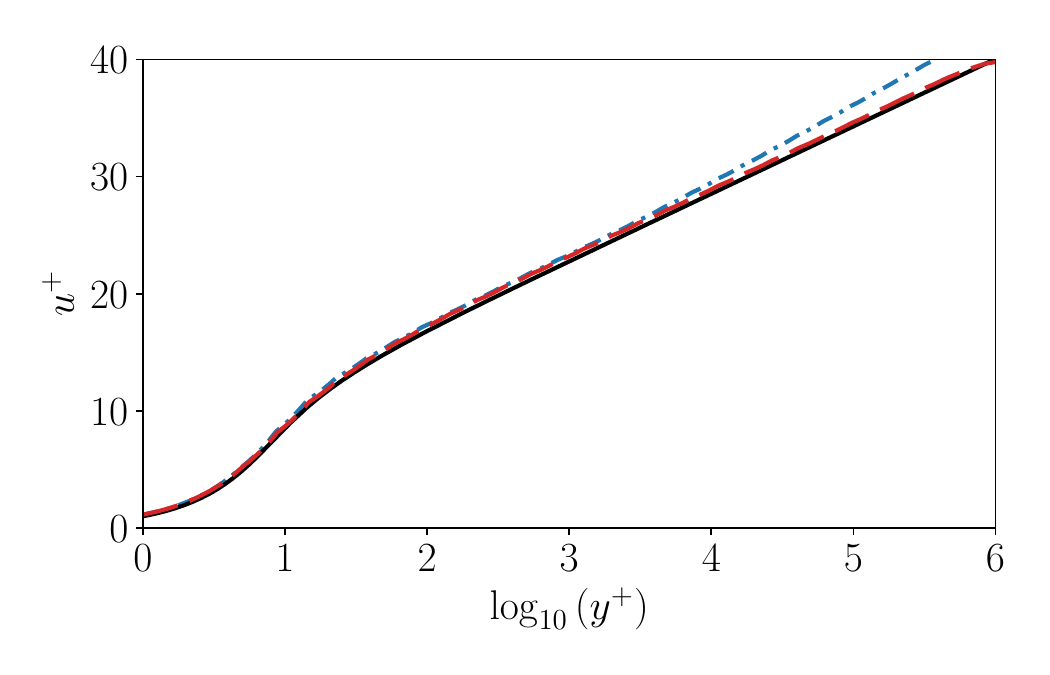}
    \caption{Plane channel: $\mathrm{Re}=8\times10^{7}$}
    \label{fig:canonical-c}
  \end{subfigure}
  \hspace{0.01\textwidth}
  \begin{subfigure}[t]{0.48\textwidth}
    \centering
    \includegraphics[width=\textwidth]{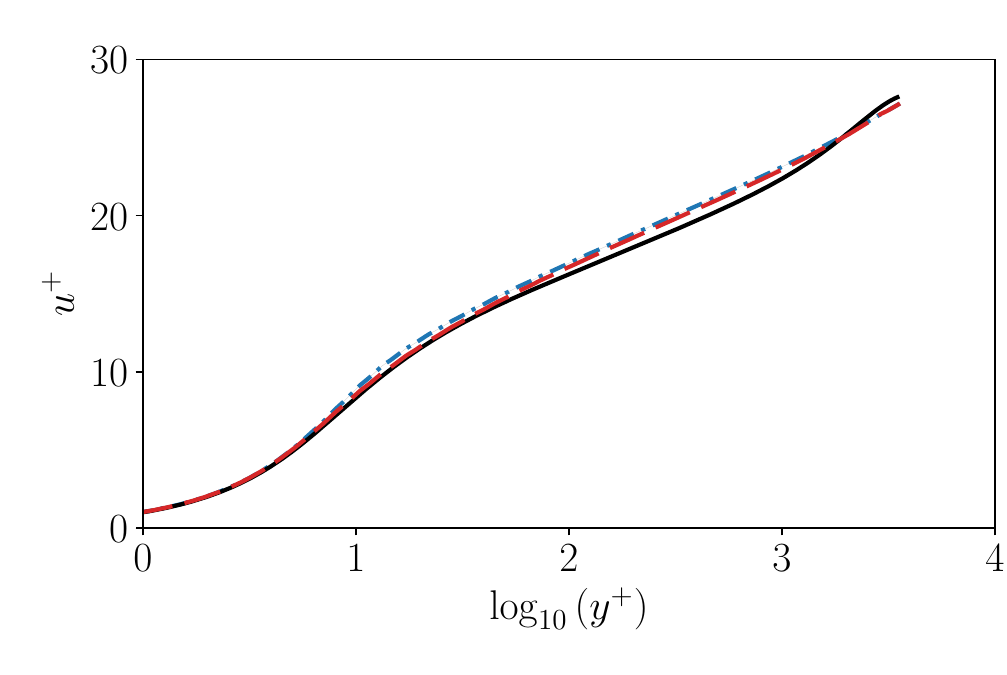}
    \caption{Zero pressure gradient flat plate}
    \label{fig:canonical-d}
  \end{subfigure}

  \caption{Performance evaluation of the \textbf{unified foundation model} in \textbf{attached boundary layers} against the ground truth and the baseline \(k\)--\(\omega\) model. Mean velocity profiles \(u^+\) are plotted for plane channel flows at (a) \(\mathrm{Re}=3300\), (b) \(\mathrm{Re}=7890\), (c) \(\mathrm{Re}=8\times10^{7}\), and (d) zero pressure gradient flat plate flow; the x-axis denotes the wall-normal coordinate $\log_{10}(y^+)$.}
  
  \label{fig:canonical-results}
\end{figure}

\begin{figure}[!htb]
  \centering

  \includegraphics[width=0.8\textwidth]{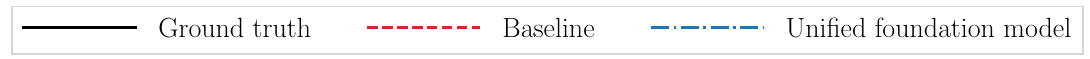}

  \begin{subfigure}[t]{0.48\textwidth}
    \centering
    \includegraphics[width=\textwidth]{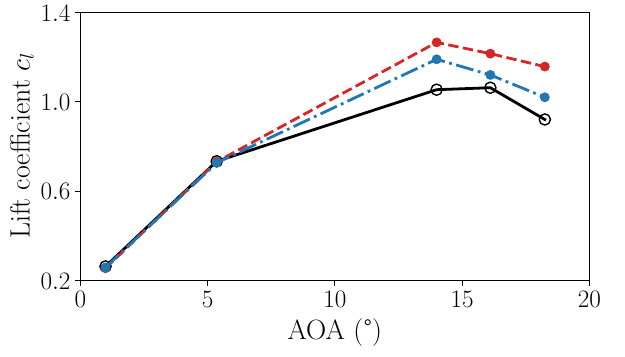}
    \caption{Lift coefficient $c_l$}
    \label{fig:airfoil-cl}
  \end{subfigure}
  \hspace{0.02\textwidth}
  \begin{subfigure}[t]{0.48\textwidth}
    \centering
    \includegraphics[width=\textwidth]{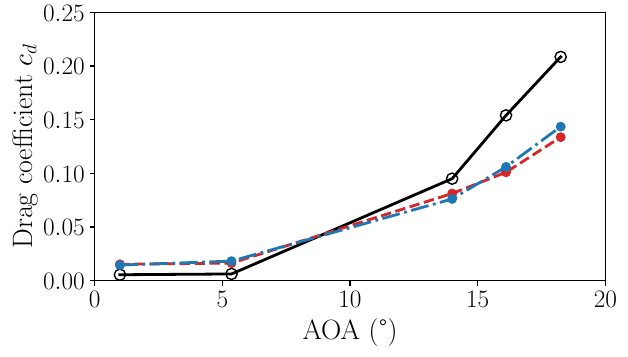}
    \caption{Drag coefficient $c_d$}
    \label{fig:airfoil-cd}
  \end{subfigure}

  \caption{Performance evaluation of the \textbf{unified foundation model} for \textbf{S809 airfoil flow} across angles of attack against the ground truth and the baseline \(k\)--\(\omega\) model. Lift coefficient \(c_l\) (left) and drag coefficient \(c_d\) (right) are shown as functions of the angle of attack (AOA).}
  
  \label{fig:cd_cl_comparison}
\end{figure}

The unified foundation model matches the baseline \(k\)--\(\omega\) model in eight cases of attached boundary layers, demonstrating good accuracy and generalization across Reynolds numbers and geometries. We evaluate the model on channel flows at \(\mathrm{Re} = 3300, \, 7890\), \text{and} \(8 \times 10^7\), on a flat plate flow, with results shown in Fig.~\ref{fig:canonical-results}. We also test the model on two S809 airfoil cases with low angles of attack (\(\text{AOA} = 1^\circ\) and \(5^\circ\)) and two NACA 0012 airfoil cases with low angles of attack (\(\text{AOA} = 10^\circ\) and \(15^\circ\)), with results shown in Fig.~\ref{fig:cd_cl_comparison} and Table~\ref{tab:naca0012_lift}. For the four airfoil cases at low angles of attack, where the flow remains fully attached, all models perform similarly.

The unified foundation model accurately reproduces near-wall mean velocity profiles across canonical wall-bounded flows and Reynolds numbers. Profiles of \(u^+\) versus \(y^+\) (log scale) for DNS data, the baseline \(k\)--\(\omega\) model, and the unified foundation model are shown in Fig.~\ref{fig:canonical-results}. Because the \(u^+\)–\(y^+\) relation reflects momentum diffusion from the wall, it provides a sensitive benchmark for turbulence models. At \(\mathrm{Re}=3300\), it follows DNS in the viscous sublayer (\(y^+ \lesssim 5\)) and buffer layer (\(5 \lesssim y^+ \lesssim 30\)), and it maintains agreement in the logarithmic region (\(30 \lesssim y^+ \lesssim 300\)), as shown in Fig.~\ref{fig:canonical-results}a. At \(\mathrm{Re}=7890\), the predicted profile remains consistent with DNS and exhibits behavior similar to the baseline model over the entire wall-normal range, as shown in Fig.~\ref{fig:canonical-results}b. For the plane channel flow at \(\mathrm{Re}=8\times10^{7}\), the velocity profile lies predominantly in the logarithmic region beyond the immediate near-wall layer. In this regime, the unified foundation model preserves the correct logarithmic slope but slightly overpredicts the mean velocity relative to DNS over most of the wall-normal range, as shown in Fig.~\ref{fig:canonical-results}c. In the zero pressure gradient flat plate flow, the unified foundation model captures the expected boundary-layer development and produces velocity profiles comparable to those of the baseline \(k\)--\(\omega\) model, indicating robustness beyond the channel-flow configuration, as shown in Fig.~\ref{fig:canonical-results}d.

These results highlight the unified foundation model’s strong generalizability. The model is not trained on any of these cases, yet it captures geometry-agnostic and physically grounded features of near-wall turbulence. This behavior arises from enforcing physical consistency within the turbulence model rather than case-specific training, enabling transfer across attached boundary layer flows.

\begin{figure}[!htb]
  \centering

  \includegraphics[width=0.8\textwidth]{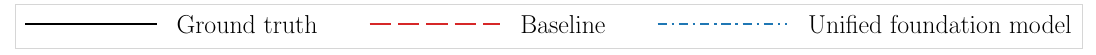}

  \begin{subfigure}[t]{0.3\textwidth}
    \centering
    \includegraphics[width=\textwidth]{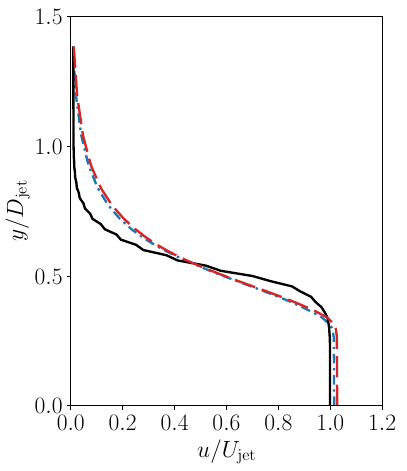}
    \caption{$x/D_{\text{jet}} = 2$}
    \label{fig:jet-a}
  \end{subfigure}
  \hspace{0.01\textwidth}
  \begin{subfigure}[t]{0.3\textwidth}
    \centering
    \includegraphics[width=\textwidth]{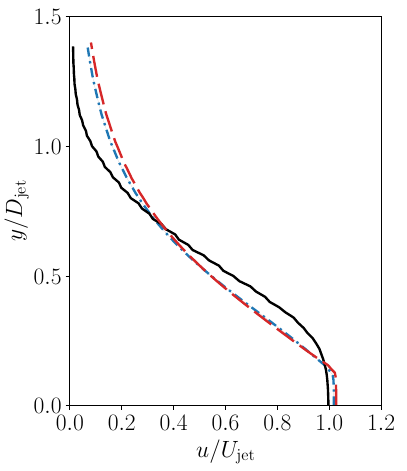}
    \caption{$x/D_{\text{jet}} = 5$}
    \label{fig:jet-b}
  \end{subfigure}
  \hspace{0.01\textwidth}
  \begin{subfigure}[t]{0.3\textwidth}
    \centering
    \includegraphics[width=\textwidth]{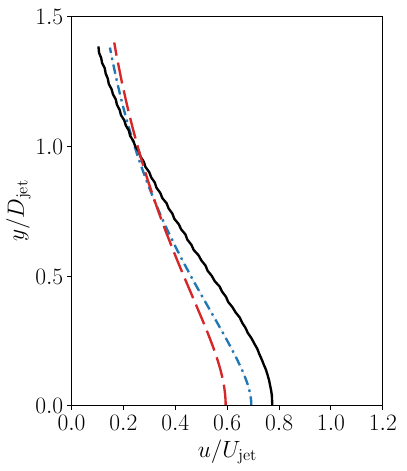}
    \caption{$x/D_{\text{jet}} = 10$}
    \label{fig:jet-c}
  \end{subfigure}

  \begin{subfigure}[t]{0.3\textwidth}
    \centering
    \includegraphics[width=\textwidth]{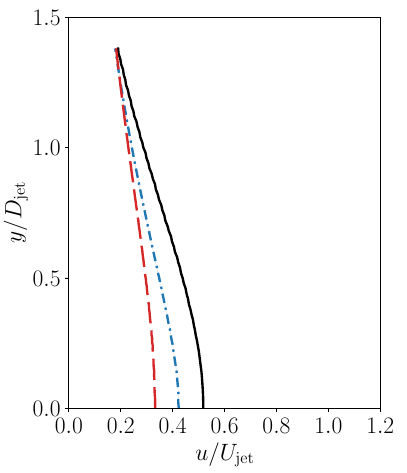}
    \caption{$x/D_{\text{jet}} = 15$}
    \label{fig:jet-d}
  \end{subfigure}
  \hspace{0.01\textwidth}
  \begin{subfigure}[t]{0.3\textwidth}
    \centering
    \includegraphics[width=\textwidth]{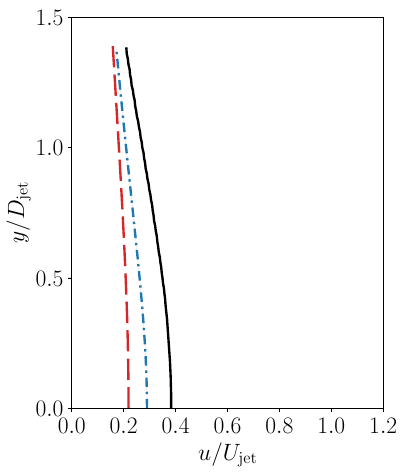}
    \caption{$x/D_{\text{jet}} = 20$}
    \label{fig:jet-e}
  \end{subfigure}
  \hspace{0.01\textwidth}
  \begin{subfigure}[t]{0.3\textwidth}
    \centering
    \includegraphics[width=\textwidth]{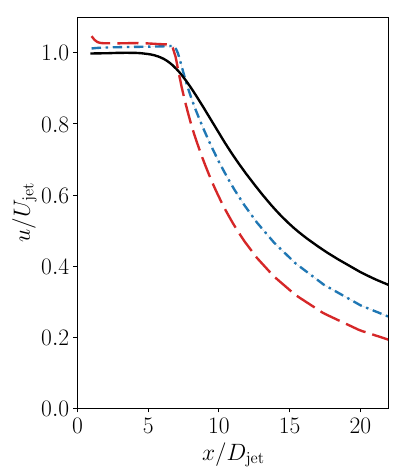}
    \caption{jet centerline}
    \label{fig:jet-f}
  \end{subfigure}

  \caption{\added[id=Author]{Performance evaluation of the \textbf{unified foundation model} in a \textbf{round jet} against the ground truth and the baseline \(k\)--\(\omega\) model. Velocity profiles $U/U_{\mathrm{jet}}$ are shown at multiple downstream locations, plotted against the normalized wall-normal coordinate $y/D_{\mathrm{jet}}$ for subfigures (a)--(e), while the normalized centerline velocity is presented as a function of the downstream distance $x_{\mathrm{center}}/D_{\mathrm{jet}}$ in (f).}}
  
  \label{fig:jet-results}
\end{figure}

\paragraph{Free-shear flow}
The unified foundation model reproduces centerline velocity decay in a round jet with higher fidelity than the baseline \(k\)--\(\omega\) model in Fig.~\ref{fig:jet-results}f. We evaluate the model on the same jet flow case used for training. Compared to high-fidelity DNS data, which show a gradual velocity decay downstream, the unified foundation model closely follows the DNS curve across the entire range. It captures jet entrainment and turbulent mixing more accurately than the baseline model, particularly in the far-field region where the baseline model typically underpredicts the decay. \added[id=Author]{The velocity profiles along the wall-normal direction, which are not directly observed during training, are also predicted with improved accuracy (Fig.~\ref{fig:jet-results}a--e). This indicates that the unified foundation model generalizes well beyond the observed quantities and provides a more consistent representation of the underlying flow field.}

\begin{figure}[!htb]
  \centering

  \includegraphics[width=0.8\textwidth]{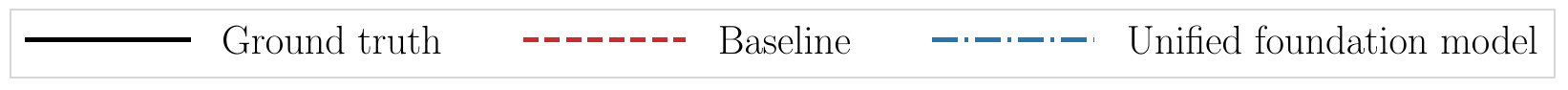}

  \begin{subfigure}[t]{0.48\textwidth}
    \centering
    \includegraphics[width=\textwidth]{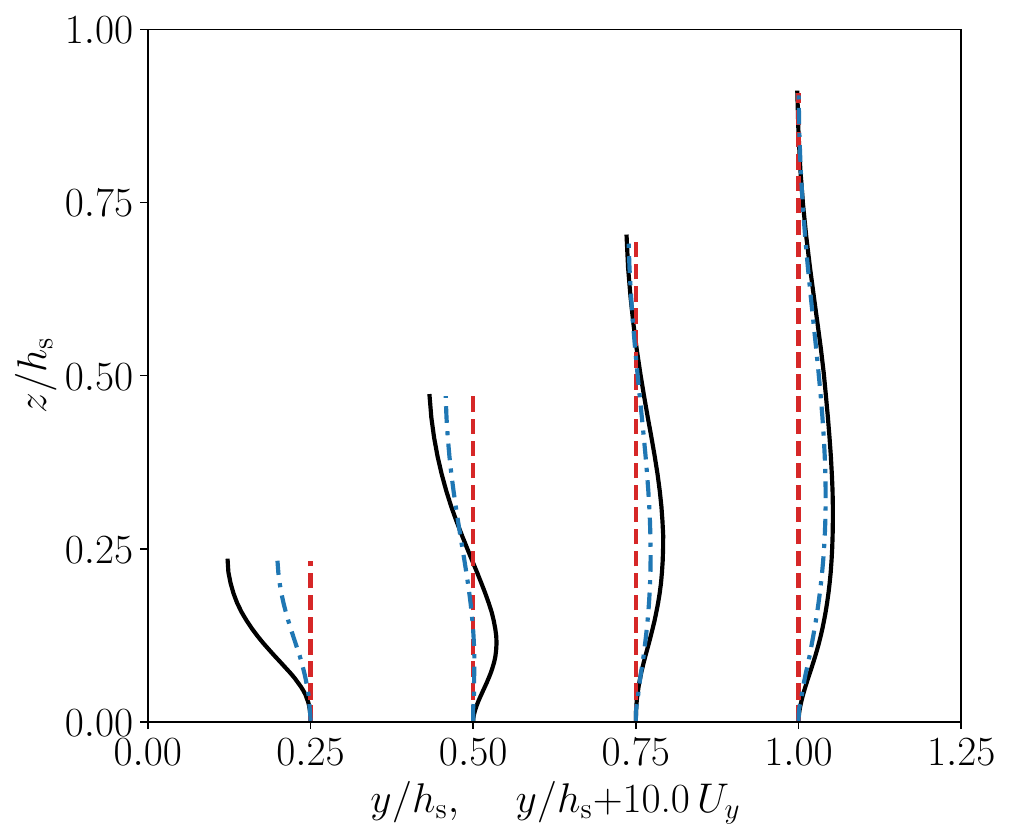}
    \caption{$\mathrm{Re}=3500$}
    \label{fig:secondary-a}
  \end{subfigure}
  \hspace{0.02\textwidth}
  \begin{subfigure}[t]{0.48\textwidth}
    \centering
    \includegraphics[width=\textwidth]{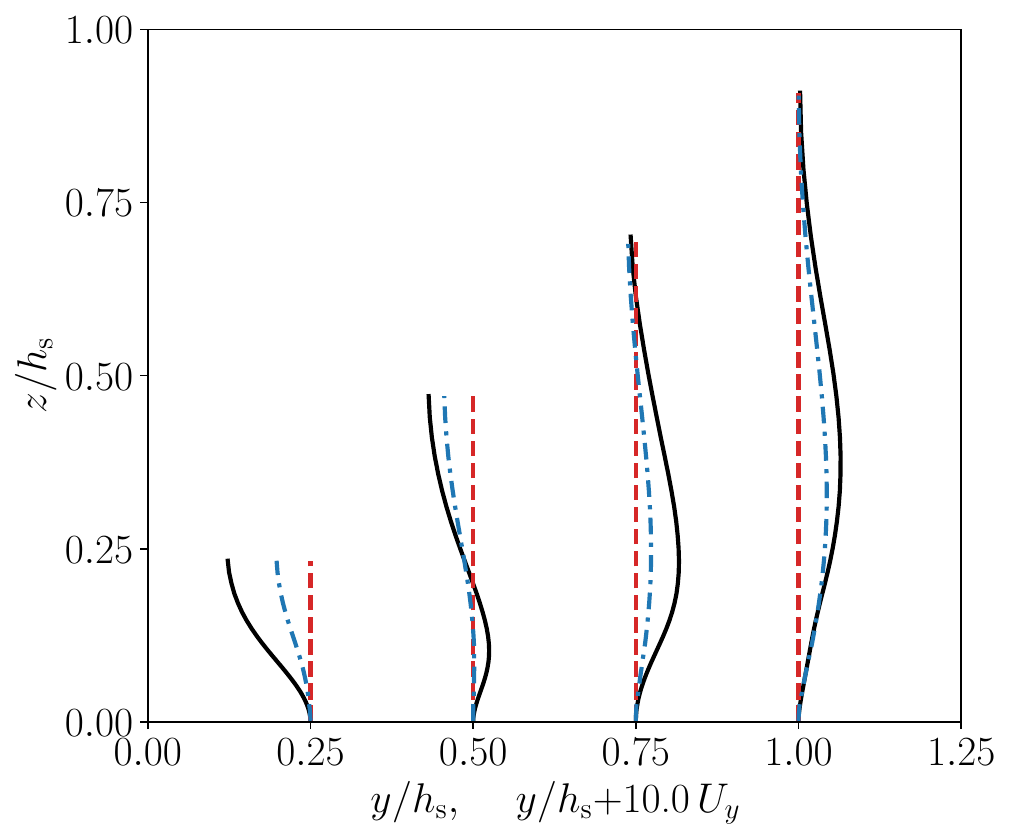}
    \caption{$\mathrm{Re}=2600$}
    \label{fig:secondary-b}
  \end{subfigure}

  \begin{subfigure}[t]{0.48\textwidth}
    \centering
    \includegraphics[width=\textwidth]{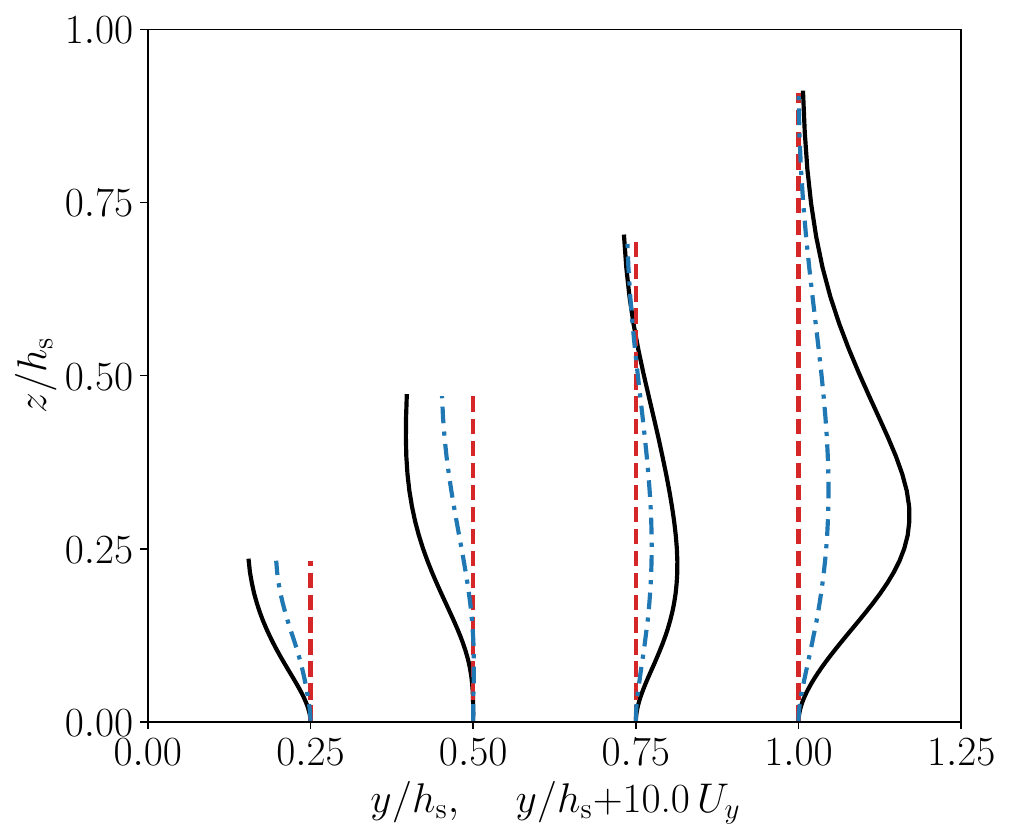}
    \caption{$\mathrm{Re}=1800$}
    \label{fig:secondary-c}
  \end{subfigure}
  \hspace{0.02\textwidth}
  \begin{subfigure}[t]{0.48\textwidth}
    \centering
    \includegraphics[width=\textwidth]{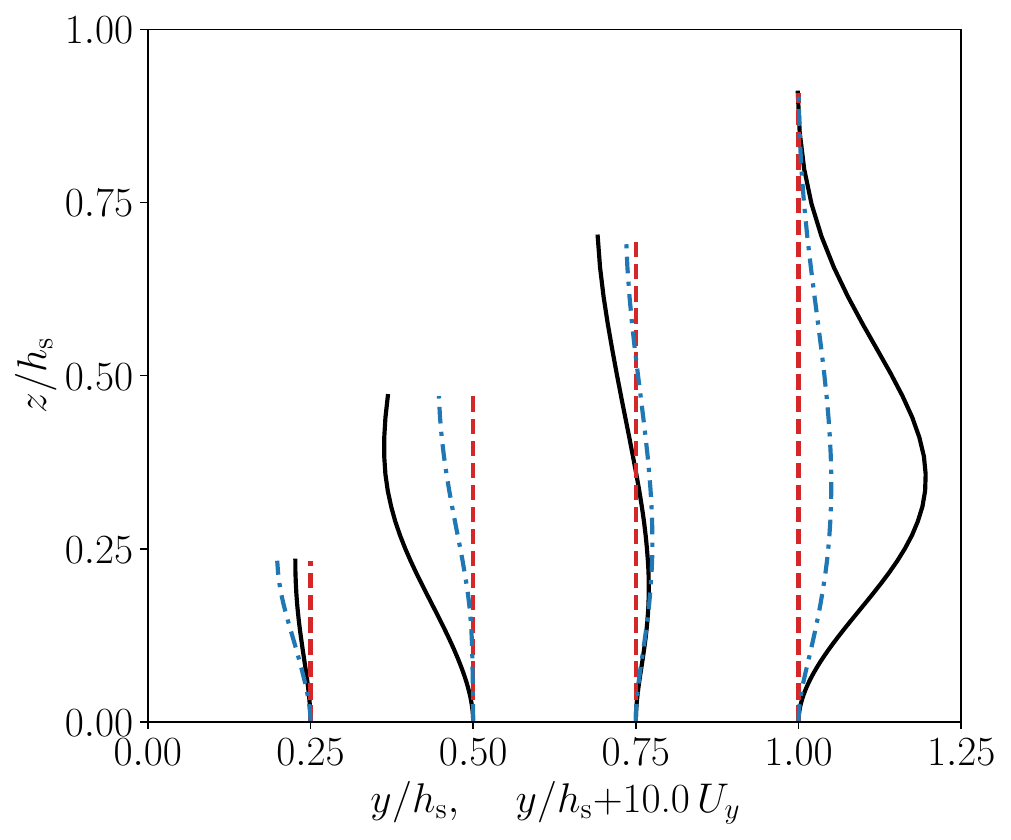}
    \caption{$\mathrm{Re}=1100$}
    \label{fig:secondary-d}
  \end{subfigure}

  \caption{Performance evaluation of the \textbf{unified foundation model} for secondary flow in \textbf{square duct configurations} against the ground truth and the baseline \(k\)--\(\omega\) model. Cross-sectional secondary velocity profiles \(U_y\) are shown at four Reynolds numbers: (a) \(\mathrm{Re}=3500\), (b) \(\mathrm{Re}=2600\), (c) \(\mathrm{Re}=1800\) and (d) \(\mathrm{Re}=1100\). Each panel plots \(U_y\) at multiple wall-normal locations \(z/h_{\mathrm{s}}\), where \(h_{\mathrm{s}}\) denotes half the duct height; horizontal offsets of \(+10.0\,U_y\) are applied for visualization.}
  
  \label{fig:secondary-flow-results-1}
\end{figure}

\begin{figure}[!htb]
  \centering

  \includegraphics[width=0.8\textwidth]{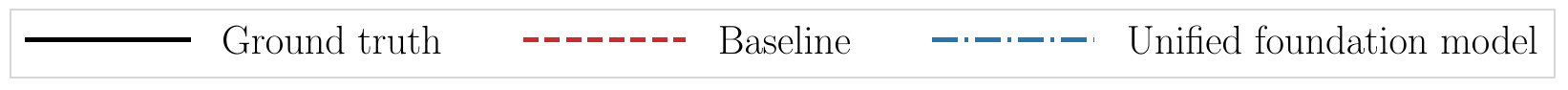}

  \begin{subfigure}[t]{0.48\textwidth}
    \centering
    \includegraphics[width=\textwidth]{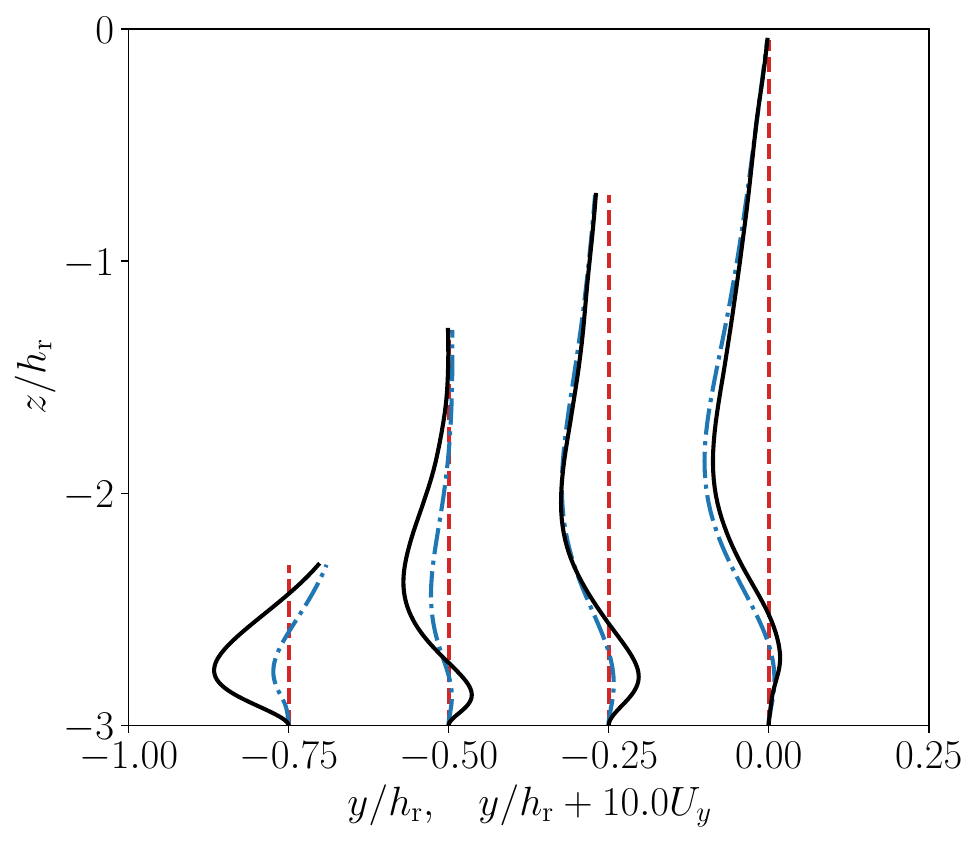}
    \caption{$\mathrm{AR}=3$}
    \label{fig:secondary2-a}
  \end{subfigure}
  \hspace{0.02\textwidth}
  \begin{subfigure}[t]{0.48\textwidth}
    \centering
    \includegraphics[width=\textwidth]{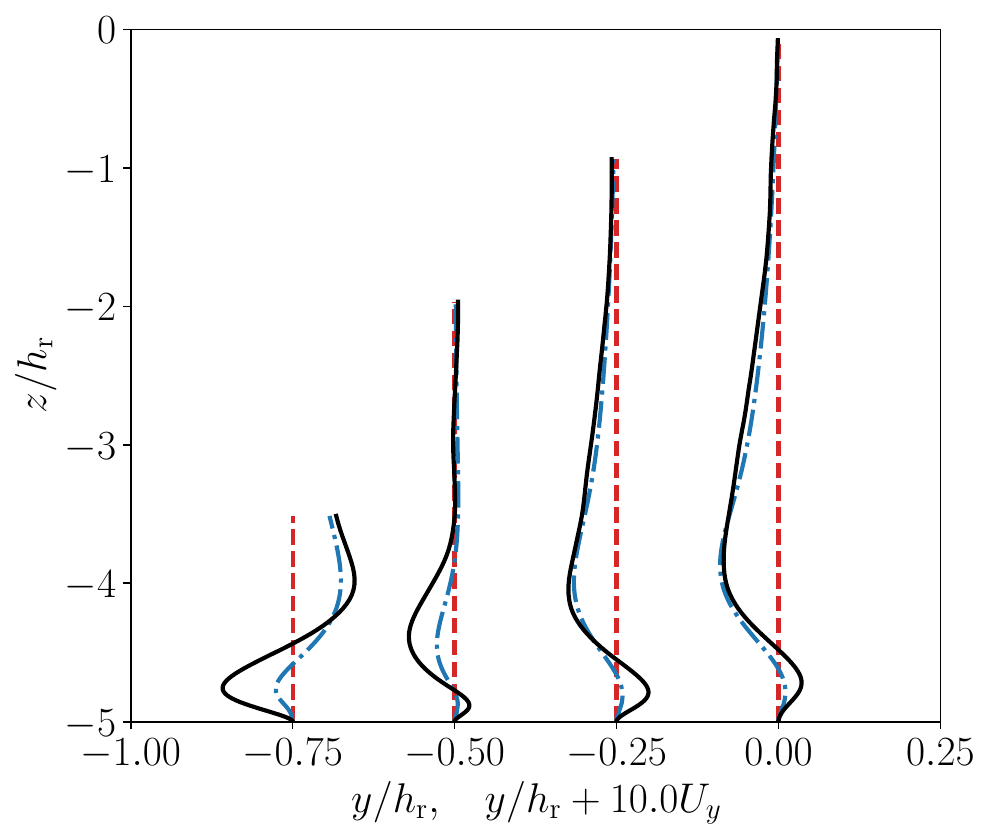}
    \caption{$\mathrm{AR}=5$}
    \label{fig:secondary2-b}
  \end{subfigure}

  \begin{subfigure}[t]{0.48\textwidth}
    \centering
    \includegraphics[width=\textwidth]{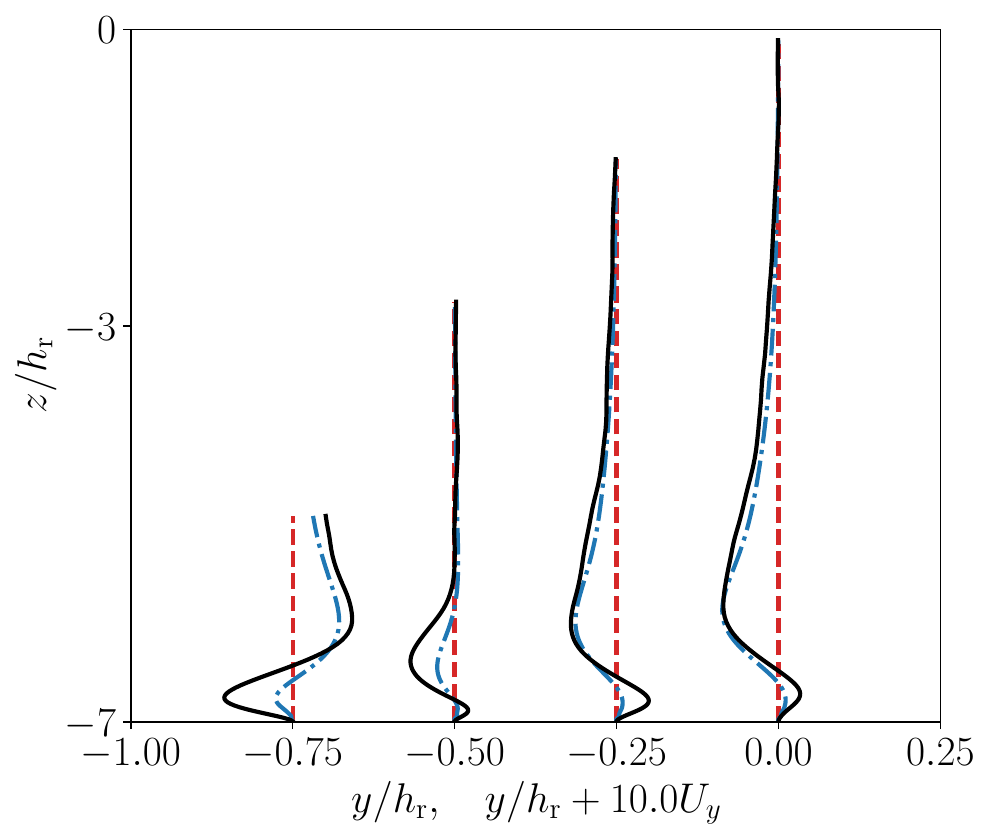}
    \caption{$\mathrm{AR}=7$}
    \label{fig:secondary2-c}
  \end{subfigure}
  \hspace{0.02\textwidth}
  \begin{subfigure}[t]{0.48\textwidth}
    \centering
    \includegraphics[width=\textwidth]{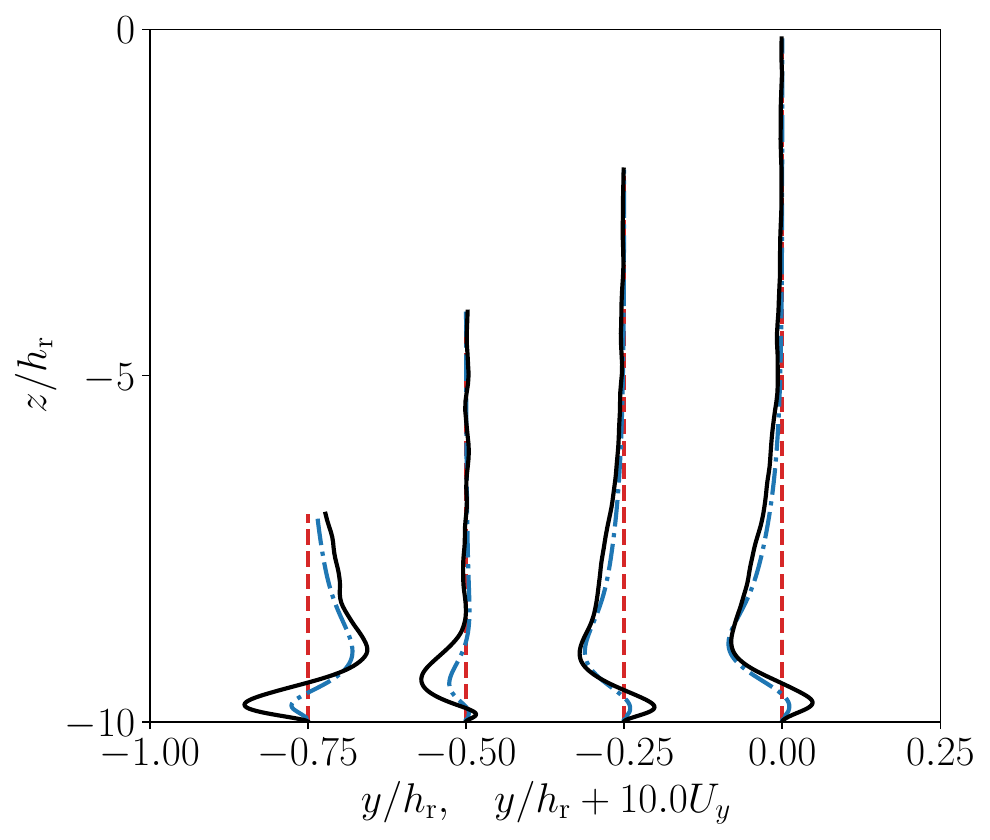}
    \caption{$\mathrm{AR}=10$}
    \label{fig:secondary2-d}
  \end{subfigure}

  \caption{Performance evaluation of the \textbf{unified foundation model} for secondary flow in \textbf{rectangular duct configurations} against the ground truth and the baseline \(k\)--\(\omega\) model. Cross-sectional secondary velocity profiles \(U_y\) are shown for four different aspect ratios (a) \(\mathrm{AR}=3\), (b) \(\mathrm{AR}=5\), (c) \(\mathrm{AR}=7\) and (d) \(\mathrm{AR}=10\). Each panel plots \(U_y\) at multiple wall-normal locations \(z/h_{\mathrm{r}}\), where \(h_{\mathrm{r}}\) denotes half the height of the rectangular duct; horizontal offsets of \(+10.0\,U_y\) are applied for visualization.}
  
  \label{fig:secondary-flow-results-2}
\end{figure}

\paragraph{Secondary flows}
The unified foundation model accurately predicts secondary flows across a wide range of duct configurations, generalizing beyond its training conditions. We evaluate the model on eight cases as shown in Fig.~\ref{fig:secondary-flow-results-1} and~\ref{fig:secondary-flow-results-2}: four square ducts at decreasing Reynolds numbers (\(\mathrm{Re} = 3500,\, 2600,\, 1800,\, 1100\)) and four rectangular ducts at increasing aspect ratios (AR = 3, 5, 7, 10). Among these, the square duct at \(\mathrm{Re} = 1100, \, 2600\) and the rectangular duct at AR = 5 appear in the training set; all other cases serve as generalization tests. For the square ducts, lower Reynolds numbers increase the sensitivity of secondary flow structures to turbulence model accuracy due to stronger viscous effects. Across all Reynolds numbers, the unified foundation model outperforms the baseline \(k\)--\(\omega\) model, maintaining closer agreement with DNS results in Fig.~\ref{fig:secondary-flow-results-1}. For the rectangular ducts, increasing the aspect ratio amplifies and distorts secondary flows, leading to anisotropic patterns that the baseline \(k\)--\(\omega\) model fails to capture. The unified foundation model captures both the amplification and spatial shifting of secondary vortices with aspect ratio, preserving their qualitative structure and quantitative magnitude in Fig.~\ref{fig:secondary-flow-results-2}. These results confirm that the model generalizes effectively to both Reynolds number and geometric extrapolation, demonstrating robustness for complex secondary flows beyond the training set.

\begin{figure}[!htb]
  \centering
  \includegraphics[width=0.8\textwidth]{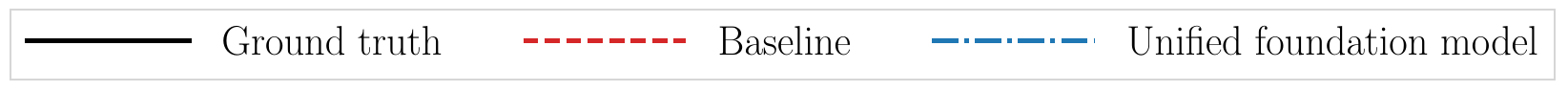}
  \begin{subfigure}[t]{0.48\textwidth}
    \centering
    \includegraphics[width=\textwidth]{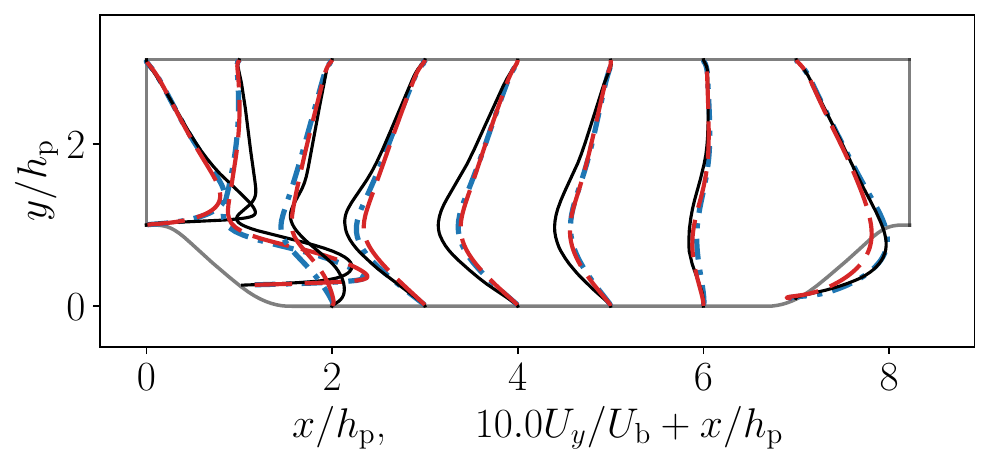}
    \caption{Periodic hill, $\alpha = 0.8$}
    \label{fig:sep-hill-a}
  \end{subfigure}
  \hspace{0.02\textwidth}
  \begin{subfigure}[t]{0.48\textwidth}
    \centering
    \includegraphics[width=\textwidth]{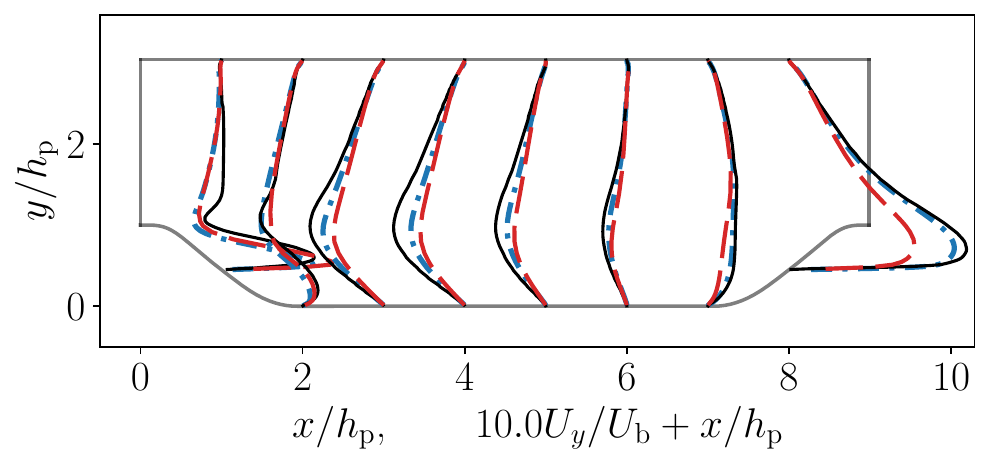}
    \caption{Periodic hill, $\alpha = 1.0$}
    \label{fig:sep-hill-b}
  \end{subfigure}
  \begin{subfigure}[t]{0.48\textwidth}
    \centering
    \includegraphics[width=\textwidth]{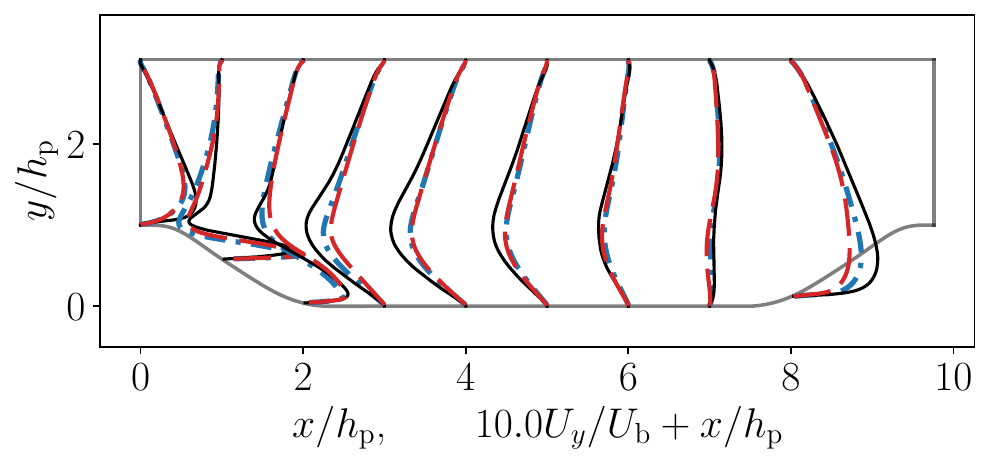}
    \caption{Periodic hill, $\alpha = 1.2$}
    \label{fig:sep-hill-c}
  \end{subfigure}
  \hspace{0.02\textwidth}
  \begin{subfigure}[t]{0.48\textwidth}
    \centering
    \includegraphics[width=\textwidth]{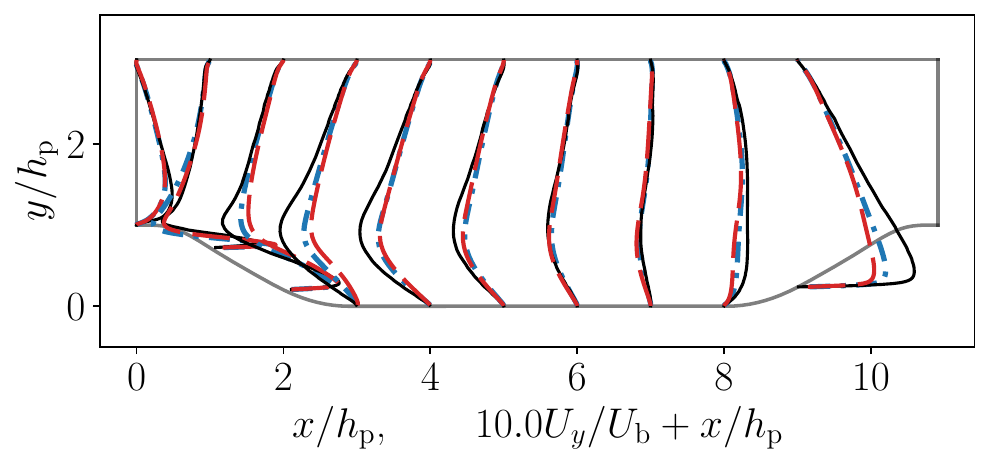}
    \caption{Periodic hill, $\alpha = 1.5$}
    \label{fig:sep-hill-d}
  \end{subfigure}
  \caption{Performance evaluation of the \textbf{unified foundation model} for \textbf{periodic hill flows}, in comparison with the ground truth and the baseline $k$--$\omega$ model. Vertical velocity profiles $U_y$, normalized by the bulk velocity $U_{\mathrm{b}}$, are shown for periodic hill configurations with slope scaling factors (a) $\alpha_{\text{p}} = 0.8$, (b) $\alpha_{\text{p}} = 1.0$, (c) $\alpha_{\text{p}} = 1.2$, and (d) $\alpha_{\text{p}} = 1.5$. The profiles are plotted as $10\,U_y/U_{\mathrm{b}} + x/h_{\text{p}}$ versus the normalized wall-normal coordinate $y/h_{\text{p}}$ at multiple streamwise locations $x/h_{\text{p}}$.}
  
  \label{fig:separation-periodic-hills}
\end{figure}

\begin{figure}[!htb]
  \centering

  \includegraphics[width=0.8\textwidth]{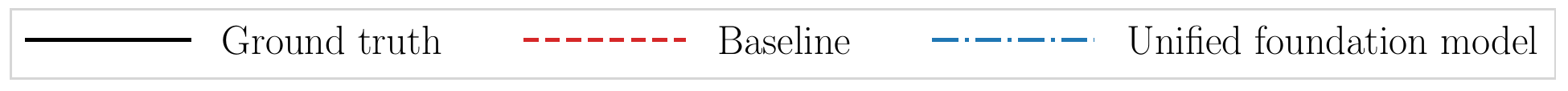}
  \begin{subfigure}[t]{0.48\textwidth}
    \centering
    \includegraphics[width=\textwidth]{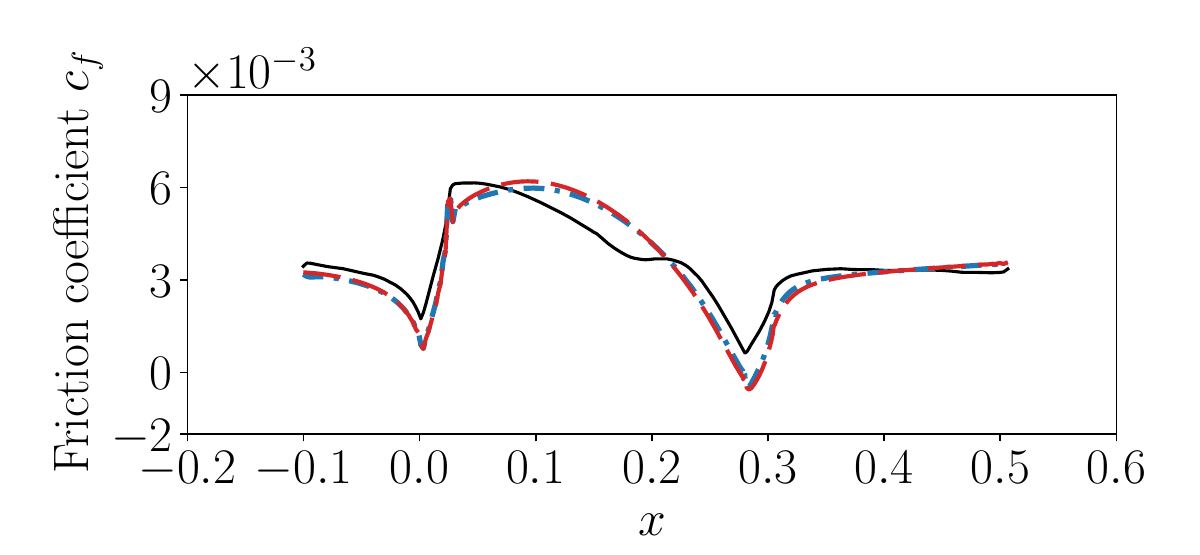}
    \caption{Bump, height = 20mm}
    \label{fig:bump-a}
  \end{subfigure}
  \hspace{0.02\textwidth}
  \begin{subfigure}[t]{0.48\textwidth}
    \centering
    \includegraphics[width=\textwidth]{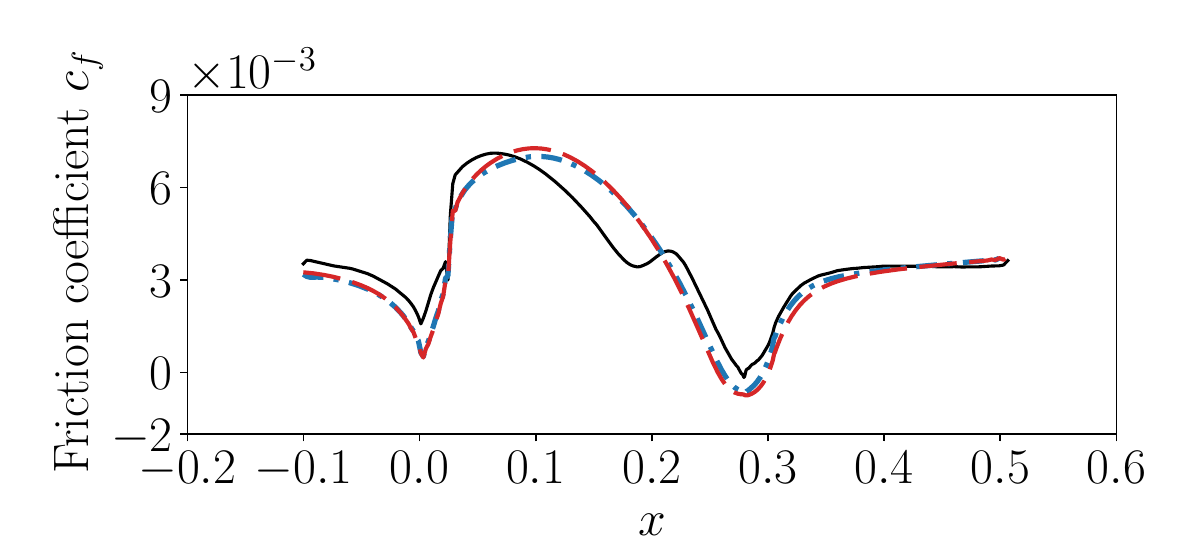}
    \caption{Bump, height = 26mm}
    \label{fig:bump-b}
  \end{subfigure}
  
  \vspace{0.01em}
  \begin{subfigure}[t]{0.48\textwidth}
    \centering
    \includegraphics[width=\textwidth]{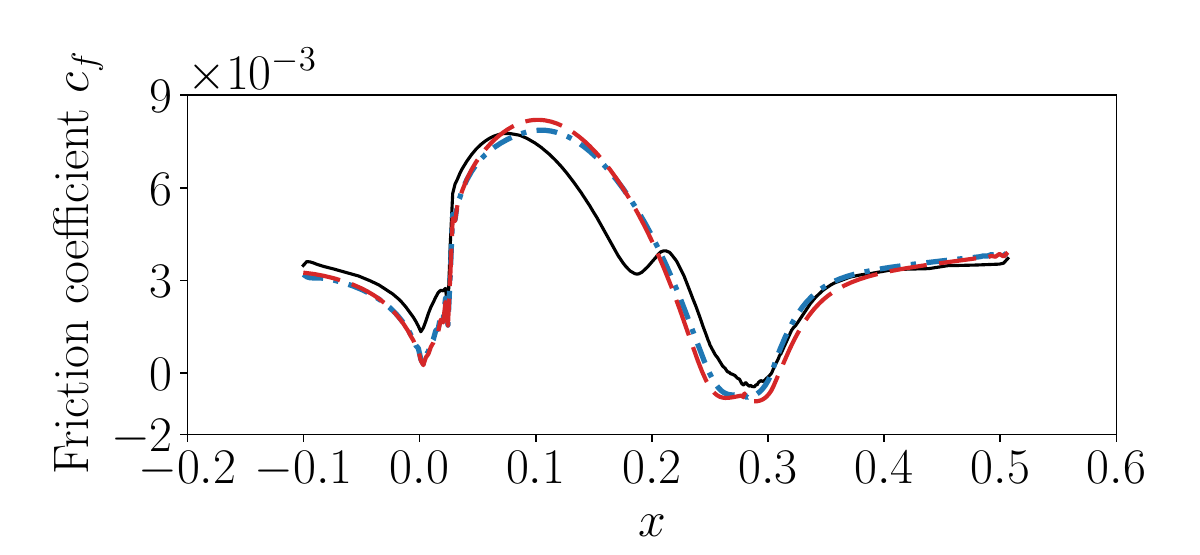}
    \caption{Bump, height = 31mm}
    \label{fig:bump-c}
  \end{subfigure}
  \hspace{0.02\textwidth}
  \begin{subfigure}[t]{0.48\textwidth}
    \centering
    \includegraphics[width=\textwidth]{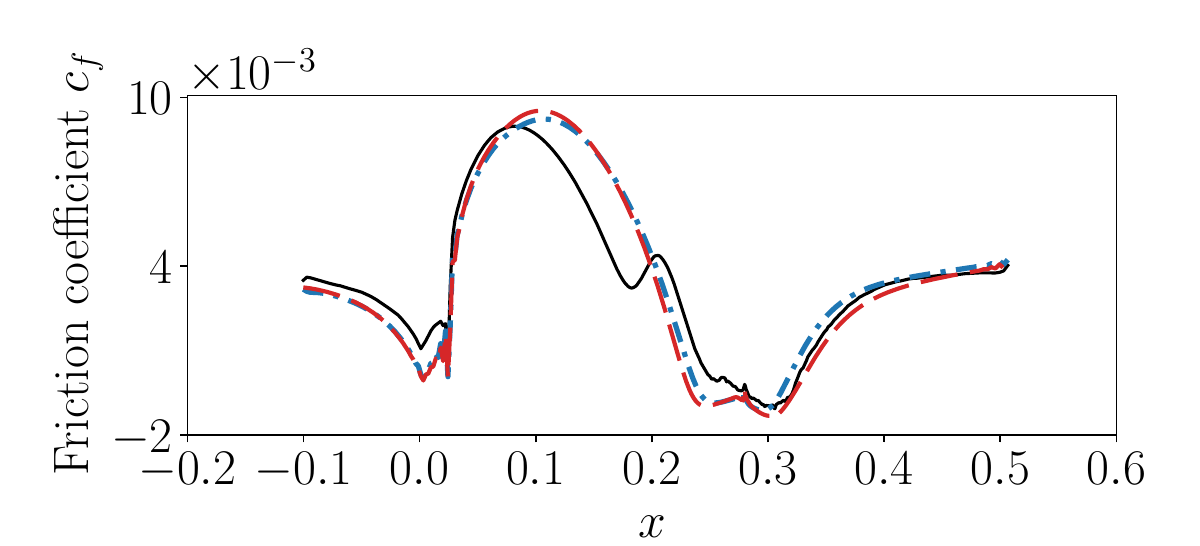}
    \caption{Bump, height = 38mm}
    \label{fig:bump-d}
  \end{subfigure}
  
  \vspace{0.01em}
  \begin{subfigure}[t]{0.48\textwidth}
    \centering
    \includegraphics[width=\textwidth]{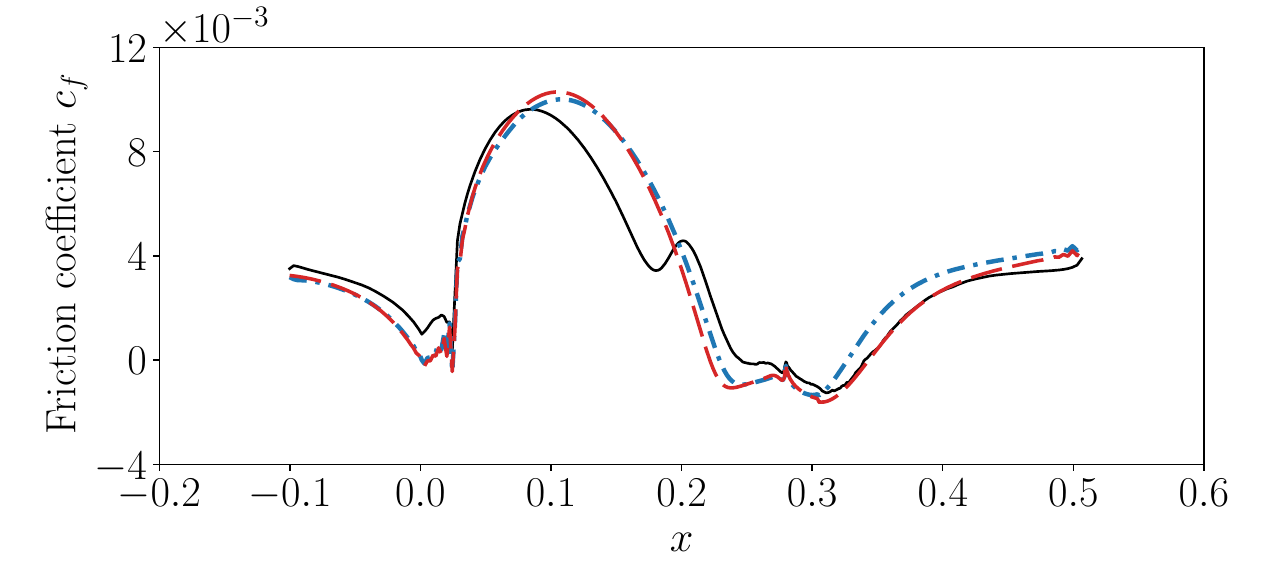}
    \caption{Bump, height = 42mm}
    \label{fig:bump-e}
  \end{subfigure}
  \caption{Performance evaluation of the \textbf{unified foundation model} for \textbf{bump flows} against the ground truth and the baseline $k$--$\omega$ model. Streamwise distributions of the wall skin-friction coefficient \(c_f\) are shown along the lower surface for bump heights of (a) \(20\,\mathrm{mm}\), (b) \(26\,\mathrm{mm}\), (c) \(31\,\mathrm{mm}\), (d) \(38\,\mathrm{mm}\), and (e) \(42\,\mathrm{mm}\). }
  
  \label{fig:separation-bump}
\end{figure}

\begin{figure}[!htb]
  \centering

  \includegraphics[width=0.8\textwidth]{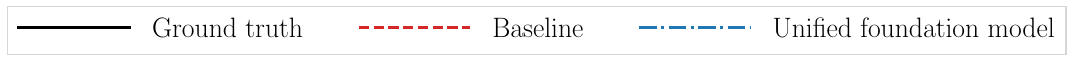}

  \begin{subfigure}[t]{0.45\textwidth}
    \centering
    \includegraphics[width=0.8\textwidth]{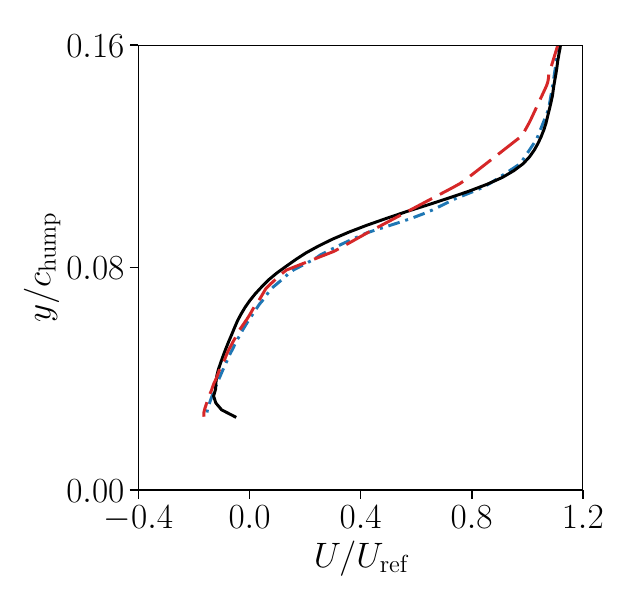}
    \caption{$x/c_{\mathrm{hump}} = 0.8$}
    \label{fig:hump-a}
  \end{subfigure}
  \hspace{0.02\textwidth}
  \begin{subfigure}[t]{0.45\textwidth}
    \centering
    \includegraphics[width=0.8\textwidth]{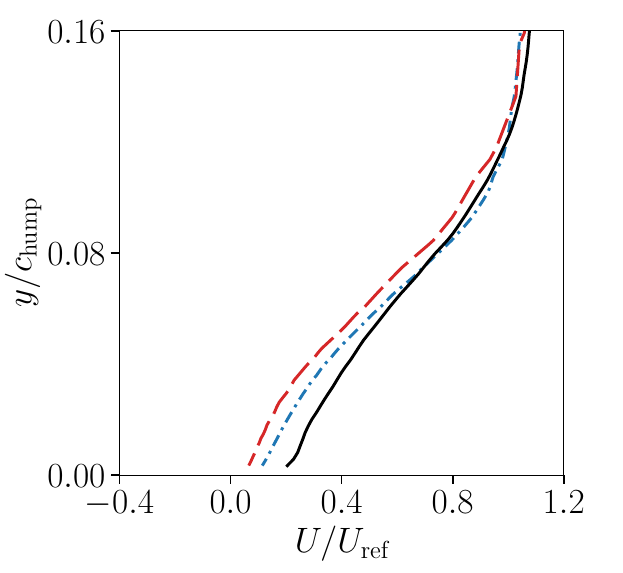}
    \caption{$x/c_{\mathrm{hump}} = 1.2$}
    \label{fig:hump-b}
  \end{subfigure}

  \caption{Performance evaluation of the \textbf{unified foundation model} in \textbf{hump flow} against the ground truth and the baseline $k$--$\omega$ model. Mean streamwise velocity profiles \(U/U_{\mathrm{ref}}\) are shown at two streamwise locations, (a) \(x/c_{\mathrm{hump}} = 0.80\) and (b) \(x/c_{\mathrm{hump}} = 1.20\), representing regions upstream and downstream of separation, respectively. The $y$-axis represents the wall-normal coordinate, normalized by $c_{\mathrm{hump}}$.}
  \label{fig:separation-hump}
\end{figure}

\begin{figure}
      \centering
      \includegraphics[width=0.72\textwidth]{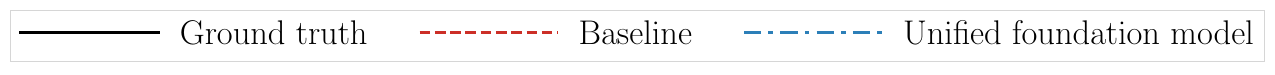}
      
  \begin{subfigure}[t]{0.43\textwidth}
    \centering
    \includegraphics[width=1\textwidth]{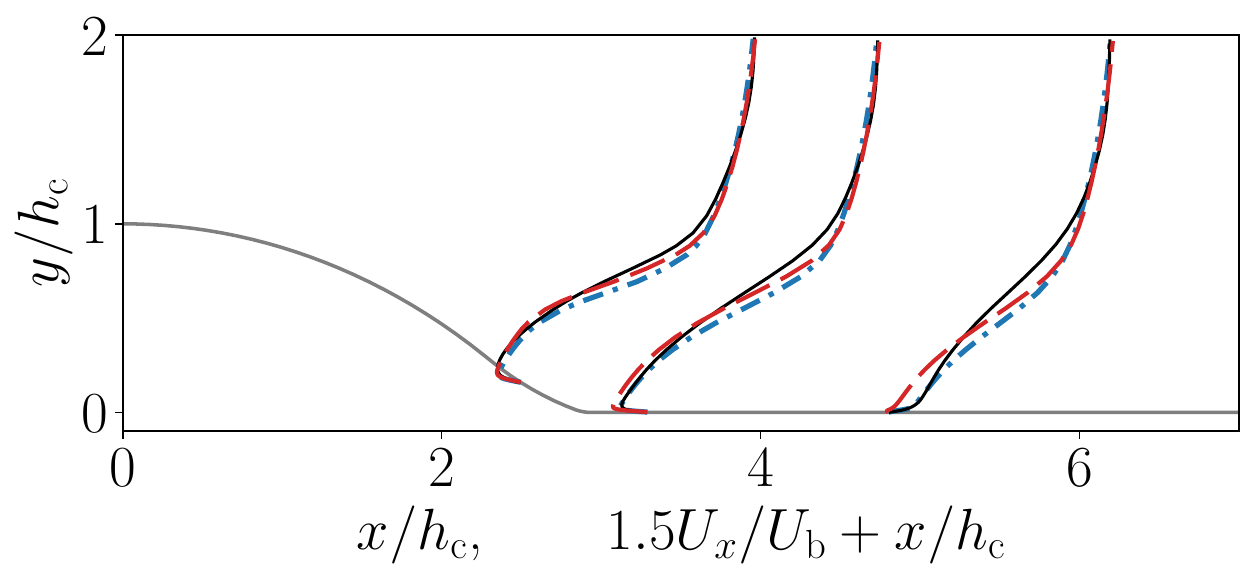}
    \caption{Streamwise velocity profiles}
    \label{fig:drag}
  \end{subfigure}
  \begin{subfigure}[t]{0.54\textwidth}
    \centering
    \includegraphics[width=1\textwidth]{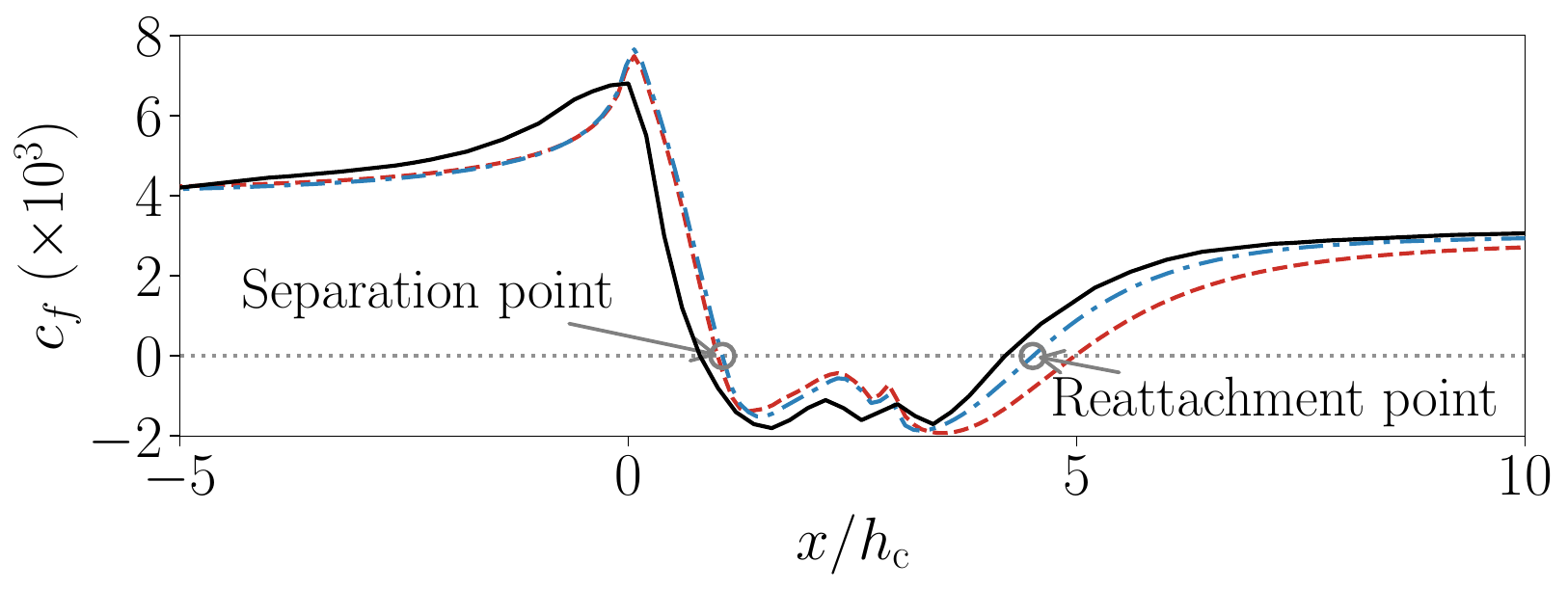}
    \caption{Friction coefficient}
  \end{subfigure}
  \caption{\added[id=R2]{Performance evaluation of the \textbf{unified foundation model} in a \textbf{curved step flow} against the ground truth and the baseline $k$--$\omega$ model. (a) Streamwise velocity profiles \(U_x\), normalized by the bulk velocity \(U_{\mathrm{b}}\), are shown at multiple streamwise locations downstream of the curved step. The profiles $U_x$ versus the wall-normal coordinate \(y\) normalized by the curved-step height \(h_\mathrm{c}\). (b) Skin-friction coefficient distributions $c_f$, are shown along the streamwise direction $x/h_{\mathrm{c}}$. Separation and reattachment points are marked using circles.}}
  \label{fig:curved-step-2}
\end{figure}

\paragraph{Separated flows}
The unified foundation model improves prediction accuracy for many separated flows. We evaluate the unified foundation model on seven types of benchmark cases: periodic hill flows with different slope scaling factors \(\alpha_{\text{p}}=0.8, 1.0, 1.2, 1.5\), flow over a bump with different bump heights \(h_{\text{b}}=20\,\mathrm{mm}, 26\,\mathrm{mm}, 31\,\mathrm{mm}, 38\,\mathrm{mm}, 42\,\mathrm{mm}\), flow over a curved step, flow over a hump, S809 airfoil cases at three high angles of attack $\text{AOA} = 14^\circ, 16^\circ, 18^\circ$ and NACA 0012 airfoil cases at one high angle of attack ($18^\circ$). Among these, the periodic hill flow with slope scaling factor \(\alpha_{\text{p}} = 1.0\), the S809 airfoil flow at \(\text{AOA} = 14^\circ\), flow over a curved step, flow over a hump and flow over a bump appear in the training set. All seven types of cases are challenging for linear eddy viscosity models because the local equilibrium assumption and Boussinesq hypothesis fail in separated regions. Test results are shown in Figs.~\ref{fig:cd_cl_comparison}--\ref{fig:curved-step-2} and Table~\ref{tab:naca0012_lift}. 

The unified foundation model improves predictions over the baseline $k$--$\omega$ model for strongly separated flows, particularly in airfoil and periodic hill configurations. For the S809 airfoil, the unified foundation model generalizes accurately across a wide range of angles of attack, closely following experimental trends in both lift and drag (Fig.~\ref{fig:cd_cl_comparison}). In the regime with large flow separation, the baseline model substantially overpredicts lift and underpredicts drag, failing to capture the nonlinear separation-driven transitions, especially at $\mathrm{AOA} = 18^\circ$. In contrast, the unified foundation model reproduces the sharp rise in $c_d$ and the drop in $c_l$ with higher fidelity. Similar improvements are observed for periodic hill flows with varying slope steepness (Fig.~\ref{fig:separation-periodic-hills}), where steeper slopes intensify separation and delay reattachment. Across all slopes, the unified foundation model reproduces the asymmetric development of the separated shear layer and its downstream recovery more faithfully than the baseline predictions.

For cases with incipient or moderate separation, the unified foundation model exhibits performance that is comparable to, and occasionally slightly better than, the baseline $k$--$\omega$ model. In the bump flows (Fig.~\ref{fig:separation-bump}), the predicted skin-friction coefficient distributions closely follow the baseline across all bump heights, with only minor deviations near separation and reattachment and nearly identical results for the largest bump height (Fig.~\ref{fig:separation-bump}e). In the hump case, the unified foundation model achieves higher overall accuracy than the baseline model, especially producing fuller near-wall velocity profiles and improved gradients near separation and reattachment, particularly at $x/c_{\mathrm{hump}} = 0.80$ (Fig.~\ref{fig:separation-hump}). For the curved step case, the unified model yields slightly more accurate normalized velocity profiles downstream of separation, indicating improved representation of near-wall behavior (Fig.~\ref{fig:curved-step-2}a). The skin-friction coefficient distribution, though not observed during training, shows more accurate predictions of separation and reattachment locations (Fig.~\ref{fig:curved-step-2}b), indicating improved generalization and a more consistent representation of the flow field. For the NACA 0012 airfoil, the unified foundation model achieves slightly better performance than the baseline model under attached-flow conditions ($\mathrm{AOA} = 10^\circ, 15^\circ$) as shown in Table~\ref{tab:naca0012_lift}, while at $\mathrm{AOA} = 18^\circ$ the unified foundation model modestly reduces lift overprediction relative to the baseline, suggesting limited but consistent gains in post-separation regimes.

\begin{table}[h!]
  \centering
  \renewcommand{\arraystretch}{1.3}
  \setlength{\tabcolsep}{15pt}
  \begin{tabular}{lccc}
    \specialrule{1pt}{0pt}{0pt}
    NACA 0012 cases  & Ground truth & Baseline $k$--$\omega$ model & \textbf{Unified foundation model} \\
    \specialrule{1pt}{0pt}{0pt}
    \(\text{AOA} = 10^\circ\)  & 1.08 & 1.12 & \textbf{1.11} \\
    \(\text{AOA} = 15^\circ\)  & 1.49 & 1.65 & \textbf{1.58} \\
    \(\text{AOA} = 18^\circ\)  & 1.01 & 1.75 & \textbf{1.69} \\
    \specialrule{1pt}{0pt}{0pt}
  \end{tabular}
  \caption{Performance evaluation of the \textbf{unified foundation model} for \textbf{NACA 0012 airfoil flow} against the ground truth and the baseline $k$--$\omega$ model. Lift coefficient predictions are listed for the NACA 0012 airfoil at three angles of attack (\(\mathrm{AOA} = 10^\circ,\ 15^\circ,\ 18^\circ\)).}
  \label{tab:naca0012_lift}
\end{table}

\begin{figure}[!htb]
  \centering

  \begin{subfigure}[t]{0.65\textwidth}
    \centering
    \includegraphics[width=0.8\textwidth]{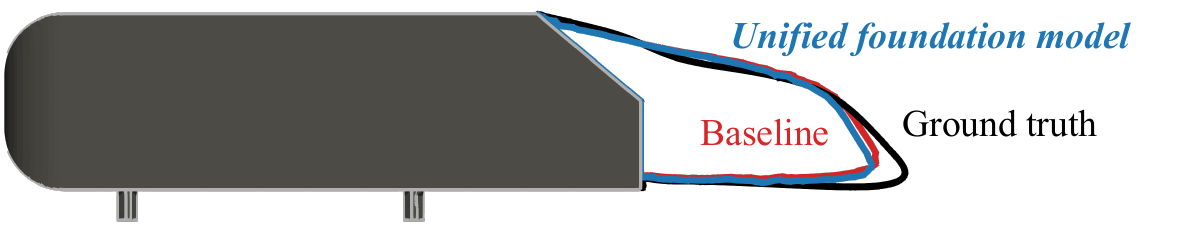}
    \caption{\added[id=Author]{Separation region comparison for generic car}}
    \label{fig:foundation-3d-car}
  \end{subfigure}

  \vspace{0.5em}

  \begin{subfigure}[t]{0.65\textwidth}
    \centering
    \includegraphics[width=0.8\textwidth]{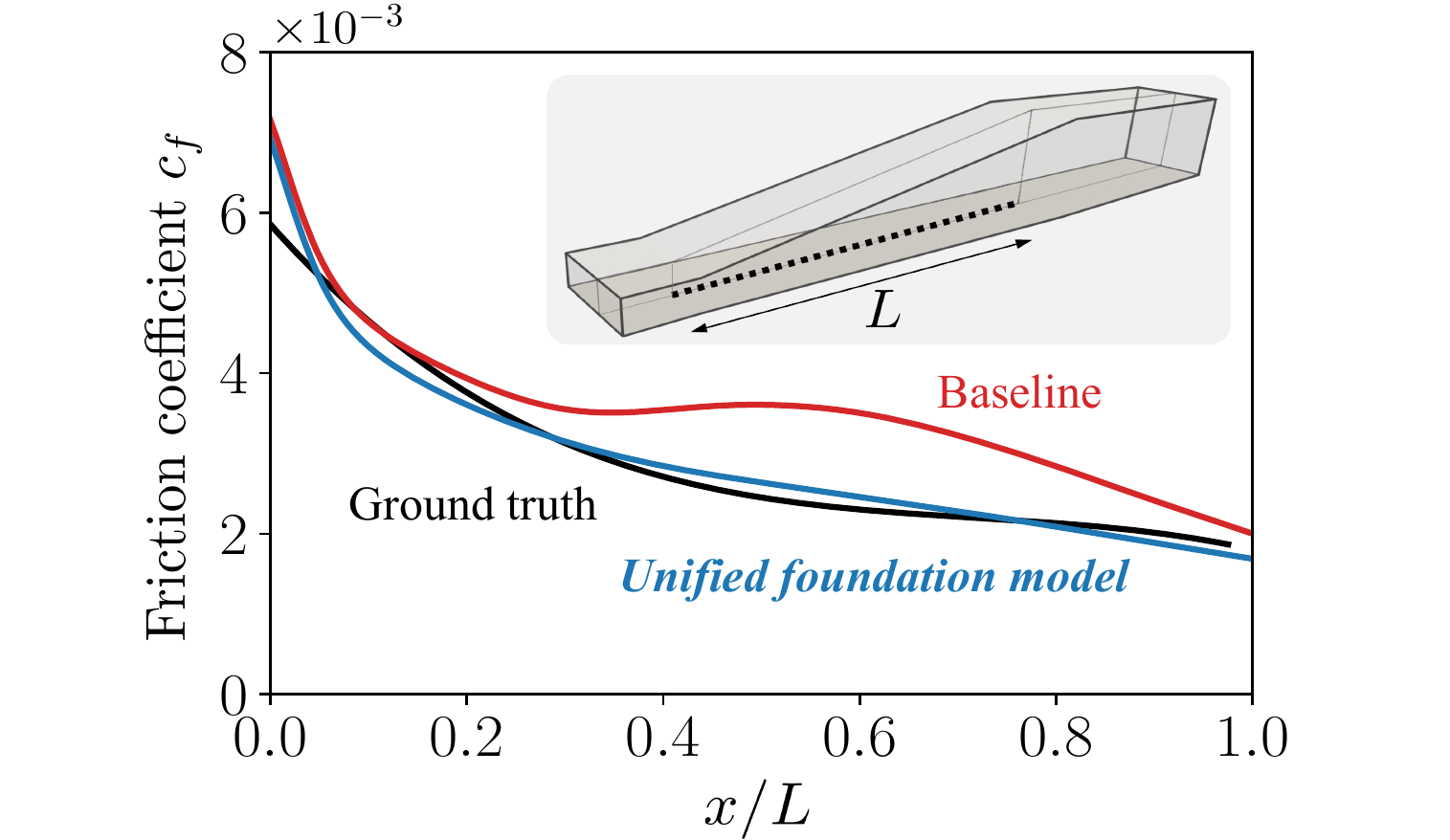}
    \caption{Friction coefficient comparison for asymmetric 3D diffuser }
    \label{fig:foundation-3d-diffuser}
  \end{subfigure}

\begin{subfigure}[t]{0.95\textwidth}
  \centering

  \begin{minipage}{0.25\textwidth}
    \centering
    \includegraphics[width=\textwidth]{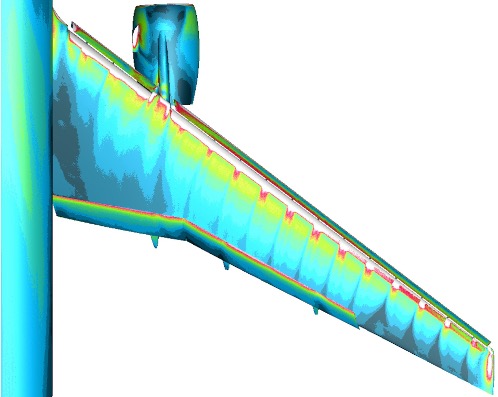}
    \\ \small Ground truth
  \end{minipage}
  \hfill
  \begin{minipage}{0.25\textwidth}
    \centering
    \includegraphics[width=\textwidth]{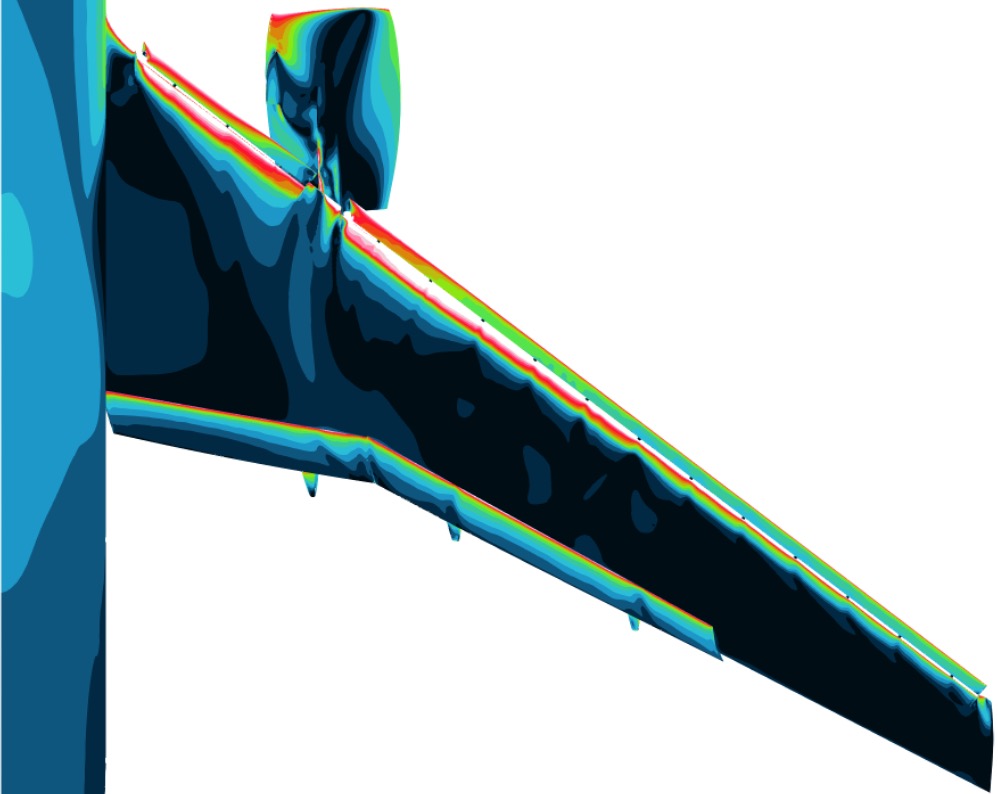}
    \\ \small \textcolor{red}{Baseline}
  \end{minipage}
  \hfill
  \begin{minipage}{0.25\textwidth}
    \centering
    \includegraphics[width=\textwidth]{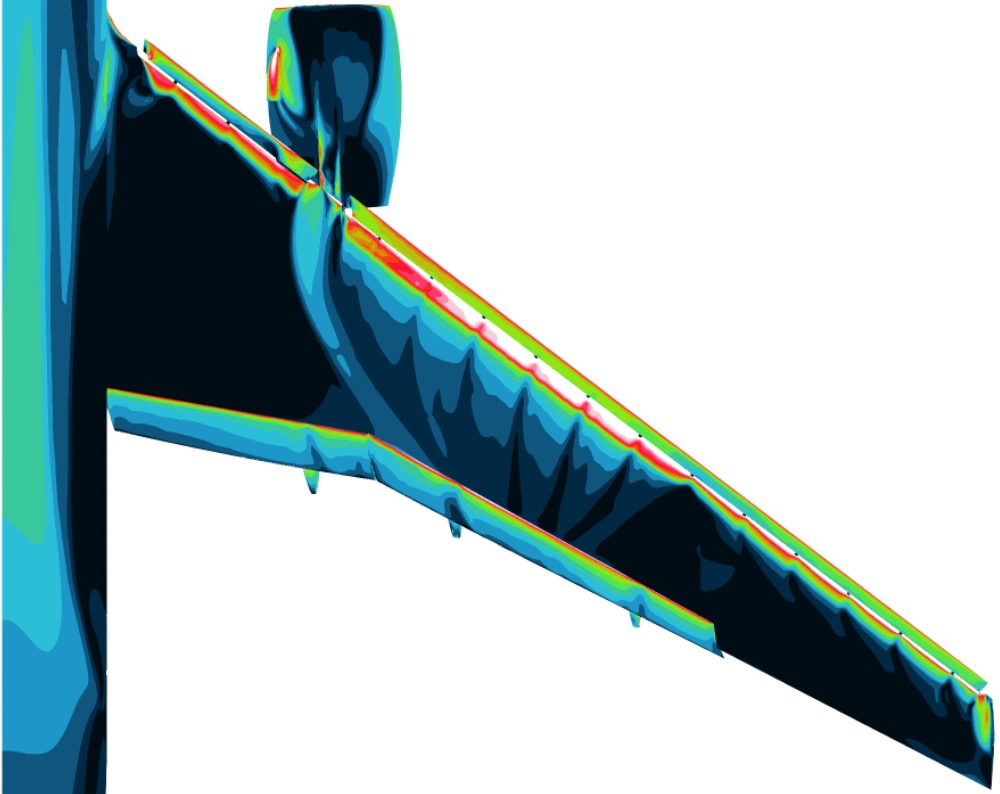}
    \\ \small \textcolor{darkblue}{\textit{\textbf{Unified foundation model}}}
  \end{minipage}  
  \hfill
  \begin{minipage}{0.07\textwidth}
    \centering
    \includegraphics[width=\textwidth]{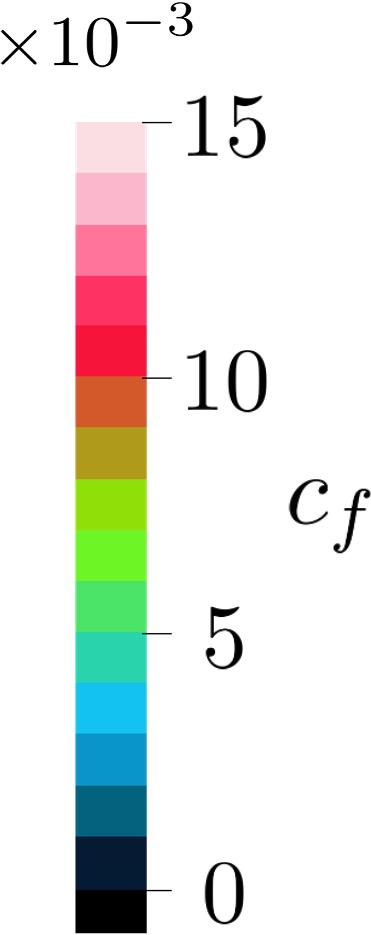}
  \end{minipage}  
  \caption{\added[id=Author]{Comparison of skin-friction coefficients on the generic aircraft predicted by different models at $\mathrm{AOA}=21.47^\circ$}}
  \label{fig:foundation-3d-aircraft}
\end{subfigure}

  \caption{Performance of the \textbf{unified foundation model} on \textbf{complex three-dimensional configurations}, including a generic car, an asymmetric diffuser, and a generic aircraft, in comparison with the baseline $k$--$\omega$ model and ground truth. Panel (a) shows the predicted separation region for the generic car, where solid contours indicate the wake and the $U_x=0$ contour delineates flow separation. Panel (b) presents the skin-friction coefficient $c_f$ for the asymmetric diffuser along the bottom-wall midline. \added[id=Author]{Panel (c) shows the surface skin-friction coefficient on the generic aircraft at $\mathrm{AOA}=21.47^\circ$ predicted by the baseline and the unified foundation model, compared with the ground truth~\cite{goc2024wind}.}}
  
  \label{fig:foundation-3d-results}
\end{figure}

\begin{figure}
  \centering
  \includegraphics[width=0.99\textwidth]{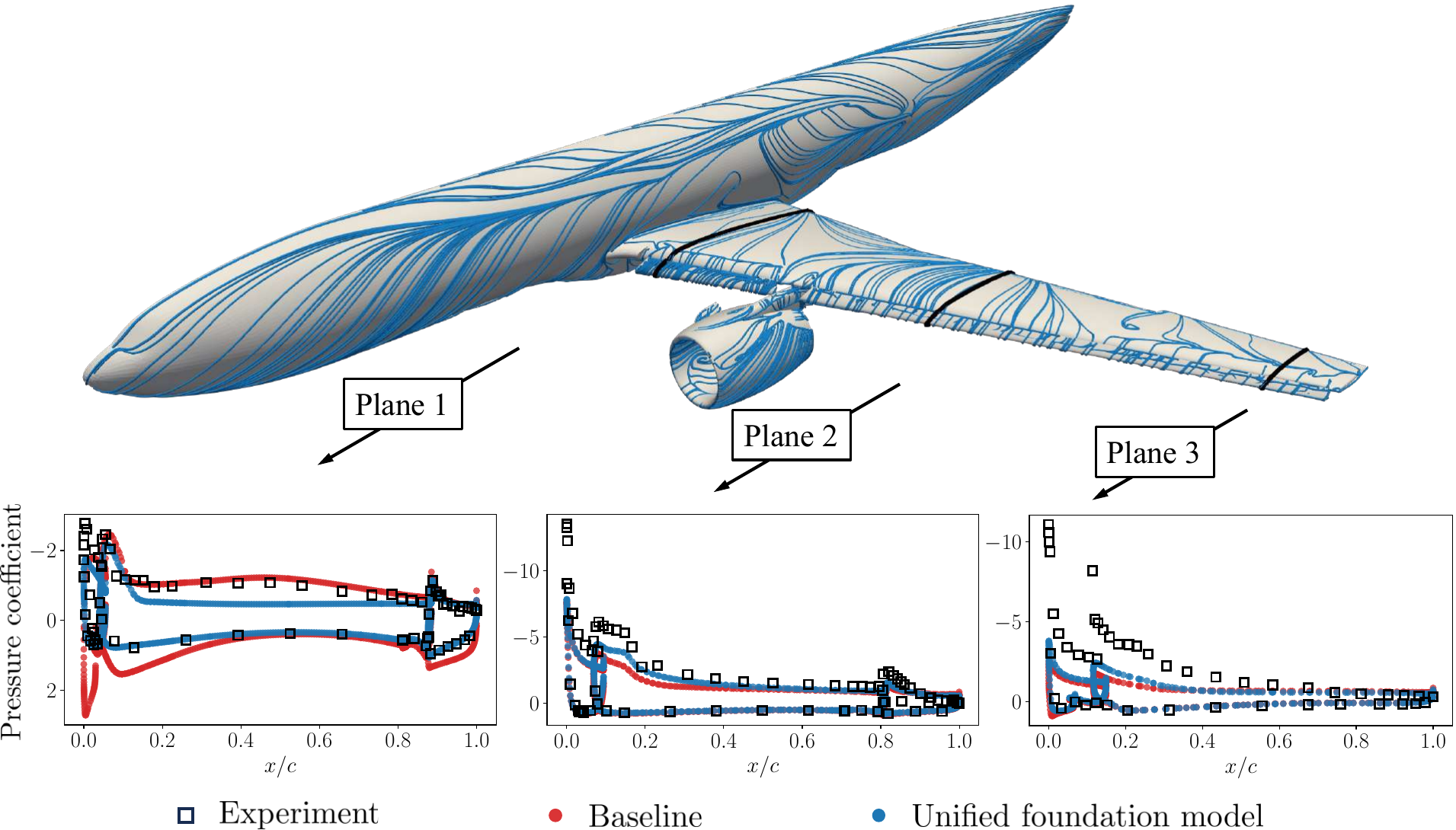}
  \caption{\added[id=Author]{Surface streamlines on the generic aircraft at $\mathrm{AOA}=21.47^\circ$ predicted by the unified foundation model, together with pressure-coefficient distributions at three spanwise sections: Plane~1 (root), Plane~2 (mid-span), and Plane~3 (tip). The baseline model predicts a dominant spanwise flow, whereas the unified model captures the mid-span flow behavior without large-scale separation, consistent with experiments~\cite{evans2020test}. This improves the pressure prediction at Plane~2 and yields slight improvement at Plane~3. At Plane~1, the unified model improves the pressure distribution on the lower surface, while discrepancies remain on the upper surface; the associated separation behavior contributes to the underprediction of lift.}}
\label{fig:crm-pressure}
\end{figure}

\paragraph{Complex 3D flows}
\deleted[id=Author]{The unified foundation model consistently outperforms the baseline $k$--$\omega$ model across both complex 3D cases shown in Fig.~\ref{fig:foundation-3d-results}.}
\added[id=Author]{The unified foundation model yields predictions that are either improved or comparable to the baseline $k$--$\omega$ model across the complex three-dimensional cases, without degrading performance (Fig.~\ref{fig:foundation-3d-results})}.
\deleted[id=Author]{In the generic car case, the unified foundation model more closely aligns with the ground truth in predicting the separated wake, although it still underpredicts the wake extent (see Fig.~\ref{fig:foundation-3d-results}a). In this case, the quadratic term $g^{(2)}$ is manually deactivated due to the model's sensitivity to this term, without noticeably affecting prediction accuracy. In contrast, the baseline model fails to capture the full separation, producing a wake region that is significantly contracted relative to the ground truth. }
\added[id=Author]{In the generic car case, the unified foundation model and the baseline model predict a similar separated wake, with nearly identical wake extent (see Fig.~\ref{fig:foundation-3d-results}a). Consistent with this observation, the predicted drag coefficients are also very similar for the two models (Table~\ref{tab:drag_coeff}). In this configuration, the unified foundation model performs on par with the baseline, without noticeable degradation. The quadratic term $g^{(2)}$ is manually deactivated due to its sensitivity, without affecting the prediction.}
In the asymmetric 3D diffuser, the unified foundation model closely follows the gradual downstream decay of the wall skin friction \(c_f\) observed in the ground truth, whereas the baseline model markedly overpredicts \(c_f\) and maintains elevated values far downstream (see Fig.~\ref{fig:foundation-3d-results}b).
\added[id=Author]{The generic aircraft serves as a benchmark configuration in the aerospace community and provides a stringent test across flow regimes. At a low angle of attack (pre-stall, $\mathrm{AOA}=7.05^\circ$), the baseline model performs well, and the unified foundation model recovers this behavior while slightly improving the predictions of lift and drag. At a high angle of attack (post-stall, $\mathrm{AOA}=21.47^\circ$), where strong separation occurs, the baseline model shows significant deficiencies, whereas the unified model improves the predictions, particularly for drag (see Table~\ref{tab:drag_lift_coeff}). The flow over the wing is also better captured, as indicated by improved skin-friction distributions (see Fig.~\ref{fig:foundation-3d-results}c) and pressure coefficients at multiple spanwise sections (see Fig.~\ref{fig:crm-pressure}). The unified model reduces the excessive spanwise flow predicted by the baseline model and recovers the mid-span flow behavior without large-scale separation, leading to improved pressure predictions at the mid-span and tip sections.
In this case, the quadratic term $g^{(2)}$ is also deactivated due to stability issues.
Despite these improvements, discrepancies remain at high angle of attack. The lift is underpredicted compared to both the ground truth and the baseline model, and the pressure distributions still deviate from the ground truth, particularly near the root section. These errors indicate that the separation is not fully captured, which limits the accuracy of the integrated force predictions.}

\begin{table}
  \centering
  \small
  \renewcommand{\arraystretch}{1.2}
  \setlength{\tabcolsep}{6pt}
  \begin{tabular}{lcccc}
    \specialrule{1pt}{0pt}{0pt}
     Angle of attack& \multicolumn{2}{c}{\textbf{$7.05^{\circ}$}} & \multicolumn{2}{c}{\textbf{$21.47^{\circ}$}} \\
    \hline

    & \textbf{\(c_d\)} & \textbf{\(c_l\)} & \textbf{\(c_d\)} & \textbf{\(c_l\)} \\

    \specialrule{1pt}{0pt}{0pt}
    Ground truth 
    & 0.16 & 1.80
    & 0.39 & 2.39 \\

    Baseline \(k\)--\(\omega\) model 
    & 0.21 & 1.77 
    & 0.73 & 2.20 \\

    Unified foundation model 
    & \textbf{0.20} & \textbf{1.81} 
    & \textbf{0.47} & \textbf{1.75} \\
    \specialrule{1pt}{0pt}{0pt}
  \end{tabular}
  \caption{\added[id=Author]{Drag and lift coefficients ($c_d$, $c_l$) for the generic aircraft predicted by the unified foundation model, compared with the ground truth and the baseline $k$--$\omega$ model.}}
  \label{tab:drag_lift_coeff}
\end{table}

\begin{figure}
  \centering
  \includegraphics[width=0.7\textwidth]{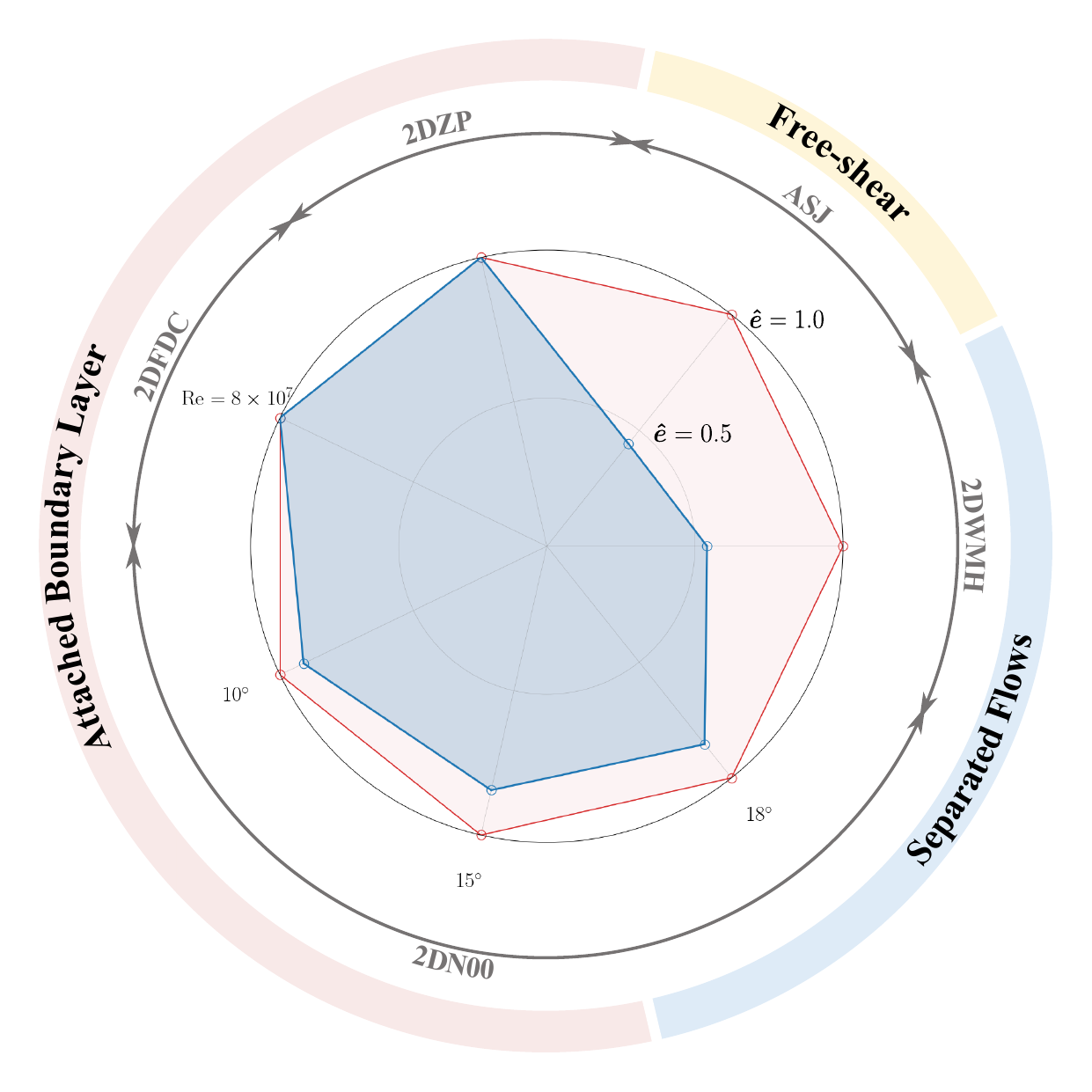}
  \caption{Performance evaluation of the \textbf{unified foundation model} on the \textbf{NASA challenge cases}. The normalized misfit $\hat e$ is shown as the radial coordinate.
  The radar chart compares $\hat e$ across canonical attached boundary layer flows, separated flows, and free-shear flows, with the baseline $k$--$\omega$ model shown for reference, corresponding to the unit red polygon with markers ($\hat{e}=1$), while the unified foundation model is shown as a blue filled polygon with markers. Across most test cases, the blue filled polygon lies inside the red polygon.
  The test cases (see Table~\ref{tab:nasa-cases}) include 2D zero-pressure-gradient flat plate (2DZP), 2D fully developed channel flow at a high Reynolds number (2DFDC), axisymmetric subsonic jet (ASJ), 2D NASA wall-mounted hump with separated flow (2DWMH), and 2D NACA 0012 airfoils (2DN00).}
  \label{fig:radar-nasa}
\end{figure}

\paragraph{Performance on NASA turbulence modeling challenge cases}
NASA turbulence modeling challenge cases span attached boundary layers, separated flows, and free-shear flows. The attached boundary layer cases include a 2D zero–pressure-gradient flat plate, a 2D fully developed channel flow, and NACA~0012 airfoil with two low angles of attack. The remaining cases comprise an axisymmetric subsonic jet, a 2D wall-mounted hump, and a NACA~0012 airfoil at high angle of attack. The name mappings for all cases are summarized in Table~\ref{tab:corresponding-na}.
The unified foundation model demonstrates improved overall performance across the NASA turbulence modeling challenge cases compared to the baseline $k$--$\omega$ model shown in Fig.~\ref{fig:radar-nasa}, indicating robust generalization within a widely adopted benchmark suite. The attached boundary layer cases (2D zero-pressure-gradient flat plate, 2D fully developed channel flow, and low-angle-of-attack NACA~0012 airfoil) are treated as verification, as baseline models already perform well in these regimes. For these flows, the unified foundation model exhibits performance comparable to the baseline 
$k$--$\omega$ model, confirming that the proposed training framework preserves established accuracy where the physics are well captured. In contrast, the unified foundation model achieves clear improvements across the remaining three cases (the axisymmetric subsonic jet, the 2D wall-mounted hump and the NACA~0012 at angle of attack of 14$^\circ$), as reflected by the reduced radial extent of the blue polygon in Fig.~\ref{fig:radar-nasa}. For the NACA~0012 airfoil at high angle of attack, i.e., $\text{AOA} = 18^\circ$, the quadratic term \(g^{(2)}\) in the constitutive relation is explicitly set to zero to ensure numerical stability. Even under this restriction, the unified foundation model remains competitive and extends predictive coverage relative to the baseline closure in the post-stall regime.

\begin{table}
\centering

\begin{tabular}{lll}
    \specialrule{1pt}{0pt}{0pt}
\textbf{Acronym} & \textbf{Case description} & \textbf{Case library label(s)} \\
    \specialrule{1pt}{0pt}{0pt}
2DZP  & 2D zero-pressure-gradient flat plate &
BL1 \\
2DFDC  & 2D fully developed channel flow at high Reynolds number &
BL2.3 \\
ASJ    & Axisymmetric subsonic jet &
FS1 \\
2DWMH  & 2D NASA wall-mounted hump separated-flow &
SP5 \\
2DN00  & 2D NACA 0012 airfoil &
BL3.3, BL3.4, SP4.4 \\
    \specialrule{1pt}{0pt}{0pt}
\end{tabular}
\caption{ \label{tab:nasa-cases}
Summary of cases in the \textbf{NASA Turbulence Modeling Challenge}~\cite{rumsey2022nasa}, with descriptions and the corresponding labels in our case library (see Table~\ref{tab:cases-setup}). 
}
\label{tab:corresponding-na}

\end{table}

\subsection{Specialist turbulence model for separated flows}

We develop a specialist turbulence model for separated flows by fine-tuning unified foundation model on benchmark canonical cases that exhibit distinct separation mechanisms. It is then evaluated both within these training categories and on a geometrically more complex three-dimensional configuration. 
\deleted[id=Author]{In this specialist model, only the linear constitutive relation is retained, with the coefficient functions $g^{(1)}$ and $\beta$ fine-tuned starting from the unified foundation model, ensuring stability while focusing model capacity on flow separation.}
\added[id=Author]{In this specialist model, we remove the quadratic term \(g^{(2)}\mathbf{T}^{(2)}\) and retain only the linear constitutive relation. The coefficient functions \(g^{(1)}\) and \(\beta\) are then fine-tuned from the unified foundation model, improving stability while focusing the model capacity on flow separation.}

\begin{figure}[!htb]
  \centering

  \begin{subfigure}[t]{0.45\textwidth}
    \centering
    \includegraphics[width=0.8\textwidth]{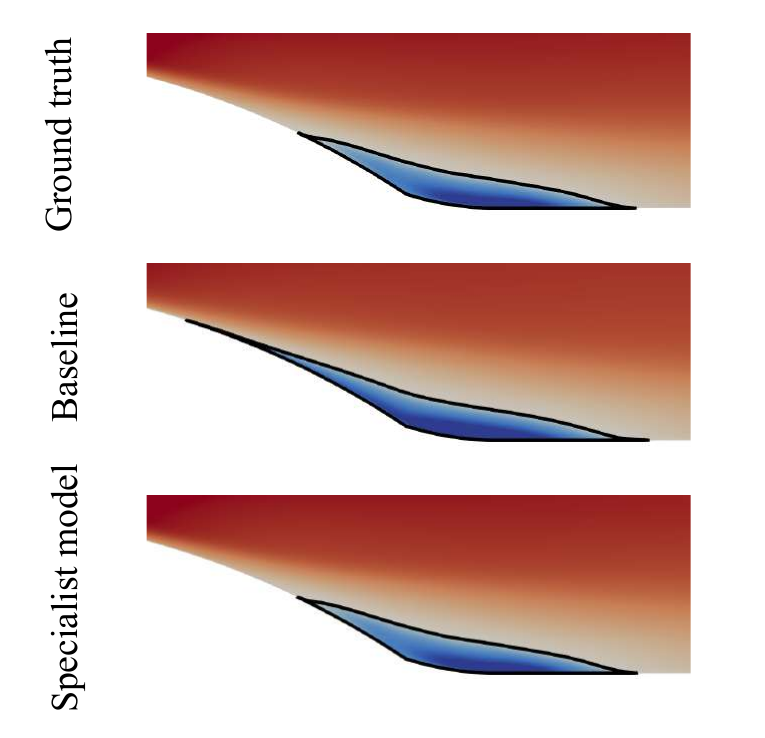}
    \caption{Bump with height 42 mm}
    \label{fig:sep-bump}
  \end{subfigure}
  \hspace{-0.08\textwidth}
  \begin{subfigure}[t]{0.45\textwidth}
    \centering
    \includegraphics[width=0.8\textwidth]{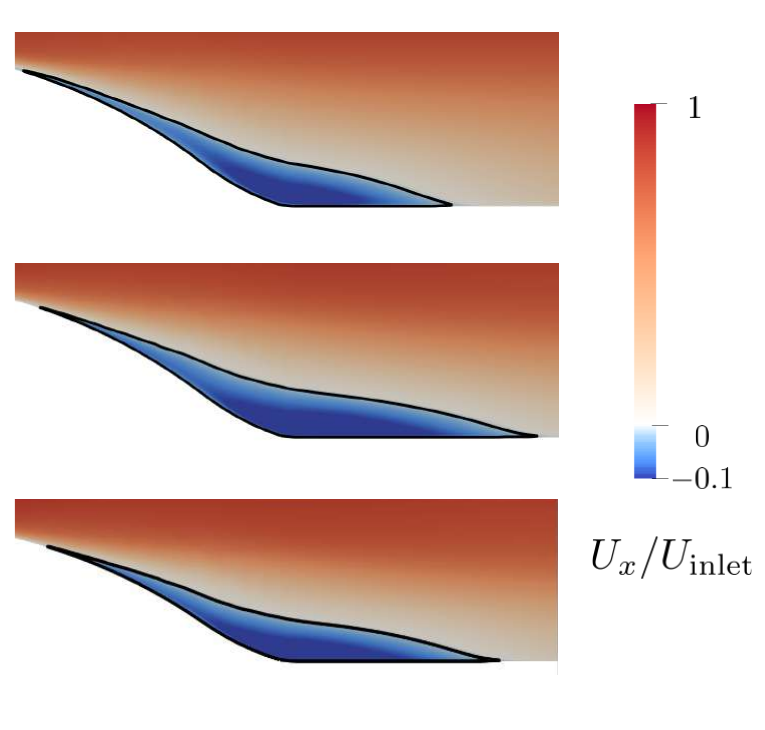}
    \caption{Curved step}
    \label{fig:sep-curved-step}
  \end{subfigure}

  \caption{Fine-tuning performance of the \textbf{specialist model for separated flows}. Separation predictions are shown for flows over (a) a \textbf{bump} with height 42~mm and (b) a \textbf{curved step}, with the ground truth, the baseline $k$--$\omega$ model, and the specialist model arranged from top to bottom. The contours represent the non-dimensional streamwise velocity $U_x/U_\text{inlet}$, with the black contour line indicating $U_x = 0$ to delineate the separated recirculation region.}
  
  \label{fig:separation_train_result}
\end{figure}

\paragraph{Fine-tuning results}
The specialist model improves separation predictions for the two fine-tuning cases (the \(42\,\mathrm{mm}\) bump and the curved step), as shown by comparisons with the ground truth and the baseline $k$--$\omega$ model in Fig.~\ref{fig:separation_train_result}. The specialist model more accurately captures both the onset and reattachment of separation, yielding separation bubbles that closely follow the ground-truth isolines. The improvement is particularly evident for the bump configuration, where the specialist model reproduces the separation extent and recovery with high fidelity. However, the baseline model substantially overpredicts the extent of the separated region, leading to delayed reattachment relative to the reference solution. 

\begin{figure}
  \centering
  \includegraphics[width=0.7\textwidth]{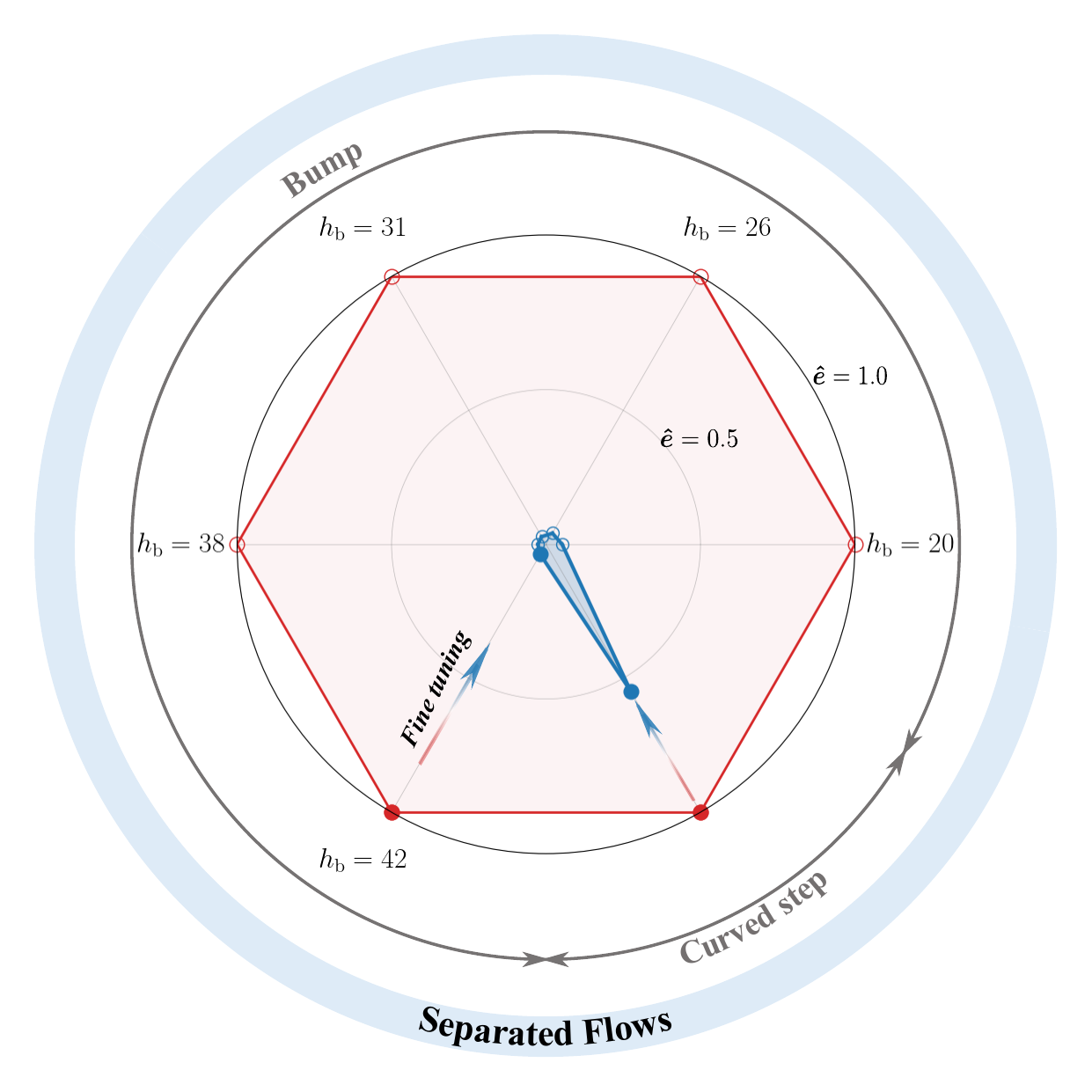}
  \caption{Performance evaluation of the \textbf{specialist model for separated flows} within the \textbf{training categories}, shown as a radar chart compared with the baseline $k$--$\omega$ model. Bumps with height from 20 to 42~mm ($h_{\text{b}} = 20$--$42$~mm) and a curved step are included. The radial coordinate represents the normalized misfit $\hat{e}$. The performance of the specialist model is shown as a blue filled polygon, while that of the baseline $k$--$\omega$ model is shown as a red polygon with markers ($\hat{e}=1$). Arrows indicate the fine-tuning cases, demonstrating additional improvements in the specialist model through further training on specific separated flows.}
  \label{fig:expert-separation-flow-testing}
\end{figure}

\paragraph{Test results on flows within fine-tuning categories}
The specialist model generalizes well to flows over bumps, consistently improving separation prediction across a range of geometries.
We evaluate this behavior for bump heights of $h_{\text{b}} = 20,\,26,\,31,$ and $38\,\mathrm{mm}$, which fall within the training categories (see Fig.~\ref{fig:case-setup-foundation-paper}), with performance summarized in Fig.~\ref{fig:expert-separation-flow-testing}.
Across all bump heights, the specialist model yields smaller misfit values than the baseline model, indicating a consistent reduction in separation-prediction error and that the results approach the single-training reference, which represents the optimized upper bound for these configurations.

\begin{figure}[!htb]
  \centering

  \begin{subfigure}[t]{0.75\textwidth}
    \centering
    \includegraphics[width=\textwidth]{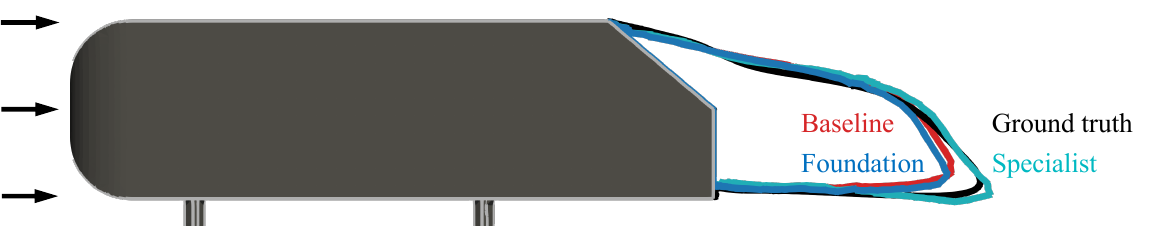}
    \caption{\added[id=Author]{Separated region comparison for generic car}}
    \label{fig:ahmed-top}
  \end{subfigure}

  \vspace{0.5em} 

  \begin{subfigure}[t]{0.725\textwidth}
    \centering
    \includegraphics[width=\textwidth]{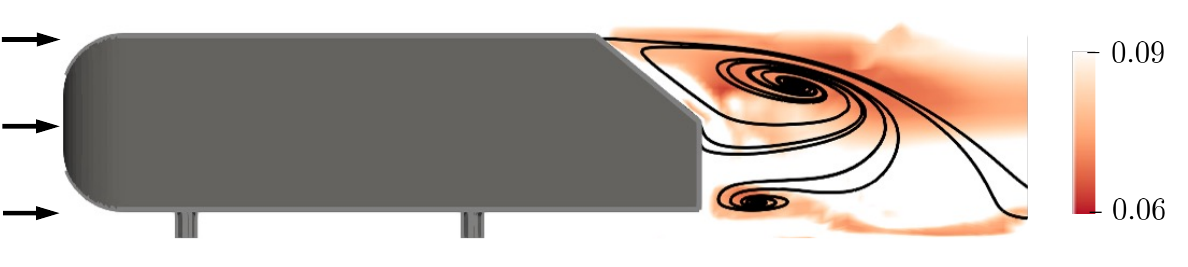}
    \caption{Streamline and $-g^{(1)}$ distribution in the wake region from specialist model}
    \label{fig:ahmed-bottom}
  \end{subfigure}

  \caption{Performance evaluation of the \textbf{specialist model for separated flows} on the \textbf{generic car} (Ahmed body), showing (a) comparison of the separated wake region, delineated by the $U_x=0$ contour, for the baseline model, unified foundation model, specialist model and ground truth; (b) streamlines and spatial distribution of the correction coefficient $-g^{(1)}$ predicted by the specialist model. The decreased values of $-g^{(1)}$ indicate a reduction relative to the standard value $C_\mu = 0.09$ used in the baseline $k$--$\omega$ model, within the wake region, as illustrated by the streamlines (solid black lines). The arrows indicate inflow direction.}
  
  \label{fig:ahmed_body_result}
\end{figure}

\paragraph{Test results on generic car}
The specialist model for separated flows exhibits improved predictive capability on the three-dimensional generic car (see Fig.~\ref{fig:ahmed_body_result}), providing more accurate drag estimates and wake separation than the baseline and unified foundation models. In this case, the drag coefficient $c_d$ is the key quantity of interest and is summarized in Table~\ref{tab:drag_coeff} for the different models. Relative to the ground truth, the specialist model provides the closest estimate of the drag coefficient\deleted[id=Author]{, although a noticeable discrepancy remains}.

The specialist model most accurately reproduces the wake structure by appropriately reducing eddy viscosity in the separated region.
Wake structures predicted by the different models are compared in Fig.~\ref{fig:ahmed_body_result}a, where the $U_x = 0$ contour delineates the separated region. The specialist model shows the best agreement with the ground truth, accurately reproducing both the size and downstream extent of the separated region. \deleted[id=Author]{The unified foundation model captures a larger wake that extends further downstream but still deviates from the ground truth, whereas the baseline model produces a markedly contracted separation bubble.}
The streamline patterns and the spatial distribution of $-g^{(1)}$ are shown in Fig.~\ref{fig:ahmed_body_result}b to examine the model behavior underlying this improvement. Regions where $-g^{(1)}$ decreases relative to its nominal value of $0.09$ are concentrated in the wake, where the eddy viscosity $\nu_t$ is locally reduced according to~\cite{zhang2023physical}
\begin{equation}
\nu_t = -\frac{g^{(1)} k}{C_{\mu} \omega},
\end{equation}
with $C_\mu = 0.09$. This localized reduction in $\nu_t$ is consistent with enhanced separation and improved agreement with the ground-truth wake structure.

\begin{table}
  \centering
  \renewcommand{\arraystretch}{1.3}
  \setlength{\tabcolsep}{15pt}
  \begin{tabular}{lc}
      \specialrule{1pt}{0pt}{0pt}
    \textbf{Model} & \textbf{Drag coefficient} \\
    \specialrule{1pt}{0pt}{0pt}
    Ground truth             & 0.248 \\
    Baseline  \(k\)--\(\omega\) model                & 0.295 \\
    Unified foundation model & 0.297 \\
    Specialist model         & \textbf{0.275} \\
    \specialrule{1pt}{0pt}{0pt}
  \end{tabular}

  \caption{\added[id=Author]{Drag coefficient evaluation of the specialist model for separated flows compared to the unified foundation model, the ground truth, and the baseline \(k\)--\(\omega\) model.}}
  \label{tab:drag_coeff}
\end{table}

In summary, the specialist model for separated flows improves prediction of the separation behavior across the training cases, placing onset and reattachment points closer to the ground truth. 
For unseen flows within the training categories, the model consistently reduces error relative to the baseline, indicating good generalization ability for closely related geometries. When applied to the generic car, the specialist model maintains improved performance, with localized corrections concentrated in the wake region that are consistent with more accurate predictions of the separation extent and associated flow structures.

\subsection{Specialist turbulence model for secondary flows with separation}

We develop a specialist turbulence model targeting secondary flows with separation and evaluate its performance on three-dimensional configurations. The model reduces the baseline $k$--$\omega$ model’s overprediction of separation, more accurately reproduces corner-vortex structures, and successfully captures the interaction between secondary flow and separation in a three-dimensional diffuser.

\paragraph{Fine-tuning results}

\begin{figure}[!htb]
  \centering

  \begin{subfigure}[t]{0.45\textwidth}
    \centering
    \includegraphics[width=0.8\textwidth]{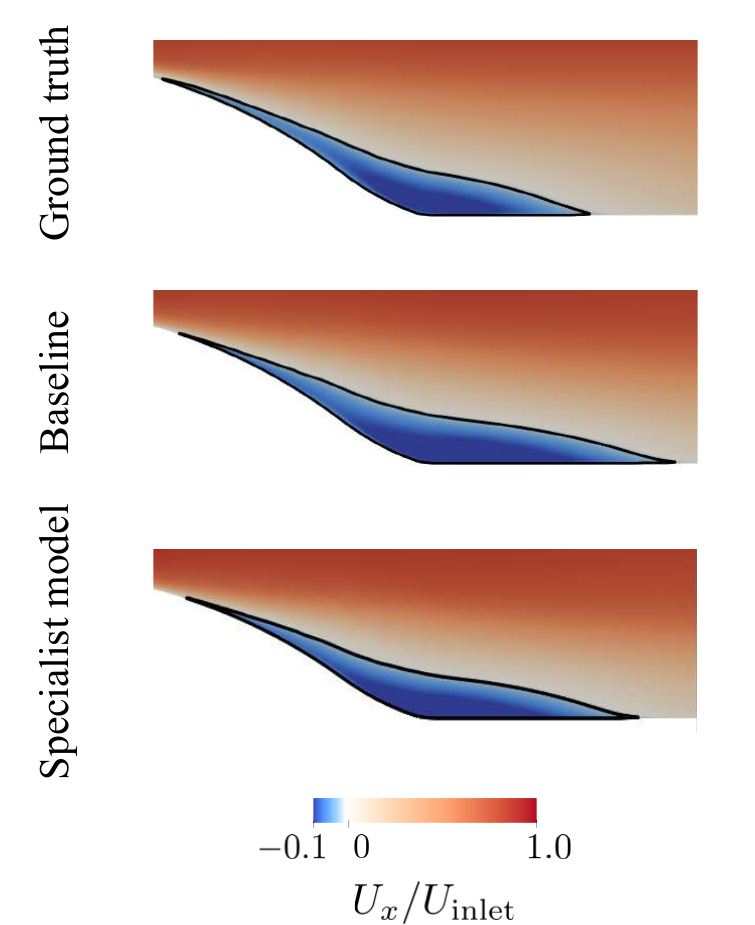}
    \caption{Curved step}
    \label{fig:curved-step}
  \end{subfigure}
  \hspace{-0.08\textwidth}
  \begin{subfigure}[t]{0.45\textwidth}
    \centering
    \includegraphics[width=0.8\textwidth]{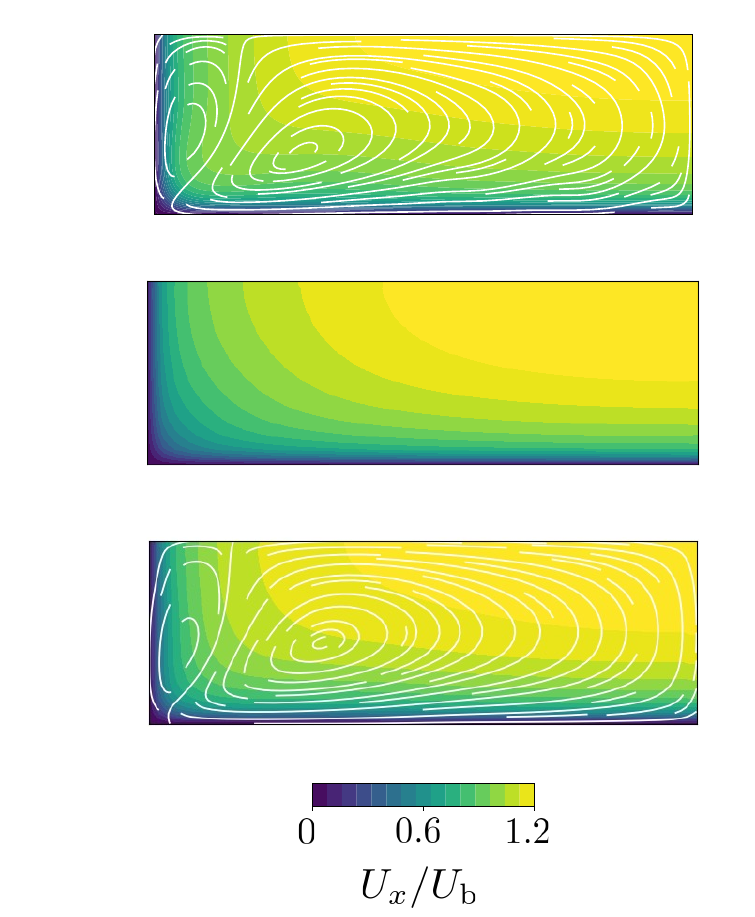}
    \caption{Rectangular duct (AR=3)}
    \label{fig:rect-duct}
  \end{subfigure}

  \caption{Fine-tuning performance of the \textbf{specialist model for secondary flows with separation} against the ground truth and the baseline $k$--$\omega$ model. Separation prediction results are shown for (a) flow over a \textbf{curved step}, and in-plane secondary flow predictions are shown for (b) flow through a \textbf{rectangular duct} with aspect ratio $\mathrm{AR}=3$. The results for the rectangular duct with aspect ratio $\mathrm{AR}=10$ are omitted for brevity. For the curved step, contours of the non-dimensional streamwise velocity $U_x/U_\text{inlet}$ are shown, where the black contour line ($U_x = 0$) marks the separation boundary and highlights differences in separation extent and reattachment location. For the rectangular duct, cross-sectional contours of $U_x/U_{\text{b}}$ overlaid with streamlines illustrate the corner-vortex structures and secondary circulations.}
  
  \label{fig:expert-secondary-flow-training}
\end{figure}

The specialist model improves predictions for both separated and secondary flows. For the curved step, which is governed by flow separation, the specialist model reduces the baseline model’s overprediction of the separated region and yields a separation extent closer to the ground truth (Fig.~\ref{fig:expert-secondary-flow-training}a).
For the rectangular duct flows, which are characterized by secondary motions, the in-plane velocity fields predicted by the specialist model, the baseline model, and the ground truth are compared in Fig.~\ref{fig:expert-secondary-flow-training}b. For the rectangular duct with aspect ratio $\mathrm{AR}=3$, the flow exhibits asymmetric corner vortices, with vortices along the horizontal walls extending further in the spanwise direction and weakening those along the vertical walls. The specialist model accurately reproduces this asymmetry and the associated vortex structure, closely matching the ground truth, whereas the baseline model fails to capture any corner vortices.

\begin{figure}
  \centering
  \includegraphics[width=0.7\textwidth]{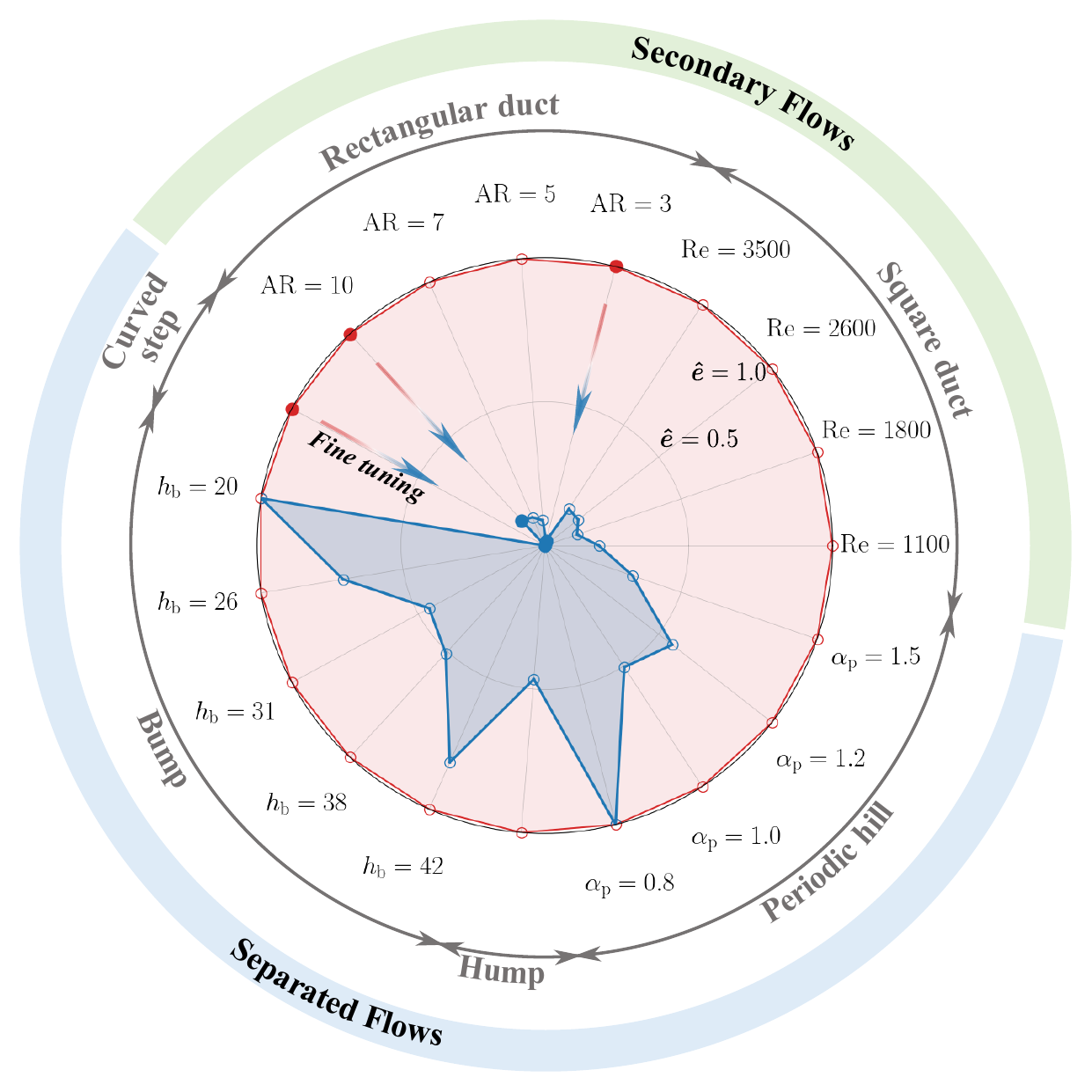}
  \caption{Performance evaluation of the \textbf{specialist model for secondary flows with separation} across \textbf{separated and secondary flows}, summarized in a radar chart and compared with the baseline $k$--$\omega$ model. The evaluated cases include canonical secondary flows (square and rectangular ducts) and separated flows (curved step, bump, periodic hill, and hump), covering a broad range of geometric and flow conditions. The radial coordinate represents the normalized prediction error $\hat{e}$, where the red polygon with markers ($\hat{e}=1$) corresponds to the baseline model and the blue filled polygon corresponds to the specialist model. Arrows indicate the fine-tuning cases, and for most remaining test cases, the blue polygon lies inside the red polygon, indicating improved performance of the specialist model.}
  \label{fig:expert-secondary-flow-testing-basic}
\end{figure}

\paragraph{Test results on flows within fine-tuning categories}
The specialist model generalizes well to separated and secondary flows beyond those included in the fine-tuning set. Its performance on these cases is summarized in Fig.~\ref{fig:expert-secondary-flow-testing-basic}. Across secondary flow configurations, i.e., square and rectangular ducts, the specialist model consistently achieves higher prediction accuracy than the unified foundation model (see Fig.~\ref{fig:radar-unified-foundation-model}). For separated flows, it generally maintains, and in some cases slightly improves upon, the accuracy inherited from the unified foundation model. Notably, the $42$~mm bump case exhibits a clear performance gain, while the periodic hill and hump cases show accuracy comparable to that of the unified foundation model.

\begin{figure}
  \centering
  \includegraphics[width=0.9\textwidth]{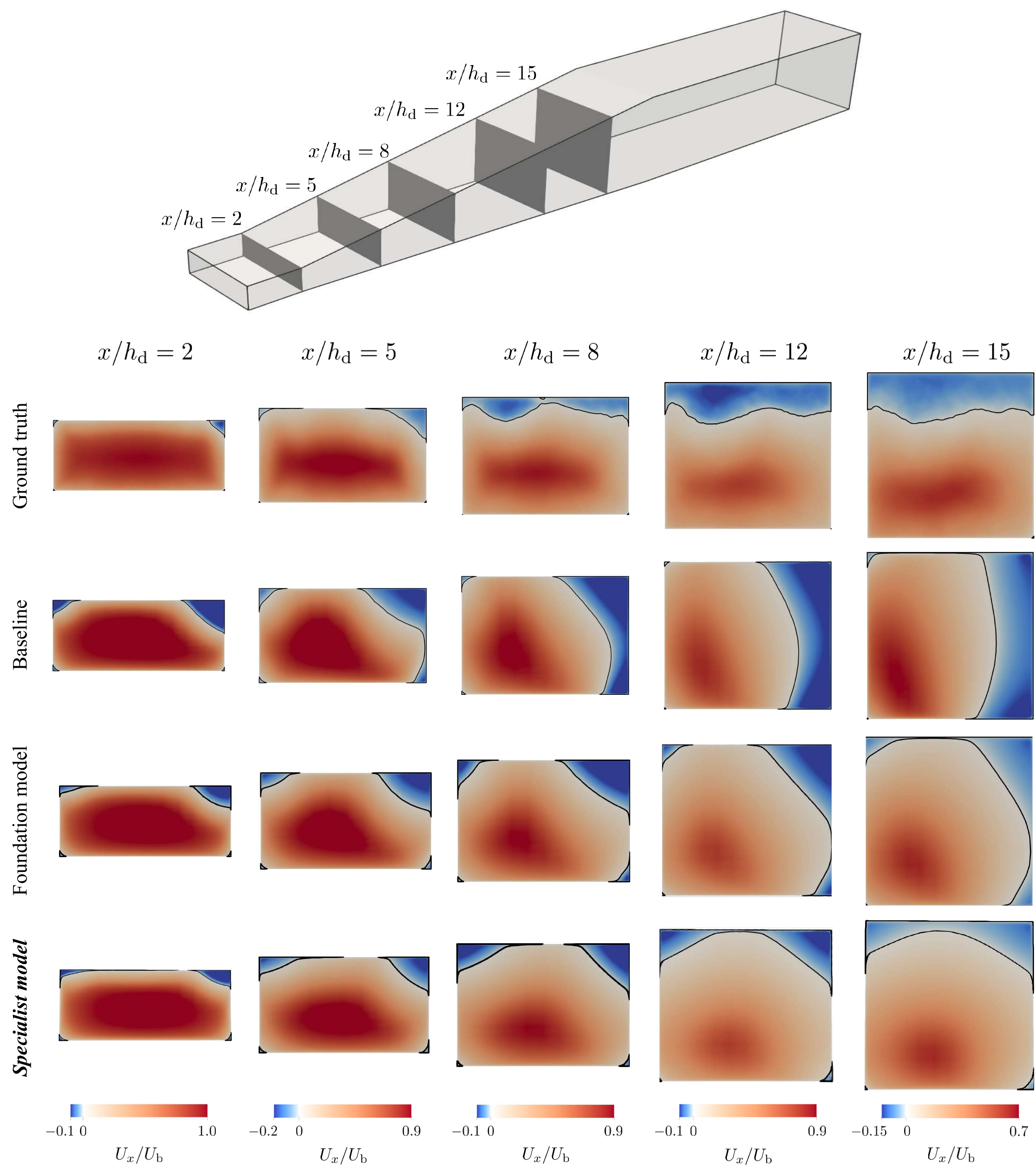}
  \caption{Performance evaluation of the \textbf{specialist model for secondary flows with separation} on the \textbf{three-dimensional diffuser} against the unified foundation model, the baseline $k$--$\omega$ model, and the ground truth (\added[id=R2]{obtained from DNS mean flow field~\cite{ohlsson2010dns}}). Separation predictions at five streamwise cross-sections are extracted at $x/h_{\text{d}} = 2$, 5, 8, 12, and 15, where $h_{\text{d}}$ denotes the diffuser inlet height. Contours represent the normalized streamwise velocity $U_x/U_{\text{b}}$, and the $U_x = 0$ isoline marks the separated regions.}
  \label{fig:expert-secondary-flow-testing}
\end{figure}

\begin{figure}[p]
  \centering
  \includegraphics[width=0.8\textwidth]{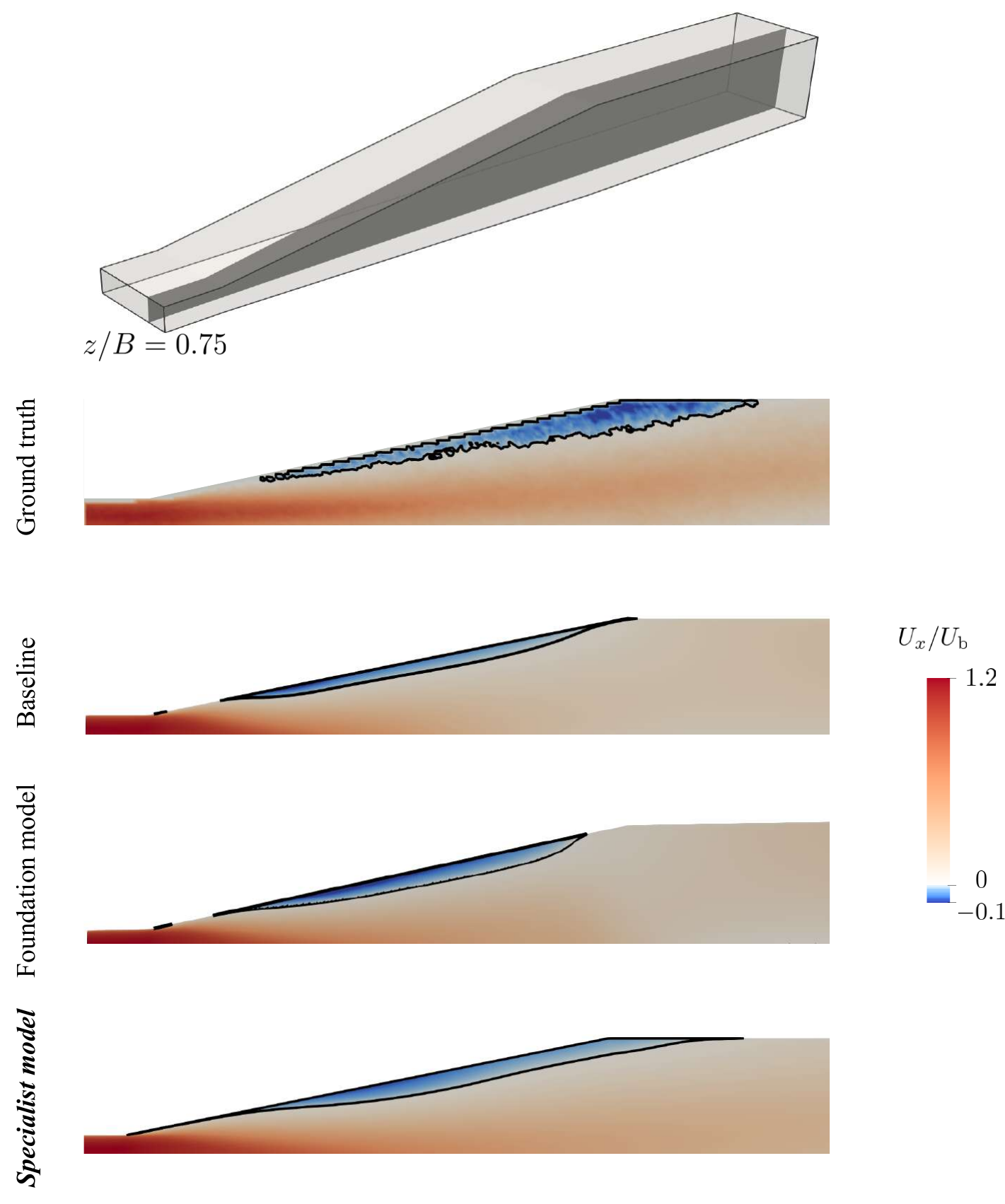}
  \caption{Performance evaluation of the \textbf{specialist model for secondary flows with separation} on the \textbf{3D diffuser} against the unified foundation model, the baseline $k$--$\omega$ model, and the ground truth (\added[id=R2]{obtained from experiments~\cite{cherry2008geometric}}). Upper-wall separations on a spanwise cross-section at $z/B = 0.75$, where $B$ is the diffuser inlet width, are plotted. Contours represent the streamwise velocity $U_x/U_{\text{b}}$, with the $U_x=0$ isoline delineating the separated region.}
  \label{fig:expert-secondary-flow-testing-2}
\end{figure}

\paragraph{Test results on three-dimensional diffuser}
The specialist model improves prediction accuracy for the asymmetric diffuser, a three-dimensional flow in which secondary flow and separation are strongly coupled. Streamwise velocity fields and separation patterns at multiple downstream locations (Fig.~\ref{fig:expert-secondary-flow-testing}) show how the specialist model corrects the baseline separation topology. In the ground truth, separation originates along the upper wall and grows downstream while remaining attached to the inclined side wall. The baseline model instead predicts extensive separation along the inclined side wall and fails to reproduce the dominant upper-wall separation. The specialist model suppresses this spurious side-wall separation and shifts the separation line to the upper wall, improving agreement with the reference topology. The unified foundation model exhibits intermediate behavior: at downstream stations ($x/h_{\text{d}}=12$ and $15$), it partially strengthens the upper-wall separation and weakens side-wall separation, but still departs from the ground-truth separation pattern. Complementary spanwise cross sections near the inclined side wall (Fig.~\ref{fig:expert-secondary-flow-testing-2}) further confirm that the specialist model best captures the downstream propagation of upper-wall separation and its reattachment at the diffuser exit, outperforming both the baseline and unified foundation models.

Overall, the specialist model for secondary flows with separation improves predictive performance across both secondary-flow and separated-flow categories and effectively captures their interaction. Its performance on the three-dimensional diffuser, including the suppression of spurious inclined side-wall separation and improved prediction of upper-wall separation, suggests improved generalization ability beyond the training categories.

\FloatBarrier
\clearpage

\FloatBarrier
\clearpage
\appendix
\onecolumn
\appendix

\renewcommand{\refname}{Supplementary References}
\putbib

\end{bibunit}

\end{document}